\documentclass[twocolumn]{aastex63}
\usepackage{amsmath}
\usepackage{enumerate}
\usepackage{enumitem}   
\usepackage{hyperref}
\usepackage{color, colortbl}
\usepackage{subfigure}
\usepackage{color}
\usepackage{courier}
\usepackage{mathtools}
\usepackage[flushleft]{threeparttable}
\usepackage{enumitem}
\usepackage{rotating}
\usepackage{tabularx}
\usepackage{tabularx}
\usepackage{rotating}
\usepackage{CJK}
\usepackage{multirow}
\usepackage{booktabs}
\usepackage{courier}
\usepackage{longtable}

\defcitealias{Sch19}{Paper~I} 	

\newcommand\clearrow{\global\let\rowmac\relax}

\newcommand{\co}{\textit{CO-only}}
\newcommand{\ha}{\textit{H$\alpha$-only}}
\newcommand{\overlap}{\textit{overlap}}
\newcommand{\Mstar}{\ensuremath{M_{\ast}}}
\newcommand{\cosightline}{\textit{CO sight lines}}
\newcommand{\hasightline}{\textit{H$\alpha$ sight lines}}
\received{}
\revised{}
\accepted{}
\submitjournal{}

\begin{document}
	

\title{The Gas–Star Formation Cycle in Nearby Star-forming Galaxies II. Resolved Distributions of CO and H$\alpha$ Emission for 49 PHANGS Galaxies}

\begin{CJK}{UTF8}{bkai}
\author{Hsi-An Pan (潘璽安)}
\affiliation{Max-Planck-Institut f\"ur Astronomie, K\"onigstuhl 17, D-69117 Heidelberg, Germany}
\affiliation{Department of Physics, Tamkang University, No.151, Yingzhuan Rd., Tamsui Dist., New Taipei City 251301, Taiwan}
\email{hapan@gms.tku.edu.tw}

\author{Eva Schinnerer}
\affiliation{Max-Planck-Institut f\"ur Astronomie, K\"onigstuhl 17, D-69117 Heidelberg, Germany}

\author{Annie Hughes}
\affiliation{CNRS, IRAP, Av. du Colonel Roche BP 44346, F-31028 Toulouse cedex 4, France}

\author{Adam Leroy}
\affiliation{Department of Astronomy, The Ohio State University, 140 West 18th Ave, Columbus, OH 43210, USA}

\author{Brent Groves}
\affiliation{International Centre for Radio Astronomy Research, The University of Western Australia, Crawley, WA 6009, Australia}

\author{Ashley Thomas Barnes}
\affiliation{Argelander-Institut f\"{u}r Astronomie, Universit\"{a}t Bonn, Auf dem H\"{u}gel 71, 53121, Bonn, Germany}

\author{Francesco Belfiore}
\affiliation{INAF – Osservatorio Astrofisico di Arcetri, Largo E. Fermi 5, I-50157, Firenze, Italy}

\author{Frank Bigiel}
\affiliation{Argelander-Institut f\"{u}r Astronomie, Universit\"{a}t Bonn, Auf dem H\"{u}gel 71, 53121, Bonn, Germany}

\author{Guillermo A. Blanc}
\affiliation{Observatories of the Carnegie Institution for Science, 813 Santa Barbara Street, Pasadena, CA 91101, USA}
\affiliation{Departamento de Astronom\'{i}a, Universidad de Chile, Camino del Observatorio 1515, Las Condes, Santiago, Chile}

\author{Yixian Cao}
\affiliation{Aix Marseille Université, CNRS, LAM (Laboratoire d’Astrophysique de Marseille), F-13388 Marseille, France}

\author{M\'{e}lanie Chevance}
\affiliation{Astronomisches Rechen-Institut, Zentrum f\"{u}r Astronomie der Universit\"{a}t Heidelberg, M\"{o}nchhofstra{\ss}e 12-14, D-69120 Heidelberg, Germany}

\author{Enrico Congiu}
\affiliation{Departamento de Astronom\'{i}a, Universidad de Chile, Camino del Observatorio 1515, Las Condes, Santiago, Chile}

\author{Daniel A. Dale} 
\affiliation{Department of Physics and Astronomy, University of Wyoming, Laramie, WY 82071, USA}

\author{Cosima Eibensteiner}
\affiliation{Argelander-Institut f\"{u}r Astronomie, Universit\"{a}t Bonn, Auf dem H\"{u}gel 71, 53121, Bonn, Germany}

\author{Eric Emsellem}
\affiliation{European Southern Observatory, Karl-Schwarzschild Stra{\ss}e 2, D-85748 Garching bei M\"unchen, Germany}
\affiliation{Univ Lyon, Univ Lyon 1, ENS de Lyon, CNRS, Centre de Recherche Astrophysique de Lyon UMR5574, F-69230 Saint-Genis-Laval, France}

\author{Christopher M. Faesi}
\affiliation{University of Massachusetts -- Amherst, 710 N. Pleasant Street, Amherst, MA 01003, USA}

\author{Simon C. O. Glover}
\affiliation{Universit\"{a}t Heidelberg, Zentrum f\"{u}r Astronomie, Institut f\"{u}r Theoretische Astrophysik, Albert-Ueberle-Str 2, D-69120 Heidelberg, Germany}

\author{Kathryn Grasha}
\affiliation{Research School of Astronomy and Astrophysics, Australian National University, Canberra, ACT 2611, Australia}

\author{Cinthya N. Herrera}
\affiliation{Institut de Radioastronomie Millim\'{e}trique (IRAM), 300 Rue de la Piscine, F-38406 Saint Martin d'H\`{e}res, France}

\author{I-Ting Ho}
\affiliation{Max-Planck-Institut f\"ur Astronomie, K\"onigstuhl 17, D-69117 Heidelberg, Germany}

\author{Ralf S. Klessen}
\affiliation{Universit\"{a}t Heidelberg, Zentrum f\"{u}r Astronomie, Institut f\"{u}r Theoretische Astrophysik, Albert-Ueberle-Str 2, D-69120 Heidelberg, Germany}
\affiliation{Universit\"{a}t Heidelberg, Interdisziplin\"{a}res Zentrum f\"{u}r Wissenschaftliches Rechnen, Im Neuenheimer Feld 205, D-69120 Heidelberg, Germany}

\author{J. M. Diederik Kruijssen}
\affiliation{Astronomisches Rechen-Institut, Zentrum f\"{u}r Astronomie der Universit\"{a}t Heidelberg, M\"{o}nchhofstra{\ss}e 12-14, D-69120 Heidelberg, Germany}

\author{Philipp Lang}
\affiliation{Max-Planck-Institut f\"ur Astronomie, K\"onigstuhl 17, D-69117 Heidelberg, Germany}

\author{Daizhong Liu}
\affiliation{Max-Planck-Institut f\"ur Astronomie, K\"onigstuhl 17, D-69117 Heidelberg, Germany}

\author{Rebecca McElroy}
\affiliation{Sydney Institute for Astronomy, School of Physics A28, The University of Sydney, NSW 2006, Australia}

\author{Sharon E. Meidt}
\affiliation{Sterrenkundig Observatorium, Universiteit Gent, Krijgslaan 281 S9, B-9000 Gent, Belgium}

\author{Eric J.\,Murphy}
\affiliation{National Radio Astronomy Observatory, 520 Edgemont Road, Charlottesville, VA 22903-2475, USA}

\author{J\'{e}r\^{o}me Pety}
\affiliation{Institut de Radioastronomie Millim\'{e}trique (IRAM), 300 Rue de la Piscine, F-38406 Saint Martin d'H\`{e}res, France}
\affiliation{Sorbonne Universit\'{e}, Observatoire de Paris, Universit\'{e} PSL, CNRS, LERMA, F-75014, Paris, France}

\author{Miguel Querejeta}
\affiliation{Observatorio Astron\'{o}mico Nacional (IGN), C/Alfonso XII, 3, E-28014 Madrid, Spain}

\author{Alessandro Razza}
\affiliation{Departamento de Astronom\'{i}a, Universidad de Chile, Camino del Observatorio 1515, Las Condes, Santiago, Chile}

\author{Erik Rosolowsky}
\affiliation{Department of Physics, University of Alberta, Edmonton, AB T6G 2E1, Canada}

\author{Toshiki Saito}
\affiliation{Max-Planck-Institut f\"ur Astronomie, K\"onigstuhl 17, D-69117 Heidelberg, Germany}

\author{Francesco Santoro}
\affiliation{Max-Planck-Institut f\"ur Astronomie, K\"onigstuhl 17, D-69117 Heidelberg, Germany}

\author{Andreas Schruba}
\affiliation{Max-Planck-Institut f\"{u}r extraterrestrische Physik, Giessenbachstra{\ss}e 1, D-85748 Garching, Germany}

\author{Jiayi Sun}
\affiliation{Department of Astronomy, The Ohio State University, 140 West 18th Ave, Columbus, OH 43210, USA}

\author{Neven Tomi\v{c}i\'c}
\affiliation{INAF-Osservatorio Astronomico di Padova, Vicolo Osservatorio 5, 35122 Padova, Italy}

\author{Antonio Usero}
\affiliation{Observatorio Astron\'{o}mico Nacional (IGN), C/Alfonso XII, 3, E-28014 Madrid, Spain}

\author{Dyas Utomo}
\affiliation{National Radio Astronomy Observatory, 520 Edgemont Road, Charlottesville, VA 22903-2475, USA}

\author{Thomas G. Williams}
\affiliation{Max-Planck-Institut f\"ur Astronomie, K\"onigstuhl 17, D-69117 Heidelberg, Germany}

\begin{abstract} 
The relative distribution of molecular gas and star formation in galaxies gives insight into the physical processes and timescales of the cycle between gas and stars.
In this work, we track the relative spatial configuration of CO and H$\alpha$ emission at high resolution in each of our galaxy targets, and use these measurements to quantify the  distributions of regions in different evolutionary stages of star formation: from molecular gas without star formation traced by H$\alpha$ to star-forming gas, and to \textsc{H\,ii} regions.
The large sample,  drawn from the Physics at High Angular resolution in Nearby GalaxieS ALMA and narrowband H$\alpha$ (PHANGS-ALMA and PHANGS-H$\alpha$) surveys, spans a wide range of stellar mass and morphological types, allowing us to investigate the dependencies of the gas-star formation cycle on global galaxy properties.
At a resolution of 150~pc, the incidence of regions in different stages shows a dependence on stellar mass and Hubble type of galaxies over the radial range probed.
Massive and/or earlier-type galaxies exhibit a significant reservoir of   molecular gas without star formation traced by H$\alpha$, while lower-mass  galaxies harbor substantial \textsc{H\,ii} regions that may have dispersed their birth clouds or formed from low-mass, more isolated clouds.
Galactic structures add a further layer of complexity to relative distribution of CO and H$\alpha$ emission.
Trends between galaxy properties and  distributions of gas traced by CO and H$\alpha$ are visible only when the observed spatial scale is $\ll$ 500~pc,  reflecting the critical resolution requirement to distinguish stages of star formation process.
\end{abstract} 

\section{Introduction}

The conversion from gas to stars is a complex process that ultimately determines many observed properties of a galaxy, such as its observed morphology at different wavelengths and stellar mass.
In star-forming galaxies, stars form through the collapse of dense cores inside giant molecular clouds (GMCs). Therefore, the rate at which stars form is determined by the properties of GMCs, such as their level of turbulence, chemical composition, strength and structure of magnetic fields, or the flux of cosmic rays \citep{Mac04,Mck07}.

\cite{Sch59} observed a tight correlation between the star formation rate (SFR) and the volume density of gas in the Milky Way. 
Later on, \cite{Ken98} showed that the SFR and gas surface densities ($\Sigma_\mathrm{SFR}$ and $\Sigma_\mathrm{gas}$) are tightly correlated on the scales of integrated galaxies, a relationship that is now known as the Kennicutt--Schmidt  relation.
Many recent studies have shown that the Kennicutt--Schmidt relation, at least when considering the surface density of molecular gas ($\Sigma_\mathrm{H_{2}}$), holds down to kpc scales, but with significant variation among galaxies \citep[e.g.,][]{Big08,Ler08,Sch11,Ler13,Mom13}.
The Kennicutt--Schmidt relation has also become a commonly-used prescription for implementing star formation in numerical simulations of galaxies \citep[e.g.,][]{Kat92,Tey02,Sch15}.

However, cloud-scale ($\sim 100$~pc) observations in the Local Group  and a few nearby star-forming galaxies reveal that the relationship between cold gas and stars is more complex.
The correlation between $\Sigma_\mathrm{SFR}$ and $\Sigma_\mathrm{H_{2}}$ develops considerable scatter when the spatial resolution is sufficiently high to spatially separate the individual elements of the surface densities: GMCs and star-forming (\textsc{H\,ii}) regions \citep[e.g.,][]{Ono10,Sch10,Kre18,Que19}.
This breakdown of the scaling relation has been attributed to the evolution of GMCs \citep{Sch10,Fel11,Kru14}.
The separation between GMCs and star formation tracers is now regularly used as an empirical probe of the cycle between gas and star formation \citep{Kaw09,Sch10,Kru14,Kru18}, including the timescale of evolutionary cycling between GMCs and star formation \citep{Kru19,Che20,Kim21} and the impact of destructive stellar feedback (e.g., photoionization, stellar winds, and supernova explosions) on the structure of interstellar medium (ISM) and future star formation \citep{Bar20,Bar21,Che21}.

Moreover, recent cloud-scale studies of extragalactic GMCs have found evidence that GMCs are diverse in their physical properties, such as surface density and dynamical state \citep{Hug13,Col14,Ros21}.
Various environmental mechanisms determine when and which pockets of the GMCs collapse,  such as galactic shear, differential non-circular motions, gas flows along and through  stellar dynamical  structures (e.g.,  bars and spiral arms), and accretion flows \citep{Kle10,Mei13,Mei18,Col18,Jef18,Jef20}.
Theoretical study predicts that these mechanisms have different timescales and cause the star formation process to vary from galaxy to galaxy and from place to place within a galaxy  \citep{Jef21}.
Therefore, to   understand how star formation works in galaxies, a large sample size is indispensable to cover a range of  galactic environments and ISM properties/conditions.

In our previous paper (\citealt{Sch19}; hereafter  \citetalias{Sch19}), we developed a simple, robust method that quantifies the relative spatial distributions of molecular gas and recent star formation, as well as the spatial-scale dependence of the relative distributions.
The method  considers the presence or absence of molecular gas traced by CO emission and star formation traced by H$\alpha$ emission in a given region (i.e., sight line or pixel) at a given observed resolution.
The method was applied to eight nearby galaxies with $\sim$ 1$\arcsec$ resolution molecular gas  observations from the Physics at High Angular resolution in Nearby GalaxieS survey (PHANGS; \citealt{Ler21a,Ler21b}) and the PdBI Arcsecond Whirlpool Survey (PAWS;  \citealt{Sch13})  that have matched resolution  narrowband H$\alpha$ observations.
However, most of the galaxies in \citetalias{Sch19} have similar global properties, they are massive, star-forming, spiral galaxies.

Given that GMC properties vary between and within galaxies, we extend this work to link the gas--star formation cycle and several secular and environmental probes.
In this paper, we apply the  method to 49 galaxies with high-resolution  CO and H$\alpha$ observations selected from PHANGS.
Galaxies in our extended sample cover a wider range in stellar mass (\Mstar) and morphology (Hubble type) compared to the galaxies in \citetalias{Sch19}.
The extended sample allows us to investigate  how the distribution of different star formation phases -- from non- or pre-star-forming gas, to star-forming clouds, and to regions forming massive stars -- depends upon global galaxy properties (i.e., \Mstar, morphology, and dynamical  structures). This is the first time that the relative distribution of molecular and ionized (H$\alpha$) gas has been quantified across a such a large and diverse sample of galaxies at high resolution (150~pc).
The  resolution of 150~pc is sufficiently high to sample individual star-forming units and to separate such regions.

This paper is organized as follows. 
In Section~\ref{sec_data}, we describe the observations of molecular gas and star formation tracers, CO and H$\alpha$, respectively.
Section~\ref{sec_method} introduces the methodology for measuring the presence or absence of different tracers.
Section~\ref{sec_results} presents the distribution of molecular gas and star formation tracers  as a function of galaxy properties and at a series of resolutions,  from our 150~pc to 1.5~kpc.
Section~\ref{sec_discussion} discusses the main results. 
The conclusions are presented in Section~\ref{sec_summary}.

\section{Data}
\label{sec_data}
PHANGS\footnote{www.phangs.org} is a multi-wavelength campaign to observe the tracers of the star formation process in a diverse but representative sample of nearby ($\lesssim$ 19 Mpc) low-inclination galaxies. 
The typical spatial resolution achieved with the multi-wavelength  observations  is $\sim$ 100~pc.
The combination of ALMA \citep{Ler21a,Ler21b},  VLT/MUSE \citep{Ems21}, narrowband H$\alpha$ (A.~Razza et al.~in~prep.), and HST \citep{Lee21} observations  yields an unprecedented view of star formation at different phases, from gas to star clusters.
The galaxies were selected to have log(\Mstar/M$_{\sun}$) $\gtrsim$ 9.75 and to be visible to ALMA,  but  with the current best approach for mass estimation,  the sample extends down to log(\Mstar/M$_{\sun}$) $\approx$ 9.3.
The galaxies are lying on or near the star-forming main sequence.
More details on the survey design and scientific motivation are presented in \cite{Ler21b}.
In this work, we focus on the molecular gas and ionized (H$\alpha$) gas observed by the PHANGS-ALMA and PHANGS-H$\alpha$ (narrowband) surveys, respectively.

\begin{longtable*}{cccccccccccc}
	\caption{Galaxy sample used in this work.} \\

		\hline
		& (a) & (b) & (c) & (d) & (e) & (f) & (g) & (h) & (i) & (j) & (k) \\
		\multirow{2}{*}{galaxy}	& dist.   &incl. & $R_{25}$& log(SFR) & log(\Mstar) & log($M_\mathrm{H_{2}}$) & log($M_\mathrm{HI}$) & \multirow{2}{*}{
		$\Delta$MS}& \multirow{2}{*}{T-type} &\multirow{2}{*}{GD spiral arms} & \multirow{2}{*}{bar} \\
		&[Mpc] & [$^{\circ}$]&[$\arcsec$]&[M$_{\sun}$~yr$^{-1}$]&[M$_{\sun}$]&[M$_{\sun}$]&[M$_{\sun}$]&&&&\\
		\hline
IC1954 & 12.0 & 57.1 & 89.8 & -0.52 & 9.6 & 8.7 & 9.0 & -0.08 & 3.3 & 0 & 1 \\
IC5273 & 14.2 & 52.0 & 91.9 & -0.28 & 9.7 & 8.6 & 9.0 & 0.08 & 5.6 & 0 & 1 \\
NGC0628 & 9.8 & 8.9 & 296.6 & 0.23 & 10.3 & 9.4 & 9.7 & 0.22 & 5.2 & 1 & 0 \\
NGC1087 & 15.9 & 42.9 & 89.1 & 0.11 & 9.9 & 9.2 & 9.0 & 0.34 & 5.2 & 0 & 1 \\
NGC1300 & 19.0 & 31.8 & 178.3 & 0.04 & 10.6 & 9.4 & 9.7 & -0.19 & 4.0 & 1 & 1 \\
NGC1317 & 19.1 & 23.2 & 92.1 & -0.4 & 10.6 & 8.9 & $\dots$ & -0.61 & 0.8 & 0 & 1 \\
NGC1365 & 19.6 & 55.4 & 360.7 & 1.22 & 10.9 & 10.3 & 9.9 & 0.79 & 3.2 & 1 & 1 \\
NGC1385 & 17.2 & 44.0 & 102.1 & 0.3 & 10.0 & 9.2 & 9.4 & 0.51 & 5.9 & 0 & 0 \\
NGC1433 & 12.1 & 28.6 & 185.8 & -0.38 & 10.4 & 9.3 & 9.3 & -0.51 & 1.5 & 1 & 1 \\
NGC1511 & 15.3 & 72.7 & 110.9 & 0.34 & 9.9 & 9.2 & 9.6 & 0.6 & 2.0 & 0 & 0 \\
NGC1512 & 17.1 & 42.5 & 253.0 & -0.03 & 10.6 & 9.1 & 9.8 & -0.25 & 1.2 & 1 & 1 \\
NGC1546 & 17.7 & 70.3 & 111.2 & -0.11 & 10.3 & 9.3 & 8.7 & -0.17 & -0.4 & 1 & 0 \\
NGC1559 & 19.4 & 65.4 & 125.6 & 0.55 & 10.3 & 9.6 & 9.5 & 0.51 & 5.9 & 0 & 1 \\
NGC1566 & 17.7 & 29.5 & 216.8 & 0.64 & 10.7 & 9.7 & 9.8 & 0.32 & 4.0 & 1 & 1 \\
NGC2090 & 11.8 & 64.5 & 134.6 & -0.5 & 10.0 & 8.7 & 9.4 & -0.34 & 4.5 & 1 & 0 \\
NGC2283 & 13.7 & 43.7 & 82.8 & -0.35 & 9.8 & 8.6 & 9.5 & -0.04 & 5.9 & 1 & 1 \\
NGC2835 & 12.4 & 41.3 & 192.4 & 0.06 & 9.9 & 8.8 & 9.3 & 0.26 & 5.0 & 0 & 1 \\
NGC2997 & 14.1 & 33.0 & 307.7 & 0.64 & 10.7 & 9.8 & 9.7 & 0.34 & 5.1 & 1 & 0 \\
NGC3351 & 10.0 & 45.1 & 216.8 & 0.04 & 10.3 & 9.1 & 8.9 & -0.01 & 3.1 & 0 & 1 \\
NGC3511 & 13.9 & 75.1 & 181.2 & -0.08 & 10.0 & 9.0 & 9.1 & 0.09 & 5.1 & 0 & 1 \\
NGC3596 & 11.0 & 25.1 & 109.2 & -0.56 & 9.6 & 8.7 & 8.8 & -0.12 & 5.2 & 1 & 0 \\
NGC3626 & 20.0 & 46.6 & 88.3 & -0.63 & 10.4 & 8.6 & 8.9 & -0.76 & -0.8 & 0 & 1 \\
NGC3627 & 11.3 & 57.3 & 308.4 & 0.57 & 10.8 & 9.8 & 9.0 & 0.21 & 3.1 & 1 & 1 \\
NGC4207 & 15.8 & 64.5 & 45.1 & -0.71 & 9.7 & 8.7 & 8.6 & -0.32 & 7.7 & 0 & 0 \\
NGC4254 & 13.0 & 34.4 & 151.1 & 0.47 & 10.4 & 9.9 & 9.7 & 0.4 & 5.2 & 1 & 0 \\
NGC4293 & 15.8 & 65.0 & 187.1 & -0.24 & 10.4 & 9.0 & 7.7 & -0.37 & 0.3 & 0 & 0 \\
NGC4298 & 13.0 & 59.2 & 76.1 & -0.48 & 9.9 & 9.2 & 9.0 & -0.23 & 5.1 & 0 & 0 \\
NGC4321 & 15.2 & 38.5 & 182.9 & 0.53 & 10.7 & 9.9 & 9.4 & 0.23 & 4.0 & 1 & 1 \\
NGC4424 & 16.2 & 58.2 & 91.2 & -0.53 & 9.9 & 8.4 & 8.3 & -0.28 & 1.3 & 0 & 0 \\
NGC4457 & 15.0 & 17.4 & 83.8 & -0.5 & 10.4 & 9.0 & 8.4 & -0.58 & 0.3 & 1 & 0 \\
NGC4496A & 14.9 & 53.8 & 101.2 & -0.21 & 9.5 & 8.6 & 9.2 & 0.31 & 7.4 & 0 & 1 \\
NGC4535 & 15.8 & 44.7 & 244.4 & 0.31 & 10.5 & 9.6 & 9.6 & 0.14 & 5.0 & 0 & 1 \\
NGC4540 & 15.8 & 28.7 & 65.8 & -0.77 & 9.8 & 8.6 & 8.5 & -0.46 & 6.2 & 0 & 1 \\
NGC4548 & 16.2 & 38.3 & 166.4 & -0.27 & 10.7 & 9.2 & 8.8 & -0.55 & 3.1 & 1 & 1 \\
NGC4569 & 15.8 & 70.0 & 273.6 & 0.13 & 10.8 & 9.7 & 8.9 & -0.23 & 2.4 & 1 & 1 \\
NGC4571 & 14.0 & 32.7 & 106.9 & -0.57 & 10.0 & 8.9 & 8.7 & -0.4 & 6.4 & 0 & 0 \\
NGC4689 & 15.0 & 38.7 & 114.6 & -0.39 & 10.1 & 9.1 & 8.6 & -0.31 & 4.7 & 0 & 0 \\
NGC4694 & 15.8 & 60.7 & 59.9 & -0.89 & 9.9 & 8.3 & 8.6 & -0.63 & -1.8 & 0 & 0 \\
NGC4731 & 13.3 & 64.0 & 189.7 & -0.31 & 9.4 & 8.6 & 9.4 & 0.24 & 5.9 & 1 & 1 \\
NGC4781 & 11.3 & 59.0 & 111.2 & -0.34 & 9.6 & 8.8 & 9.2 & 0.08 & 7.0 & 0 & 1 \\
NGC4941 & 15.0 & 53.4 & 100.7 & -0.38 & 10.2 & 8.7 & 8.4 & -0.32 & 2.1 & 0 & 1 \\
NGC4951 & 15.0 & 70.2 & 94.2 & -0.49 & 9.8 & 8.6 & 9.0 & -0.18 & 6.0 & 0 & 0 \\
NGC5042 & 16.8 & 49.4 & 125.6 & -0.23 & 9.9 & 8.8 & 9.0 & -0.01 & 5.0 & 0 & 1 \\
NGC5068 & 5.2 & 35.7 & 224.5 & -0.55 & 9.3 & 8.4 & 8.8 & 0.07 & 6.0 & 0 & 1 \\
NGC5134 & 19.9 & 22.7 & 81.3 & -0.37 & 10.4 & 8.8 & 8.9 & -0.47 & 2.9 & 0 & 1 \\
NGC5530 & 12.3 & 61.9 & 144.9 & -0.48 & 10.0 & 8.9 & 9.1 & -0.31 & 4.2 & 0 & 0 \\
NGC5643 & 12.7 & 29.9 & 157.4 & 0.39 & 10.2 & 9.4 & 9.1 & 0.4 & 5.0 & 0 & 1 \\
NGC6300 & 11.6 & 49.6 & 160.0 & 0.27 & 10.4 & 9.3 & 9.2 & 0.18 & 3.1 & 0 & 1 \\
NGC7456 & 15.7 & 67.3 & 123.3 & -0.59 & 9.6 & 9.3 & 8.7 & -0.16 & 6.0 & 0 & 0 \\
\hline
\multicolumn{12}{@{}p{\linewidth}}{\footnotesize Note -- (a) Distance \citep{Ana21}.  (b) Inclination \citep{Lan20}. (c) Optical radius from the Lyon-Meudon Extragalactic Database (LEDA).  (d) \& (e): SFR and \Mstar\ \citep{Ler21b}. (f) Aperture-corrected total molecular gas mass  based on the PHANGS-ALMA observations \citep{Ler21a}. (g) Atomic gas mass from LEDA.  (h) Offset from the star-forming main sequence $\Delta$MS \citep{Cat18,Ler21b}. (i) Hubble type from LEDA.  (j) \& (k)  Presence ($=$ 1) and absence ($=$ 0) of grand-design spiral arms and stellar bar \citep{Que21}.}
	\label{tab_galaxy_sampe_props}
\end{longtable*}	
		
\subsection{CO Images: PHANGS-ALMA}
\label{sec_data_co}
The 90 PHANGS-ALMA galaxies were observed in \mbox{CO(2--1)} using the ALMA \mbox{12-m} and \mbox{7-m}  arrays,  and  total-power antennas.
The data were imaged in CASA \citep{Mcm07} version~5.4.0.
We use the spectral line cubes delivered in the internal data release version~3.4.
The  data  have  native  spatial  resolutions  of $\sim$ 25--180~pc, depending on the source distance.
The typical 1$\sigma$ noise  level is $\sim$ 0.3~K per 2.5~km~s$^{-1}$ channel,  but varies slightly between galaxies.
We  use  the  ``broad mask'' integrated intensity maps.
These  maps  include most CO emission (98\% with a 5 -- 95th percentile range of 73 -- 100\%) in the cube, meaning that they have high completeness.
For full details of the sample, observing and reduction processes, and final data products see \cite{Ler21a}.

 We create maps of molecular gas surface density ($\Sigma_\mathrm{H_{2}}$) by applying a radially-varying  CO-to-H$_{2}$ conversion factor ($\alpha_\mathrm{CO}$) to the CO integrated intensity map,  following the method described in \cite{Sun20a}. 
We briefly summarize the steps here.

Many studies have shown that $\alpha_\mathrm{CO}$ increases with decreasing metallicity ($Z$) \citep[e.g.,][]{Wil95,Ari96,Ler11,Sch12}. 
Our adopted radially-varying $\alpha_\mathrm{CO}$ takes into account the  radial metallicity gradient of galaxies.
The metallicity at one effective radius ($R_\mathrm{e}$) in each galaxy is predicted according to the global \Mstar\ and the global \Mstar--$Z$ relation reported by \cite{San19} based on the \cite{PP04} metallicity calibration.
Then the  $Z$ at 1$R_\mathrm{e}$ is extended to cover the entire galaxy assuming a universal radial metallicity gradient of $-0.1$~dex $R_\mathrm{e}^{-1}$ \citep{San14}. 
Finally, $\alpha_\mathrm{CO}$ at each galactocentric radius is calculated via  the  relation determined by \cite{Acc17}:
\begin{equation}
	\label{eq_z_alphaco}
	\alpha _{\mathrm{CO}}=4.35\, Z{}'^{-1.6}\, M_{\odot }\: \mathrm{pc}^{-2}\, (\mathrm{K\: km\: s^{-1}})^{-1},
\end{equation}
where $Z{}'$ is  the local gas-phase abundance normalized to the solar value ($12 + \log[{\rm O}/{\rm H}]) = 8.69$; \citealt{Asp09}).
Since $\alpha_\mathrm{CO}$  is defined for the $^{12}\mathrm{CO}(J=1\rightarrow0)$ transition, we apply a constant
$^{12}\mathrm{CO}(J=2\rightarrow1)$  to $^{12}\mathrm{CO}(J=1\rightarrow0)$  brightness temperature ratio of $R_{21} = 0.65$. We do not account for galaxy to galaxy  (and also inside a galaxy)  variations in this ratio, which are typically  $\sim$ 0.1~dex \citep{Ler13,Yaj21,denB21,Ler21c}.
  We test our results against using a constant Galactic  $\alpha_\mathrm{CO}$ and discuss the choice of  $R_{21}$ in Appendix~\ref{sec_appendix_alphaco} and \ref{sec_appendix_r21}, respectively.

\subsection{\texorpdfstring{H$\alpha$}{Halpha} Images: PHANGS-\texorpdfstring{H$\alpha$}{Halpha}}
\label{sec_ha_data}
To create maps of recent star formation in our PHANGS galaxies, we obtained $R$-band and H$\alpha$-centered narrowband imaging for our sample. 
The 65 PHANGS-H$\alpha$ galaxies were observed by the Wide Field Imager (WFI) instrument at the MPG-ESO \mbox{2.2-m} telescope at the La Silla Observatory or by the Direct CCD at the Ir\'{e}n\'{e}e du Pont \mbox{2.5-m} telescope at the Las Campanas Observatory.
Among the 65 galaxies, 32 were observed by the \mbox{2.2-m} telescope and 36 by du Pont telescope, including three galaxies that were observed by both instruments.
For galaxies with repeated observations, we use the observation that has the best spatial resolution. 
The field of view (FoV) of WFI and du Pont observations are $34\arcmin \times 33\arcmin$ and $8.85\arcmin \times 8.85\arcmin$, respectively.
Full details of the observations, data reduction, and map construction can be found in A.~Razza et al.\ (in prep.).
The images used in this work correspond to the internal release version 1.0 of the PHANGS-H$\alpha$ survey.
The main steps are summarized here (see also  \citetalias{Sch19}).

The data frames were astrometrically and photometrically calibrated using Gaia DR2 catalogs \citep{Gaia18} cross-matched to all stars in the full FoV of the images. 
Typical seeing for the data is $\sim$ 1$\arcsec$ and the final astrometric accuracy is $\lesssim$ 0.1\arcsec.
The sky background is computed in each exposure by masking all the sources $>$ 2$\sigma$ above the sigma-clipped mean, including an elliptical area around the galaxies based on the galaxy geometric parameters.
A 2D plane is then fit to this background and subtracted, with this process occurring for each exposure frame.
Each background-subtracted frame is then combined using inverse-variance weighting.

Then the stellar continuum is subtracted from the combined images.
The flux scale is determined using the median  of  the  flux  ratios  for  a  selection  of  non-saturated  stars  that are matched between the H$\alpha$ and the $R$-band images.
Using this flux ratio as a basis, we obtain a first estimate of the H$\alpha{+}$[\textsc{N\,ii}] flux by subtracting the  $R$-band image from the H$\alpha{+}$[\textsc{N\,ii}]+continuum image.

However, the blended H$\alpha{+}$[\textsc{N\,ii}] line also contributes to the  $R$-band data. 
Using the estimated H$\alpha{+}$[\textsc{N\,ii}] image we determine the H$\alpha{+}$[\textsc{N\,ii}] contribution to the $R$-band image. 
We subtract this estimated H$\alpha{+}$[\textsc{N\,ii}]  contamination from the $R$-band image and iterate this process until successive continuum estimates differ by less than 1\%. 
Then we subtract this continuum estimate to obtain a flux-calibrated line (H$\alpha{+}$[\textsc{N\,ii}]) image.

We correct the measured H$\alpha$ flux for the loss due to the filter transmission, using the spectral shape of the narrowband filter and the position of the H$\alpha$ line within the filter. 
We also correct for the contribution of the [\textsc{N\,ii}] lines at $654.8$ and $658.3$~nm to the narrowband filter flux, assuming  a uniform [\textsc{N\,ii}]/H$\alpha$ ratio of 0.3. 
This value is derived from high-spectral resolution observations of  \textsc{H\,ii} regions in NGC~0628 with the VLT/MUSE instrument,with a typical scatter of $\pm$ 0.1 \citep{Kre16,San21}.
We treat this as a characteristic spectrum for all of our targets, but note possible variation in [\textsc{N\,ii}]/H$\alpha$ as a source of uncertainty.
Finally, we correct all images for foreground Galactic extinction using  \cite{SchFin11}, who assumes a \cite{Fit99} reddening law with $R_{V}$ = 3.1.

\subsection{Sample Selection from PHANGS} 
The sample of galaxies used in this work is a subset of the full PHANGS-ALMA and PHANGS H$\alpha$   observations.
Since our main analysis is performed at a fiducial resolution of 150~pc, the selected galaxies are required to be detected in both CO and H$\alpha$, and that a physical resolution better than 150~pc is achieved for \emph{both} observations.
Moreover, we only include galaxies that had been observed by all ALMA arrays (i.e., 12-m+7-m+total power) by the time of the internal data release v3.4.
No additional cut (e.g. on \Mstar) is applied to the sample, besides the selection criteria that are inherited from the parent sample (see above).
This results in a sample of 51 galaxies. 
Among these,   NGC~2566 has many foreground stars that impact the reliability of the H$\alpha$ data and NGC~6744 has incomplete ALMA coverage. 
These two galaxies are therefore also excluded from our analysis, resulting in a final sample  of 49 galaxies.
Global properties of the sample are presented  in Table~\ref{tab_galaxy_sampe_props}.

\begin{figure*}
	\centering
	\includegraphics[width=0.9\textwidth]{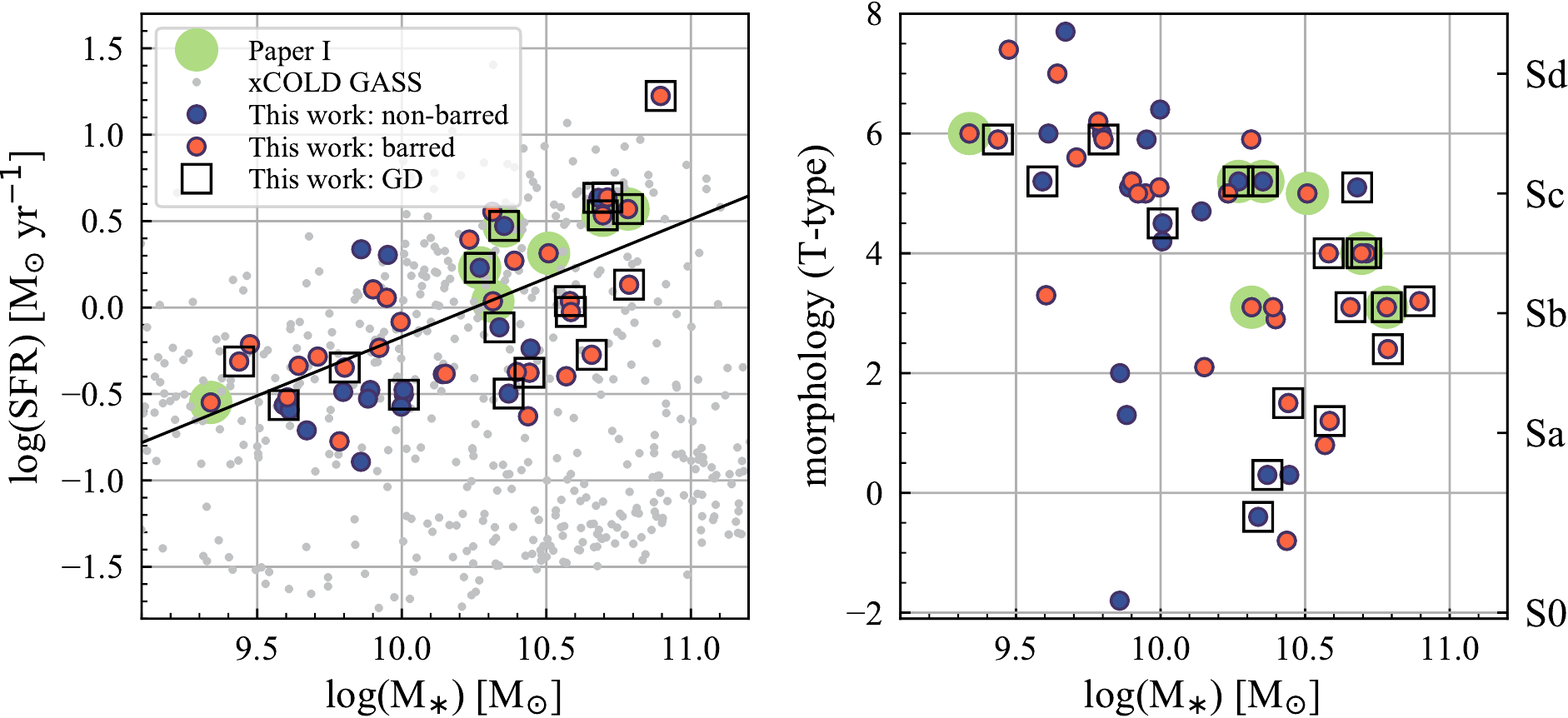}  
	\caption{Left: Integrated star formation rate (SFR) versus stellar mass (\Mstar).  Gray dots represent local galaxies in the xCOLD GASS survey \citep{Sai17}.  Large colored circles represent the PHANGS galaxies used in this work. Galaxies with a bar are shown in red, while galaxies without a bar are shown in blue. Further, galaxies with grand-design (GD) spiral arms are marked by open squares.  The  integrated SFR and \Mstar\ of PHANGS galaxies  are derived from GALEX and WISE \citep{Ler19,Ler21b}. The black line represents the local star-forming main sequence \citep{Cat18,Ler19,Ler21b}. Right: Hubble type versus \Mstar. The symbols are the same as in the left panel. Note that the Hubble T-type  considers ellipticity  and  strength of spiral arms, and does not reflect the presence or absence of the  bar. Galaxies already included in the sample of \citetalias{Sch19} are highlighted by a green circle.}
	\label{fig_sample}
\end{figure*}

The left panel of Figure~\ref{fig_sample} shows the SFR--\Mstar\ relation for our  sample overlaid on   a sample of local galaxies from the  xCOLD GASS survey (grey circles; \citealt{Sai17}).
The  integrated SFR and \Mstar\  are derived based on GALEX and WISE \citep{Ler19,Ler21b}.
The line in the figure represents the local star-forming main sequence derived by \cite{Ler19}.
There are roughly equal numbers of galaxies above and below the main sequence.
The offset from the main sequence ($\Delta$MS) spans $\pm$ 0.8~dex ($\sim$ a factor of 6).   
Galaxies already included in the sample of \citetalias{Sch19} are highlighted by a green circle (NGC~0628, NGC~3351, NGC~3627, NGC~4254, NGC~4321, NGC~4535, and NGC~5068).

We  further classify our galaxies based on the presence of bar and grand-design  spiral arms.
In Figure~\ref{fig_sample}, blue and red circles denote non-barred and barred galaxies, respectively, while the galaxies with grand-design spiral arms are marked by open squares.
Information on the galactic structures is provided in Table~\ref{tab_galaxy_sampe_props}.
We define a galaxy as barred if a bar component was implemented in the PHANGS environmental masks \citep{Que21} (\texttt{morph$\_$bar$\_$flag} in the PHANGS sample table version~1.5). 
These bar identifications mostly follow \cite{Her15} and \cite{Men07}, with some modifications based on the multi-wavelength and kinematic information available in PHANGS.
For spiral arms, we adopt the flags \texttt{morph$\_$spiral$\_$arms} (i.e., grand-design spiral arms) from the PHANGS sample table, which comes from visual inspection of multi-wavelength data by four PHANGS collaboration members.
Strictly speaking, the  \texttt{morph$\_$spiral$\_$arms} flag in the sample table indicates whether the environmental masks include spiral masks or not. 
It is generally true that we implemented spiral arms mostly for grand-design spirals (and did not attempt to do so for flocculent arms). 
However, in some cases, e.g. due to inclination, we found that the spiral mask was not reliable even though the galaxy shows clear spiral arms and was classified as grand-design by \cite{But15}.
Therefore,  our classification does not always agree with arm classifications from the literature (e.g., \citealt{But15} for S4G\footnote{S4G: Spitzer Survey of Stellar Structure in Galaxies}).

Morphology classification is presented by Hubble morphological T-type in this work.
The  T-type values for S0, and Sa -- Sd galaxies are approximately -2, 1, 3, 5, and 7, respectively.  
Note that T-type  considers ellipticity  and  strength of spiral arms, but does not reflect the presence or absence of the bar.
The right panel of Figure~\ref{fig_sample} displays the Hubble type of our target galaxies as a function of \Mstar.
The Hubble type of the galaxies in our sample ranges from $-1.8$ to 7.7 (approximately equivalent  to S0--Sd).
Our sample shows the expected trend:  earlier types (i.e., smaller Hubble type values) are generally more massive \citep[e.g.,][]{Kel14,Gon15,Lai16}, but the  correlation is rather poor at the high-mass end of  our sample of $\log(\Mstar/\mathrm{M}_{\sun}) > 10$.
In this work, we use the term ``earlier'' to denote  galaxies with lower values of Hubble type, but note that our  working sample does not contain elliptical galaxies; the earliest-type galaxy in our sample is NGC~4694 with Hubble type of $-$1.8 ($\sim$ S0).

\section{Methodology}
\label{sec_method}
This section introduces the method used to quantify the relative distribution of molecular gas traced by CO emission and recent star formation traced by H$\alpha$ emission.
The method is identical to that used in \citetalias{Sch19}, with some changes in tuning parameter values.

\subsection{\texorpdfstring{H$\alpha$}{Halpha}: Filtering Out Emission from Diffuse Ionized Gas}
\label{sec_ha_filtering}
Our analysis uses H$\alpha$ as a tracer for the location of recent high-mass star formation. 
However, H$\alpha$ not only arises from \textsc{H\,ii} regions surrounding the massive stars that ionize them, but also the larger scale diffuse ionized gas (DIG). 
To correctly correlate the sites of star formation with  molecular gas, our analysis must remove this diffuse component.
DIG is warm (⁠$\sim$ 10$^{4}$~K) and low density ($<$ 0.1 cm$^{-3}$) gas found in the ISM of galaxies which seems similar to the warm ionized medium observed in the Milky Way  \citep[see the review by][]{Haf09}.
The energy sources of DIG  are still not well understood. 
Spectral features, such as the emission-line ratios and ionizing spectrum, of DIG are different from those of \textsc{H\,ii} regions powered by massive young stars \citep[e.g.,][]{Hoo03,Bla09,Zha17,Tom17,Tom19}, indicating the presence of additional sources of ionization. 
Since DIG constitutes a substantial fraction of the H$\alpha$ flux in star-forming galaxies \citep[e.g.,][]{Oey07,Tom21}, one must remove the DIG contribution from the H$\alpha$ fluxes when using H$\alpha$ as a star-formation tracer.

Following the approach utilized in \citetalias{Sch19}, a   two-step unsharp masking technique is used to remove the DIG from the H$\alpha$ images,
We first identify diffuse emission on scales larger than \textsc{H\,ii} regions, and then we take into account higher levels of DIG contribution and clustering of \textsc{H\,ii} regions that are often found in galactic structures \citep[e.g., spiral arms,][]{Kre16}.
More specifically, the following steps are undertaken to remove DIG in the original H$\alpha$ images.
\begin{enumerate}
	\item \emph{Unsharp mask with a kernel of 200~pc}. We  smooth our original image with a Gaussian kernel with FWHM size of 200~pc, slightly larger than the largest \textsc{H\,ii} regions \citep{Oey03,Azi11,Whi11}.  
	Then we subtract this smoothed image from the original image. We identify initial \textsc{H\,ii}  regions as the parts of the map still detected at high signal-to-noise in this filtered map. Specifically, first of all, peaks above  5$\sigma$  are identified, and then the mask is expanded to contain all   connected regions that are above 3$\sigma$. \textsc{H\,ii}  regions are identified as pixels enclosed within the masks. 
	\item \emph{Subtract a scaled version of the initial \textsc{H\,ii} regions from the DIG map.} We subtract a scaled version of the \textsc{H\,ii} regions identified in the previous step from the original map. The scaling factor is an arbitrary choice, but  we do not want to over-subtract at this stage. A scaling factor of 0.1 is adopted in this work.
	\item  \emph{Unsharp mask with a kernel of 400~pc}. We smooth our \textsc{H\,ii} region-subtracted image with a Gaussian kernel that has FWHM of 400~pc,  larger than that in Step~1. This scale is set such to detect higher levels of DIG contribution and clustering of \textsc{H\,ii} regions. Then, we subtract this smooth version of the image from the original image. We identify our final set of \textsc{H\,ii} regions  in this filtered map using the same S/N criteria in Step 1.
\end{enumerate}

 On average, the DIG removal process removes $\sim$65\%  of the H$\alpha$ emission  across the sample,  consistent with the DIG fractions of  PHANGS galaxies measured by different approaches \citep{Che20,Bel21}.
 Moreover, our mean DIG fraction is in good agreement with the mean DIG fraction ($\sim$ 60\%) derived from H$\alpha$-surface brightness-based method for the 109 nearby star-forming galaxies in \citet{Oey07}.  
 A similar mean DIG fraction is also suggested for CALIFA (The Calar Alto Legacy Integral Field Area Survey) galaxies based on integral-field-spectroscopy (IFS)-based \textsc{H\,ii}/DIG separator \citep{Lac18}. The DIG fractions of our sample galaxies are provided in Table \ref{tab_dig_frac}.
We note that the tuning parameters adopted in this work are different from what was used in \citetalias{Sch19}.
The choice of parameters in this work is optimized to reproduce  \textsc{H\,ii} regions  identified with our PHANGS-MUSE IFS H$\alpha$-line images which have similar spatial resolution  \citep{San21}. 
Changing the adopted kernel sizes has only a minor impact on the number of  \textsc{H\,ii} regions but does affect their sizes.
Therefore,  a key assessment of the performance of the DIG removal strategy is to avoid  \textsc{H\,ii} region overgrowth.
This can be evaluated using emission-line ratios accessible via spectroscopic  observations: for example, the [\textsc{N\,ii}]/H$\alpha$ and  [\textsc{S\,ii}]/H$\alpha$ ratios are higher in the DIG relative to \textsc{H\,ii} regions \cite[e.g.,][]{Hoo99,Bla09,Kre16,Tom17,Tom21}.
The full catalog of \textsc{H\,ii} regions identified in our narrowband H$\alpha$ maps, including a detailed description of how we verified the narrowband \textsc{H\,ii} regions using PHANGS-MUSE spectroscopic information and the dependence of DIG fraction on galaxy properties will be presented in a forthcoming paper (H.-A.~Pan et al.\ in prep.).
Our results remain qualitatively unchanged when using the tuning parameters  in \citetalias{Sch19}, as discussed in Appendix~\ref{sec_appendix_impact_um}.
In this work, we assume all the H$\alpha$ emission surviving from the DIG removal process is from \textsc{H\,ii} regions, and  the contribution from other powering sources, such as AGN, post-AGB stars, and shocks, are statistically minor in the analysis. 
Separating these sources from  \textsc{H\,ii} regions rely on emission-line diagnostics and therefore spectroscopic observations.

Here we note two important caveats of our  DIG removal process. 
We use a signal-to-noise threshold when identifying \textsc{H\,ii} regions during the unsharp masking step.
Since the noise and native resolution of the input H$\alpha$ images vary, the effective H$\alpha$ surface brightness threshold applied to our fiducial maps therefore also varies, 
  corresponding to SFR surface densities of $\sim$ 10$^{-3}$ -- 10$^{-2}$ M$_{\sun}$~yr$^{-1}$~kpc$^{-2}$ depending on the galaxy target\footnote{We adopt Equation (6) in \cite{Cal07} for the relation between SFR and H$\alpha$ emission, which assumes a Kroupa initial mass function.}.
For a point source at the native resolution of our H$\alpha$  data,  the effective sensitivity limits in terms of H$\alpha$ surface brightness threshold applied to the fiducial maps  corresponds to  \textsc{H\,ii} region luminosities ($\log (L_\mathrm{\textsc{H\,ii}\,region}^\mathrm{sensitivity} / {\rm erg \, s^{-1}}$)) between $36.7$ and $38.4$, with most ($\sim$80\%) being between 37 and 38.
The $L_\mathrm{\textsc{H\,ii}\,region}^\mathrm{sensitivity}$ for our galaxies are listed in Table \ref{tab_res_sen}.
The \textsc{H\,ii} region sensitivity limits are comparable to the turn-over point of the \textsc{H\,ii} region luminosity function measured by narrowband H$\alpha$ imaging in the literature \citep[e.g.,][]{Bra06,Oey07}, but we may miss the low-luminosity \textsc{H\,ii} regions detected by  optical integral field units  \citep{Kre16,Rou18,San21}, which are unavailable at the necessary resolution for the bulk of the galaxies in our sample.
Therefore, we may underestimate the number of sight lines with H$\alpha$ emission (see Appendix~\ref{sec_appendix_ha_sens}  for a detailed discussion on the impact of H$\alpha$ sensitivity).
Moreover, our  H$\alpha$-line images are not corrected  for dust attenuation, thus the maps may miss the most heavily embedded regions.
Since our main analysis focuses mostly on the location (rather than the amount) of  massive star formation, we consider internal extinction as a secondary issue.
However, for some analysis based on flux, we may underestimate H$\alpha$ flux for the regions where CO (and therefore dust)  is present.

\begin{table}[]
	\centering
	\caption{Fraction (\%) of diffuse ionized gas (DIG) inside $<$ 0.6~R$_{25}$ for each galaxy in our sample. On average, the DIG removal process (Section \ref{sec_ha_filtering}) removes $\sim$65\%  of the H$\alpha$ emission across the sample.}
	\label{tab_dig_frac}
	\begin{tabular}{cccccc}
		\hline
		\multirow{2}{*}{galaxy}	 & DIG &\multirow{2}{*}{galaxy}	 & DIG&\multirow{2}{*}{galaxy}	 & DIG\\
		          & [\%] & &[\%] && [\%] \\
		\hline
		IC1954 & 70 & NGC2997 & 46 & NGC4569 & 62 \\
		IC5273 & 73 & NGC3351 & 62 & NGC4571 & 68 \\
		NGC0628 & 51 & NGC3511 & 65 & NGC4689 & 77 \\
		NGC1087 & 49 & NGC3596 & 52 & NGC4694 & 91 \\
		NGC1300 & 66 & NGC3626 & 86 & NGC4731 & 65 \\
		NGC1317 & 80 & NGC3627 & 64 & NGC4781 & 74 \\
		NGC1365 & 60 & NGC4207 & 74 & NGC4941 & 84 \\
		NGC1385 & 49 & NGC4254 & 50 & NGC4951 & 80 \\
		NGC1433 & 69 & NGC4293 & 88 & NGC5042 & 92 \\
		NGC1511 & 64 & NGC4298 & 58 & NGC5068 & 52 \\
		NGC1512 & 65 & NGC4321 & 58 & NGC5134 & 88 \\
		NGC1546 & 79 & NGC4424 & 87 & NGC5530 & 66 \\
		NGC1559 & 62 & NGC4457 & 81 & NGC5643 & 71 \\
		NGC1566 & 48 & NGC4496A & 55 & NGC6300 & 58 \\
		NGC2090 & 74 & NGC4535 & 78 & NGC7456 & 87 \\
		NGC2283 & 57 & NGC4540 & 69 & & \\
		NGC2835 & 41 & NGC4548 & 87 & & \\
		\hline
	\end{tabular}
\end{table}

\begin{table*}[]
	
		\centering
		\caption{ Parameters of H$\alpha$ and CO observations. H$\alpha$ res. and CO res. denote the native physical resolution of H$\alpha$ and CO observations. $L_\mathrm{\textsc{H\,ii}\,region}^\mathrm{sensitivity}$ is the effective sensitivity limits in terms of \textsc{H\,ii} region luminosity  at the native resolution (Section \ref{sec_ha_filtering}). The sensitivity of CO observation is represented by 1$\sigma$ $\Sigma_\mathrm{H_{2}}$ at 150~pc resolution (Section \ref{sec_co_threshold}). }
		\label{tab_res_sen}
		\begin{tabular}{ccccc|ccccc}
			\hline
\multirow{2}{*}{galaxy} & H$\alpha$ res. & CO res. & $L_\mathrm{\textsc{H\,ii}\,region}^\mathrm{sensitivity}$ & 1$\sigma$ $\Sigma_\mathrm{H_{2}}$ & \multirow{2}{*}{galaxy}&H$\alpha$ res. & CO res. & $L_\mathrm{\textsc{H\,ii}\,region}^\mathrm{sensitivity}$ & 1$\sigma$ $\Sigma_\mathrm{H_{2}}$ \\
 & [pc] & [pc] & [log(erg~s$^{-1}$)] & [M$_{\sun}$~pc$^{-2}$] & & [pc] & [pc] & [log(erg~s$^{-1}$)] & [M$_{\sun}$~pc$^{-2}$] \\
\hline
IC1954 & 88 & 91 & 37.6 & 0.9 & NGC4293 & 53 & 88 & 36.7 & 1.5 \\
IC5273 & 77 & 120 & 37.4 & 0.8 & NGC4298 & 71 & 105 & 37.3 & 1.0 \\
NGC0628 & 41 & 53 & 37.0 & 1.5 & NGC4321 & 47 & 121 & 37.0 & 2.1 \\
NGC1087 & 69 & 123 & 36.9 & 1.8 & NGC4424 & 81 & 89 & 37.7 & 1.7 \\
NGC1300 & 73 & 95 & 36.8 & 3.1 & NGC4457 & 91 & 80 & 37.8 & 2.2 \\
NGC1317 & 74 & 147 & 37.3 & 1.6 & NGC4496A & 73 & 90 & 37.2 & 1.3 \\
NGC1365 & 58 & 130 & 36.9 & 2.4 & NGC4535 & 87 & 119 & 37.6 & 1.6 \\
NGC1385 & 85 & 105 & 36.9 & 2.6 & NGC4540 & 77 & 104 & 37.3 & 2.8 \\
NGC1433 & 74 & 62 & 37.1 & 1.6 & NGC4548 & 73 & 132 & 37.1 & 1.0 \\
NGC1511 & 84 & 107 & 37.8 & 0.9 & NGC4569 & 89 & 128 & 37.3 & 0.7 \\
NGC1512 & 67 & 90 & 36.8 & 1.4 & NGC4571 & 85 & 79 & 37.3 & 1.8 \\
NGC1546 & 125 & 114 & 37.6 & 0.7 & NGC4689 & 97 & 85 & 37.4 & 1.9 \\
NGC1559 & 129 & 117 & 37.7 & 1.5 & NGC4694 & 73 & 88 & 37.9 & 1.3 \\
NGC1566 & 62 & 104 & 36.9 & 2.0 & NGC4731 & 61 & 98 & 37.2 & 0.5 \\
NGC2090 & 52 & 73 & 36.9 & 1.0 & NGC4781 & 52 & 72 & 37.5 & 0.9 \\
NGC2283 & 54 & 87 & 37.5 & 1.5 & NGC4941 & 94 & 115 & 37.4 & 0.7 \\
NGC2835 & 56 & 50 & 37.2 & 1.7 & NGC4951 & 83 & 91 & 37.6 & 0.8 \\
NGC2997 & 64 & 92 & 36.8 & 1.3 & NGC5042 & 84 & 107 & 37.7 & 1.0 \\
NGC3351 & 56 & 70 & 37.4 & 1.4 & NGC5068 & 32 & 24 & 37.4 & 2.0 \\
NGC3511 & 75 & 121 & 37.4 & 0.4 & NGC5134 & 91 & 118 & 37.8 & 1.7 \\
NGC3596 & 63 & 65 & 37.4 & 3.0 & NGC5530 & 65 & 66 & 37.5 & 1.2 \\
NGC3626 & 148 & 114 & 38.4 & 2.5 & NGC5643 & 73 & 75 & 37.6 & 1.6 \\
NGC3627 & 80 & 86 & 37.3 & 1.3 & NGC6300 & 60 & 60 & 37.2 & 1.7 \\
NGC4207 & 70 & 93 & 37.6 & 2.0 & NGC7456 & 84 & 127 & 37.4 & 0.4 \\
NGC4254 & 59 & 107 & 37.2 & 3.1 & & & & & \\
		\hline
\end{tabular}
\end{table*}

\subsection{CO: Applying a Physical Threshold}
\label{sec_co_threshold}
The CO images are treated using a similar scheme. 
We clip the CO images  \emph{at our best-matching resolution of 150~pc}   using a   $\Sigma_\mathrm{H_{2}}$  threshold  of 10~M$_{\sun}$~pc$^{-2}$ accounting for galaxy inclination.
This corresponds to a 3$\sigma$ $\Sigma_\mathrm{H_{2}}$ sensitivity of our CO map with the lowest  sensitivity  at this spatial scale (Table  \ref{tab_res_sen}).
The applied threshold value is lower than the threshold used in \citetalias{Sch19} (i.e., 12.6 M$_{\sun}$~pc$^{-2}$) due to the lower sensitivity of the PAWS M51 observations.
To have a data sample with homogeneous observational properties, M51 is not included in this work.
Our results remain qualitatively unchanged if different  $\Sigma_\mathrm{H_{2}}$  clipping values are adopted (see Appendix~\ref{sec_appendix_co_threshold}).

\subsection{Measuring Sight Line Fractions}
\label{sec_measure_sight lines}
First of all,  the thresholded H$\alpha$  maps are  regridded to match the pixel grid of the CO maps since the FoV of our ALMA observations is considerably smaller than that of narrowband observations.  
We convolve each thresholded  CO and H$\alpha$ image by a Gaussian to a succession of  resolutions, ranging from our highest common resolution of 150~pc to 1500~pc, in steps of 100~pc.
Then we clip the low-intensity emission in the convolved images. 
For each convolved image, we blank the faintest sight lines that collectively contribute 2\% of the total flux in the image to suppress convolution artifacts.
The results remain robust to small variations (1 -- 4\%) of this threshold.

Finally, we measure the presence or absence of the two tracers at each resolution in a FoV  extending to 0.6~$R_{25}$, which is the largest radial extent probed by our data in all galaxies,  corresponding to $\sim$ 6.4~kpc on average (5 -- 22 kpc, mostly $<$ 15 kpc).
We divide each  sight line (pixel)  within 0.6~$R_{25}$ in the thresholded and artifact-clipped CO and H$\alpha$ images into one of four categories:
\begin{itemize}
	\item \co: only CO emission is present
	\item \ha: only H$\alpha$ emission is present
	\item \overlap: both CO and H$\alpha$ emission are present
	\item \emph{empty}: neither CO nor H$\alpha$ emission is present\footnote{We note that the \emph{empty} pixels in the filtered maps may contain DIG and/or CO emission with surface density below the applied threshold in the original maps.}.
\end{itemize}
The fraction of sight lines with (i.e., \co$+$\ha$+$\overlap) and without (i.e., \emph{empty}) any emission are given in Table \ref{tab_sight_line_empty}.
At the highest common resolution of 150~pc, the median fraction of sight lines without any emission within 0.6~$R_{25}$ of our galaxies is as high as 70\%, ranging from 13 -- 97\%.
Moreover, the fractions of \emph{empty} sight lines decrease with increasing spatial scales (Figure \ref{fig_appendix_spatial_scale}, see also \citealt{Pes21}). 
The  distributions of empty pixels among galaxies is  a potentially interesting diagnostic of ISM evolution and host-galaxy properties. We defer a detailed analysis of the statistics relating to empty pixels to a future investigation, since the main focus of this paper is the impact of host galaxy properties and observing scale on the relative distribution of  molecular and ionized gas.

We measure the fraction of sight lines and the fraction of flux in each region type, i.e., \co, \ha, and \overlap, at each resolution.
All galaxies in our sample have non-zero fractions for the three  region types at the highest common resolution of  150~pc  (Appendix~\ref{sec_appendix_individual_galaxy}).
Since we do not consider the sight lines where neither CO nor H$\alpha$ emission is present, the sum of \co, \ha, and \overlap\ sight lines  is 100\%.
We also define \cosightline\ as regions that are classified as either \co\ or \overlap\ (i.e., sight lines with CO emission, regardless of whether they are associated with H$\alpha$ emission  or not), while \hasightline\ are defined as regions of \ha\ or  \overlap\ (i.e., sight lines with  H$\alpha$ emission, regardless of whether they are associated with CO emission).

\begin{table*}
	\centering
		\caption{Fraction of sight lines with (and without) any CO or H$\alpha$ emission inside $<$ 0.6~R$_{25}$ for our sample galaxies, measured at 150~pc resolution.}
		\label{tab_sight_line_empty}
		\begin{tabular}{|cc|cc|cc|cc|cc|}
			\hline
IC1954 & 48(52) & NGC1512 & 8(92) & NGC3596 & 46(54) & NGC4496A & 24(76) & NGC4941 & 20(80) \\
IC5273 & 33(67) & NGC1546 & 30(70) & NGC3626 & 7(93) & NGC4535 & 19(81) & NGC4951 & 30(70) \\
NGC0628 & 27(73) & NGC1559 & 50(50) & NGC3627 & 31(69) & NGC4540 & 40(60) & NGC5042 & 9(91) \\
NGC1087 & 58(42) & NGC1566 & 22(78) & NGC4207 & 40(60) & NGC4548 & 11(89) & NGC5068 & 31(69) \\
NGC1300 & 14(86) & NGC2090 & 40(60) & NGC4254 & 71(29) & NGC4569 & 15(85) & NGC5134 & 21(79) \\
NGC1317 & 12(88) & NGC2283 & 46(54) & NGC4293 & 3(97) & NGC4571 & 32(68) & NGC5530 & 42(58) \\
NGC1365 & 13(87) & NGC2835 & 29(71) & NGC4298 & 87(13) & NGC4689 & 45(55) & NGC5643 & 49(51) \\
NGC1385 & 39(61) & NGC2997 & 36(64) & NGC4321 & 41(59) & NGC4694 & 10(90) & NGC6300 & 48(52) \\
NGC1433 & 16(84) & NGC3351 & 22(78) & NGC4424 & 11(89) & NGC4731 & 10(90) & NGC7456 & 14(86) \\
NGC1511 & 37(63) & NGC3511 & 39(61) & NGC4457 & 26(74) & NGC4781 & 53(47) & & \\
	\hline
		\end{tabular}
\end{table*}

\section{Results}
\label{sec_results}
\subsection{CO and \texorpdfstring{H$\alpha$}{Halpha} Fractions at 150~pc Resolution }
\label{sec_results_fractions}

\begin{figure*}[!ht]
	\begin{center}
		\includegraphics[scale=0.65]{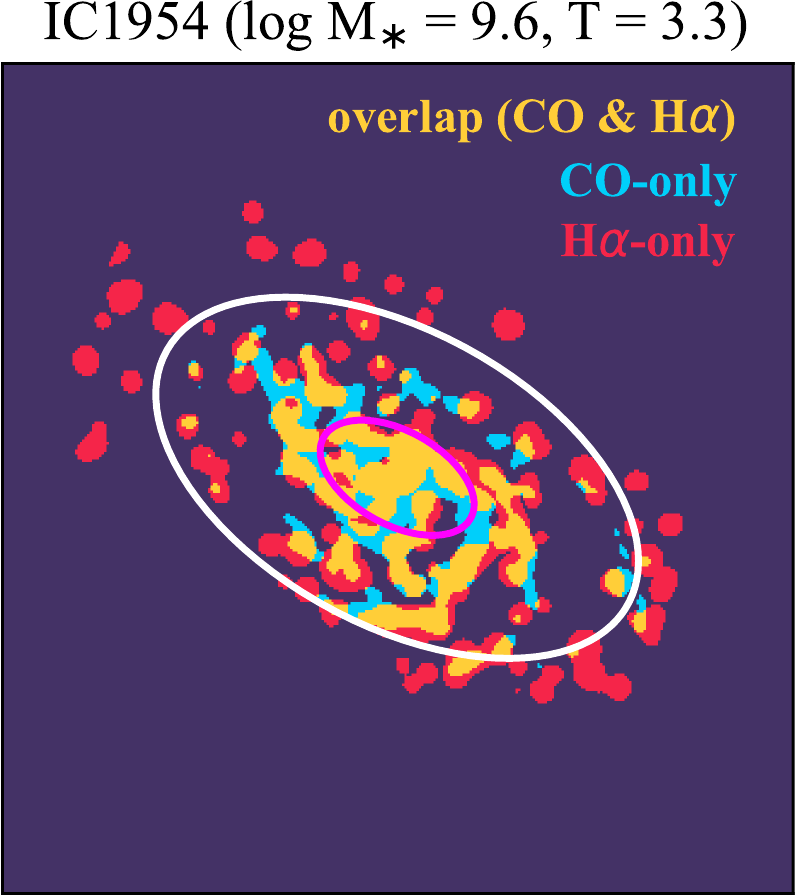}
		\includegraphics[scale=0.65]{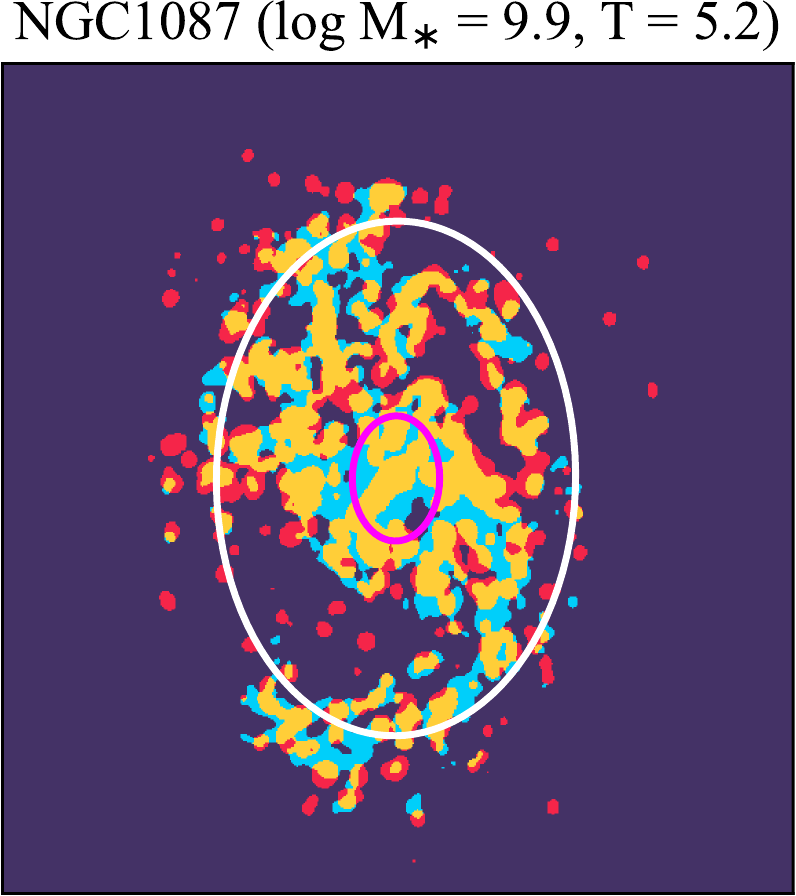}
		\includegraphics[scale=0.65]{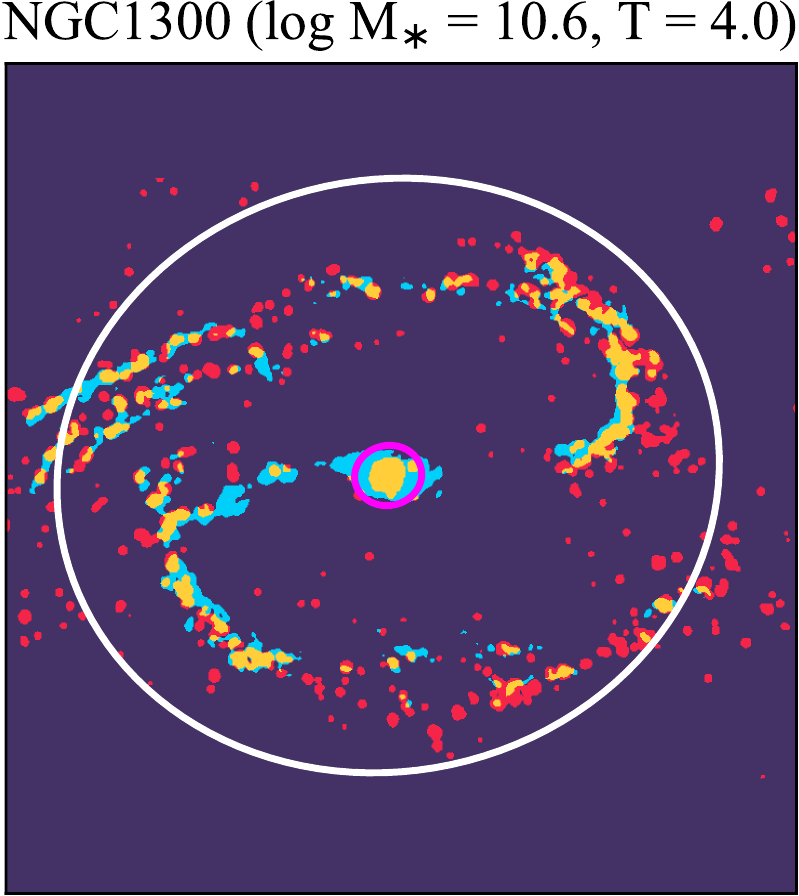}\\
		\vspace{10pt}
		\includegraphics[scale=0.65]{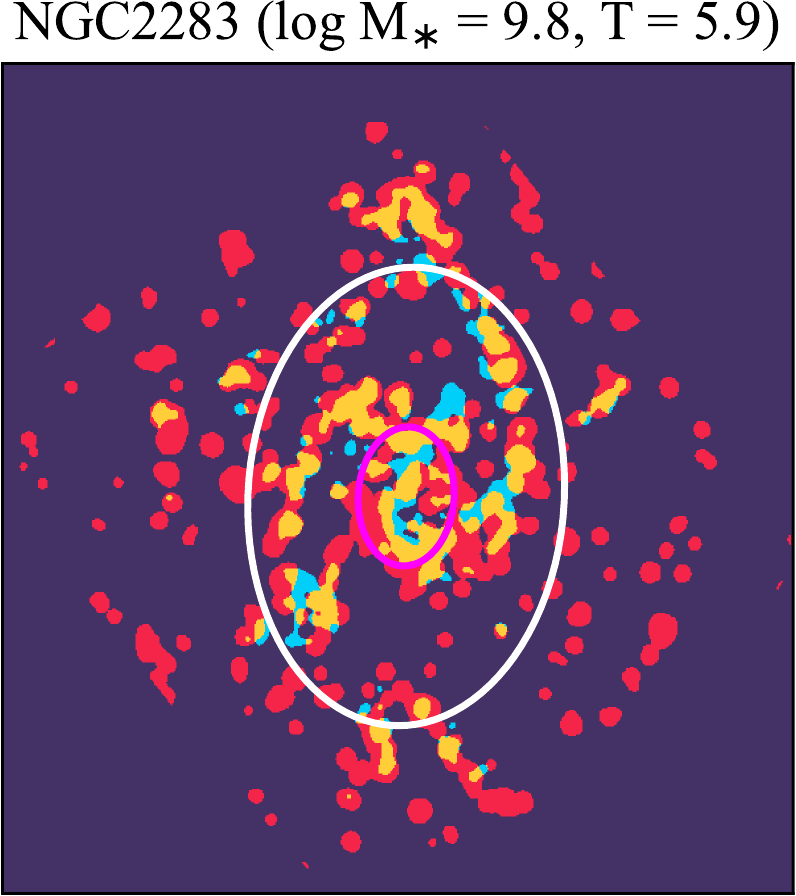}
		\includegraphics[scale=0.65]{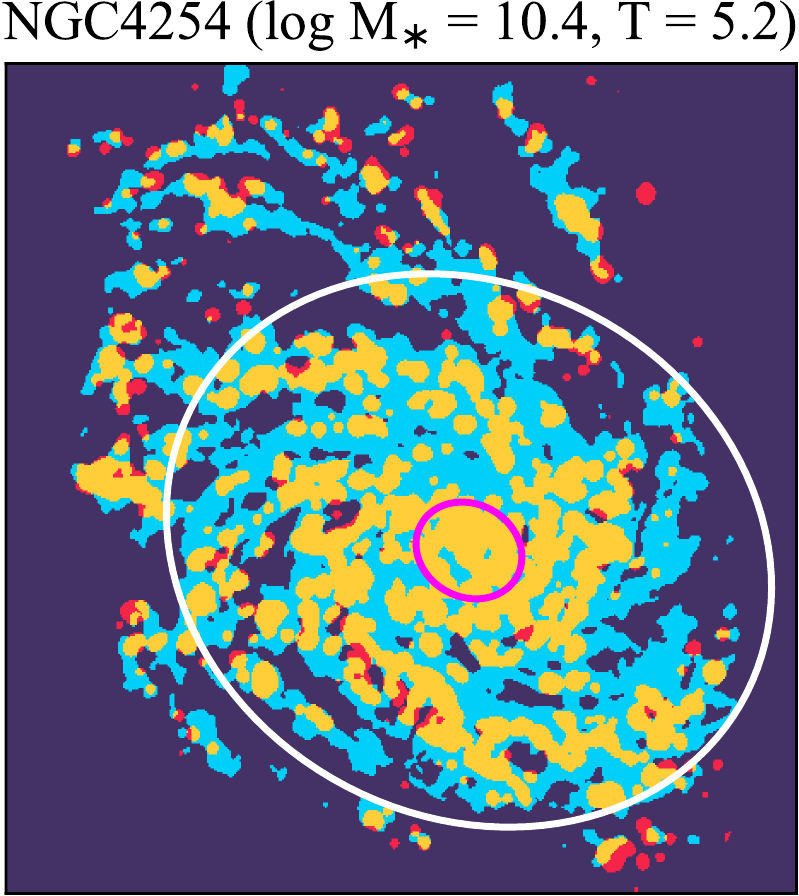}
		\includegraphics[scale=0.65]{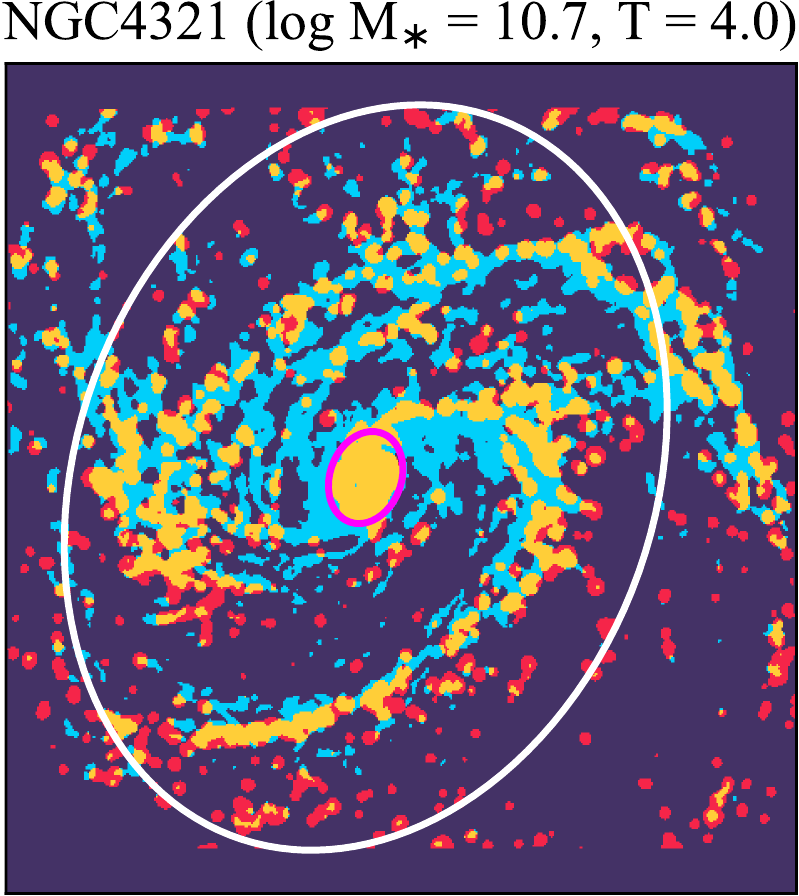}\\
		\vspace{10pt}
		\includegraphics[scale=0.65]{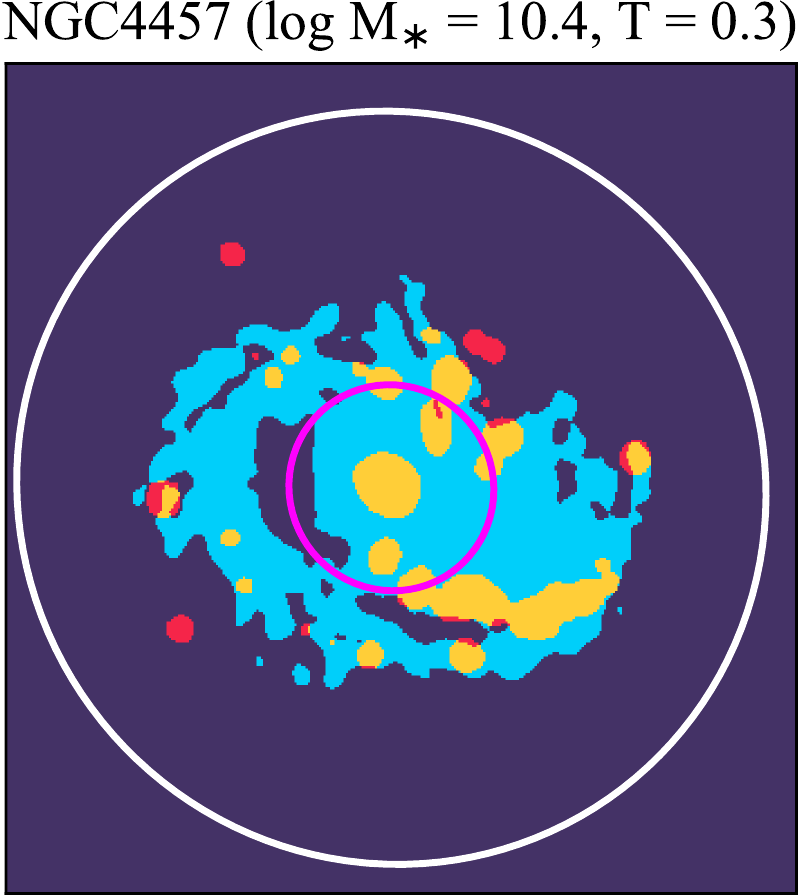}
		\includegraphics[scale=0.65]{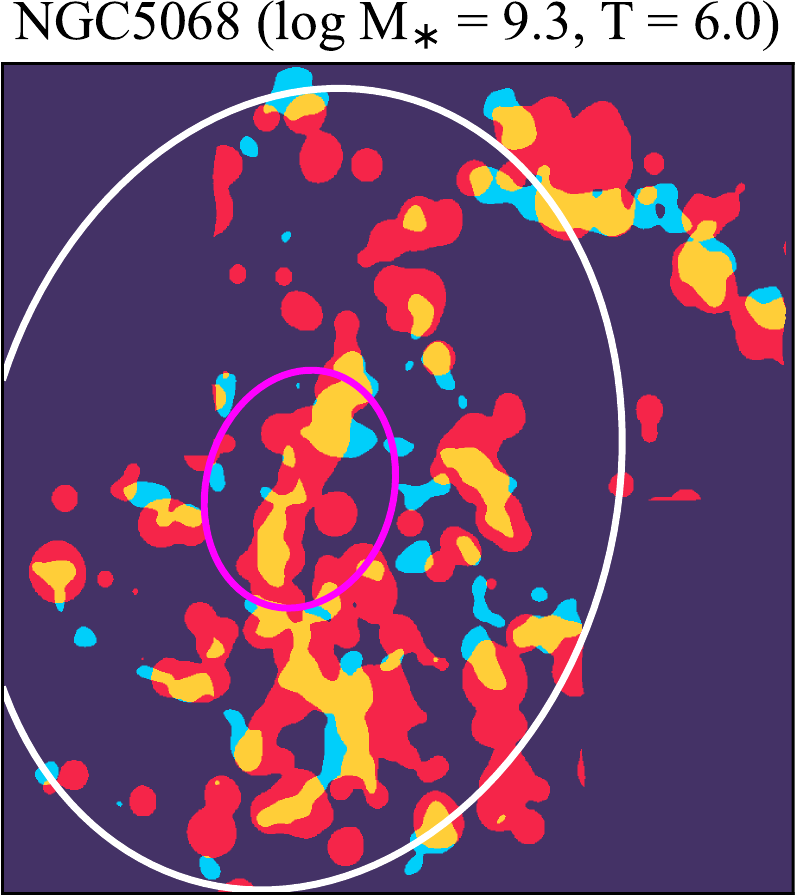}
		\includegraphics[scale=0.65]{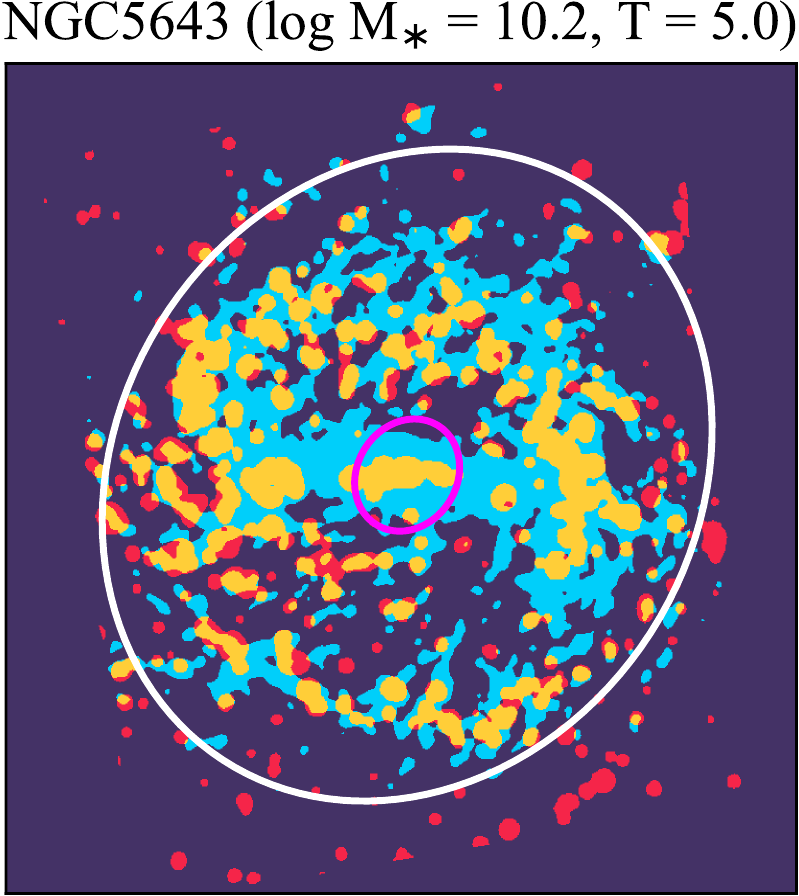}
	\end{center}
	\vspace{-15pt}
	\caption{Examples of the spatial distribution of different sight lines.  Galaxy maps show the regions of \co\ (blue), \ha\ (red), and \overlap\ (yellow) sight lines at  150~pc resolution.  The inner ellipses (magenta) mark the central region, defined as the central 2~kpc in deprojected diameter. The outer ellipses (white) indicate the 0.6$R_{25}$ regions where we measure the global sight line fractions. The \Mstar\ (in unit of solar mass in log scale) and Hubble type of each galaxy are given in the top of each panel.
	}
	\label{fig_pie_example}
\end{figure*}

There are significant galaxy-to-galaxy variations in the CO and H$\alpha$ distributions.
Figure~\ref{fig_pie_example} presents some examples of the distribution of different sight line categories at 150~pc resolution (maps for the full sample are provided in  Appendix~\ref{sec_appendix_individual_galaxy}).
The blue, red, and yellow regions denote \co,  \ha, and  \overlap\ sight lines, respectively.
We define the galactic \emph{center} as the region within 1~kpc (i.e., 2~kpc in diameter) of the galaxy nucleus.
The region that we define as the  center is indicated in each panel as a magenta  ellipse in Figure~\ref{fig_pie_example}, while the region that we use to measure the global sight fraction  is indicate as a white  ellipse (i.e., 0.6~$R_{25}$).

\begin{figure*}
	\centering
	\includegraphics[width=0.95\textwidth]{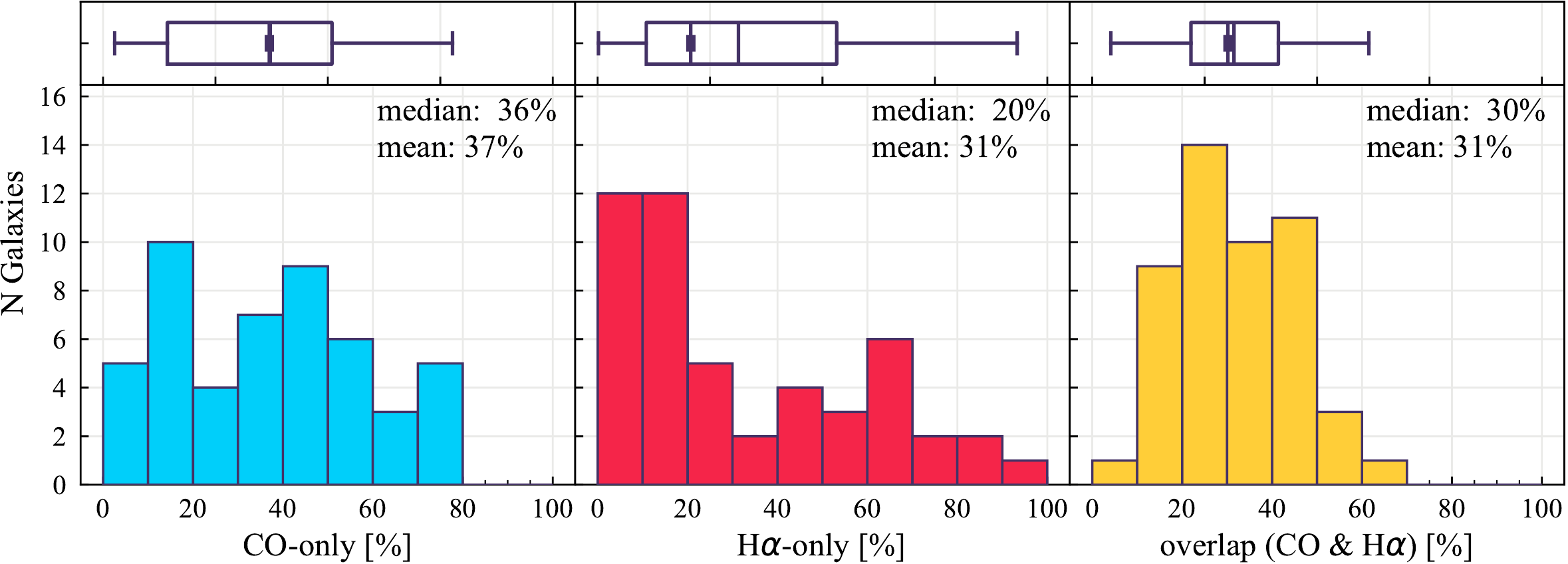}  
	\caption{Distribution of global \co\ (left), \ha\ (middle), and \overlap\ (right) sight line fractions at resolution of 150~pc. Corresponding box plots are shown at the top of each panel.
		The boxes show the interquartile ranges (IQR; the Q1/25th percentile to Q3/75th percentile), and the horizontal whiskers extend to Q1 $-$ 1.5$\times$IQR  and  Q3 $+$ 1.5$\times$IQR.
		The inner vertical belt-like symbol and line in the boxes represent the median and mean of the distribution, respectively; the values are also given in the upper-right of each panel. Substantial galaxy-to-galaxy variations  are seen for all sight line categories.} 
	\label{fig_global_hist_0150pc_thresh}
\end{figure*}

The histograms of Figure~\ref{fig_global_hist_0150pc_thresh}, from left to right, show the distribution of   \co, \ha, and \overlap\ fractions within the fiducial FoV at 150~pc scale, respectively. 
The boxplots shown at  the top of each panel summarize the statistics for the sight line fractions.
The sight line fractions for  each individual galaxy are provided in Table~\ref{tab_fractions} of Appendix~\ref{sec_appendix_individual_galaxy}).
The median and mean sight line fractions are given in the upper-right corner of the panels. 
In the rest of the paper, we will  use the median as the measure of central tendency because the mean is more sensitive to extreme values. 
The mean values  are given in the relevant figures and tables  for reference.

We find a wide range of \co\ fraction in our sample from 3 to 78\%,  with a median of 36\%.
While the \ha\ fraction peaks at the lower end ($<$ 20\%), \ha\ sight lines show a wider range of spatial  coverage than the \co\ sight lines, from nearly 0 to almost 95\%. 
The median \ha\ fraction is 20\%.
The \overlap\ region  exhibits a narrower range  than the \co\ or \ha\ regions, and shows a preference for  intermediate values from 20 to 50\%, with a median of 30\%.

In terms of the relative frequency of the three types of sight lines, our results are qualitatively consistent  with the conclusions of  \citetalias{Sch19} based on a smaller sample of only eight galaxies.
However,  \citetalias{Sch19} reports considerably higher median values for \co\ and \overlap\ sight lines (42\% for \co\ and 37\% for \overlap) and slightly higher median for \ha\ (23\%).
We ascribe this difference to the  combined effect of different thresholds for CO and H$\alpha$ images and sample composition (e.g., \Mstar\ distribution; see Figure \ref{fig_sample} and  next subsection).

\subsubsection{Trends with Host Galaxy Properties}
\label{sec_global_galprops}

To explore the potential origin of the galaxy-to-galaxy variation in the spatial distributions of CO and H$\alpha$, we first  compute the Spearman rank correlation coefficients  between the  sight line fractions and various host-galaxy and observational properties.
The  correlation coefficients  are given in Table~\ref{tab_sight line_cc_galprops}. 
In this work, a significant correlation is defined as a correlation coefficient  of absolute value greater than 0.3.

The strongest correlations are found with \Mstar\ and Hubble type for \co\ and \ha\ fractions.
The fractions of \co\ and \ha\ regions are  moderately correlated with  \Mstar, with correlation coefficients of 0.53 and $-0.44$, respectively.
The \co\ and \ha\ fractions also correlate with Hubble type, with correlation coefficients of  $-0.59$ and 0.38, respectively.
In contrast to the \co\ and \ha\ regions, \overlap\ fractions show no significant correlation with \Mstar\ and Hubble type in terms of correlation coefficients,   $-$0.01 and 0.24, respectively.

\begin{figure*}
	\centering
	\includegraphics[scale=0.85]{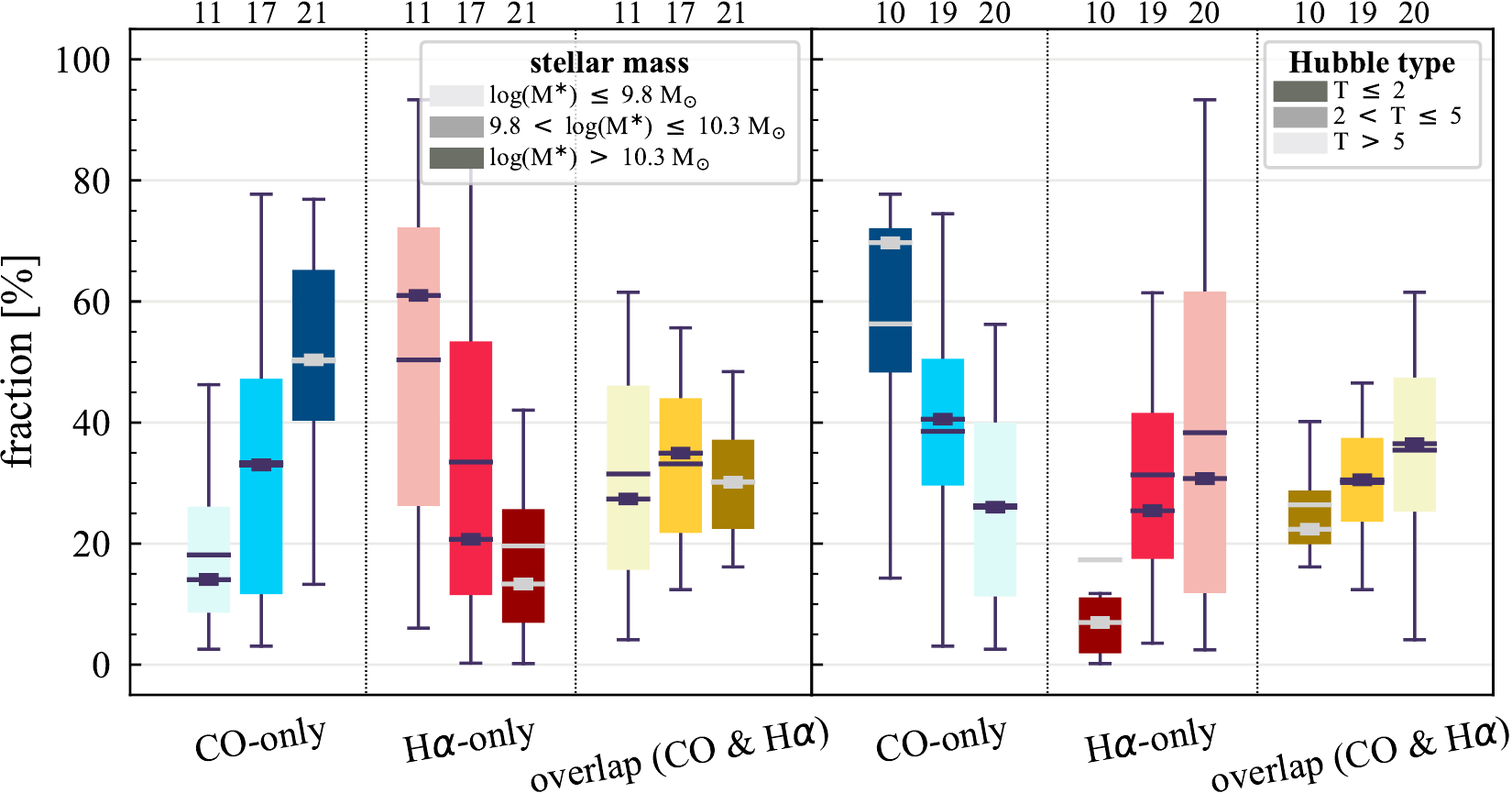}
	
	\caption{Variations of the global sight line fractions at  150~pc resolution   as a function of \Mstar\ (left) and Hubble type (right). For a given type of sight line, the color  darkness of the boxplots resembles increasing \Mstar\   (from left to right) or decreasing Hubble type value (from right to left). The number of galaxies in each \Mstar\ and Hubble type bin are shown in the top of the plots. Symbols of the boxplot are the same as in Figure~\ref{fig_global_hist_0150pc_thresh}.  The \co\ and \ha\ sight line fractions are correlated with \Mstar\ and Hubble type, while the \overlap\ fractions are less sensitive to galaxy properties.
	}  
	\label{fig_global_boxes}
\end{figure*}

To further visualize the dependence of the sight line fractions on \Mstar\ and Hubble type, Figure~\ref{fig_global_boxes} shows boxplots of sight line fractions as a function of \Mstar\ (left) and Hubble type  (right).
Galaxies are divided into three groups according to their \Mstar\ or Hubble type.
The darker colors indicate increasing \Mstar\ or decreasing Hubble type value.
The median and mean sight line fractions for a given \Mstar\ or  Hubble type are given in Table~\ref{tab_sight line_vs_galprops}.

The left panel of Figure~\ref{fig_global_boxes} shows a tendency for more massive galaxies to have higher \co\ fractions.
The median   \co\ fractions increase from 14\% to 33\% and to 50\%  from our lowest to highest \Mstar\ bin. This dependency partially explains the higher median \co\ fractions in \citetalias{Sch19} because that sample is largely dominated by galaxies with $\log(\Mstar/\mathrm{M}_{\sun}) > 10.2$.
An opposite trend is exhibited for \ha\ fractions, with the median  fraction decreasing gradually from   61\% to 21\% and to 13\%. 
Moreover, the  two lower \Mstar\ bins reveal a  larger scatter in \ha\ fractions than for the  highest \Mstar\ bin, while the opposite trend is observed for \co\ sight lines.
We note that the median \ha\ fraction in our highest-\Mstar\ bin is lower than the median \ha\ fraction of galaxies with similar mass in \citetalias{Sch19} because  the \textsc{H\,ii}  regions in this work are generally smaller than that in \citetalias{Sch19}. This is driven by the different kernel sizes used in the unsharp masking technique to remove emission associated with the DIG.
The median  \overlap\ fractions remain at a nearly constant value as a function of \Mstar\  (27\% to 35\% and to 30\%), but the scatter in \overlap\ fraction  decreases with increasing \Mstar.

The trends with Hubble type and \Mstar\ are consistent in the sense that late type galaxies tend to be less massive (right panel of Figure~\ref{fig_global_boxes}).
The three Hubble type bins in the right panel of Figure~\ref{fig_global_boxes} roughly correspond to earlier types than Sab ($\mathrm{T} \leq 2$), around Sb--Sc (2 $<$ $\mathrm{T}$ $\leq$ 5), and later than Sc ($\mathrm{T} > 5$).
Unlike for \Mstar,  the \overlap\ fraction  shows an increasing trend toward the later-type galaxies. 
However,  the differences in the \overlap\ fraction between the different galaxy types are still significantly smaller than that for \co\ and \ha\ fractions, and the correlation coefficient (0.20) indicates a non-significant correlation.

We use partial rank correlation to examine whether the dependence of   \co\ and \ha\ fractions on Hubble type  is entirely due to the correlation with \Mstar\  or the other way around.
The partial rank correlation coefficient $r_{12,3}$ measures strength of the correlation between $x_{1}$ and $x_{2}$ when excluding the effect of $x_{3}$. 
 The partial rank correlation can be computed based on the Spearman rank correlation coefficient between the three variables as follows
 \begin{equation}
 r_{12,3}=\frac{r_{12}-r_{13}r_{23}}{\sqrt{(1-r_{13}^{2})(1-r_{23}^{2})}},
 \label{eq_partialcc}
 \end{equation}
where  $r_{ij}$ denoting the correlation between variables $i$ and $j$. 
Using the rank correlation coefficients in Table \ref{tab_sight line_cc_galprops} and Equation (\ref{eq_partialcc}), the partial rank correlations between \co\ and \ha\ with \Mstar\ become 0.35 and $-$0.32, respectively when Hubble type is controlled.
The correlations between  \co\ and \ha\ with Hubble type  are $-$0.45 and 0.21  while holding \Mstar.
 The  partial correlation coefficients between  these two sight line fractions with \Mstar\ and Hubble type are lower than that of the bivariate coefficients.
We therefore conclude that the correlations between  \co\ and \ha\ fractions and both \Mstar\ and Hubble  type are physical in nature, but the correlation between \Mstar\ and Hubble type may come between them.
Such dependencies of sight line fractions (\co\ and \ha)  on \Mstar\ and Hubble type have also been hinted at by the small (8) galaxies sample in \citetalias{Sch19}.

We also compute the correlation coefficients for the sight line fractions with other host-galaxy  and observational properties: galaxy distance,  optical size  ($R_{25}$), disk inclination, native resolution and effective sensitivity of the  H$\alpha$ ($\log (L_\mathrm{\textsc{H\,ii}\,region}^\mathrm{sensitivity}$ in Section \ref{sec_ha_filtering}) and CO (1$\sigma$ $\Sigma_\mathrm{H_{2}}$ at 150~pc resolution) observations, specific SFR (sSFR $=$ SFR/\Mstar), and offset from the star-forming main-sequence ($\Delta$MS) (Table~\ref{tab_sight line_cc_galprops}).
Scatter plots of the sight line fractions as a function of all the properties we explore in this section are shown in Appendix~\ref{sec_appendix_obs_impact}.
Galaxies with lower \Mstar\ are generally more nearby in our sample, caused by a potential sample-selection bias.
Therefore, the  dependence  of  sight line fractions on distance, sensitivity, and resolution might be a result of this selection effect.
In principle, sight line fractions could correlate with DIG fraction, in the sense that removing a higher fraction of H$\alpha$ flux would lead to a higher \co\ fraction and lower  \ha\ and \overlap\ sight lines.
Such a dependence is seen in terms of correlation coefficients, but only for  \co\ (0.31) and \overlap\ regions ($-$0.37).
The sight line fractions show no significant correlation with other galaxy and observational properties we explore.

 The \co\ fraction shows a correlation with  sSFR (-0.32).
 At the same time, sSFR is correlated with \Mstar, in the sense that along the star-forming main sequence, galaxies with higher \Mstar\ tend to have lower sSFR \citep{Bri04,Sal07}.
We checked the partial correlation of \co\ fraction and sSFR taking \Mstar\ as the control variable. 
The correlation between \co\ fraction and sSFR no longer exists ($-$0.17) when \Mstar\ is controlled for, but the correlation between  \co\ and \Mstar\ still holds while controlling for the effect of sSFR  (0.47). 
This suggests that the correlation with sSFR is an outcome of the dependence on \Mstar.
There is no correlation between the sight line types and $\Delta$MS, which we discuss further in Section~\ref{sec_discussion_sf}.

\begin{table}
	\centering
	\caption{Spearman correlation coefficients between sight line fractions at 150~pc resolution and global properties. The significant correlations, which we define as $\lvert \mathrm{coefficient} \rvert$ $\geq$ 0.3, are highlighted in bold-face. Scatter plots for each pair of variables are presented in Appendix~\ref{sec_appendix_obs_impact}.}
		\label{tab_sight line_cc_galprops}
	\begin{tabular}{lccc}
	\hline
	\	& \co\   &\ha\ & \overlap \\
	\hline
	galaxy properties  & & &\\
	\hline
	$M_*$ & {\bf 0.53} & {\bf $-$0.44} & $-$0.01 \\
	Hubble type & {\bf $-$0.59} & {\bf 0.38}& 0.24\\
	distance & {\bf 0.39} &  $-$0.28& $-$0.09\\
	$R_{25}$ & 0.08 &0.02 &$-$0.17\\
	inclination & $-$0.09 &0.15& $-$0.16\\
	DIG fraction & {\bf0.31} & $-$0.08& {\bf $-$0.37}\\
	\hline
     observations  & & &\\
     \hline
	effective H$\alpha$ sensitivity & {\bf 0.30}&$-$0.18& $-$0.14\\ 
	H$\alpha$ native resolution & {\bf 0.38} &$-$0.25 &$-$0.13\\
	effective CO sensitivity & 0.25 & {\bf $-$0.39} & {\bf 0.36}\\
	CO native resolution & 0.29& $-$0.25& 0.02\\
	\hline
	star formation  & & &\\
	\hline
	sSFR & {\bf $-$0.32} & 0.17 & 0.20 \\
	$\Delta$MS & $-$0.15& 0.02 &0.21\\
	\hline

\end{tabular}
\end{table}

\begin{table*}[]
	\centering
	\caption{Median (mean) sight line fractions at the 150~pc spatial scale for different stellar mass and Hubble type bins.  The median (mean) stellar mass for each Hubble type bin are provided in the bottom row.  }
	\label{tab_sight line_vs_galprops}
	\begin{tabular}{lccc}
		\hline
		& log(\Mstar/M$_{\sun}$) $\leq$ 9.8 & \;9.8 $<$ log(\Mstar/M$_{\sun}$) $\leq$ 10.3 & \;log(\Mstar/M$_{\sun}$) $>$ 10.3 \\
		\hline
		\co\ [\%]                                   & 14 (18)                                             & 33 (33)                                                            & 50 (50)                                                 \\
		\ha\ [\%]                              & 61 (50)                                             & 21 (33)                                                            & 13 (20)                                                 \\
		\overlap\ [\%]                               & 27 (32)                                             & 35 (33)                                                            & 30 (30)                                                 \\
		\hline                        
		& T $\leq$ 2                                       & 2 $<$ T $\leq$ 5                                          & T $>$ 5                                        \\
		\hline                        
		\co\ [\%]                                    & 70 (56)                                             & 41 (39)                                                            & 26 (26)                                                 \\
		
		\ha\ [\%]                                    & 7 (17)                                              & 25 (31)                                                            & 31 (38)                                                 \\
		\overlap\ [\%]                               & 22 (26)                                             & 31 (30)                                                            & 36 (35)                                                 \\
		\hline
		log(\Mstar/M$_{\sun}$) & 10.4 (10.3)                                         & 10.4 (10.4)                                                        & 9.8 (9.9)                      \\
		\hline                        
	\end{tabular}
\end{table*}

\subsubsection{Radial Distribution of CO and \texorpdfstring{H$\alpha$}{Halpha} Sight Lines}

\begin{figure*}
	\centering
	\includegraphics[scale=0.6]{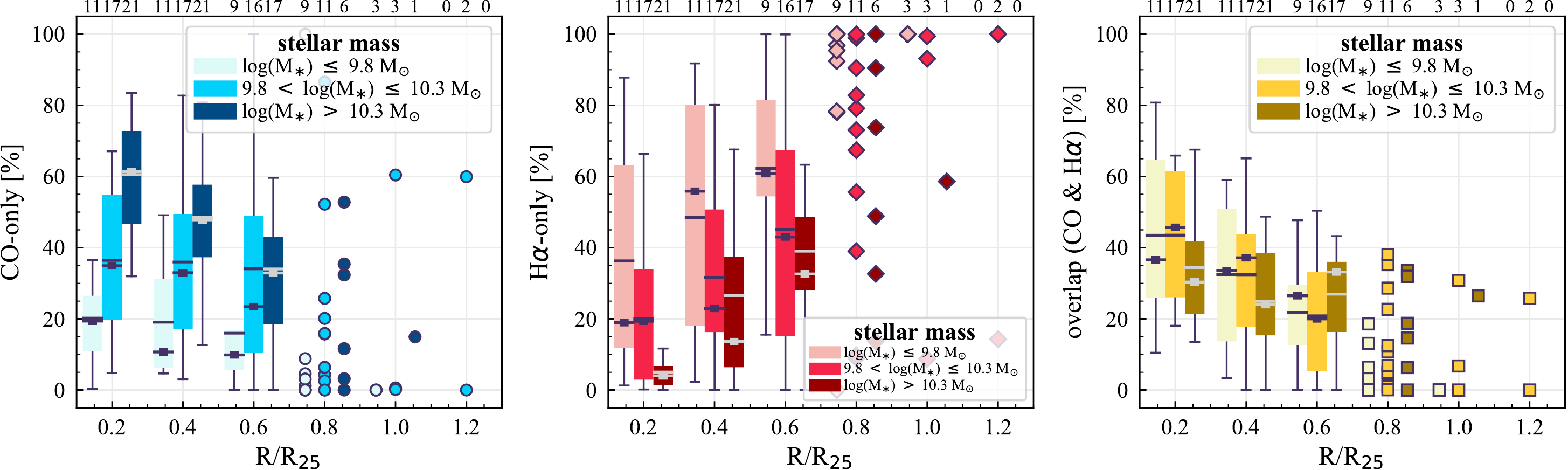}\\
	\vspace{10pt}
	\includegraphics[scale=0.6]{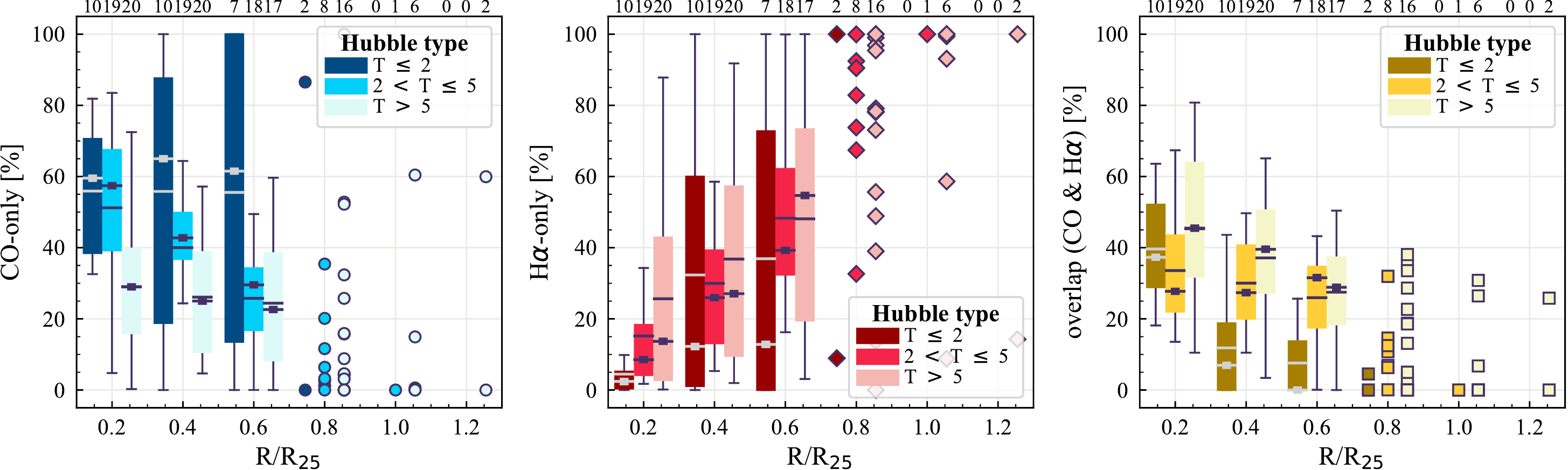}
	\caption{Radial profiles of \co\ (left), \ha\ (middle), and \overlap\ (right) sight lines for galaxies with different global properties at 150~pc resolution. Top row: Radial profiles of  sight line fractions  from  0.2 to 1.2 times $R_{25}$ stacked in bins of stellar mass. Line and color styles are the same as in Figure~\ref{fig_global_boxes}. The number of galaxies in each radial bin are shown in the top of the plots. Since the number of galaxies with maximum radius of $> 0.6R_{25}$  is low, we show the sight line fractions of each individual galaxy at these radii by symbols rather than boxplots. The color coding of the symbols is the same as for the boxplots at $R < 0.6 R_{25}$. Bottom row: Radial profiles of sight line fractions for galaxies in the three Hubble type bins. All  sight line categories show a strong radial dependence.   These trends observed for global sight line fractions  in Figure \ref{fig_global_boxes} are almost preserved radially from the center out to 0.6~$R_{25}$ (corresponding to $\sim$ 6~kpc on average). 
	}  
	\label{fig_radial_all}
\end{figure*}

We quantify the  radial trends of \co, \ha, and \overlap\ fractions (from left to right) in Figure~\ref{fig_radial_all}.
Here,  boxplots showing the galaxy distributions for each of the sight line fractions   are shown as a function of deprojected galactocentric radius normalized to $R_{25}$ in  annuli of width 0.2~$R_{25}$. 
For each galaxy, we only compute its radial sight line fractions out to the maximum radius of complete azimuthal \mbox{[0,\,2$\pi$]} coverage. 
For each radial bin, the light  to dark  boxplots represent the distributions for the lowest (later) to highest (earlier) bins of \Mstar\ (Hubble type). 
The three boxes at a given radius are offset by 0.05~$R_{25}$ on the plot for clarity.
The number of galaxies in each radial bin is indicated above each panel.
Some galaxies have maximum complete radius  up to 1.2~$R_{25}$. 
For reference, we show the sight line fractions of each individual galaxy at these radii using symbols rather than  boxplots. 
Note that the data points at $R$ $>$ 0.6~$R_{25}$ regime  are dominated by large spiral galaxies.
Due to the biased sample and low number statistics, data at $>$ 0.6~$R_{25}$ are not included in our discussion.
The color coding of each symbol is the same as for the boxplots at $R$ $<$ 0.6~$R_{25}$.

The sight line fractions show a strong radial dependence.
\co\ sight lines decrease with increasing radius and the fractions of \ha\ sight lines increase with radius.
The ordering between  sight line fractions and \Mstar\  is  observed in each radial bin at $R \lesssim 0.6~R_{25}$, suggesting that the dependence (or lack of dependence)  of the total sight line fractions on \Mstar\ and Hubble type in Figure~\ref{fig_global_boxes} is driven by the  local trends at all radii.
The median  \co\ fractions at $R \lesssim 0.6~R_{25}$ is at least doubled when moving from the lowest to the highest \Mstar\ bins.
The radial profiles of \ha\ sight lines also show a clear ranking with \Mstar\ at $R \lesssim 0.6~R_{25}$, increasing from the highest to the lowest \Mstar\ bins.
The differences between \Mstar\ bins are considerably smaller for \overlap\ regions,  but it can be seen that the radial profile of \overlap\ sight lines is  shallower for the highest \Mstar\ bin than the two lower \Mstar\ bins.
This is at least partially due to the fact that lower mass galaxies generally have lower \co\ fraction at larger radii than high mass galaxies; given that the \overlap\ regions appear to be embedded in the \co\ regions (Figure~\ref{fig_pie_example}),   the chance to have \overlap\ sight lines at large radii of lower mass galaxies is small.

The observed correlation between the sight line fractions and Hubble type in Figure~\ref{fig_global_boxes} is also seen in most of the radial bins, but the  rankings are not as obvious as for \Mstar, partially due to lower number statistics for the earliest bin.

\subsection{Trends with Galactic Structure}
\label{sec_global_structures}
Molecular  gas is preferentially formed or collected efficiently in galactic structures such as bars and spiral arms.
Since the distribution of molecular gas subsequently determines the potential sites of star formation, it is natural to expect that the distributions of CO and H$\alpha$ emission are also regulated by galactic structures.

\begin{figure*}
	\centering
	\includegraphics[scale=0.8]{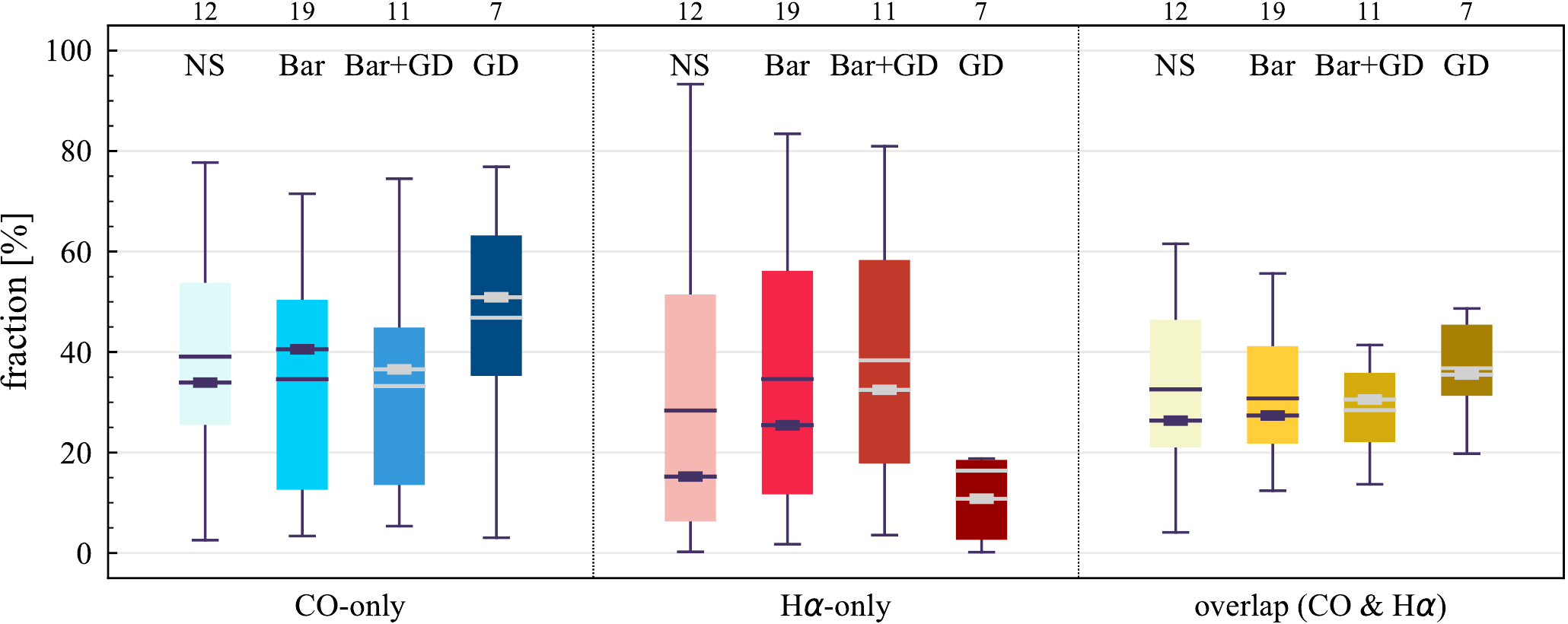}
	\caption{Sight line fractions at 150~pc resolution for galaxies without structures (bar or grand-design spiral arms; NS), galaxies with a bar but without grand-design spiral arms (Bar), galaxies with both a bar and grand-design spiral arms (Bar+GD), and galaxies with grand-design spiral arms but no bar (GD). GD  exhibits a distinct sign of higher \co\ and \overlap\ fractions and lower \ha\ fractions than the other populations. 
	}  
	\label{fig_structures}
\end{figure*}

We classify our target galaxies into four groups according to the presence and absence of bar and grand-design (GD) spiral arms: 
\begin{enumerate}
	\item \emph{no structures (NS):} galaxies without bar and GD spiral arms (e.g., galaxies with flocculent/multiple arms are in this category)
	\item \emph{Bar:} galaxies with a bar but no GD spiral arms
	\item \emph{Bar+GD:} galaxies with a bar and grand-design spiral arms
	\item \emph{GD:} galaxies with grand-design spiral arms but without a bar.
\end{enumerate}
The number of galaxy in groups (1) to (4) are 12, 19, 11, and 7, respectively.
The statistics of sight line fractions for each category at 150~pc resolution are provided in Table~\ref{tab_structures}; the corresponding boxplots  are shown in Figure~\ref{fig_structures}. 
While the sight line fractions for NS, Bar, and Bar+GD span over a similar range, GD galaxies exhibits a distinct sign of higher \co\ and \overlap\ fractions and lower \ha\ fractions than the other populations.  
Since GD galaxies have a lower median \Mstar\ (log(\Mstar/M$_{\sun}$) $=$ 10.3) than the Bar+GD galaxies (10.7) (Table \ref{tab_structures}), the differences in \co\ and \ha\ fractions between GD and Bar+GD are opposite to what one would expect if \Mstar\ is the dominant driver of the sight line fractions, and points to the potential importance of galactic structure on regulating the star formation process.

\begin{figure}
	\includegraphics[scale=0.48]{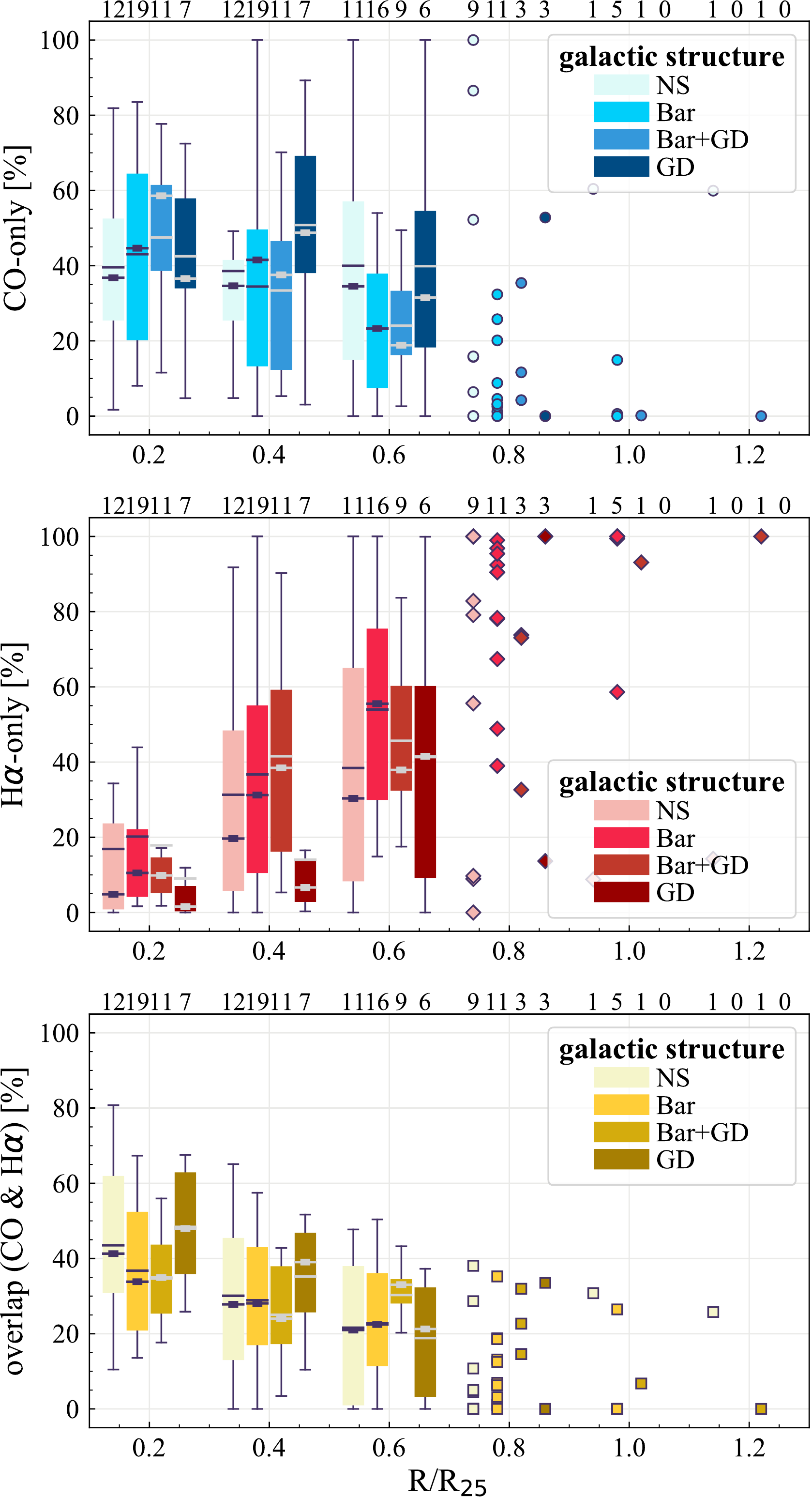}
	\caption{Different radial sight line profiles for galaxies with different galactic structures at 150~pc resolution.  The abbreviations are as follows: NS -- galaxies without structures (bar or grand-design spiral arms); Bar -- galaxies with a bar but without grand-design spiral arms; Bar+GD -- galaxies with both bar and grand-design spiral arms; and GD -- galaxies with grand-design spiral arms only. The plot style is analogous to Figure~\ref{fig_radial_all}. The figure implies  the importance of galactic dynamics in regulating star formation
	}  
	\label{fig_rad_structures}
\end{figure}

Figure~\ref{fig_rad_structures} presents the radial sight line fractions for each structure type at 150~pc resolution.
The bar length in our  galaxy sample ranges  from $\sim$ $0.1 {-} 0.9$~$R_{25}$,  with most bar lengths around $0.1 {-} 0.5$~$R_{25}$.
The median bar length of Bar+GD galaxies (0.3~$R_{25}$) is slightly longer than that of Bar galaxies (0.2~$R_{25}$).
Galaxies with a bar (Bar and Bar+GD) visually show stronger radial dependence of \co\ sight lines than galaxies without a bar (NS and GD),   in the sense that their median \co\ fraction   gradually decreases with increasing radius.
The opposite trend is observed for \overlap\ regions.
Moreover, Figure~\ref{fig_rad_structures} suggests  that the high total \co\ fraction in GD galaxies (Figure~\ref{fig_structures}) is due to the increased fraction at $\sim$ $0.4 {-} 0.6$~$R_{25}$.
On the other hand, the low total \ha\ fraction can be attributed to a lack of   \ha\  regions at $< 0.4~R_{25}$.

In summary, the results in this section show that, in addition to global galaxy properties,  galactic dynamics add a further layer of complexity to the distribution of CO and H$\alpha$ emission. 
We note that the FoV covering fraction of galactic structures could affect the sight line fractions. 
For example, while bars are fully covered by our FoV, we may miss the outer part of some GD spiral arms.
This analysis could be taken further by counting the sight lines within individual fully-sampled structures, but that is beyond the scope of this paper.
We refer the reader to \cite{Que21} for a comprehensive empirical characterisation of the molecular gas and star formation properties in different galactic environments of PHANGS galaxies.

\begin{table*}[]
	\centering
	\caption{Median (mean) sight line fractions and stellar mass  for galaxies with different structures at 150~pc resolution. The median (mean) stellar mass for galaxies with different structures are given in the bottom row.  Number of galaxy in NS, Bar, Bar+GD, and GD are 12, 19, 11, and 7, respectively.}
\label{tab_structures}
	\begin{tabular}{lcccc}
		\hline
		& no structure (NS) & bar only (Bar)    & bar and GD spiral arms (Bar+GD) & GD spiral arms only (GD) \\
				\hline
		\co                                              & 34 (39)       & 41 (35)     & 37 (33)     & 51 (47)       \\
		\ha                                              & 15 (28)       & 25 (35)     & 32 (38)     & 11 (16)       \\
		\overlap                                         & 26 (33)       & 27 (31)     & 31 (28)     & 36 (37)       \\
				\hline
		log(\Mstar/M$_{\sun}$) & 9.9 (9.9)     & 10.0 (10.0) & 10.7 (10.5) & 10.3 (10.2)   \\
		
		\hline
	\end{tabular}
\end{table*}

\subsection{CO and \texorpdfstring{H$\alpha$}{Halpha} Flux in Single-Tracer and Overlap Regions}
\label{sec_number_vs_flux}

In this section, we explore whether  there is any difference between regions where both tracers are observed (\overlap) and regions where  only one tracer is observed (\co\ and \ha).
We estimate the  fractional contribution of \co\ (i.e., only one tracer is observed) and \overlap\ (two tracers are observed;  \emph{CO-overlap}) to the  total  \cosightline, and the fractional contribution of \ha\ and \overlap\ (\emph{H$\alpha$-overlap})  to the total \hasightline.
In other words, the sum of \co\ and \emph{CO-overlap} is normalized to 100\%, and so is the sum of  \ha\ and  \emph{H$\alpha$-overlap}.
To compare with the results based on the number of sight lines (our default fraction), the corresponding fractions for \emph{flux}  are also estimated.

\begin{figure*}
	\begin{center}
		\subfigure[]{\label{fig_number_vs_flux}\includegraphics[width=0.48\textwidth]{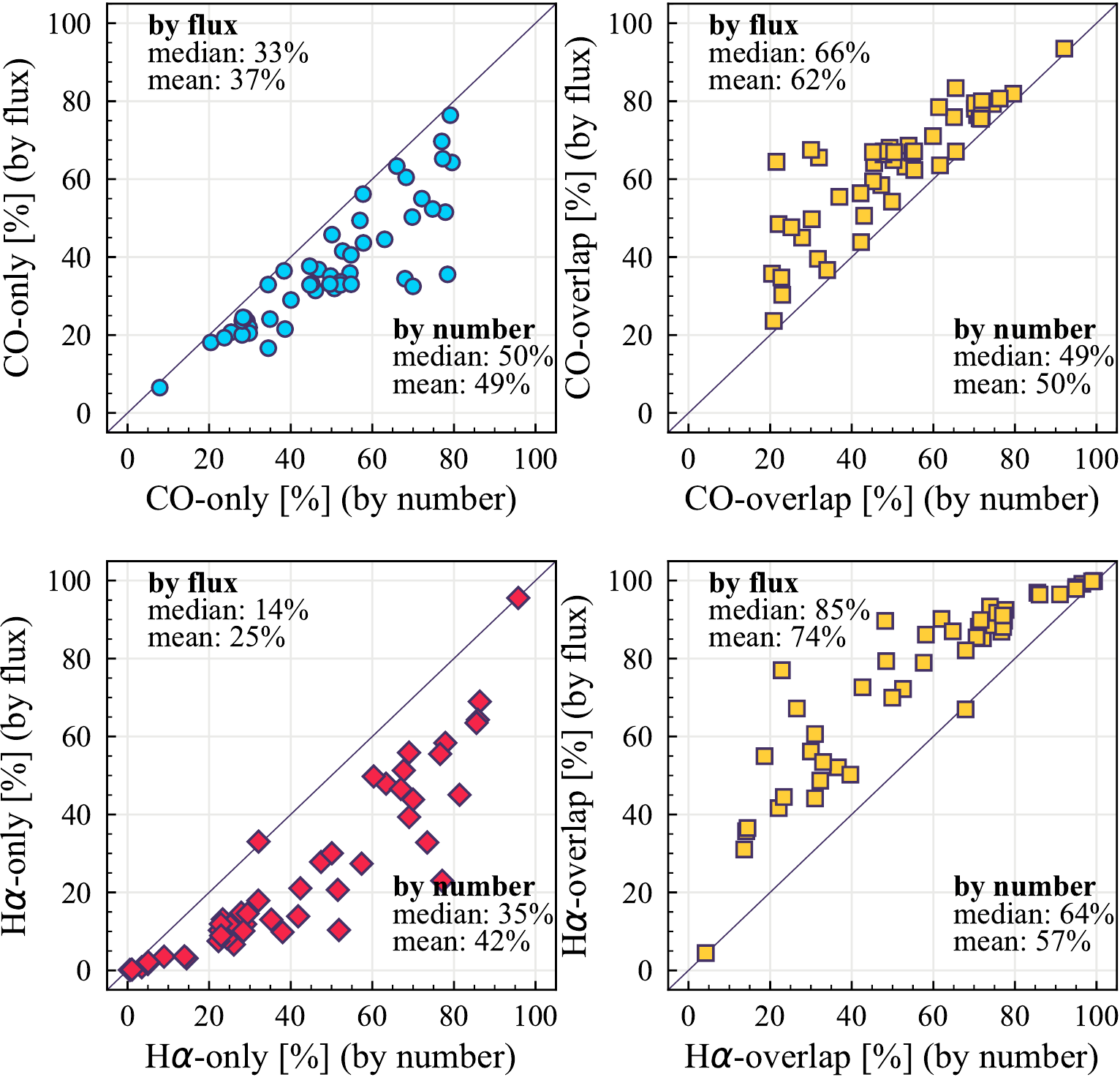}}
		\subfigure[]{\label{fig_number_vs_flux_nocenter}\includegraphics[width=0.48\textwidth]{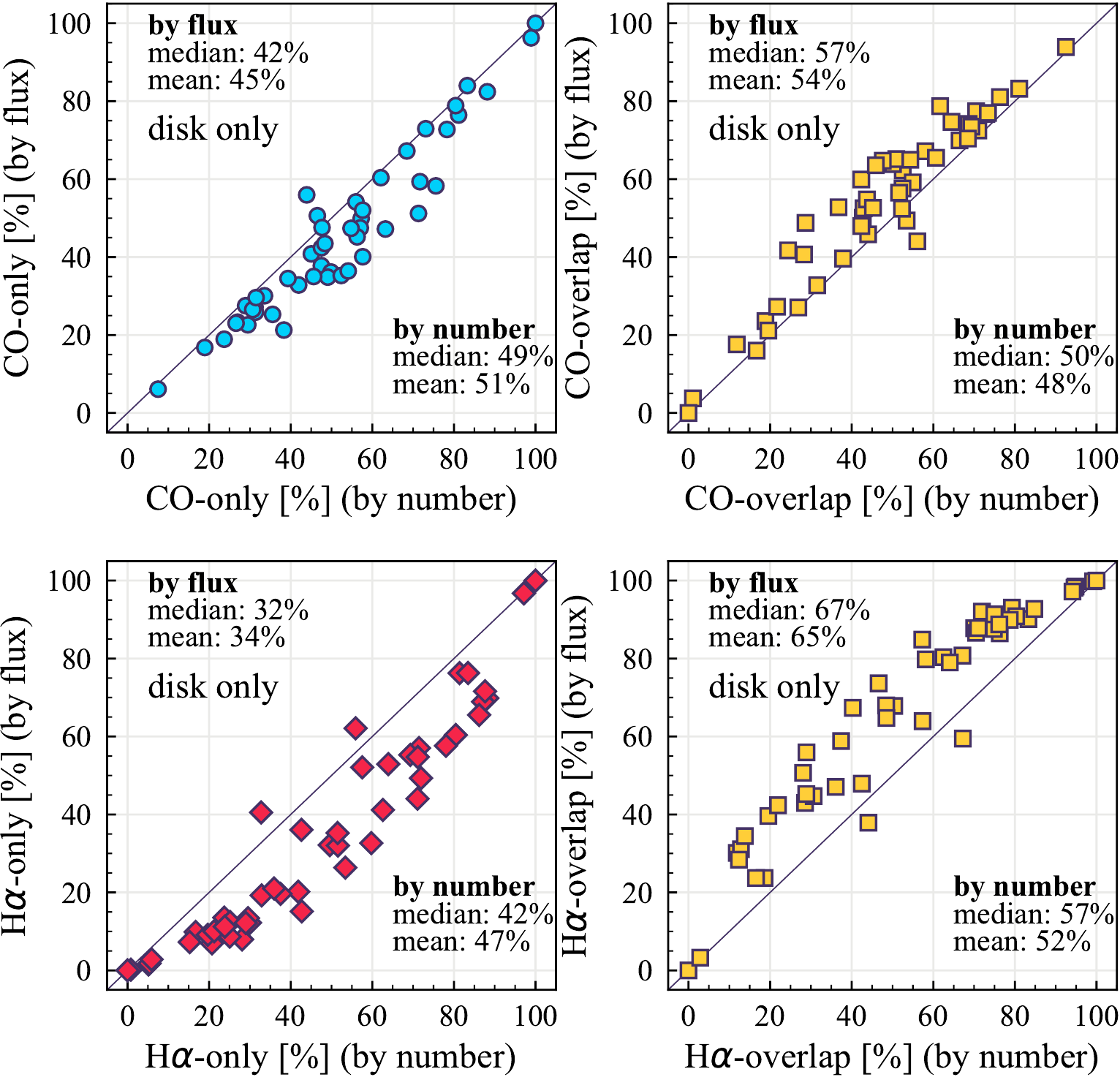}}
	\end{center}
	\caption{Comparison of  the fractions of sight line ($x$-axis) and flux ($y$-axis)   per tracer at 150~pc scale. 
	We estimate the fractional  contribution of \co\ (i.e., only one tracer is observed) and \overlap\ regions (two tracers are observed;  \emph{CO-overlap}) to the  total  number of sight lines with CO and total CO flux, and the fractional contribution of \ha\ and \overlap\ regions (\emph{H$\alpha$-overlap})  to the total H$\alpha$ sight line and flux.
	In other words, the sum of \co\ and \emph{CO-overlap} is normalized to 100, and so is the sum of  \ha\ and  \emph{H$\alpha$-overlap}. Specifically,  for each data point, the values on the $x$- and $y$-axis are calculated based on exactly the same pixels (sight lines), but the $x$-axis shows their fractional  contribution to all sight lines of the tracer and the $y$-axis shows their fractional  contribution to the  total flux of the tracer.	The solid line indicates the one-to-one correlation. Panels~(a) show the comparisons within the fiducial field of view of the 150~pc resolution images, and panels~(b) present the results excluding the  central 1~kpc (radius) regions. Overall, the fluxes  of CO and H$\alpha$ emission are higher in \overlap\ regions where emission from \emph{both} tracers are observed compared to regions where \emph{only one} tracer is observed. 
	}
	\label{fig_number_vs_flux_all}
\end{figure*}

Figure~\ref{fig_number_vs_flux} shows the comparison of sight line fractions with  flux fractions. 
Specifically, for each data point, the values on the $x$- and $y$-axis are calculated based on exactly the same pixels (sight lines), but the $x$-axis shows their fractional  contribution to the total sight line of the tracer and the $y$-axis shows their fractional  contribution to the  total flux of the tracer.
The black  line indicates a one-to-one correlation.

The median sight line fractions of  \co\ and  \emph{CO-overlap}  are approximately equal ($x$-axes in the upper panels of Figure~\ref{fig_number_vs_flux}), but the latter contributes  a larger portion to the overall flux (66\%; $y$-axis of upper-right panel).
Nonetheless,   \co\ regions still contribute one third of the CO flux  (33\%; $y$-axis of upper-left panel).
The difference between the fractions by number  of sight lines and by flux is  larger for  H$\alpha$, as shown in Figure~\ref{fig_number_vs_flux} (lower panels).
Taking all the pixels in our galaxies, the  \ha\ accounts for 36\% of the area of H$\alpha$-emitting regions, but  contributes only $\sim$ 14\% of the H$\alpha$ flux.
On the other hand,  \emph{H$\alpha$-overlap} contributes 85\% to the  total H$\alpha$ flux, and it is higher than the  64\%   sight line fraction.
Since the  \emph{H$\alpha$-overlap} regions are by definition co-spatial with CO-emitting molecular gas, they likely suffer from  dust attenuation that we do not account for in the processing of our H$\alpha$ maps due to the lack of extinction tracers (Section \ref{sec_ha_data}). 
Therefore, the true flux contribution of  \emph{H$\alpha$-overlap}  is  probably higher.

While covering only a small area,  galaxy centers  often substantially contribute  to the total flux \citep{Que21}. Since the flux in the central region of galaxies is not necessarily associated with star formation,  we therefore repeat the analysis while excluding the central 2~kpc in deprojected diameter (Figure~\ref{fig_number_vs_flux_nocenter}).
For  CO, the agreement between area and flux is much tighter; the deviation from the one-to-one line for high CO fractions almost vanishes, implying that galactic centers drive the difference.
On the other hand, H$\alpha$ fractions appear less dominated by the centers. This is at least partially due to higher extinction present in the centers.
The trend of higher flux in  \emph{H$\alpha$-overlap} regions than in  \ha\ regions persists even when excluding the centers.

\begin{figure*}
	\begin{center}
\includegraphics[width=0.8\textwidth]{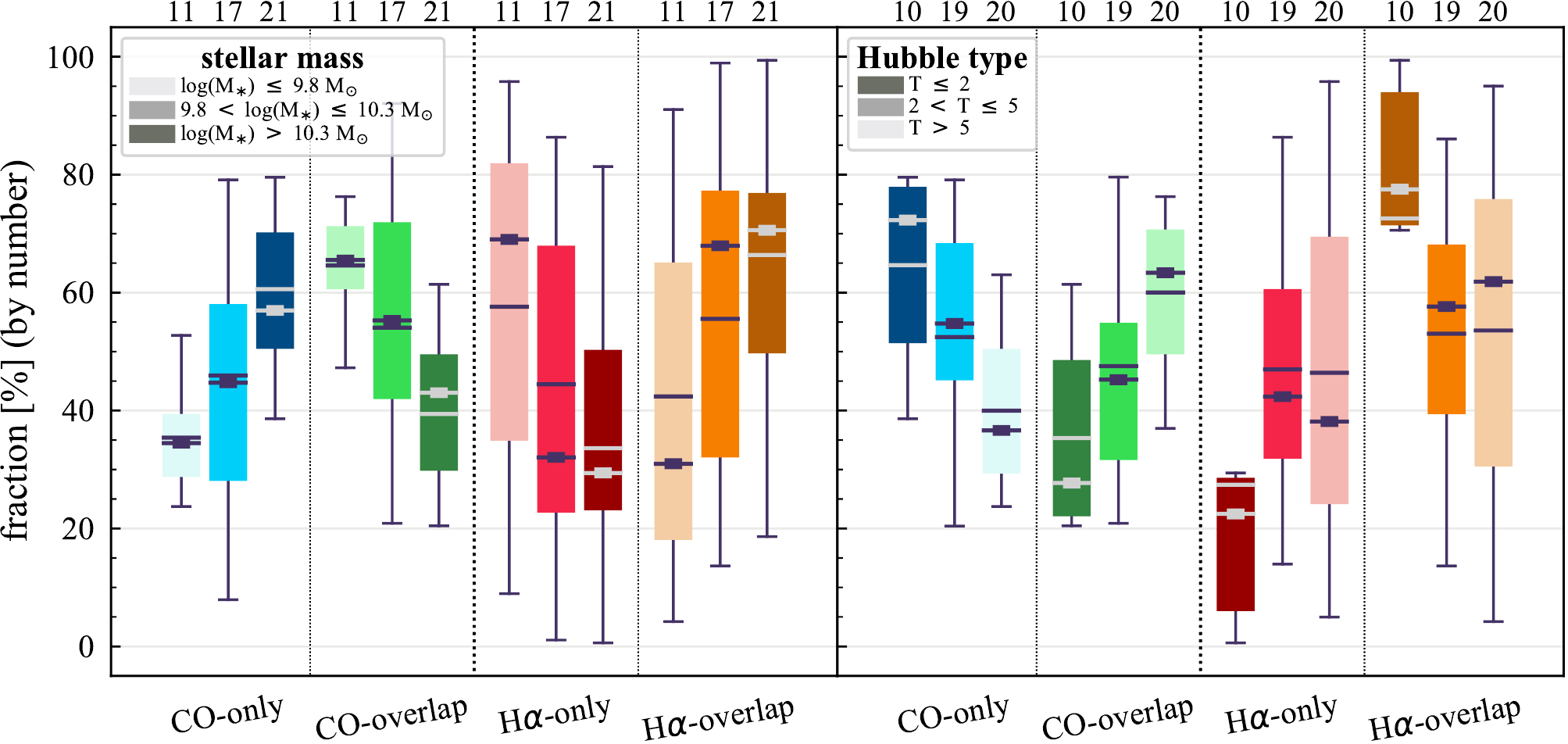}\\
\vspace{5pt}
\includegraphics[width=0.8\textwidth]{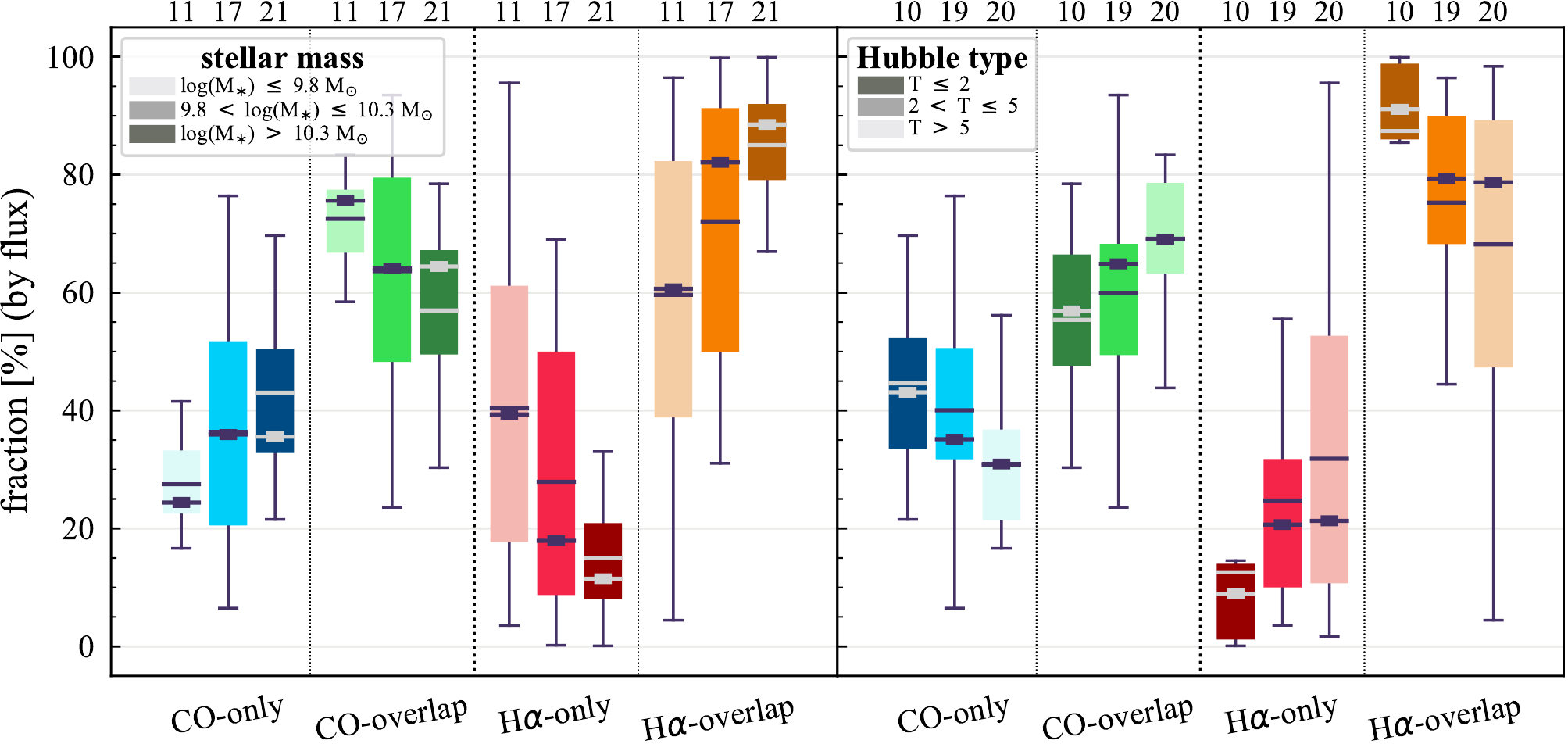}
	\end{center}
	\caption{Fractional contribution to the number of sight lines (top row) and flux (bottom row) per tracer for different  stellar mass bins (left column) and  Hubble types (right column) at 150~pc resolution. For each panel, CO and H$\alpha$ are shown on the left-hand and right-hand sides, respectively.  \co\,  \emph{CO-overlap},  \ha\, and  \emph{H$\alpha$-overlap} are shown in blue, green, red, and orange, respectively.   As in previous figures, the color darkness of boxes increases with increasing \Mstar\ and decreasing Hubble type value. For a given galaxy in a given \Mstar\ or Hubble type bin, the sum of values in the blue and green boxes (i.e., CO without and with H$\alpha$) is normalized to 100, and the sum in red and orange ones (i.e., H$\alpha$ without and with CO) is also normalized to 100.  It is true for all  \Mstar\ and Hubble type bins that \overlap\ regions contribute to a larger proportion of CO and H$\alpha$ flux than regions with only one type of emission.
	}
	\label{fig_global_coHa_sight lines}
\end{figure*}

Figure~\ref{fig_global_coHa_sight lines} compares the fractions by number  of sight lines (upper panel) and flux (lower panel) for galaxies in different \Mstar\ and Hubble type bins.
The distributions of \co\ and  \emph{CO-overlap} are shown by blue and green boxes, respectively, while   \ha\ and  \emph{H$\alpha$-overlap} are shown by red and orange boxes.
For each \Mstar\ or Hubble type bin (indicated by the darkness of the boxes),  the sum of a data point in the blue (red) box and the corresponding data point in the green (orange) box is normalized to 100.
The median and mean for each  \Mstar\ and Hubble type bin are summarized in Table~\ref{tab_COHa}.
The trends with \Mstar\ and Hubble type are the same for sight lines and  flux, but the difference among the \Mstar\ and Hubble type bins are sightly larger when considering  flux  instead of number of sight lines.
For all  \Mstar\ and Hubble type bins, the \overlap\ regions contribute to a larger proportion of CO and H$\alpha$ flux than regions with only one type of emission. 

Interestingly,   \co\ becomes dominant in the highest \Mstar\ galaxies, occupying $\sim$ 60\% of the CO-emitting regions. 
However, they are  almost as equivalently low in flux contribution as other populations, implying a generally (i.e., more extended distributed) low H$_{2}$ surface density in the highest \Mstar\ galaxies in our sample.
The same feature is seen for the earliest type galaxies, hinting that star-formation ceases to prevail over a significant area of a galaxy while gas remains there.
However, we cannot rule out that this result arises from our methodology.  Some \co gas in the low-mass galaxies may not pass our threshold due to its intrinsically low $\Sigma_\mathrm{H_{2}}$, while in higher mass galaxies, their \co gas is slightly brighter than our threshold. This would potentially add many CO-emitting sight lines, but very little flux. 
Such a possibility again highlights the  differences in molecular gas properties among galaxies with different \Mstar\ and Hubble type.

In summary,  at 150~pc spatial scale, the fluxes  of CO and H$\alpha$ emission are higher in \overlap\ regions where emission from \emph{both} tracers is observed compared to regions where \emph{only one} tracer is observed,  consistent with the finding of \citetalias{Sch19}.
This trend holds for galaxies with different \Mstar\ and Hubble type.
Nonetheless, the contribution from  regions with only one tracer (\co\ and \ha)  to the total flux remain substantial for most systems.

\begin{table*}[]
	\centering
	\caption{Median (mean) fractions of sight lines and flux per tracer for different stellar mass and Hubble type bins at 150~pc resolution.  Regular and bold fonts denote fraction of number of sight lines and flux, respectively. Note that for each individual galaxy,  the sum of \co\ and \emph{CO-overlap} is normalized to 100, and so is the sum of  \ha\ and  \emph{H$\alpha$-overlap}  (see the text for details).}
\label{tab_COHa}
	\begin{tabular}{lccc}
		\hline
		& log(\Mstar/M$_{\sun}$) $\leq$ 9.8 & \;9.8 $<$ log(\Mstar/M$_{\sun}$) $\leq$ 10.3 & \;log(\Mstar/M$_{\sun}$) $>$ 10.3 \\
		\hline
		& \multicolumn{3}{c}{sight line \% median (mean) {\bf flux \% median (mean)}}   \\
				\hline
		   \co\                    & 34 (35)  {\bf 24 (28)}                  & 45 (46) {\bf 36 (36)}                         & 57 (61)  {\bf 36 (43)}                                                 \\
		\emph{CO-overlap}                            & 66 (65)  {\bf 76 (72)}                   & 55 (54) {\bf 64 (64)}                        & 43 (39)  {\bf 64 (57)}                                                 \\
		 \ha\                       & 69 (58)  {\bf 39 (41)}                   & 32 (44)  {\bf 18 (28)}                        & 29 (34)  {\bf 11 (15)}                                                 \\
		 \emph{H$\alpha$-overlap}                            & 31 (42)  {\bf 61 (59)}                   & 68 (56)  {\bf 82 (72)}                         & 71 (66)  {\bf 89 (85)}                                                 \\
				\hline
		& T  $\leq$ 2                                       & 2 $<$ T   $\leq$ 5                                          & T  $>$ 5                                        \\
			\hline
	& \multicolumn{3}{c}{sight line \% median (mean) {\bf flux \% median (mean)}}   \\
				\hline
		\co\                       & 72 (65)  {\bf 43 (45)}                   & 55 (52)  {\bf 35 (40)}                         & 37 (40) {\bf 31 (31)}                                              \\
		\emph{CO-overlap}                            & 28 (35)  {\bf 57 (55)}                  & 45 (48)  {\bf 65 (60)}                        & 63 (60  {\bf 69 (69)}                                             \\
		 \ha\                       & 22 (27)  {\bf 09 (13)}                   & 42 (47)  {\bf 21 (25)}                        & 38 (46)  {\bf 21 (32)}                                                \\
		 \emph{H$\alpha$-overlap}                           & 78 (73)  {\bf 91 (87)}                    & 58 (53)  {\bf 79 (75)}                       & 62 (54)  {\bf 79 (68)}                                                \\
				\hline
	\end{tabular}
\end{table*}

\subsection{Distributions of CO and \texorpdfstring{H$\alpha$}{Halpha} as a Function of Spatial Scale}
\label{sec_global_resolution}
\begin{figure*}
	\centering
	\includegraphics[scale=0.57]{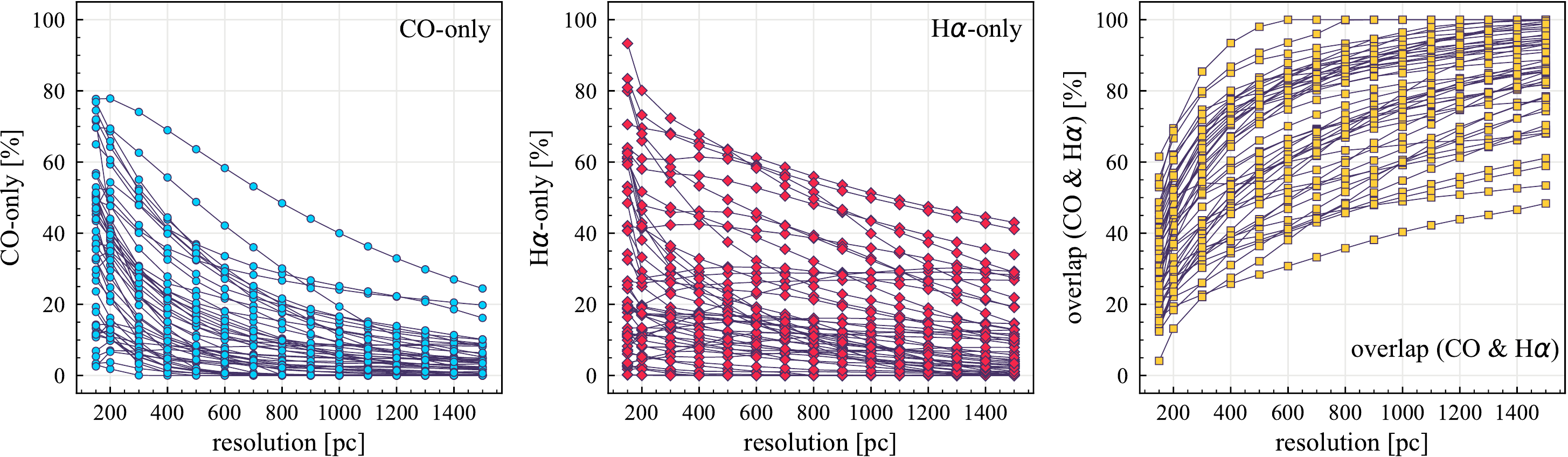}
	\caption{Fractions of sight lines as a function of spatial scale (observing resolution) from 150~pc to 1.5~kpc for each individual galaxy (i.e., one line per galaxy). From left to right, the panels show the variation of \co\ regions, \ha\ regions, and CO and H$\alpha$ \overlap\ regions, respectively.   The variations of \co\ and \overlap\ sight line fractions with spatial scale are rather uniform across the sample, while  the relation between \ha\ fractions and spatial scale  is more diverse.} 
	\label{fig_res_allgal}
\end{figure*}

We investigate the impact of spatial scale on the distributions of CO and H$\alpha$ emission.
Figure~\ref{fig_res_allgal} shows the sight line fractions for individual galaxies  as a function of spatial scale  from 150~pc to 1.5~kpc.
For most of the  galaxies, their \co\ sight lines decrease to $\lesssim$ 20\% at spatial scale $\gtrsim$ 800~pc, regardless of their \co\ fractions at spatial scale of 150~pc.
The \overlap\ regions  substantially increase and become the dominant sight lines when resolution is degraded.
This is the case for all galaxies in our sample, and the vertical ordering of \overlap\ fractions among the galaxies is almost maintained until 1.5~kpc resolution.
At the lowest resolution we consider, more than half of the regions are populated by both CO and H$\alpha$ emission in most galaxies.
While the variations of \co\ and \overlap\ sight line fractions with spatial scale are rather uniform across the sample,  the relation between \ha\ fractions and spatial scale  is more diverse.
Specifically,  galaxies with low \ha\ fractions at 150~pc scale  exhibit  a low, roughly constant fraction  toward large spatial scale  (lower resolution);   galaxies with the highest \ha\ fractional percentages at the best 150~pc scale decrease   rapidly toward low resolutions; and some galaxies show increasing  \ha\ fractions with decreasing resolutions.
For all sight line categories, the variations with spatial scale become less evident at $>$500~pc resolution. 
The  flattening point determines the critical resolution at which we stop resolving the CO and H$\alpha$ distributions.

\begin{figure*}
	\centering

	\includegraphics[scale=0.57]{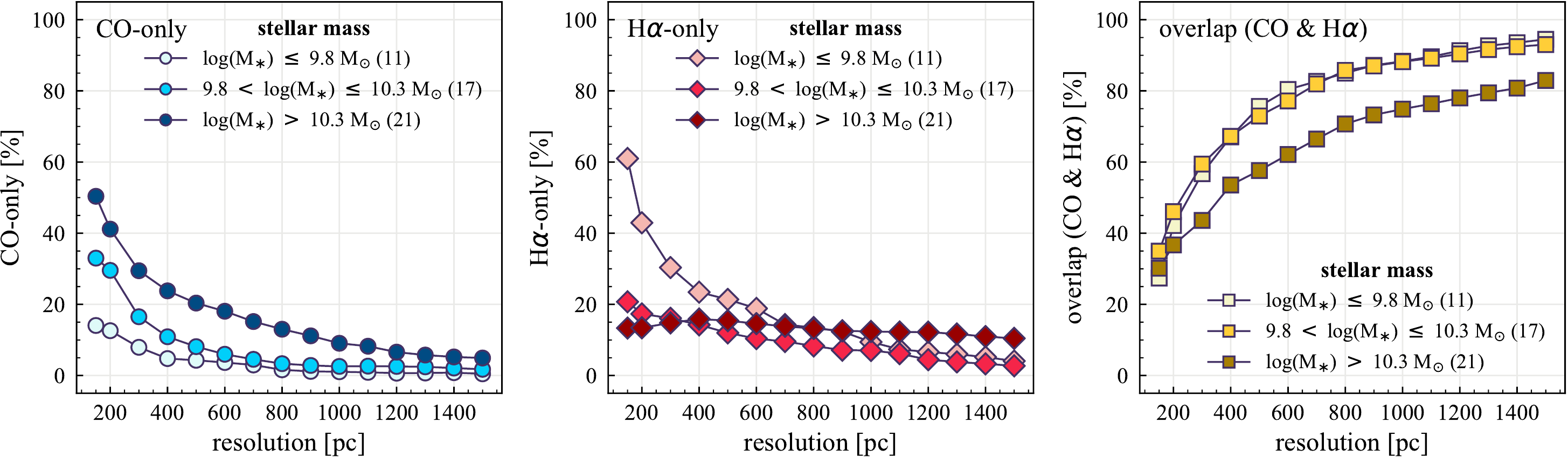}\\
	\vspace{5pt}
	\includegraphics[scale=0.57]{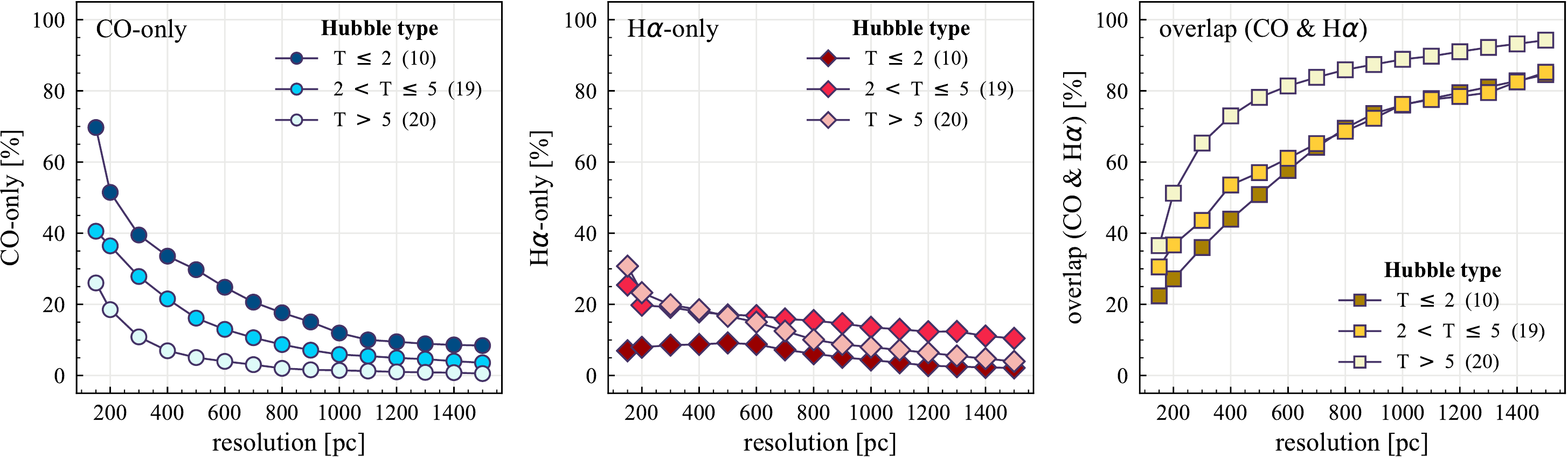}
	\caption{Fractions of sight lines as a function of spatial scale (observing resolution)  in different stellar mass (top row) and Hubble type  (bottom row) groups. From left to right, the three columns show the variation of \co\ regions, \ha\ regions, and CO and H$\alpha$ \overlap\ regions.  The figures demonstrate the important role that observing resolution can play when characterizing the distribution of CO and H$\alpha$ emission and their dependence on host galaxy properties.} 
	\label{fig_res_all}
\end{figure*}

Figures~\ref{fig_res_all}  sheds light on the nature of the different \ha\ versus spatial scale relation.
These figures are analogous to Figure~\ref{fig_res_allgal}, except that now galaxies are binned by their \Mstar\ and Hubble type.
The high global  \ha\  fractions at 150~pc scale, which tend to be relatively isolated toward the outer parts of  low-\Mstar\ and/or later-type galaxies, become quickly contaminated by other types of sight lines in the inner regions when the resolution is lowered. In other words, we see that the \co\ and \overlap\ regions increase in size and expand toward outer disks when the resolution decreases, e.g., NGC~2090, NGC~2835, and NGC~4951 in Figure \ref{fig_appendix_spatial_scale}), leading to a rapid decrease of the \ha\ fraction as a function of increasing spatial scale.
On the other hand, the \ha\ fraction of galaxies with  low-\ha\   are less sensitive to resolution.
They tend to be higher-\Mstar\ galaxies.
Their \ha\ sight lines populate both outer and/or inner disks (e.g., inter-arm regions, NGC~1300 and NGC~4321 in Figure \ref{fig_pie_example} and NGC~2997 and NGC~3627 in Figure \ref{fig_appendix_spatial_scale}).
Whether a galaxy's \ha\ fractions increases or decreases with  resolution  depends on the relative distribution of gas traced by CO and H$\alpha$.

By contrast, \co\ sight lines  vary relatively uniformly as a function of spatial scale among different galaxy populations.
The profiles show a clear ranking with \Mstar\ and Hubble type at spatial scale $<$ 500~pc.
At spatial scale $\gtrsim$ 500~pc, the dependence of  \co\ fraction on \Mstar\ and Hubble type becomes less pronounced.
While  we find no strong dependence of \overlap\ regions with \Mstar\ at  resolution of 150~pc (Figure~\ref{fig_global_boxes}),  Figure~\ref{fig_res_all} shows that galaxies in the highest \Mstar\ bin tend to have lower \overlap\ fractions  when the spatial scales are larger than $\sim$ 300~pc. 
On the other hand, the trend with Hubble type at 150~pc resolution  only holds when  the spatial scale is smaller than $\sim$ 500~pc.

In summary, the  results of this section demonstrate the important role that spatial scale can play when characterizing the distribution of CO and H$\alpha$ emission and their dependence on host galaxy properties.
The trend between sight line fractions and spatial scale was also observed in \citetalias{Sch19} for individual galaxies, here we further show 
 that the resolution dependence depends on galaxy type and the underlying high resolution CO and H$\alpha$ emission structure, indicating that there may be no simple (universal) prescription to infer the physical connection between gas and star formation from kpc-scale measurements. 

\section{Discussion}
\label{sec_discussion}

We have analyzed a sample of 49 resolution-matched CO and H$\alpha$ maps, which trace molecular gas and high-mass star formation, respectively.
At the best resolution we consider, 150~pc,  we find that the distributions of both CO and H$\alpha$ emission depend  on galaxy stellar mass and Hubble type (Section~\ref{sec_results_fractions}).
Specifically, the \co\ fractions increase with stellar mass and earlier Hubble type, while the converse is seen for  \ha\ fractions. The fraction of \overlap\ regions remains roughly constant with both quantities.

Galactic structures act as an additional factor controlling the distribution of the CO and H$\alpha$ emission (Section~\ref{sec_global_structures}).
GD galaxies exhibits a distinct sign of higher \co\ and \overlap\ fractions and lower \ha\ fractions than the other populations; galaxies with a bar (Bar and Bar+GD) visually show stronger radial dependence of \co\ sight lines than galaxies without a bar (NS and GD).

However, probing the dependence  of CO and H$\alpha$ distributions on galaxy properties requires observations with resolution high enough to distinguish between regions where only one tracer is observed and regions where both tracers are observed (Section~\ref{sec_global_resolution}).
Our results also show that, at 150~pc resolution, both CO and H$\alpha$ tend to have higher flux in regions where both CO and H$\alpha$ are found (\overlap), than in regions  where only a single tracer (\co\ and \ha) can be found (Section~\ref{sec_number_vs_flux}).

\subsection{\co\ Sight Lines}
\label{sec_discussion_quiescent_gas}

We find that galaxies in our sample contain a substantial reservoir of  \co\ molecular gas not associated with optical tracers of high-mass star formation (or above SFR surface densities of $\sim$ 10$^{-3}$ -- 10$^{-2}$ M$_{\sun}$~yr$^{-1}$~kpc$^{-2}$ depending on the galaxy target).
Our result is qualitatively consistent with studies of Local Group galaxies.
In these galaxies (the Small and Large Magellanic Clouds (SMC and LMC) and M33), about 20--50\% of GMCs are not associated  with \textsc{H\,ii} regions or young clusters\footnote{Note that one should not compare the fraction of non-star-forming ``GMCs'' in the Local Group galaxies with our ``sight line'' fractions directly due to the different counting methods, i.e.,  object-based or pixel-based approaches. Direct comparison is only possible if we assume that  GMCs have a fixed size, which is unlikely to be true \citep[e.g.,][]{Hug13,Col14,Ros21}.} \citep[e.g.,][]{Miz01,Eng03,Kaw09,Gra12,Cor17}.
Our results  further reveal that these starless clouds are not restricted to lower mass spiral and irregular galaxies, as in the Local Group, but are  observed across the whole range of the galaxy population.

\emph{Non-star-forming gas:} The sensitivity of PHANGS-ALMA is able to detect GMCs with  mass of $\gtrsim$ 10$^{5}$ M$_{\odot}$; moreover,  the \co\ sight lines are found at all surface densities from the adopted threshold to a few thousands of  M$_{\odot}$ pc$^{-2}$.
Massive star formation is certainly  expected to proceed in these relatively high-mass and high density regions. 
This implies that  part of the \co\ gas consists of non-star-forming clouds; the gas is unable to form stars because of its intrinsic properties. 
For example, molecular gas in some \co\ regions might be a diffuse, dynamically hot component  \citep{Pet13} that is not prone to star formation, or may be analogous to the gas  in the centers of early-type (elliptical) galaxies that seems not to be forming stars \citep{Cro11,Dav14}. Nonetheless, we note that  although the non-star-forming gas does not currently participate in the local on-going star-formation cycle, it may participate in star formation at some point in the future, i.e. made possible by relocating to a different, favorable site in the galactic potential that prompts a change in its dynamical state and/or organization, for example.


\emph{Low-mass star formation:} It is possible that  high-mass star formation is suppressed in the \co\ regions,  forming stars that are not massive enough to produce detectable  H$\alpha$ emission. Such molecular clouds have been found in the Large Magellanic  Cloud \citep{Ind08}.  

\emph{Embedded star formation:} Massive stars may be formed in part of the \co\ gas, but their H$\alpha$ emission is obscured by dust. However, the  embedded phase is relatively short, lasting only for a few to several Myr  \citep{Kim21}, and therefore may not account for  all \co\ regions.

\emph{Pre- and/or post-star formation:} The \co\ gas might be in the process of collapsing or may be remnant molecular gas dispersed from previous star-forming sites by stellar feedback  (e.g., photoionization, stellar winds, and supernova explosions).

Distinguishing these scenarios requires the analysis of multi-wavelength data, such as line widths and surface densities of molecular clouds, dense gas tracers, better tracers of obscured star formation (e.g.\ infrared emission), and extinction tracers, but such an analysis is beyond the scope of this paper. 
Future James Webb Space Telescope (JWST) observations  will also provide crucial insight into the complex processes of star formation and the nature of our \co\ sight lines.

In Section \ref{sec_global_resolution}, we saw that the observed sight line fractions depend on spatial scale. 
We repeated our analysis for a subsample of 17 galaxies for which our observations achieve a common 90~pc resolution to test whether the fraction of \co\ sight lines in galaxies is larger at even higher physical resolution.
A~CO threshold of 13~M$_{\sun}$~pc$^{-2}$ is adopted, corresponding to the 3$\sigma$ of the lowest sensitivity of these galaxies at 90~pc resolution.
The \co\ fractions in all galaxies show an increase by $\sim$14\% (median) as the resolution improves from 150 to 90~pc, while the  fraction of  \ha\ and \overlap\ regions for the 90~pc maps decreases by $\sim$  4\% and  8\%. This suggests that  there remains a non-negligible fraction of  \co\ gas that is not well resolved at our fiducial scale of 150~pc. 
 If we increased the resolution even more, e.g., to 10~pc, we might expect to find even more \co\ sight lines, but testing this will require higher resolution ALMA observations.
At some point, such observations will highly resolve individual clouds or other star-forming structures, and we might even detect that individual regions within a molecular cloud remain  quiescent (e.g., genuinely non-star-forming or pre-star-forming) while stars already form elsewhere. 
This is not yet the case for our data, however.

\subsection{Effect of Galactic Dynamics}
\label{sec_discussion_structures}
Both bar and grand-design spiral arms are known to stabilize the gas against collapse and thus star formation under certain circumstances  \citep{Rey98,Zur04,Ver07,Mei13}.
However, while we indeed observe a higher fraction of \co\ sight lines in grand-design spiral galaxies (Figure~\ref{fig_structures}), the  \co\ fractions of Bar and Bar+GD are  comparable to NS galaxies.
It is probably because we do not consider bar strength in this work which  is known to be correlated with SFR and star formation history of galaxies \citep[][]{Mar97,Car16,Kim17}.
 An alternative explanation could be that the gas distribution in barred galaxies is evolving heavily over time \citep[e.g.,][]{Don19}, leading to a wide variety in the gas distribution in the barred galaxies  seen in PHANGS \citep{Ler21b}. 

Nonetheless, our results show a possible trend for galaxies with bars (Bar and Bar+GD) to exhibit a stronger radial dependence in the fraction of \co\ sight lines (Figure \ref{fig_rad_structures}).
This may be attributed to bar-driven gas inflows which increase the gas concentrations  in  the  central  regions   \citep[][]{Sak99,She05,Sun20b}.
Moreover, the Bar and Bar+GD galaxies show a weaker radial dependence of \overlap\ fraction than the NS and GD galaxies in terms of median values.
Star-forming  complexes are often observed  at bar ends.
Although bar footprints are not necessarily forming stars, the star-forming bar ends  may smooth the profiles of the \overlap\ sight lines \citep[][]{Jam09,Beu17,Dia20}.
Finally, we note that we did not control for other trends (e.g., \Mstar) when  comparing sight line fractions between galaxies with different structures.
A very large sample  is required in order to  distinguish the effects of  global galaxy properties and galactic dynamics.
 

Some galaxies show a pronounced  offset between the different sigh line types with a sequence of \co\ to \overlap\ and to \ha\ when going from up- to downstream (assuming the spiral arms are trailing, e.g., NGC~4321 in Figure \ref{fig_pie_example} and NGC~0628, NGC~1566, NGC~2997 in Figure \ref{fig_appendix_spatial_scale}), consistent with expectations for a  spiral density wave \citep[see Figure 1 of][]{Pou16}.
These offsets are almost exclusively found in well-defined grand-design spiral arms and presumably lead in turn to  the high (or even highest) \co\ fraction in the disk (0.4 and 0.6 $R_{25}$)  of GD in Figure \ref{fig_rad_structures}, suggesting that   most of GD structures may indeed be density waves.
This demonstrates the potential of  the sight line method as a diagnostic of the relationship between ISM condition and galactic dynamics.
Detailed analysis of individual galaxies would be necessary to confirm the (dynamical) nature of our  grand-design spiral arms.

Besides, the offsets between molecular gas and star formation tracers also  allow measurement of the angular rotation velocity of a spiral pattern and the timescale for star formation \citep[e.g.,][]{Egu04,Egu09,Lou13}.
While such analyses have been restricted to small-sample studies in the past, PHANGS  allows for a systematic exploration of the spatial offset between the gas spiral arms and star-forming regions.
Some  barred galaxies also exhibit such \co\ and \overlap\ offsets along their spiral arms, e.g. NGC~4321 and NGC~3627 in Figure \ref{fig_pie_example} and  NGC~1365  in Figure \ref{fig_appendix_spatial_scale}, suggesting a   dynamical link between spiral arms and stellar bar \citep{Mei09,Hil20}.

\subsection{Sight Line Fractions and Star Formation}
\label{sec_discussion_sf}

\begin{figure*}
	\centering
	\includegraphics[scale=0.57]{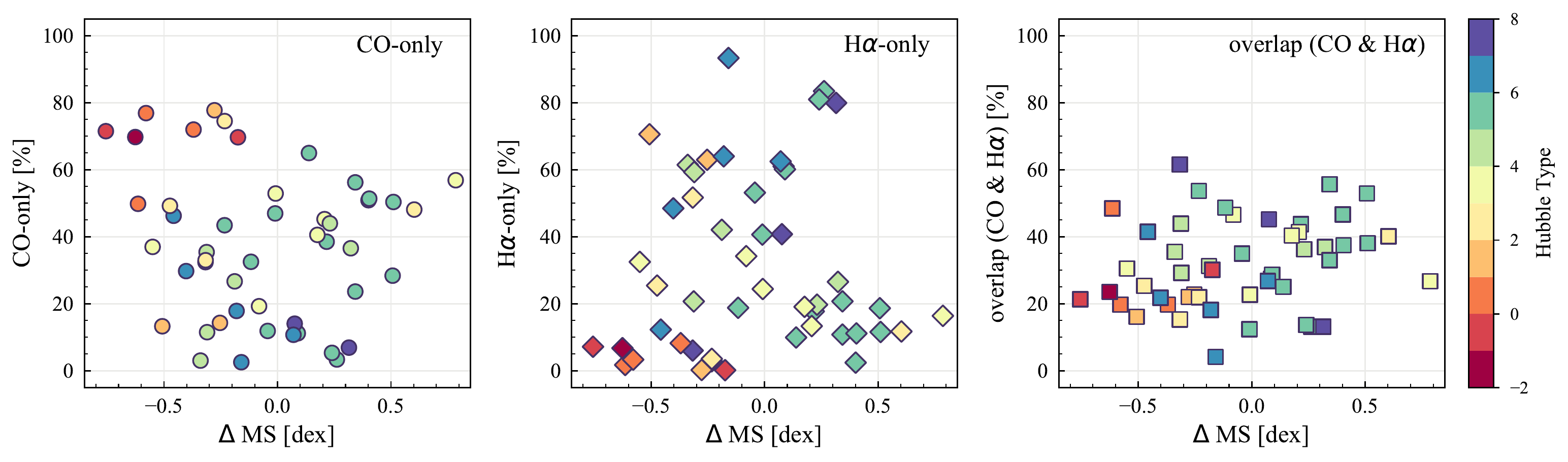}
	\caption{Fractions of sight lines at 150~pc resolution versus  $\Delta$MS, color-coded by Hubble type. From left to right, the three panels show the results for \co\ regions,\ha\ regions, and CO and H$\alpha$ \overlap\ regions, respectively.  We find no correlation between the global sight line fractions with $\Delta$MS.  Correlation coefficients of each sight line category  relative to $\Delta$MS are given in Table~\ref{tab_sight line_cc_galprops}.} 
	\label{fig_sight line_vs_deltams}
\end{figure*}

We find no  correlation between sight line fractions and star formation properties  (Section \ref{sec_global_galprops}) and no correlation with  the fractions of flux contributed by \co, \ha, \emph{CO-overlap}, and \emph{H$\alpha$-overlap} regions (correlation coefficients $\lesssim$ 0.2).
Figure~\ref{fig_sight line_vs_deltams} shows the sight line fractions against $\Delta$MS, color-coded by Hubble type.
Although there is no statistical relationship between the sight line fractions and $\Delta$MS,  galaxies with low $\Delta$MS in our sample, $\lesssim$ $-0.58$~dex or $\sim$ 4 times below the main sequence (NGC~1317, NGC~3626, NGC~4457, and NGC~4694), tend to have  high \co\ sight line fractions ($\sim$ 50--80\%).
These low-$\Delta$MS galaxies are all earlier types with Hubble type $\mathrm{T} \leq 1$ (S0--Sa). 
The spatial distribution of their \co\ regions are relatively compact inner disks, analogues to the molecular gas in elliptical galaxies \citep[e.g., ][]{Cro11,Dav14}.
By contrast, among the six  highest-$\Delta$MS  galaxies ($>$ 0.4 dex or 2.5 times above the main sequence)  in our sample, four show relatively high \co\ sight line fractions ($\gtrsim$ 50\%; NGC~1365, NGC~1559, NGC~4254, and NGC~5643).
All these high-\co\ and high-$\Delta$MS galaxies have grand-design spiral arms and/or a bar; moreover,  their \co\ sight lines follow well these galactic structures, implying a dynamic origin of the high \co\ fractions.
Although both galaxies with the highest- and lowest-$\Delta$MS in our sample show  substantial CO-emitting regions not associated with star formation, the spatial distribution of their \co\ regions is markedly different, potentially pointing to different underlying causes for the suppressed star formation in these regions.

The other two highest-$\Delta$MS galaxies (NGC~1385 and NGC~1511) have  lower, but not necessarily low, \co\ fractions (28 and 48\%).
Both of them happen to be peculiar systems.
In fact, four out of the six highest-$\Delta$MS galaxies  (NGC~1365, NGC~1385, NGC~1511, and NGC~4254)   show signs of interactions with other galaxies. 

In our working sample, 33 galaxies show  signs of interactions in terms of their morphology.
The median \overlap\ sight line fraction of the merger candidates (28\%) is slightly lower than that of isolated galaxies (36\%).
However, the \overlap\ sight lines in the merger candidates  contribute  a  higher fraction flux (90\%) to the total H$\alpha$ flux than for the apparently undisturbed galaxies (73\%).
Taking these numbers at face value, a given unit of star-forming region (\overlap)  in merger candidates contribute more significantly to the total SFR of a galaxy than a given unit of star-forming region in  undisturbed galaxies, \emph{assuming that all the H$\alpha$ emission is powered by star formation.}
We should note that galaxy interactions  may trigger central AGN \citep[e.g.,][]{Ell19} and shocks prevailing over the disks, which could contribute to H$\alpha$ emission.
However, we are not able to cleanly separate  \textsc{H\,ii} regions from other H$\alpha$-emitting sources when using only narrowband data.
Spectroscopic observations are necessary to confirm the differences between merger candidates and undisturbed galaxies.

\subsection{\texorpdfstring{H$\alpha$}{Halpha} Sight Lines at Large Galactocentric Radii}
\label{sec_discussion_ism}

The \ha\  sight lines are preferentially found at large galactocentric radii and even become dominant at $R$ $>$ 0.4~$R_{25}$ ($\geq$ 60\%) in low-\Mstar\ galaxies.
Moreover, the fraction of \ha\ sight lines is always higher for low-\Mstar\ galaxies than higher-\Mstar\ galaxies at all radii  (Figure~\ref{fig_radial_all}).
The lack of \cosightline\ at large radii may be due to  (1) the lack of gas and/or (2) the existence of low-$\Sigma_\mathrm{H_{2}}$  gas that drops below our applied $\Sigma_\mathrm{H_{2}}$ threshold (Section \ref{sec_co_threshold}).

To gain insight into the extent  of cold gas reservoirs in our galaxies, we compute the radial profiles of gas fractions ($f_\mathrm{gas}$) and molecular-to-atomic gas mass ratio ($R_\mathrm{mol}$) for galaxies that have spatially-resolved measurements of atomic gas and stellar mass surface densities ($\Sigma_\mathrm{HI}$  and $\Sigma_{\ast}$).
$f_\mathrm{gas}$  is defined as the ratio of the  total  gas mass ($\Sigma_\mathrm{H_{2}}$ +$\Sigma_\mathrm{HI}$)  to $\Sigma_{\ast}$, while $R_\mathrm{mol}$ is defined as the ratio between $\Sigma_\mathrm{H_{2}}$ and $\Sigma_\mathrm{HI}$.
The spatially-resolved HI data are taken from various sources in the literature, including the PHANGS-VLA (D.~Utomo et al. in prep.), VLA THINGS \citep{Wal08} and VIVA \citep{Chu09} surveys, and VLA archive.
The spatially-resolved $\Sigma_{\ast}$ are measured  based on Spitzer IRAC 3.6$\mu$m  or WISE 3.4$\mu$m \citep{Ler19,Ler21b}.
The typical resolution of $\Sigma_\mathrm{HI}$ and $\Sigma_{\ast}$ measurements is 1 -- 2 kpc. 
Since we are interested in the general trend of the $f_\mathrm{gas}$ and $R_\mathrm{mol}$ distributions, high spatial resolution is not needed for this purpose.
In total, for 28 and 32 galaxies we can compute their radial $f_\mathrm{gas}$ and $R_\mathrm{mol}$, respectively.
For this analysis, we rely on radial measurements at matched kpc  resolution  from  the  PHANGS  multi-wavelength  database presented in \cite{Sun20b} and J.~Sun et al. (in prep.).

\begin{figure*}
	\centering
	\includegraphics[scale=0.76]{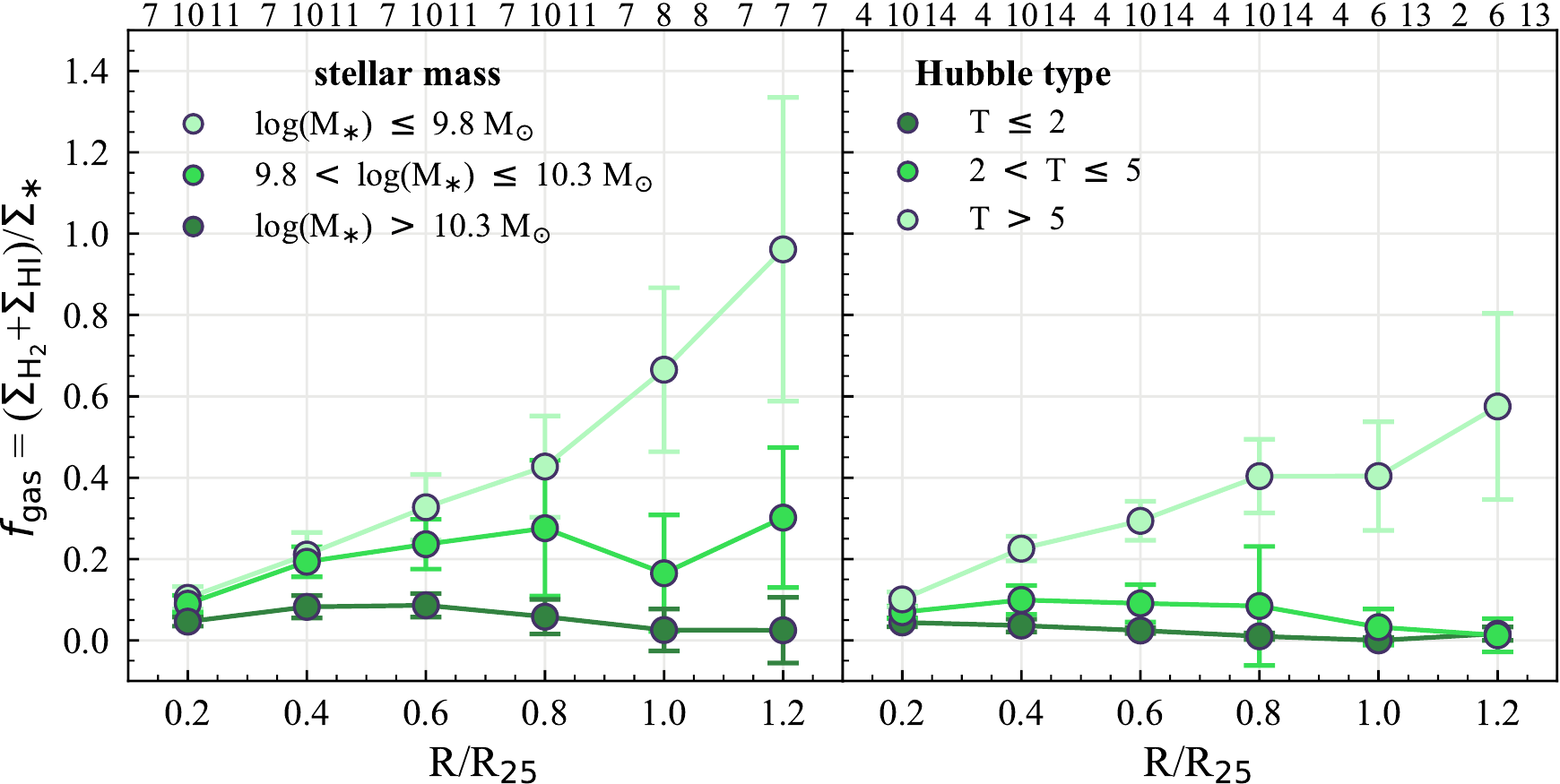}\\
	\vspace{10pt}
	\includegraphics[scale=0.76]{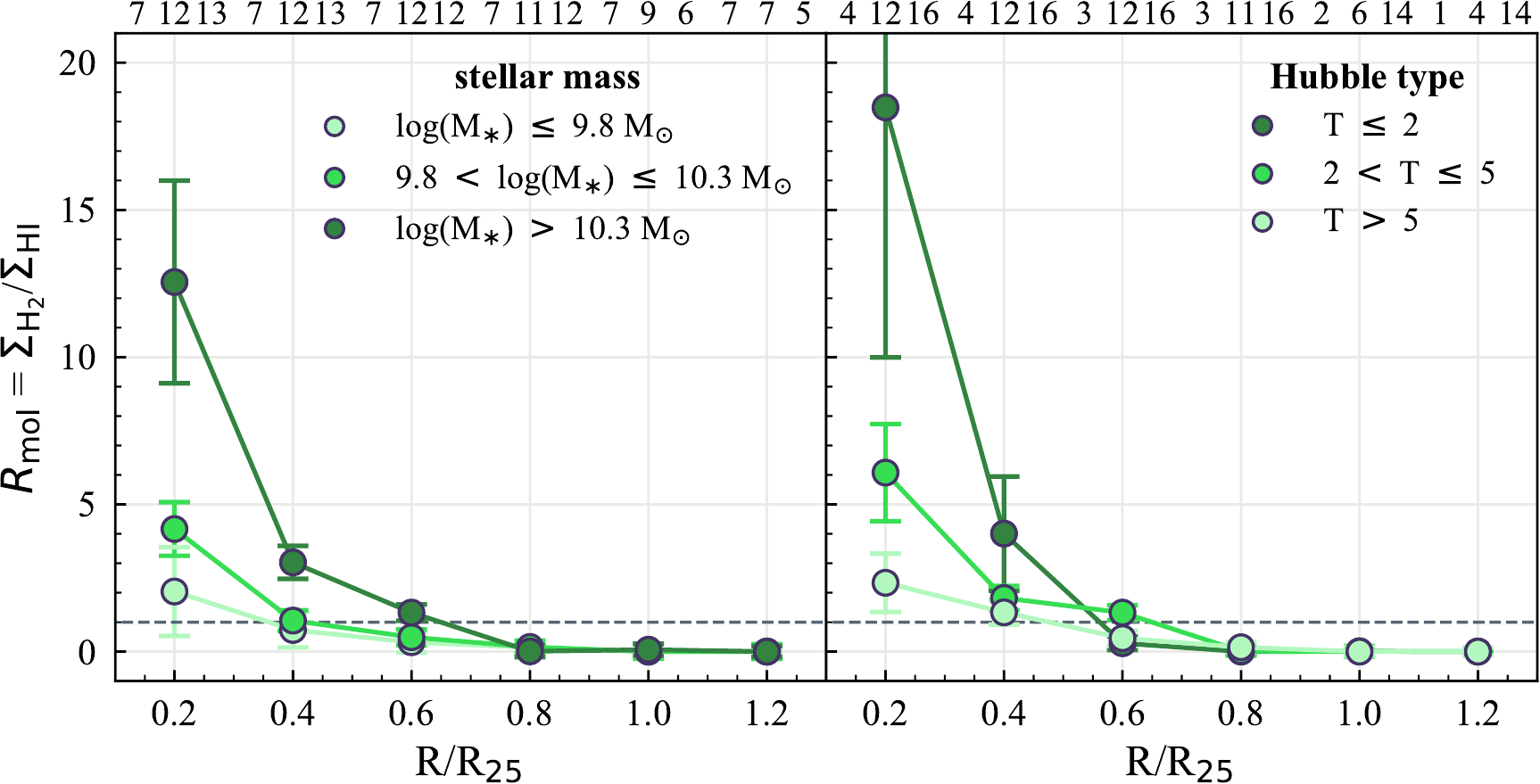}
	\caption{Radial ISM properties for different bins of stellar mass (left) and Hubble type (right). Top and bottom rows show the radial gas fraction  ($f_\mathrm{gas}$) and molecular-to-atomic gas mass ratio ($R_\mathrm{mol}$), respectively. The dashed lines in the bottom panels indicate $R_\mathrm{mol}$ $=$ 1.0. The symbol darkness is proportional to \Mstar\ or Hubble type. The error bars represent the error of the mean. The figure is created for the sub-sample of galaxies that have spatially-resoled \textsc{H\,i} and stellar mass measurements.  At $R > 0.4~R_{25}$ regime  where \ha\ regions dominate the sight lines in lower mass galaxies,  there is still a significant reservoir of gas  with respect to stellar mass ($f_\mathrm{gas}$ $\gtrsim$ 0.2), but the  gas   is  predominantly atomic ($R_\mathrm{mol}$ $<$ 1).} 
	\label{fig_radial_gas_props}
\end{figure*}

Figure~\ref{fig_radial_gas_props} shows the radial  $f_\mathrm{gas}$ (upper row) and $R_\mathrm{mol}$ (lower row) for galaxies with different \Mstar\ (left) and Hubble type (right). 
Our sample shows a gradual decrease of $f_\mathrm{gas}$ with increasing \Mstar\ at all radii.
Moreover, $R_\mathrm{mol}$ increases with  \Mstar\ (when looking at $R$ $\lesssim$ 0.6~$R_{25}$).
The trends with Hubble types are consistent in the sense that later type galaxies are less massive.
The results in  Figure~\ref{fig_radial_gas_props} are in good agreement with \cite{Sai11,Sai16} based on integrated measurements for a large sample of galaxies

Furthermore, Figure~\ref{fig_radial_gas_props} shows  that the    high-\ha\ regime  ($R > 0.4 R_{25}$) of low mass galaxies still harbors a significant reservoir of gas  with respect to stellar mass ($f_\mathrm{gas}$ $\gtrsim$ 0.2), but the  gas   is  predominantly atomic ($R_\mathrm{mol}$ $<$ 1).
Therefore, it is likely that there are molecular clouds in the outer atomic-dominated, high-\ha\ regions, but their $\Sigma_\mathrm{H_{2}}$ is low and below our applied threshold.

For galaxies with log(\Mstar/M$_{\odot}$) $>$ 10.3, around 70--98\% of the total CO emission (both median and mean are $\sim$ 90\%) are included in  our analysis of sight line fractions (i.e., $\Sigma_\mathrm{H_{2}}$ $>$ 10 M$_{\sun}$~pc$^{-2}$), while the  fraction of CO emission above our applied threshold decreases to $\sim$ 40--90\% (both median and mean are  $\sim$ 65\%) for  log(\Mstar/M$_{\odot}$) $<$ 10.3. 
We also observe a stronger variation in the \ha\ fractions for low-\Mstar\ galaxies when lowering CO threshold while keeping H$\alpha$ threshold fixed.
These   imply a prevalence of lower mass molecular clouds ($\sim 10^4 {-} 10^5$~M$_{\sun}$) in lower-mass galaxies and  significant galaxy-to-galaxy variations in their molecular cloud properties \citep[e.g.,][]{Hug13,Schruba19,Sun20a,Sun20b,Ros21}.
The galaxy-to-galaxy variations in cloud properties  also leads to the  correlations, albeit relatively weak, between the sight line fractions and CO effective sensitivity in Table \ref{tab_sight line_cc_galprops}.
Moreover, the  gas depletion time of molecular gas ($\tau_\mathrm{dep}$) is found to anti-correlate with galaxy \Mstar, with lower-\Mstar\ galaxies showing  shorter $\tau_\mathrm{dep}$ even after accounting for the metallicity dependence of $\alpha_\mathrm{CO}$ \citep{Uto18}.
The $\tau_\mathrm{dep}$-\Mstar\ relation also causes substantial H$\alpha$ emission not associated with molecular gas in low-\Mstar\ galaxies.

Finally, dissociation is presumably more efficient in  low-$\Sigma_\mathrm{H_{2}}$ environments  due to less dust shielding \citep[e.g.,][]{Wol10}.
Therefore,  CO emission is preferentially seen in high-extinction regions (e.g., inner galactic disks).
The need for high extinction to form  CO may lead to the consequence that the radial profiles of CO   are steeper than that of H$\alpha$ \citep[e.g.,][]{Ler08}.
Since $\Sigma_\mathrm{H_{2}}$  decreases rapidly with increasing galactocentric radii, the choice of  $\Sigma_\mathrm{H_{2}}$ threshold has a significant impact on the \overlap\ fractions  at large galactocentric radii, especially for low-\Mstar\ galaxies, whose $\Sigma_\mathrm{H_{2}}$ are systematically lower \citep[e.g.,][]{Hug13,Sun20b}.

\subsection{Relative Timescale of the Gas-Star Formation Cycle}
\label{sec_molgas_lifetime}
 If we assume that all  the \co\ sight lines are pre-star-forming clouds, the three categories of regions we define roughly sample the evolutionary sequence of star-forming regions, in the sense that a  cold molecular gas cloud (\co) evolves into a star-forming molecular cloud (\overlap), and then to a region of (exposed) massive young stars (\ha) where the molecular gas has been largely dispersed and/or dissociated. 
In this scenario, the areal fractions that we measure are an approximate reflection of the time spent by a star-forming region in each of these different evolutionary phases.
Similar frameworks   (i.e., counting the number of GMCs and  \textsc{H\,ii} regions)  have been applied previously to  estimate the   duration for which the molecular cloud traced by CO emission is visible  in the LMC  \citep{Kaw09}, constrain the cloud life-cycle in M33  \citep{Gra12,Cor17}, and determine the timescales for dense molecular clumps to evolve from being starless to star-forming in the Milky Way \citep{Bat17}.

The evolution of molecular clouds provides constraints on the mechanisms triggering or halting star formation at a specific location in a galaxy.
In \citetalias{Sch19}, we apply a simple version of the approach  to estimate the  duration that molecular gas traced by CO emission  is visible ($t_{\mathrm{gas}}$):
\begin{equation}
	t_{\mathrm{gas}}=t_{\mathrm{H\alpha}}\times \frac{f_{\textrm{CO-only}}+f_{\mathrm{overlap}}}{f_{\textrm{H$\alpha$-only}}+f_{\mathrm{overlap}}}=t_{\mathrm{H\alpha}}\times f_{\mathrm{scale}},
	\label{equ_fscale}
\end{equation}
where $t_{\mathrm{H\alpha}}$ represents the duration that H$\alpha$ emission from \textsc{H\,ii} regions is visible. $f_{\textrm{CO-only}}$, $f_{\textrm{H$\alpha$-only}}$, and $f_{\mathrm{overlap}}$ denote the fraction of \co, \ha, and \overlap\ sight lines, respectively.
Then $f_{\mathrm{scale}}$ represents the scaling factor to translate from the fiducial timescale, here $t_{\mathrm{H\alpha}}$, to $t_{\mathrm{gas}}$\footnote{We note that the cloud visibility time $t_{\mathrm{gas}}$ is  different from the dynamic timescale or the depletion time. A comparison of various relevant timescales, such as cloud visibility time,  free fall time, crossing time, and the  characteristic timescale for star formation regulated by galactic dynamical processes has been discussed in \cite{Che20} for  a subset of  PHANGS galaxies. A further discussion will also be presented in the upcoming paper by J.~Sun et al. (in prep.).}.
 We emphasize that, by applying Equation~\eqref{equ_fscale}, we are implicitly assuming that our individual pixels are discrete star-forming units and all the \co\ sight lines contain pre-star-forming clouds (see Section \ref{sec_discussion_quiescent_gas}).

\begin{figure}
	\centering
	\includegraphics[scale=0.85]{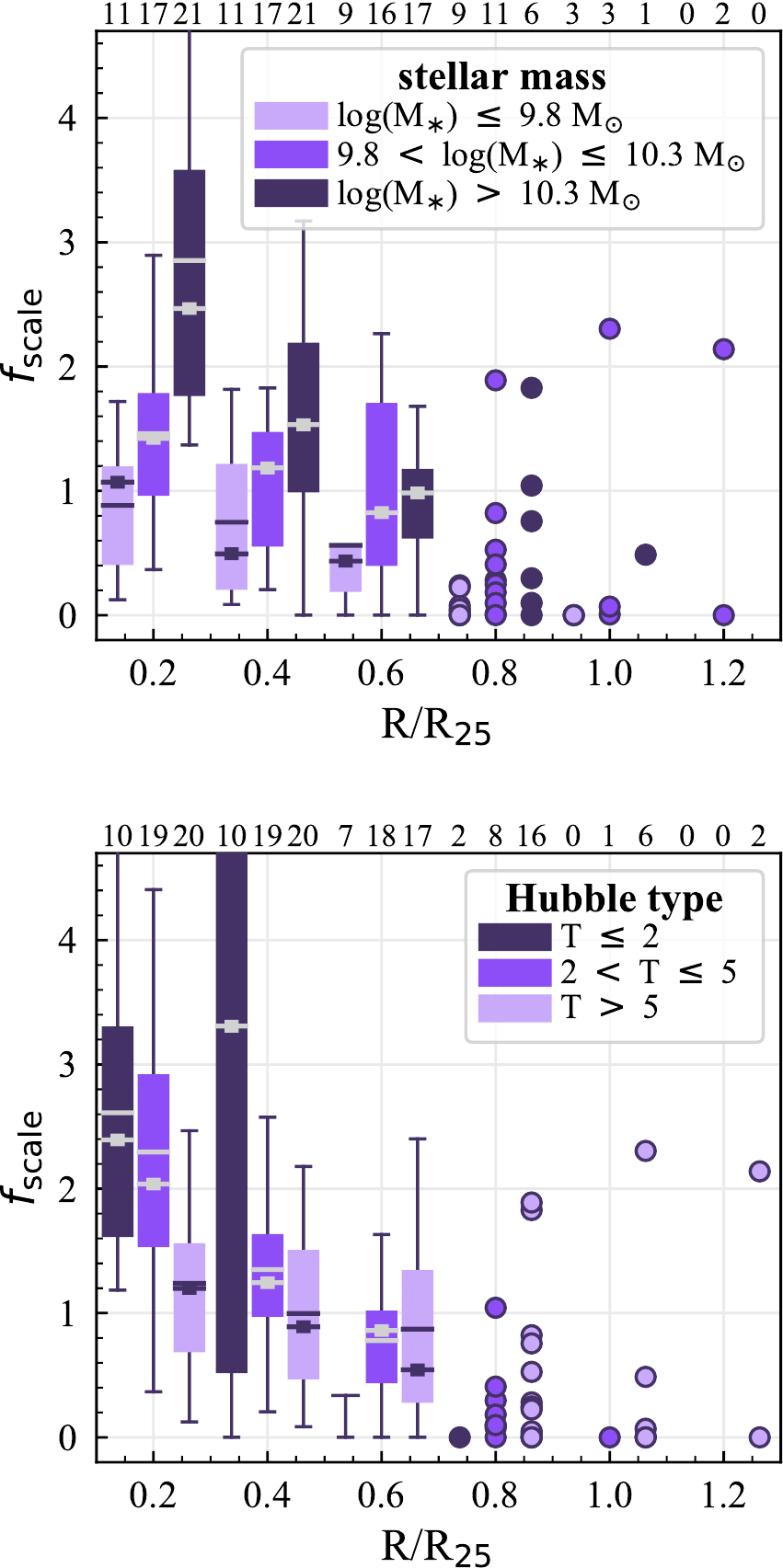}
	\caption{Radial distribution of $f_\mathrm{scale}$ for different bins of \Mstar\ (upper panel) and Hubble type (lower panel) at 150~pc resolution. $f_{\mathrm{scale}}$ represents,  to first order,  the scaling factor to translate from the fiducial timescale  $t_{\mathrm{H\alpha}}$ (the visibility time of H$\alpha$ emission of \textsc{H\,ii} regions; $\sim$ 5 -- 10 Myr) to the lifetime for a cold gas structure (see the main text for details). The plot style is the same as in  Figure~\ref{fig_radial_all}. There is a ranking of cloud lifetime or $f_{\mathrm{scale}}$ along \Mstar, but the trend with Hubble type is not as strong as for \Mstar.} 
	\label{fig_fscale}
\end{figure}

Figure~\ref{fig_fscale} shows the  radial trend in $f_{\mathrm{scale}}$ averaged in bins of \Mstar\ (upper panel) and Hubble type (lower panel)\footnote{$f_{\mathrm{scale}}$ has to be calculated in radial bins or in any  region-by-region manner,  otherwise $f_{\mathrm{scale}}$  is heavily determined by the  bright CO in the inner part and bright H$\alpha$ in the outer part, but those regions do  not form part of the same evolutionary cycle.}.
There is a ranking of  $f_{\mathrm{scale}}$ along {\Mstar} where molecular clouds in high-\Mstar\ galaxies tend to have  longer $f_{\mathrm{scale}}$  than clouds in low-\Mstar\ over the radial range probed.
Given that typical estimates for  $t_{\mathrm{H\alpha}}$ are 5--10 Myr \citep[e.g.,][]{Ler12,Ken12,Hay20}, the  $t_{\mathrm{gas}}$ of our sample at 150~pc scale, therefore, decreases from $\sim$ 5 --25 Myr for the high-\Mstar\ galaxies, to $\sim$ 5 --15 Myr for the intermediate-\Mstar\ galaxies, and to a few  Myr for the low-\Mstar\ galaxies in our sample \emph{when applying a CO threshold of 10~M$_{\sun}$~pc$^{-2}$}.
$f_{\mathrm{scale}}$ also decreases from earlier- to later-type spiral galaxies, but the trend is not as strong as for \Mstar.

\begin{figure*}
	\centering
	\includegraphics[scale=0.7]{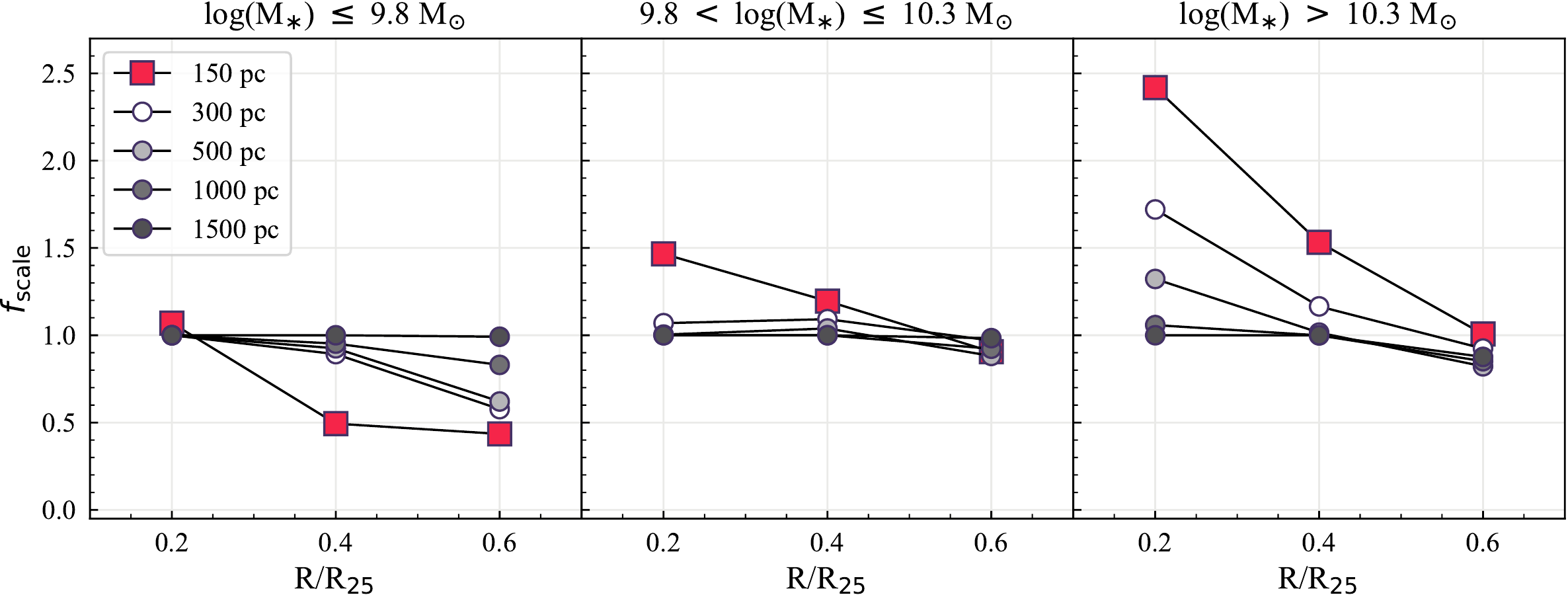}
	\caption{Median radial $f_\mathrm{scale}$ ($\propto$ cloud lifetime, at least to first order) for different \Mstar\ as a function of spatial scale (150, 300, 500, 1000, and 1500 pc).  In each panel, the red square  represents  $f_\mathrm{scale}$  at our best-matching 150~pc resolution. $f_\mathrm{scale}$  at other spatial scales is shown by circles, where increasing color darkness resembles increasing spatial scale.  At  spatial scale  $\leq$ 300~pc, there is a visible difference in  $f_{\mathrm{scale}}$  between galaxy populations, but the  differences between galaxies become considerably smaller at spatial scale of $>$ 300~pc  due to the significant decrease in \co\ sight lines and increase in \overlap\ sight lines.} 
	\label{fig_fscale_res}
\end{figure*}

The dependence of  $f_{\mathrm{scale}}$  on host galaxy properties suggests  the potential importance of environment for regulating star formation (see also \citetalias{Sch19} and \citealt{Che20}).
However,   the evolution of star-forming regions is only visible when the spatial scale is close enough to the typical region separation length between molecular clouds and \textsc{H\,ii} regions.
Figure~\ref{fig_fscale_res} shows the dependence of radial $f_{\mathrm{scale}}$  on spatial scale as a function of \Mstar.
Light to dark gray circles denote  $f_{\mathrm{scale}}$ at different  spatial scales of  300, 500, 1000, and 1500~pc, while $f_{\mathrm{scale}}$  at  150~pc resolution is indicated by red squares. 
At  spatial scales  $\leq$ 300~pc, there is a distinct difference in  $f_{\mathrm{scale}}$  between galaxy populations, as parameterized by \Mstar.
The  differences between galaxies become considerably smaller at spatial scale of $>$ 300~pc  due to the significant decrease in \co\ sight lines and increase in \overlap\ sight lines, indicating  a critical resolution requirement to resolve  the evolution of individual star-forming regions.
The critical spatial scale of  300~pc is consistent with the characteristic separation length between independent clouds or star-forming regions reported by \cite{Che20}.

While there is a clear dependence on spatial scale,  the derived $t_{\mathrm{gas}}$ at 150~pc resolution are in reasonable agreement  within a factor of a few (two to three) in most radial bins  with the  cloud \emph{lifetime} ($t_{\mathrm{GMC}}$) during which CO emission is visible estimated by \cite{Che20} for seven of our galaxies  using the  statistical method developed by  \cite{Kru18}. 
We present a direct comparison of  the radial variation of our  $t_{\mathrm{gas}}$ with $t_{\mathrm{GMC}}$ from \cite{Che20} in Appendix~\ref{sec_appendix_comparison_Che20}.
The $t_{\mathrm{gas}}$ measured at 150~pc resolution also agrees well with the estimates of   cloud (CO) visibility time based on  counting the number of GMCs with and without  \textsc{H\,ii} regions \cite[e.g.,][]{Kaw09}.

Finally, there are several systematic differences in cloud properties and uncertainties to bear in mind when using Equation (\ref{equ_fscale}) to estimate the visibility time of molecular clouds traced by CO.
First of all, more massive galaxies typically have larger mid-plane ISM pressures, which leads to higher molecular gas surface densities and thus larger GMC sizes traced by CO.
On the other hand, in a high-pressure environment, \textsc{H\,ii} regions might  be smaller.
Therefore, the number of CO and H$\alpha$ sight lines may  reflect not only the variation of cloud visibility time, but also the intrinsic differences in GMC and   \textsc{H\,ii} region properties among galaxies.
Moreover, our measurements of \hasightline\  (Section \ref{sec_ha_filtering})  is  affected by  internal extinction and non-\textsc{H\,ii} powering mechanisms.
In principle, the identification of \textsc{H\,ii} sight lines could be refined by  the use of optical IFS observations.
Similarly, the fractions of \cosightline\  depend on the applied surface density threshold (Section \ref{sec_co_threshold}).
Though our filtering methods are verified by visual inspection of the  filtered CO and H$\alpha$ images and comparing with the \textsc{H\,ii} regions identified in the PHANGS-MUSE images, the impact of methodology remains a potential source of bias (see Appendix \ref{sec_appendix_assumptions}).
Finally,  molecular clouds and \textsc{H\,ii} regions may have not yet been fully resolved by our 150~pc resolution as discussed in Section \ref{sec_discussion_quiescent_gas}.
A handful of our galaxies have PHANGS-ALMA CO and PHANGS-MUSE H$\alpha$ images with a resolution of $\sim$50~pc, comparable to the size of GMCs and \textsc{H\,ii} regions (e.g., NGC~0628, NGC~2835, and NGC~5068). At such high resolution, counting  number of objects and sight lines should become almost identical, and thus would provide a more robust estimate on cloud (CO) visibility time and even the actual lifetime of molecular clouds.
In summary, in addition to cloud visibility time, the region sizes traced by CO and H$\alpha$ emission,  the ratio between the resolution and the region separation length, and the data processing strategy may  also contribute to the derived $f_{\mathrm{scale}}$.

\subsection{Implication for the Kennicutt--Schmidt Relation}
\label{sec_discuss_ks}
Many studies have shown that the  Kennicutt--Schmidt relation between the surface densities of molecular gas and  SFR is  tight with an index $\sim$ 0.7--1.4 on kpc scales \citep[e.g.,][]{Big08,Ler08,Sch11,Ler13,Mom13}.
These findings resonate with our analysis, where \overlap\ sight lines dominate the maps when spatial scale is approximately or greater than 1~kpc; in other words, maps of molecular gas and star formation tracers become very similar.

The picture becomes more complex when the resolution increases. 
When the resolution is high enough to separate molecular clouds and star-forming regions, molecular gas and star formation surface densities become loosely correlated \citep[e.g.,][]{Bla09,Ono10,Kre18,Pes21} or even anti-correlated \citep[e.g.,][]{Sch10} because the two components no longer coincide at all times.
This spatial separation between different evolutionary stages of the star formation process is also evident in our  sight line maps at the  resolution of 150~pc (Figure \ref{fig_pie_example} and Figure \ref{fig_appendix_spatial_scale}).
The loosely (anti-)correlated molecular gas and SFR tracers  lead to an increasing scatter in the Kennicutt--Schmidt relation at small spatial scales, as a result of incomplete sampling of different evolutionary stages of the gas and star formation cycle \citep[e.g.,][]{Sch10,Fel11,Ler13,Kru14}.

Our results at 150~pc resolution  reveal an important dependence between host galaxy properties and the relative distribution of molecular gas and star formation tracers to each other. 
Namely, how molecular gas and SFR tracers relate (or do not relate) to one another on the Kennicutt--Schmidt plane depends on the host galaxy properties, such as \Mstar\ and Hubble type.
However,  any relation between the sight line fractions and host galaxy properties seen at 150~pc resolution is gradually diminished as resolution decreases, and becomes non-identifiable when at resolutions coarser than $\sim$ 500~pc.
Galactic structures   add a further complexity to the Kennicutt--Schmidt relation as the relation varies among galactic environments \citep{Pan17,Pes21,Que21}.
Since the contribution of the different environments varies as a function of galactic radius, the impact of environments could be transferred to the Kennicutt--Schmidt relation when averaging the environments.

Moreover, if the relative distributions of \co\ and \ha\  sight lines are complex at high resolution, we  might  be missing important $\tau_\mathrm{dep}$ variations (i.e., slope and intercept of  Kennicutt--Schmidt relation) and change in  cloud life cycle/time variations (i.e., scatter of the relation) when using  Kennicutt--Schmidt relation alone as a diagnostics of star formation process because the relation,  by definition, only sees  the  \overlap\ regions.

\section{Summary}
\label{sec_summary}

The main goal of this study is to investigate how global galaxy properties affect the radial distribution of various stages in the star formation cycle  using an unprecedented large sample of 49 star-forming main sequence disk galaxies (Figure~\ref{fig_sample}).
We compare high  resolution  ($\sim$ 1$\arcsec$)  observations of CO line emission and narrowband H$\alpha$ maps of  nearby galaxies selected from the PHANGS surveys (Section~\ref{sec_data}).

We adopt a simple and reproducible method developed in \citetalias{Sch19}  to quantify  the relative spatial distributions of molecular gas and recent star formation (Section~\ref{sec_method}).
The method  has taken into account the contribution of diffuse ionized gas to the H$\alpha$ emission and the metallicity dependence of CO-to-H$_{2}$ conversion factor when identifying star-forming regions and molecular clouds.

We classify each sight line (i.e., pixel) at each resolution according to the overlap between the tracers: \co, \ha, and \overlap\ (CO and  H$\alpha$)  (Figure~\ref{fig_pie_example}).
These three categories can be translated  into the following star formation phases: \co\ --    molecular gas currently not associated with star formation traced by H$\alpha$,  \overlap\ --  star-forming molecular clouds, and \ha\ -- regions of young massive stars.
We investigate whether the fractions of the different categories of sight lines vary with galaxy properties (stellar mass and Hubble type), galactocentric radius, and the presence of bars or grand-design spiral structure.
We also measure the sight line fractions at different resolutions ranging from 150~pc to 1.5~kpc.
The best common resolution (150~pc) is sufficiently high to sample individual star-forming unit and to separate such regions.
A summary of the main  results presented in this paper is as follows.

\begin{enumerate}
	
	\item At our best-matching resolution of 150~pc for our sample,  a median of 36\% of  detected sight lines in a galaxy are dominated by CO emission alone. The molecular gas surface densities of these \co regions are not necessarily low, ranging from our applied threshold (10 M$_{\sun}$~pc$^{-2}$; see Section~\ref{sec_method}) to a few thousands  M$_{\sun}$~pc$^{-2}$.
	This implies  that there is a substantial fraction of molecular gas in galaxies that is currently \textit{not} associated with young high-mass star formation traced by non-DIG optical tracer. 
	Statistically, the second most common category are \overlap\ regions where both CO and H$\alpha$ emission coincide,  accounting for a median  30\% of the sight lines at 150~pc resolution. 
	The  \overlap\ sight lines show the least variation from galaxy to galaxy.
	The \ha\ sight lines are less common than the other two categories,  with a median fraction of 20\%, but also exhibit the largest galaxy-to-galaxy variations.
	The rank of the median sight line fractions  is consistent with \citetalias{Sch19}  which used  only eight galaxies
	 (Section~\ref{sec_results_fractions} and Figure~\ref{fig_global_hist_0150pc_thresh}).

	\item At 150~pc resolution, we find strong correlations between the sight line fractions (\co\ and \ha) and global galaxy properties.
	Such dependencies had already been hinted at by  \citetalias{Sch19} which analyzed a small subset of our sample; in this work, we  quantify the dependencies.
	Specifically, the  fraction of \co\  sight lines  within the fiducial field of view increases gradually with increasing \Mstar\ and also increases gradually from later to earlier type spiral galaxies. The opposite trend is observed for \ha\ sight lines.   The fraction of \overlap\ regions is insensitive to \Mstar, but increase toward later types. These trends observed for global sight line fractions  are almost preserved radially from the center out to 0.6~$R_{25}$ (corresponding to $\sim$ 6~kpc). 
	Our results at 150~pc resolution suggest that the relationship between molecular gas and SFR tracers in the Kennicutt--Schmidt plane depends on host-galaxy properties (Sections~\ref{sec_results_fractions} and~\ref{sec_discuss_ks}, Figures~\ref{fig_global_boxes} and~\ref{fig_radial_all}).

	\item  In addition to \Mstar\ and Hubble type, we also classify galaxies according to the presence of a stellar bar and/or grand-design spiral arms. Galaxies without these structures,  galaxies with a stellar bar only, and galaxies with both a bar and grand-design spiral arms exhibit broadly similar sightline fractions. Galaxies with grand-design spiral arms but no stellar bar, however, show a distinct signature of higher \co\ and \overlap\ fractions and lower \ha\ fraction than the other populations. 
	Moreover,  galaxies with a bar show a stronger (weaker) radial dependence of \co\ (\overlap) sight lines than galaxies without a bar. These results suggest that galactic dynamics further contributes to organizing the spatial distribution of CO and H$\alpha$ emission separately within galaxies (Sections~\ref{sec_global_structures} and~\ref{sec_discussion_structures}, Figures~\ref{fig_structures} and~\ref{fig_rad_structures}).

	\item Comparing the fractions of pixels (our ``default'' approach)  with the fractions of \emph{flux} shows that \overlap\ regions tend to have higher CO and H$\alpha$  intensities compared to regions that emit CO or  H$\alpha$ alone. 
	Yet the flux traced by \emph{CO-only} and \emph{H$\alpha$-only} regions cannot be neglected, since they still contribute a median of 33\% (mean: 37\%) and 14\% (mean: 25\%) of the galaxy's total CO and  H$\alpha$ flux, respectively.
	The result is consistent with the finding of \citetalias{Sch19}. (Section~\ref{sec_number_vs_flux}, Figures~\ref{fig_number_vs_flux_all} and~\ref{fig_global_coHa_sight lines}).
	
	\item The sight line fractions show a strong dependence on spatial scale (resolution), confirming the finding of \citetalias{Sch19}. Specifically,  \co\ and \ha\ regions  rapidly vanish as spatial scale increases. Therefore, any relation between the sight line fractions and galaxy properties (\Mstar\ and Hubble type) are only evident when the resolution is $\ll$ 500~pc  (Section~\ref{sec_global_resolution}, Figures~\ref{fig_res_allgal} and~\ref{fig_res_all}).
	
	\item We find no correlation between the global sight line fractions with specific star formation rate and the offset from the star-forming main sequence ($\Delta$MS). Nonetheless, galaxies with the highest- or the lowest-$\Delta$MS in our sample both show  significant molecular gas reservoirs that appear not to be associated with star formation. However, the spatial distribution of their \co\ sight lines  are different, pointing to different underlying causes of their high \co\ fractions (Section~\ref{sec_discussion_sf} and Figure~\ref{fig_sight line_vs_deltams}).

	\item    \ha\ regions  tend to be found in atomic gas dominated regions in low-\Mstar\ systems. It is very likely that lower mass molecular clouds exist in these regions, but their $\Sigma_\mathrm{H_{2}}$  drops below our applied threshold, adding further evidence for   prominent galaxy-to-galaxy variation in molecular cloud properties, in line with previous studies \citep[e.g.,][]{Hug13,Sun20b,Ros21} (Section~\ref{sec_discussion_ism} and Figure~\ref{fig_radial_gas_props}).

	\item We  estimate the  the duration for which the molecular cloud traced by CO emission is visible  following the statistical approach in  \citetalias{Sch19}.  There is a ranking of cloud visibility time with \Mstar\ where molecular clouds in high-\Mstar\ galaxies tend to have a longer  visibility time than clouds in low-\Mstar\ over the radial range probed. The trend is related to the fact that molecular clouds in high-\Mstar\ galaxies tend to have higher molecular mass surface densities. However, the  differences between galaxies become considerably smaller when the spatial scale is larger than 500~pc  due to a significant decrease in \co\ sight line and increase in \overlap\ sight line, indicating   a critical resolution that is required to resolve  the evolution of individual star-forming regions. 	We also note several systematic differences in cloud properties and uncertainties to bear in mind when using Equation (\ref{equ_fscale}) to estimate the visibility time of molecular clouds (Section~\ref{sec_molgas_lifetime} and Figures~\ref{fig_fscale} and~\ref{fig_fscale_res}).
\end{enumerate}

The methodology presented in this paper  offers a simple, physically motivated, and reproducible approach for quantifying the relative distribution of molecular gas traced by CO emission and HII regions traced by H$\alpha$ emission.
Several caveats related with the use of this approach should be kept in mind, including the  choices of  CO-to-H$_{2}$ conversion factor, CO(2-1)-to-CO(1-0) brightness temperature ratio,  unsharp masking parameters, and H$\alpha$ and CO($\Sigma_\mathrm{H_{2}}$) thresholds.
These factors are discussed in detail  in  Appendix \ref{sec_appendix_assumptions}.
Although our results remain robust when  accounting for the impact of these factors, care should be taken when interpreting the results based on our sight line method.
Moreover, since our main analysis focuses mostly on the location (rather than the amount) of   massive star formation, we consider internal extinction as a secondary issue.
Nonetheless, there is room for improvement with  the use of optical IFS data which allows for simultaneous correction for internal extinction and tools for distinguishing  H$\alpha$ emission emitted from various powering sources.
A straightforward next step would be to test our results with large, diverse, deep ($\log L_\mathrm{H\alpha}$ $\approx$ 36~erg~s$^{-1}$), and high resolution ($\leq$ 100~pc) IFS sample.
Other follow-up studies (utilizing a large IFS sample) include the detailed investigation of the nature of \co\ and \ha\ regions, the absolute timescale of each star formation phase, and the dependence of these properties on global galaxy properties and galactic dynamics.

We thank the anonymous referee for constructive comments that improved the paper.
This work was carried out as part of the PHANGS collaboration.

HAP acknowledges funding from the European Research Council (ERC) under the European Union's Horizon 2020 research and innovation programme (grant agreement No. 694343), and the Ministry of Science and Technology (MOST) of Taiwan under grant 110-2112-M-032-020-MY3.

ES, PL, DL, RMcE, TS, FS, and TGW  acknowledge funding from the European Research Council (ERC) under the European Union's Horizon 2020 research and innovation programme (grant agreement No. 694343).

AH was supported by the Programme National Cosmology et Galaxies (PNCG) of CNRS/INSU with INP and IN2P3, co-funded by CEA and CNES, and by the Programme National “Physique et Chimie du Milieu Interstellaire” (PCMI) of CNRS/INSU with INC/INP co-funded by CEA and CNES.

The work of AKL, JS, and DU is partially supported by the National Science Foundation under Grants No. 1615105, 1615109, and 1653300.

AB and FB acknowledge funding from the European Research Council (ERC) under the European Union's Horizon 2020 research and innovation programme (grant agreement No.726384/Empire).

MC gratefully acknowledges funding from the Deutsche Forschungsgemeinschaft (DFG) through an Emmy Noether Research Group, grant number KR4801/1-1 and the DFG Sachbeihilfe, grant number KR4801/2-1, and from the European Research Council (ERC) under the European Union's Horizon 2020 research and innovation programme via the ERC Starting Grant MUSTANG (grant agreement number 714907).

EC acknowledges support from ANID project Basal AFB-170002.

CE acknowledges funding from the Deutsche Forschungsgemeinschaft (DFG) Sachbeihilfe, grant number BI1546/3-1.

CMF is supported by the National Science Foundation under Award No. 1903946 and acknowledges funding from the European Research Council (ERC) under the European Union's Horizon 2020 research and innovation programme (grant agreement No. 694343).

SCOG and RSK acknowledge financial support from the German Research Foundation (DFG) via the collaborative research center (SFB 881, Project-ID 138713538) “The Milky Way System” (subprojects A1, B1, B2, and B8). They also acknowledge funding from the Heidelberg Cluster of Excellence ``STRUCTURES'' in the framework of Germany’s Excellence Strategy (grant EXC-2181/1, Project-ID 390900948) and from the European Research Council via the ERC Synergy Grant ``ECOGAL'' (grant 855130). 

JMDK gratefully acknowledges funding from the Deutsche Forschungsgemeinschaft (DFG) in the form of an Emmy Noether Research Group (grant number KR4801/1-1) and the DFG Sachbeihilfe (grant number KR4801/2-1), and from the European Research Council (ERC) under the European Union's Horizon 2020 research and innovation programme via the ERC Starting Grant MUSTANG (grant agreement number 714907).

JP acknowledges support from the Programme National “Physique et Chimie du Milieu Interstellaire” (PCMI) of CNRS/INSU with INC/INP co-funded by CEA and CNES.

MQ acknowledges support from the research project PID2019-106027GA-C44 from the Spanish Ministerio de Ciencia e Innovaci\'on.

ER acknowledges the support of the Natural Sciences and Engineering Research Council of Canada (NSERC), funding reference number RGPIN-2017-03987.

AU acknowledges support from the Spanish funding grants PGC2018-094671-B-I00 (MCIU/AEI/FEDER) and PID2019-108765GB-I00 (MICINN).

This paper makes use of the following ALMA data: \linebreak
ADS/JAO.ALMA\#2012.1.00650.S, \linebreak 
ADS/JAO.ALMA\#2013.1.01161.S, \linebreak 
ADS/JAO.ALMA\#2015.1.00925.S, \linebreak 
ADS/JAO.ALMA\#2015.1.00956.S, \linebreak 
ADS/JAO.ALMA\#2017.1.00392.S, \linebreak 
ADS/JAO.ALMA\#2017.1.00886.L, \linebreak 
ADS/JAO.ALMA\#2018.1.01651.S. \linebreak 
ALMA is a partnership of ESO (representing its member states), NSF (USA) and NINS (Japan), together with NRC (Canada), MOST and ASIAA (Taiwan), and KASI (Republic of Korea), in cooperation with the Republic of Chile. The Joint ALMA Observatory is operated by ESO, AUI/NRAO and NAOJ.

This paper includes data gathered with the 2.5 meter du Pont located at Las Campanas Observatory, Chile, and data based on observations carried out at the MPG 2.2m telescope on La Silla, Chile.

\appendix
\renewcommand\thefigure{\thesection.\arabic{figure}}   
\restartappendixnumbering 

\section{Impact of Methodology}
\label{sec_appendix_assumptions}
\restartappendixnumbering
Here we discuss the potential impact of methodology on our results, including  the choice of CO-to-H$_{2}$ conversion factor, CO(2-1)-to-CO(1-0) ratio, unsharp masking parameters, and H$\alpha$ and CO thresholds.
Figure~\ref{fig_test_assumptions} summarizes the  sight line fractions for each individual galaxy  based on different  methodologies.
Bar graphs with darker colors make use of the default unsharp masking parameters, adopted $\Sigma_\mathrm{H_{2}}$ threshold and $\alpha_\mathrm{CO}$  described in Section \ref{sec_data} and \ref{sec_method}, while 
bar graphs with lighter colors demonstrate the impact of these three assumptions.

\subsection{CO-to-\texorpdfstring{H$_2$}{H2} Conversion Factor}
\label{sec_appendix_alphaco}
We test whether our results are sensitive to the employed $\alpha_\mathrm{CO}$ conversion factor by comparing the sight line fractions based on the metallicity- and radius-dependent $\alpha_\mathrm{CO}$ (default in this work; see \ref{sec_data_co}) and the frequently used, constant Galactic $\alpha_\mathrm{CO}$ of 4.35 M$_{\sun}$ pc$^{-1}$ (K km s$^{-1}$)$^{-1}$ \citep{Bol13}.
The metallicity- and radius-dependent $\alpha_\mathrm{CO}$ values tends to be lower than the Galactic value at  small galactocentric radii, and higher than the Galactic value at  large radii.
Therefore, for a given H$_{2}$ surface density threshold,  applying the Galactic $\alpha_\mathrm{CO}$  often increases the number of \cosightline\ (\co\ and \overlap)  at  small galactocentric radii and decreases the number of  \cosightline\ at large galactocentric radii.
The resulting impact on the sight lines fractions thus depends on the radial   distribution of CO and H$\alpha$ emission.
The different $\alpha_\mathrm{CO}$ prescriptions translate into a typical variation of sight line fractions of $\pm$10\% (mostly within 5\%).
The median and mean differences are within $\pm$ 1\% for all types of sight lines.
We again repeat our analysis on the dependence of sight line fraction  for  \Mstar, Hubble type, and spatial scale, using the sight line fractions estimated based on the Galactic $\alpha_\mathrm{CO}$.
We conclude that our results are robust against the choice of $\alpha_\mathrm{CO}$.
The sight line fractions based on the Galactic $\alpha_\mathrm{CO}$ are presented in column~Q  in Figure~\ref{fig_test_assumptions}.

	\subsection{CO (2-1)-to-CO(1-0) Ratio}
	\label{sec_appendix_r21}
	In this work, we adopt a single $^{12}$CO(2-1)-to-$^{12}$CO(1-0) brightness temperature ratio of $R_{21}$~=~0.65 to   all our sample galaxies. 
	We do not account for galaxy to galaxy variations in  $R_{21}$.
	Recent studies of nearby galaxies show that $R_{21}$ for individual galaxies is around 0.5 to 0.7, and the typical scatter is $\sim$~0.1 dex within individual galaxies \citep[e.g.,][]{Yaj21,denB21,Ler21c}. 
	Physically, $R_{21}$ may increase with SFR and gas density  because the higher-$J$ transition becomes brighter when gas is warm and/or dense \citep[e.g.,][]{Sak94,Sak97,Yaj21,denB21,Ler21c}.
	In other words, assuming a single $R_{21}$ may result in overestimation of $\Sigma_\mathrm{H_{2}}$ in \overlap\ regions (given that CO and H$\alpha$ emission  are likely to co-exist in  high-SFR and/or high-density regions).
	Nonetheless, we do not expect  $\Sigma_\mathrm{H_{2}}$ in the \overlap\ regions to drop below the  CO threshold (10 M$_{\sun}$ pc$^{-2}$) even when accounting for the varying $R_{21}$.
	Moreover, as shown in Figure \ref{fig_number_vs_flux}(top),  the sight line ratio between \co\ and  \emph{CO-overlap} regions is $\sim$ 1:1, while their flux ratio is $\sim$ 1:2, suggesting that  \emph{CO-overlap} regions are on average two times brighter than \co\ regions. On the other hand, the typical $\sim$~0.1~dex scatter of $R_{21}$ corresponds to a $\sim$~25\% difference in flux.  Therefore the difference between the    sight line ratio and the flux ratio is unlikely to disappear even if   $R_{21}$ variation were taken into account. Taken together, we conclude that	 our results  should remain valid for the typical scatter in $R_{21}$ (note again that our  analysis focuses  on the location, rather than the amount, of molecular clouds and    star formation).

\subsection{Unsharp Masking Parameters}
\label{sec_appendix_impact_um}
As described in Section~\ref{sec_method}, the adopted unsharp masking parameters are optimized to  reproduce the \textsc{H\,ii} regions detected in PHANGS-MUSE H$\alpha$ images  \citep{San21}.
We have also identified  two additional sets of parameters which also  reasonably well reproduce the \textsc{H\,ii} regions identified in  PHANGS-MUSE.
The second best  parameters  (UM$_\mathrm{2nd\_best}$) have a  200 pc kernel for  Step 1 in Section \ref{sec_ha_filtering}, a scaling factor of 0.33 for Step 2, and a kernel size of 400 pc for Step 3.
The  third best (UM$_\mathrm{paperI}$) is the set of parameters used in \citetalias{Sch19}: 300 pc for Step 1, 0.33 for Step 2, and 750 pc for Step 3.

We test how sight line fractions relate to the DIG removal process by repeating the sight line classification using \textsc{H\,ii} region maps created with the other two sets of unsharp masking parameters.
Overall, we find that the derived sight line fractions are similar among the three sets of parameters.
The median and mean differences in any sight line fraction for any two sets of parameters are always $<$ 5\%. 
For reference, the sight line fractions estimated based on different unsharp masking parameters   are presented in Figure~\ref{fig_test_assumptions} column~D (this work), column~R (UM$_\mathrm{2nd\_best}$), and column~S (UM$_\mathrm{paperI}$).
We also repeat all analyses using the sight line fractions estimated based on  UM$_\mathrm{2nd\_best}$ and  UM$_\mathrm{paperI}$.
Our results are robust against  the choice of different unsharp masking parameters, as far as they can reasonably reproduce the \textsc{H\,ii} region properties  identified in the VLT/MUSE IFU data.
The  detailed description of how we verified the narrowband \textsc{H\,ii} regions using PHANGS-MUSE spectroscopic information will be presented in a forthcoming paper (H.-A.~Pan et al.\ in prep.).

\begin{figure*}[h]
	\centering
	\includegraphics[scale=0.52]{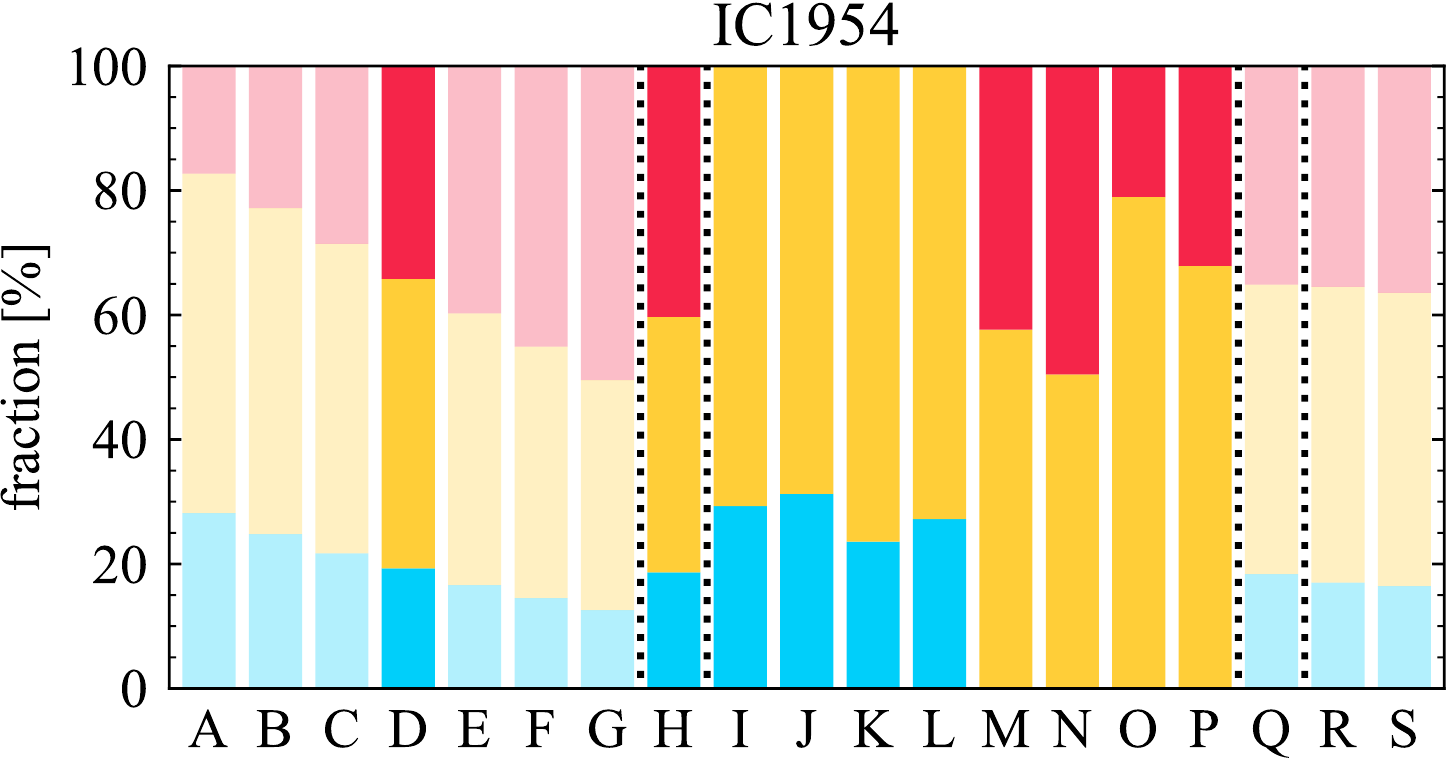}
	\includegraphics[scale=0.52]{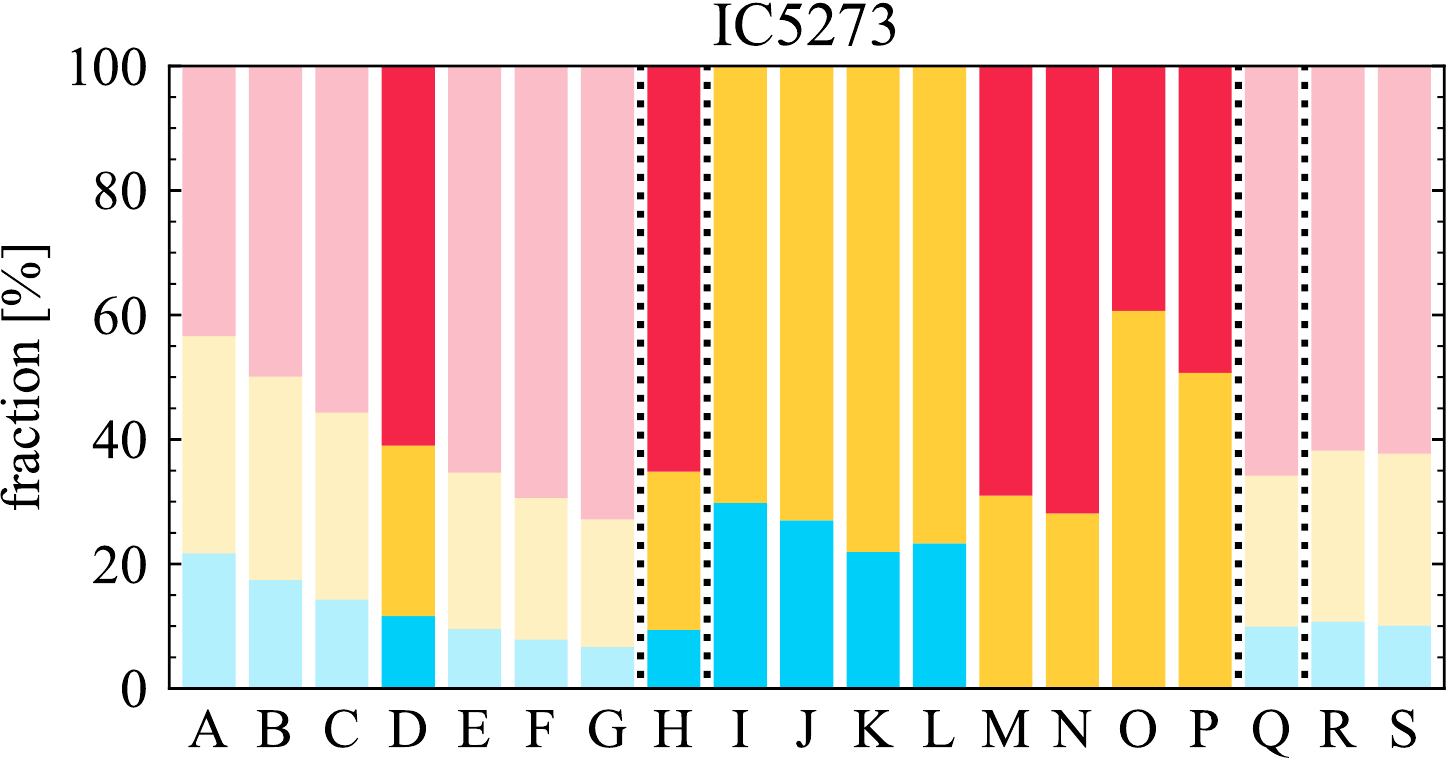}\\ 
	\vspace{10pt}
	\includegraphics[scale=0.52]{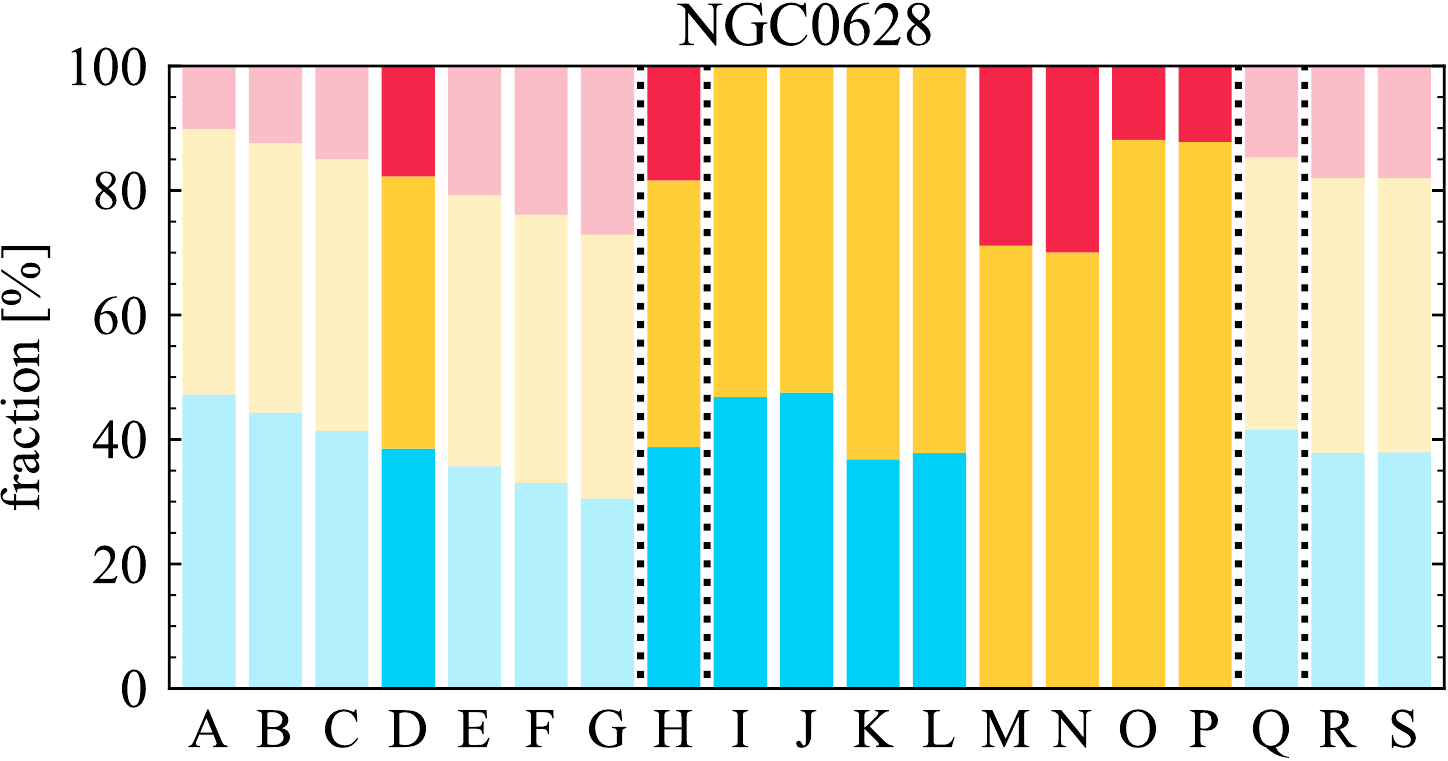}
	\includegraphics[scale=0.52]{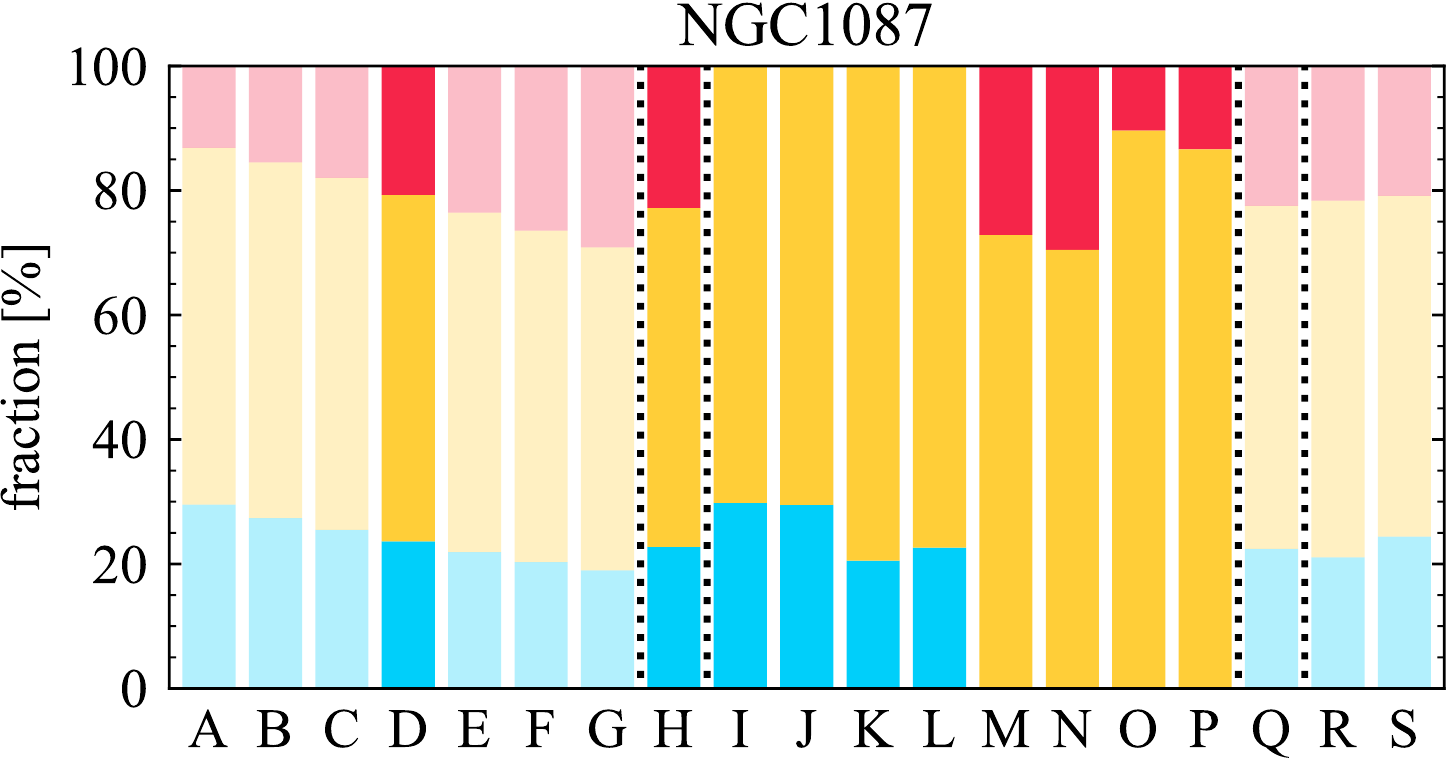}\\
	\vspace{10pt}
	\includegraphics[scale=0.52]{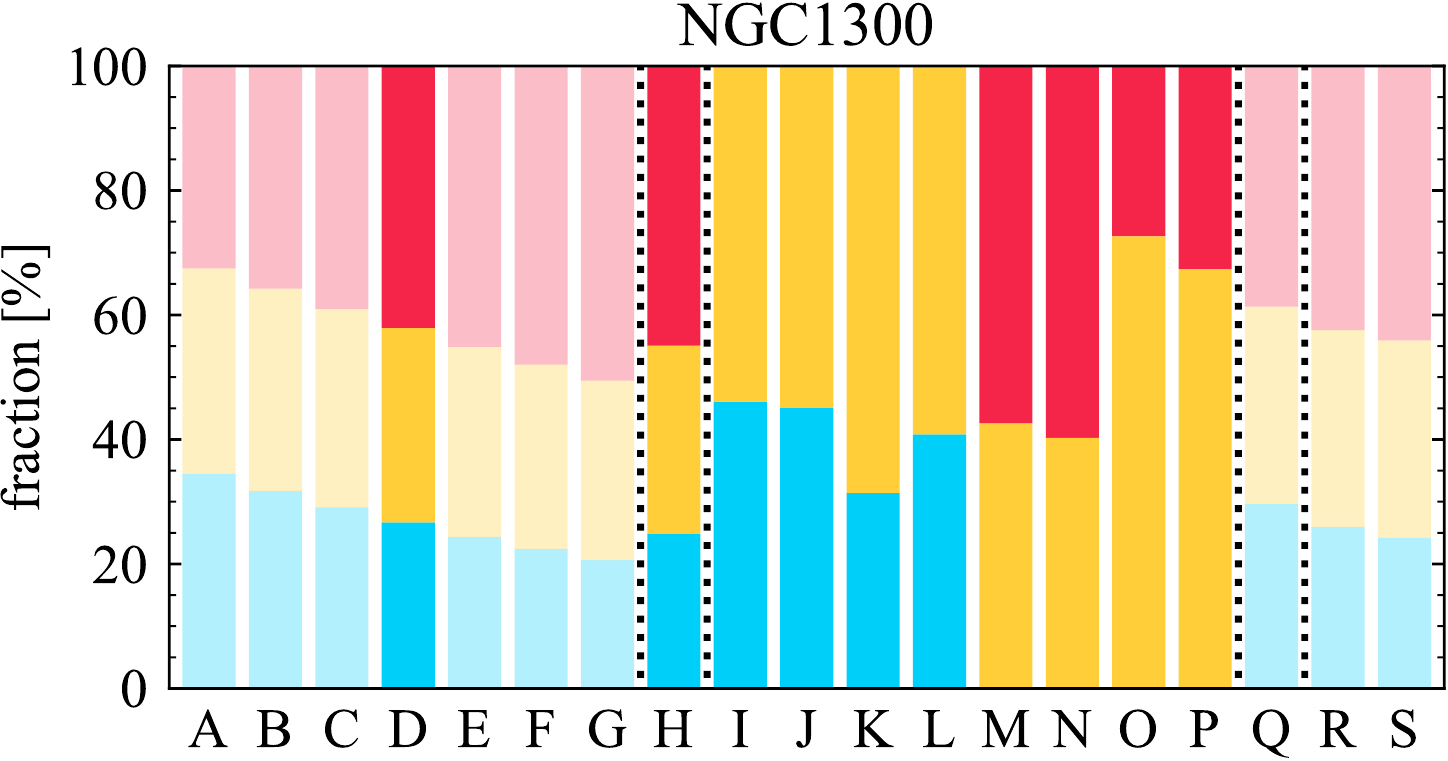}
	\includegraphics[scale=0.52]{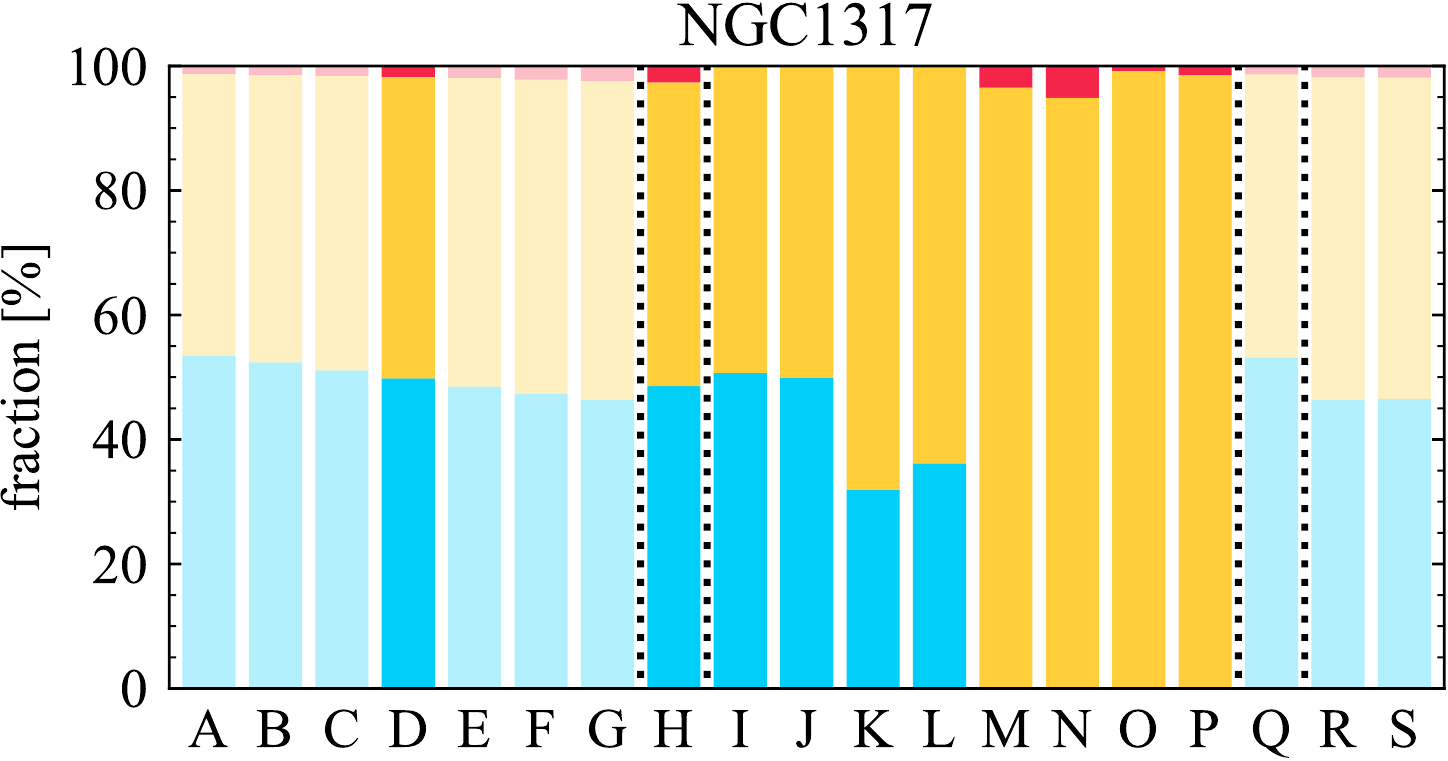}\\
	\vspace{10pt}
	\includegraphics[scale=0.52]{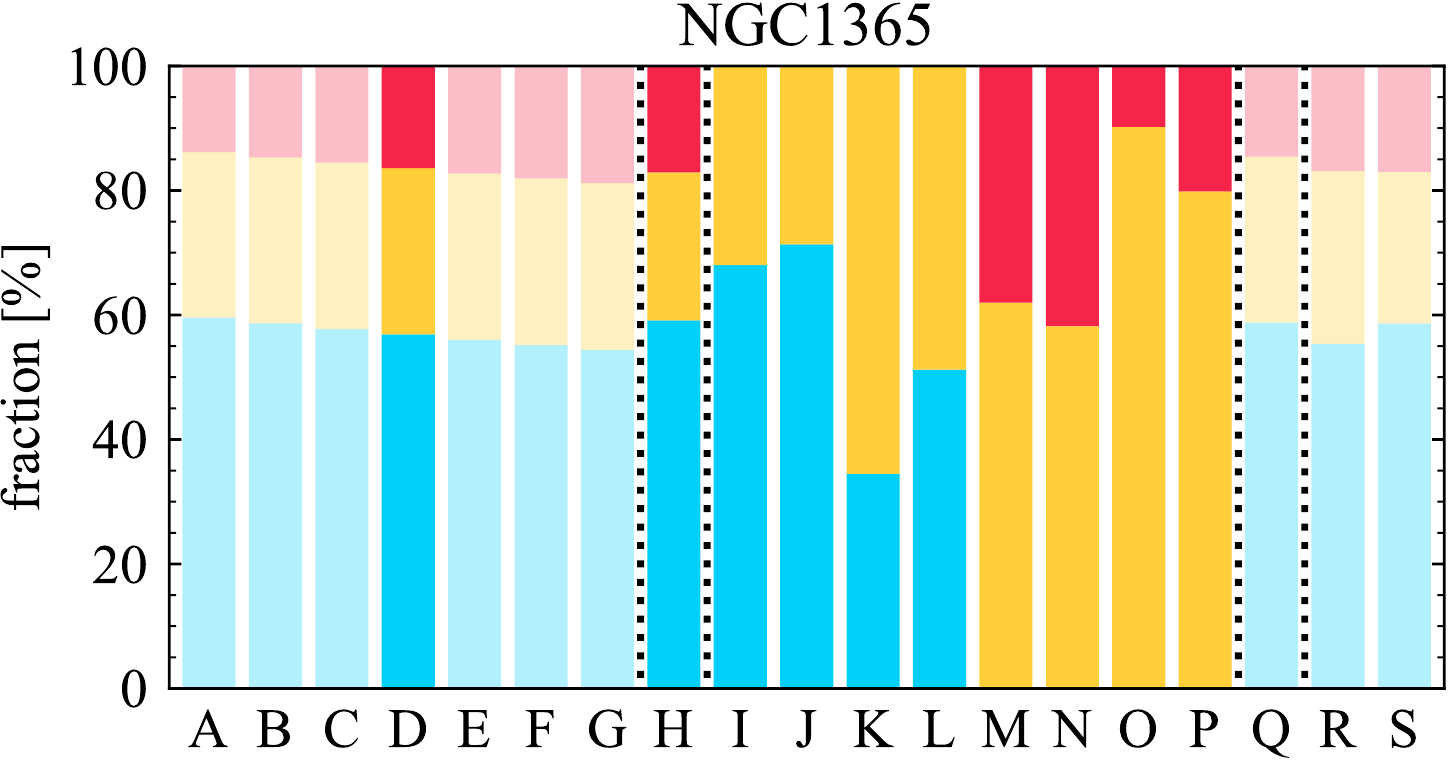}
	\includegraphics[scale=0.52]{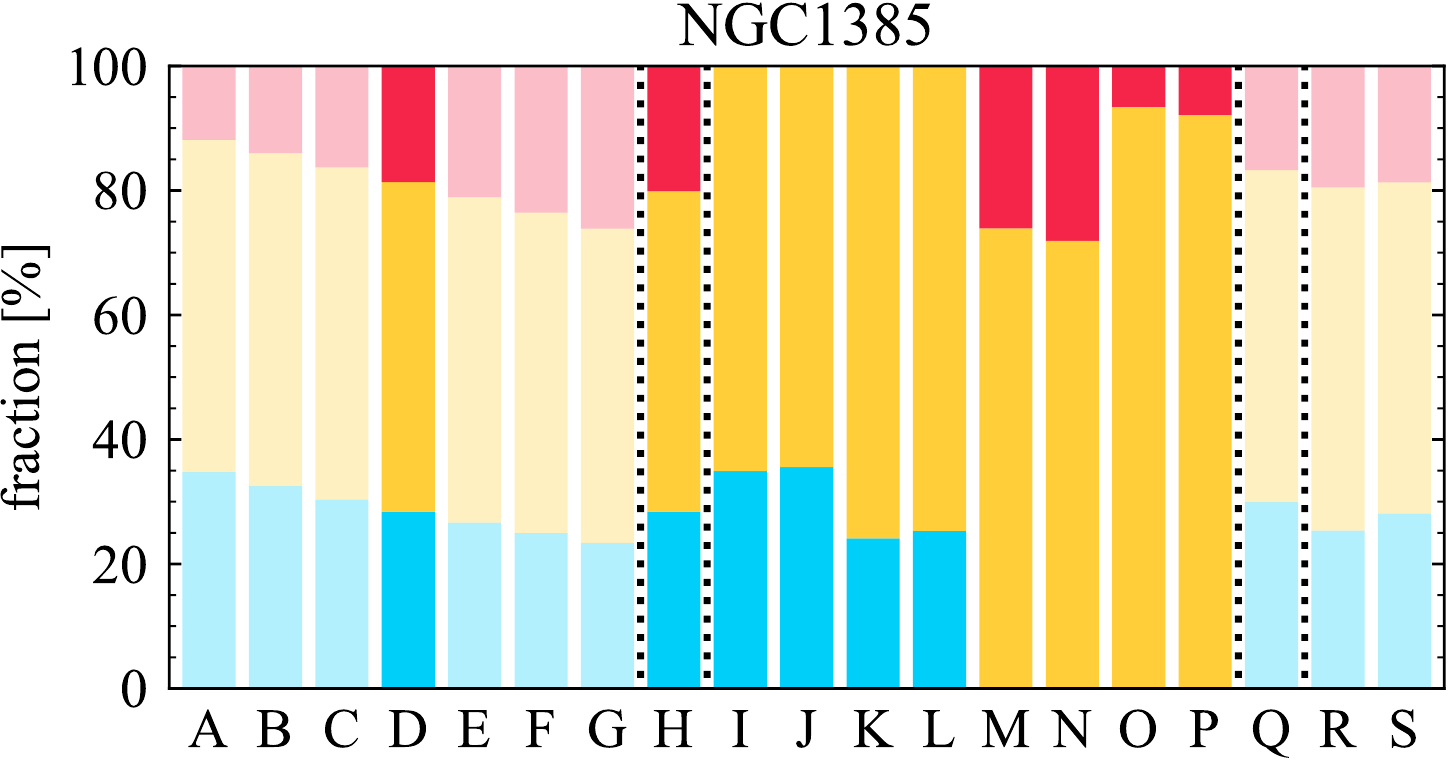}
	
	\caption{
		Bar graphs summarizing the impact of  methodology and assumptions on \co\ (blue), \ha\ (red), and \overlap\ (yellow) sight lines for individual galaxy at 150~pc resolution.
		Bar graphs with darker colors make use of the default unsharp masking parameters, adopted H$_{2}$ threshold, and $\alpha_\mathrm{CO}$ conversion factor, while 
		bar graphs with lighter colors demonstrate the impact of these three assumptions. 
		Here we show
		A--G: the \emph{number} of sight lines for \emph{our FoV} with H$_{2}$ threshold of 7--13 M$_{\sun}$~pc$^{-2}$, respectively; 
		H:  the \emph{number} of sight lines for \emph{disk} with default H$_{2}$ threshold of 10 M$_{\sun}$~pc$^{-2}$;
		I \& J: relative contribution of \emph{number} of  CO sight lines (\co\ and \overlap) for \emph{our FoV} and \emph{disk}; 
		K \& L:  relative contribution of \emph{CO flux} of  CO sight lines (\co\ and \overlap) for \emph{our FoV} and \emph{disk};
		M \& N: relative contribution of \emph{number} of  H$\alpha$ sight lines (\ha\ and \overlap) for \emph{our FoV} and \emph{disk}; 	
		O \& P:  relative contribution of \emph{H$\alpha$ flux} of  H$\alpha$ sight lines (\ha\ and \overlap) for \emph{our FoV} and \emph{disk};	
		Q:  three \emph{number} of sight lines for \emph{our FoV} with  Galactic $\alpha_\mathrm{CO}$.
		R \& S: 	 the \emph{number} of sight lines for \emph{our FoV} with unsharp masking parameters UM$_\mathrm{2nd\_best}$ and  UM$_\mathrm{Paper\,I}$, respectively. The black dotted lines are used to guide the eye.
		Sight line fractions in the column D are the default fractions used for the main analysis  in this work at 150~pc resolution.
	}
	\label{fig_test_assumptions}
\end{figure*}

\addtocounter{figure}{-1}
\begin{figure*}
	\centering
	\includegraphics[scale=0.52]{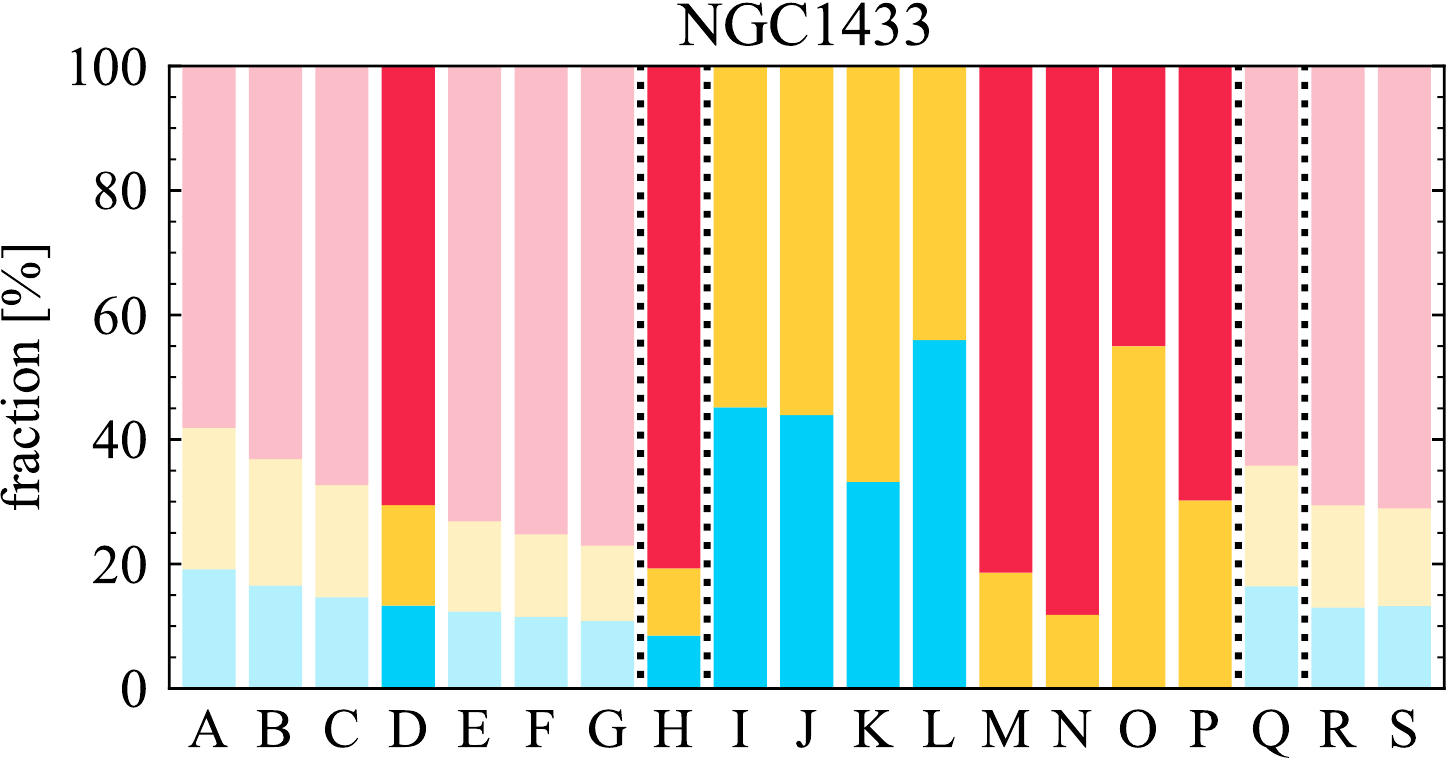}
	\includegraphics[scale=0.52]{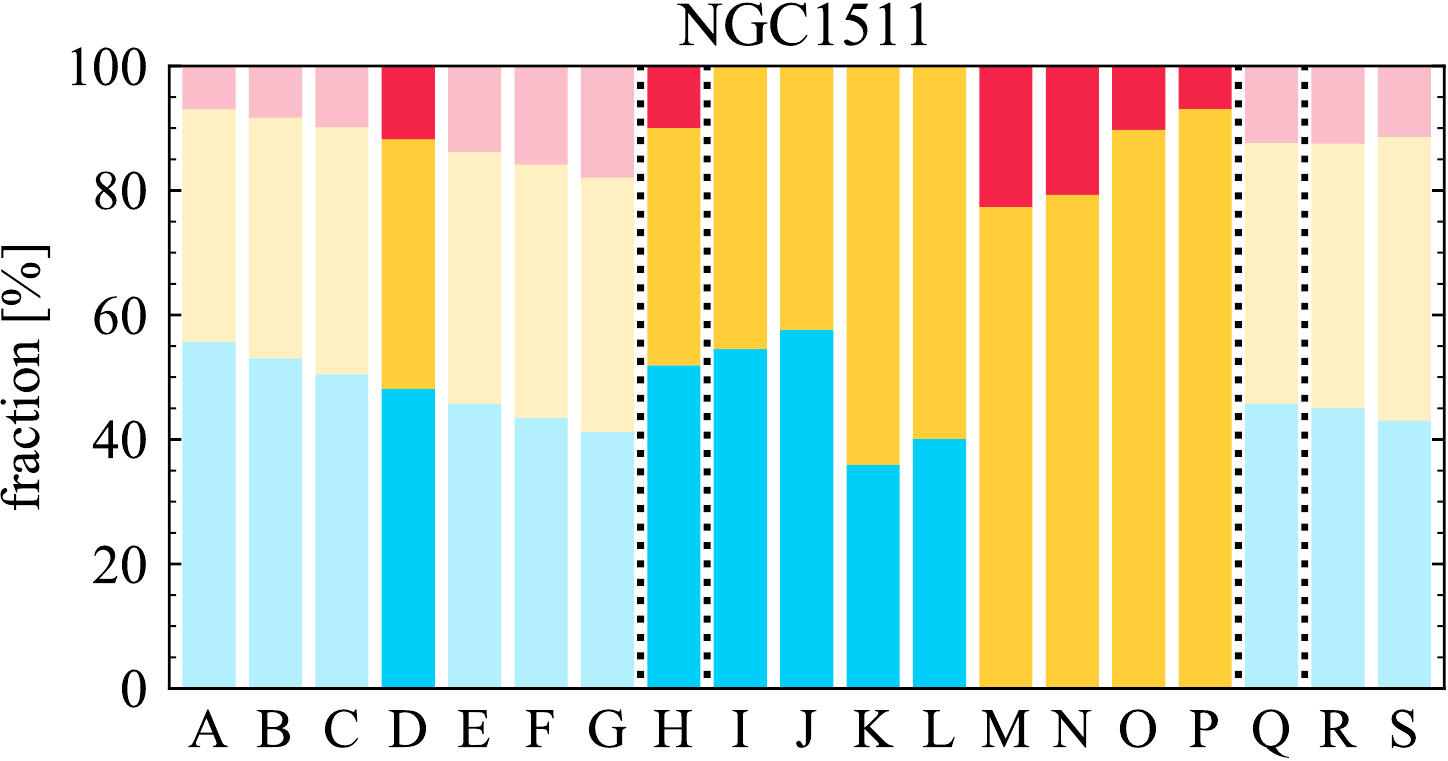}\\
	\vspace{10pt}
	\includegraphics[scale=0.52]{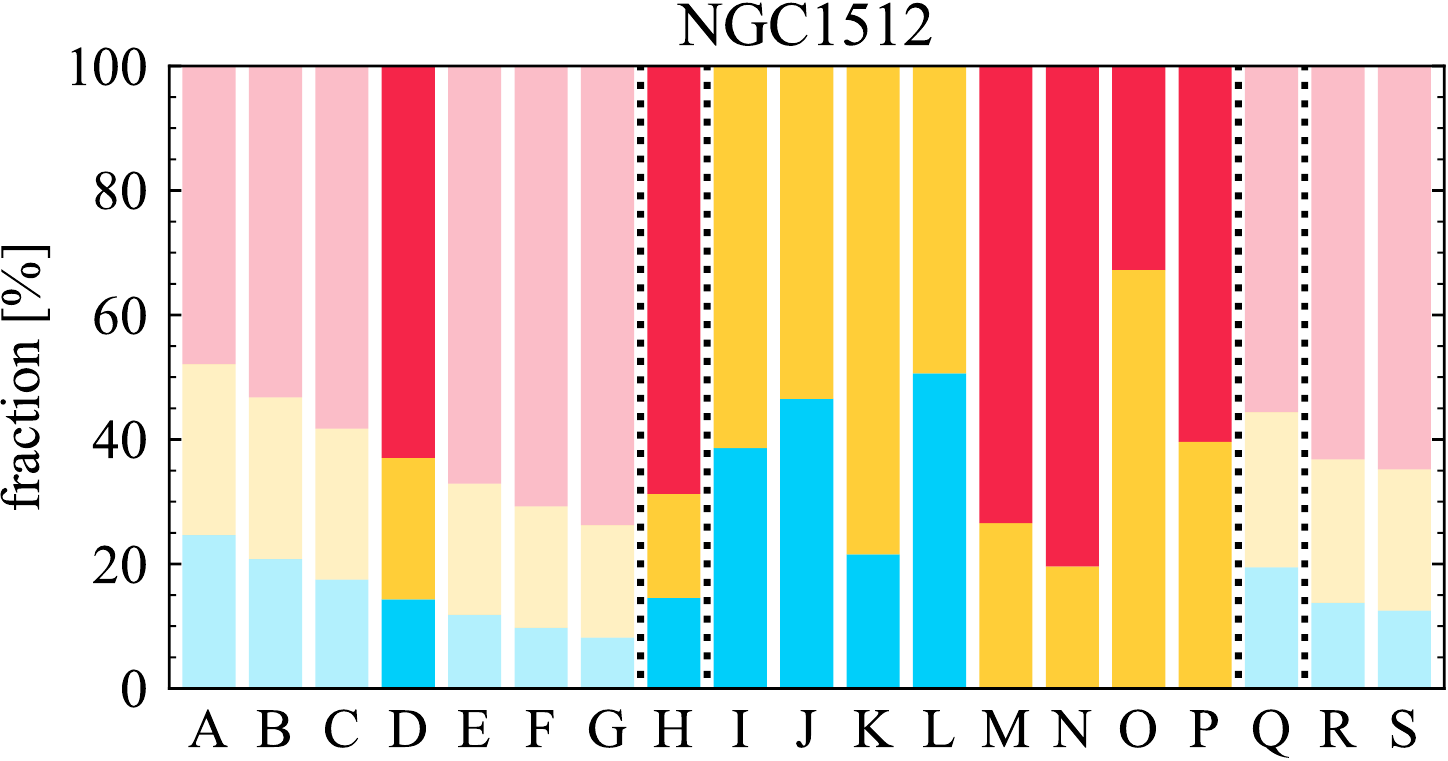}
	\includegraphics[scale=0.52]{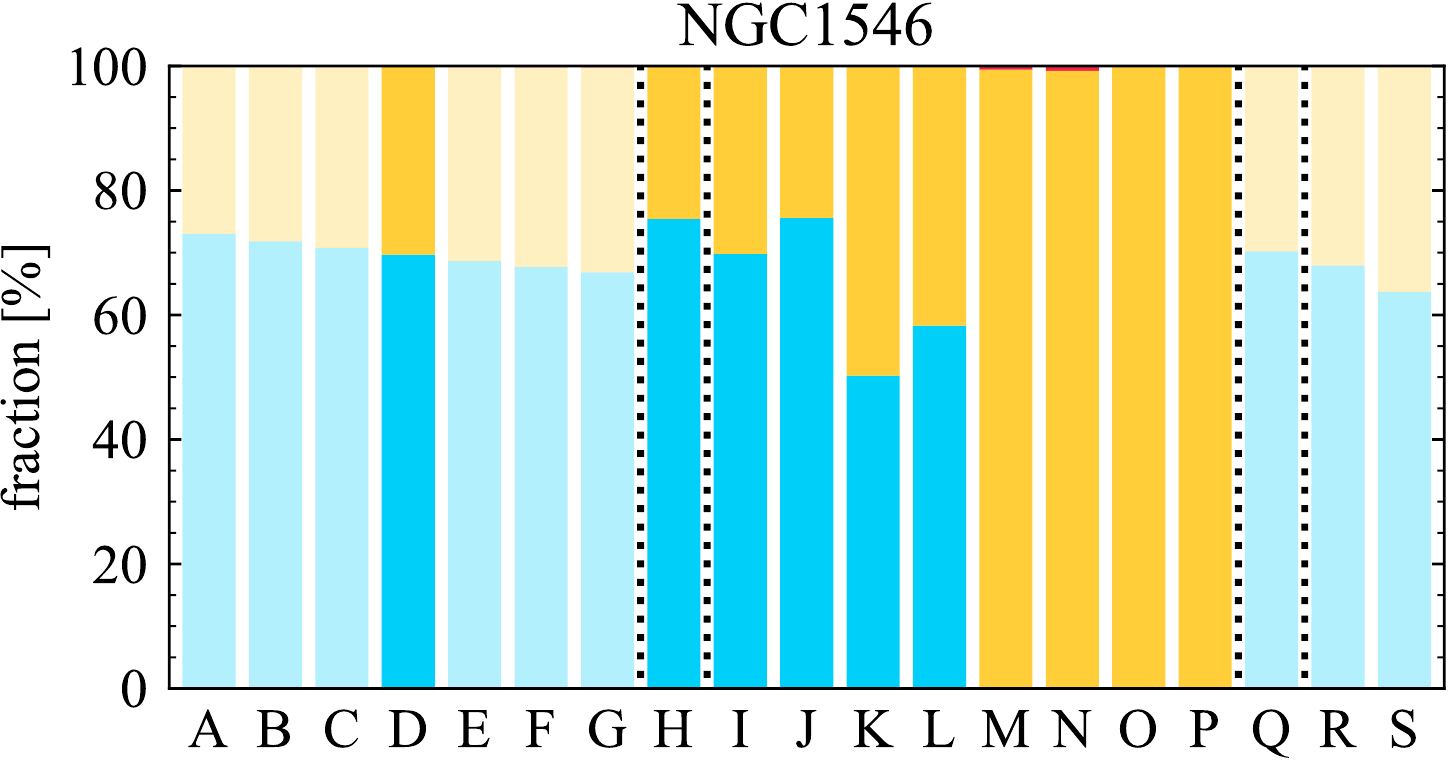}\\
	\vspace{10pt}
	\includegraphics[scale=0.52]{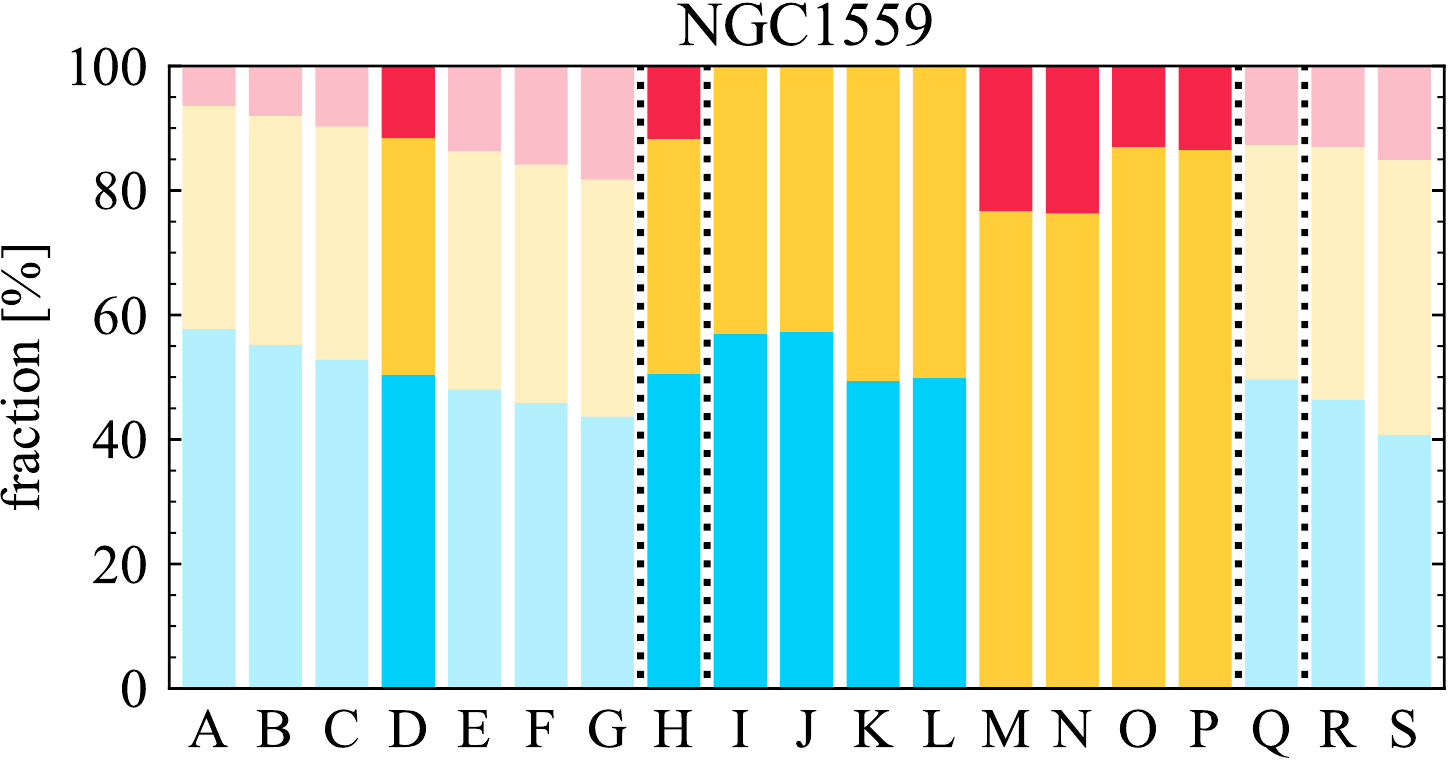}
	\includegraphics[scale=0.52]{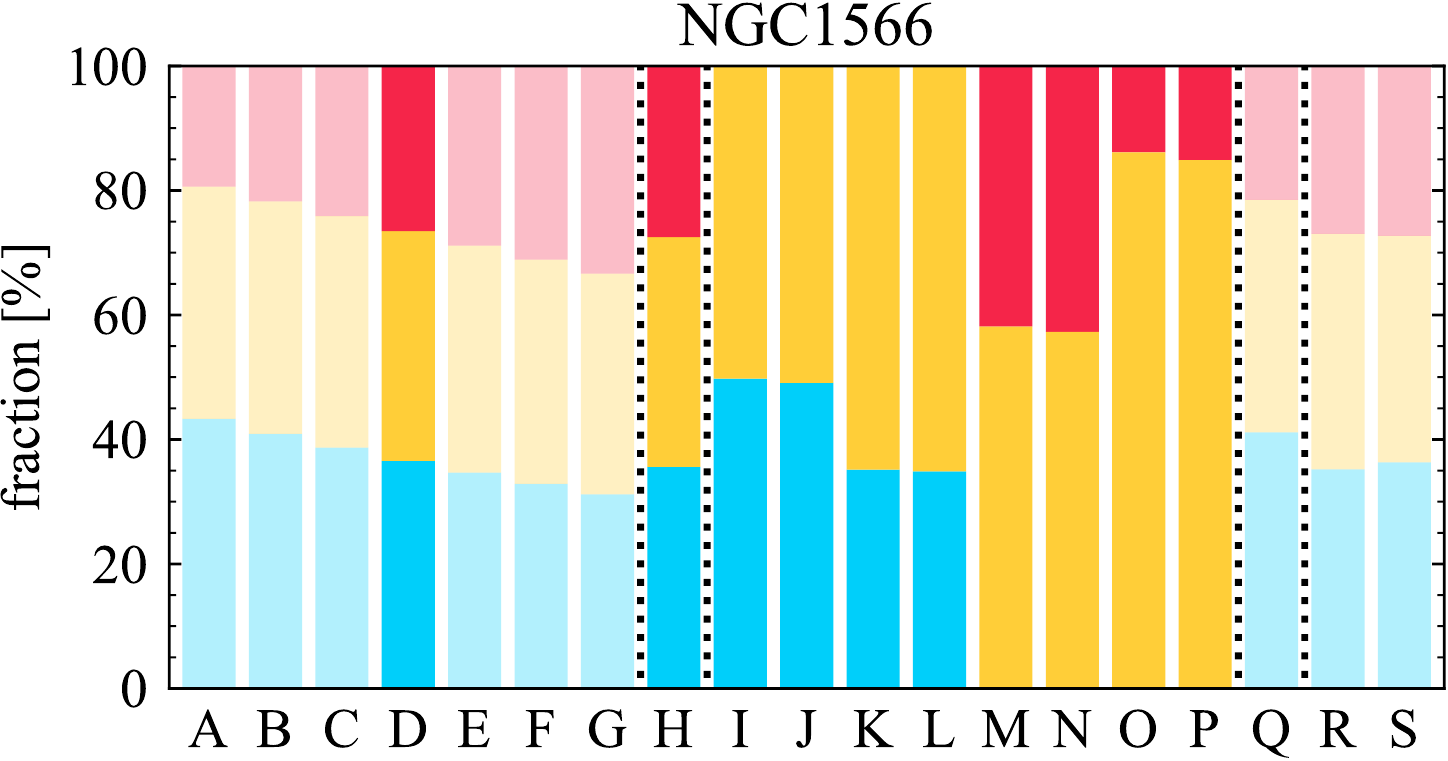}\\
	\vspace{10pt}
	\includegraphics[scale=0.52]{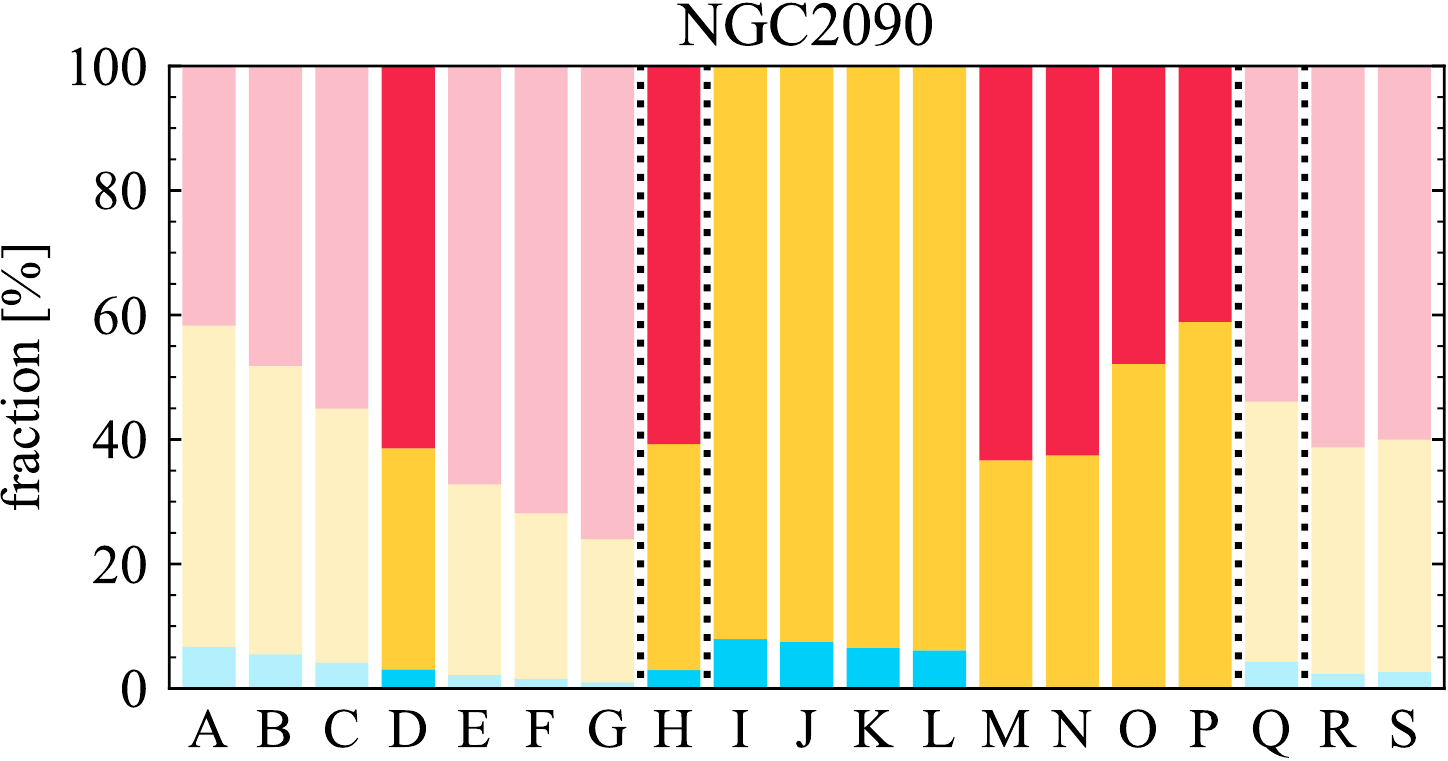}
	\includegraphics[scale=0.52]{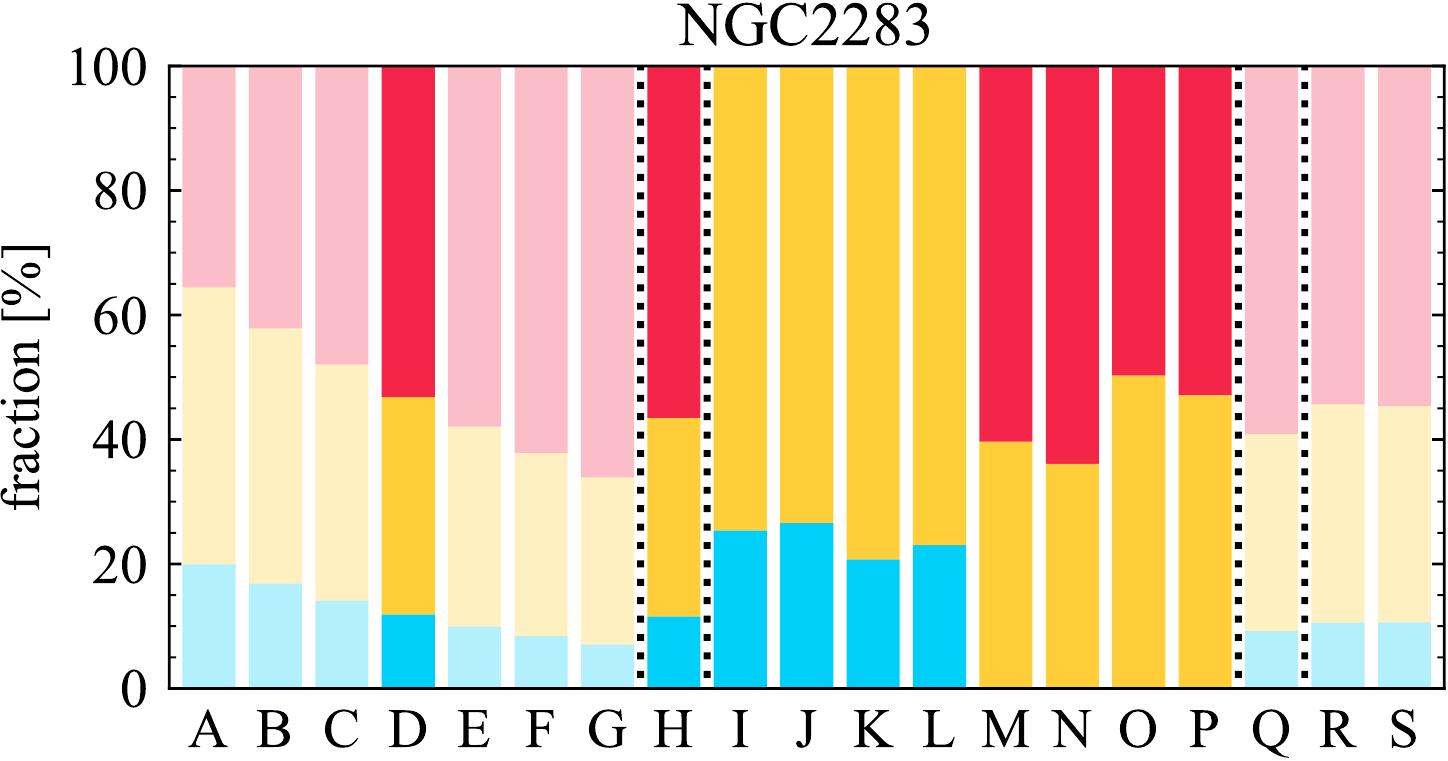}\\
	\vspace{10pt}
	\includegraphics[scale=0.52]{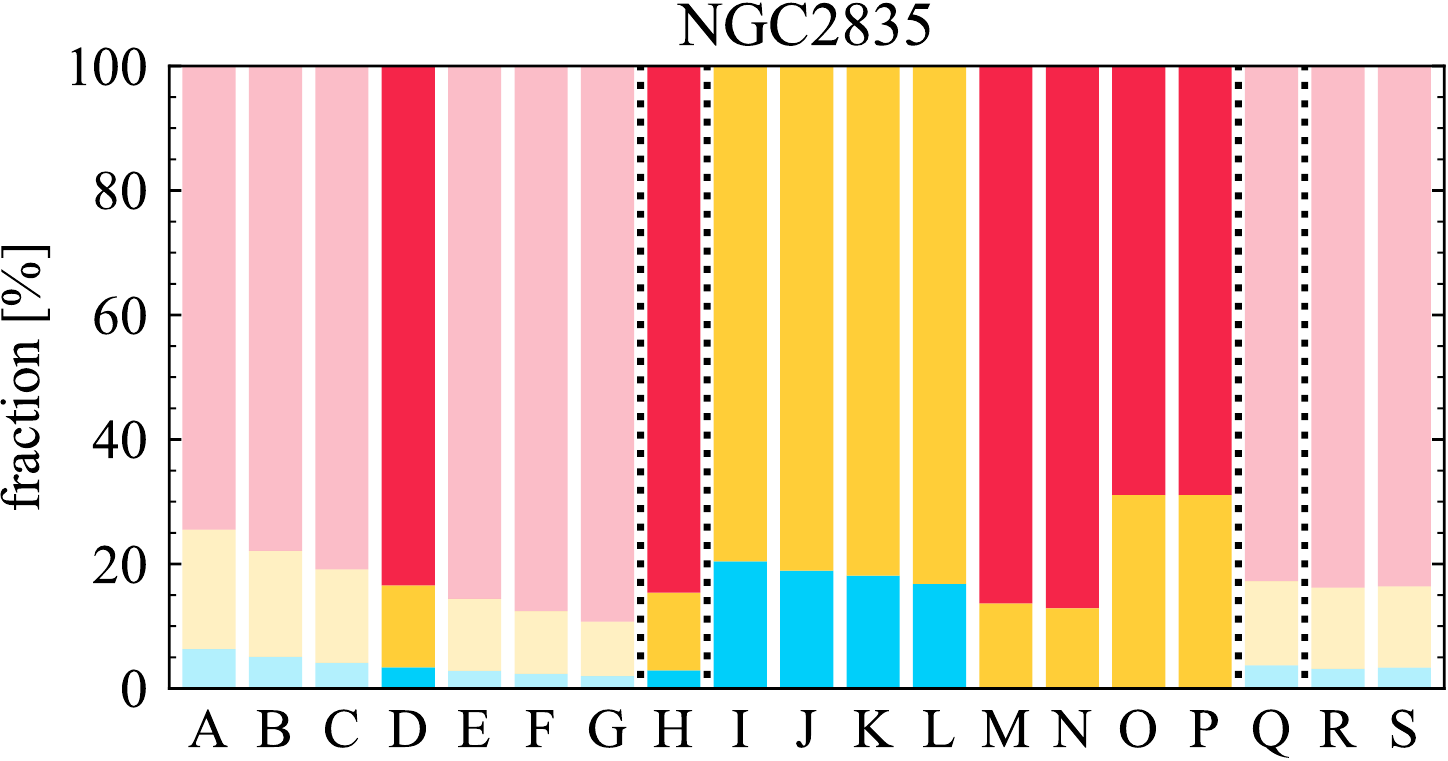}
	\includegraphics[scale=0.52]{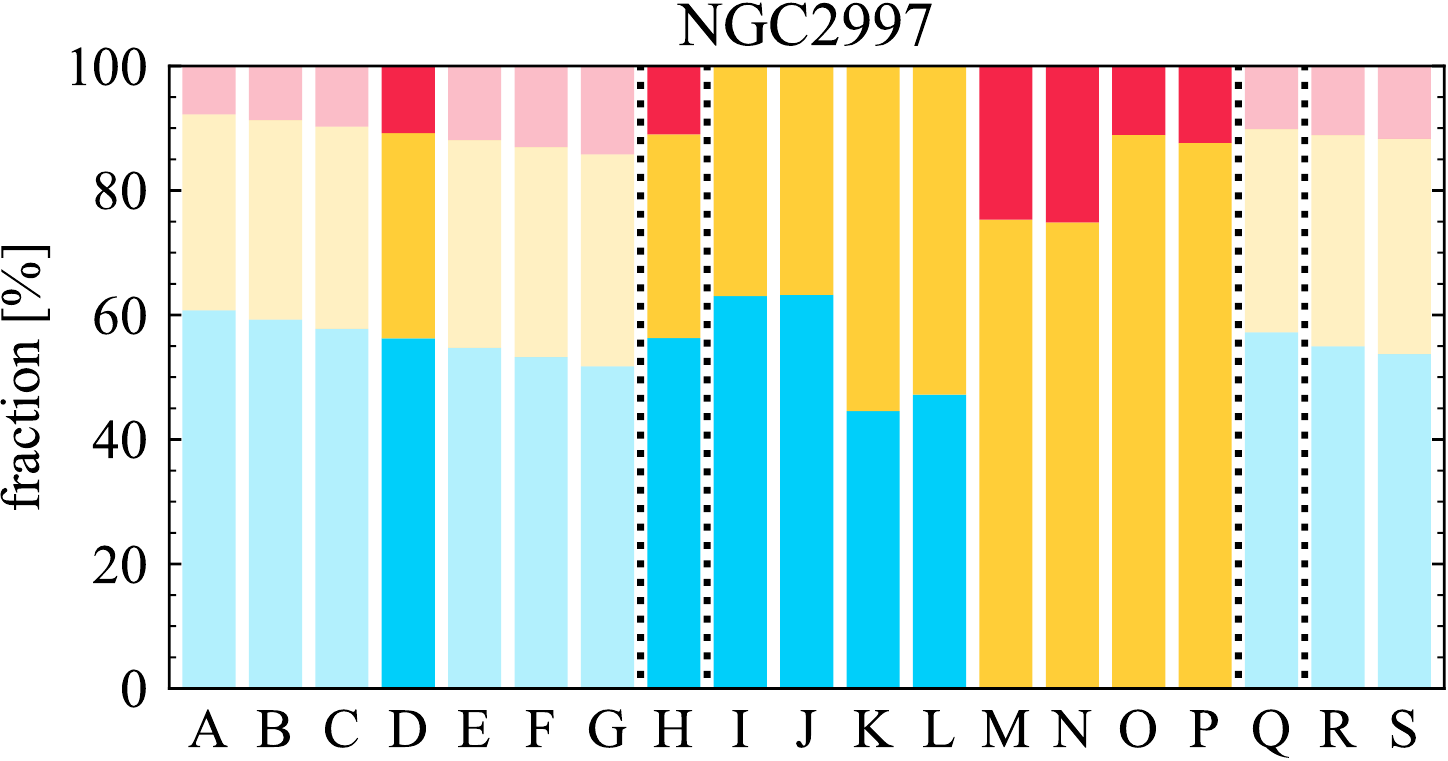}
	\caption{Continued.} 
\end{figure*}

\addtocounter{figure}{-1}
\begin{figure*}
	\centering
	\includegraphics[scale=0.52]{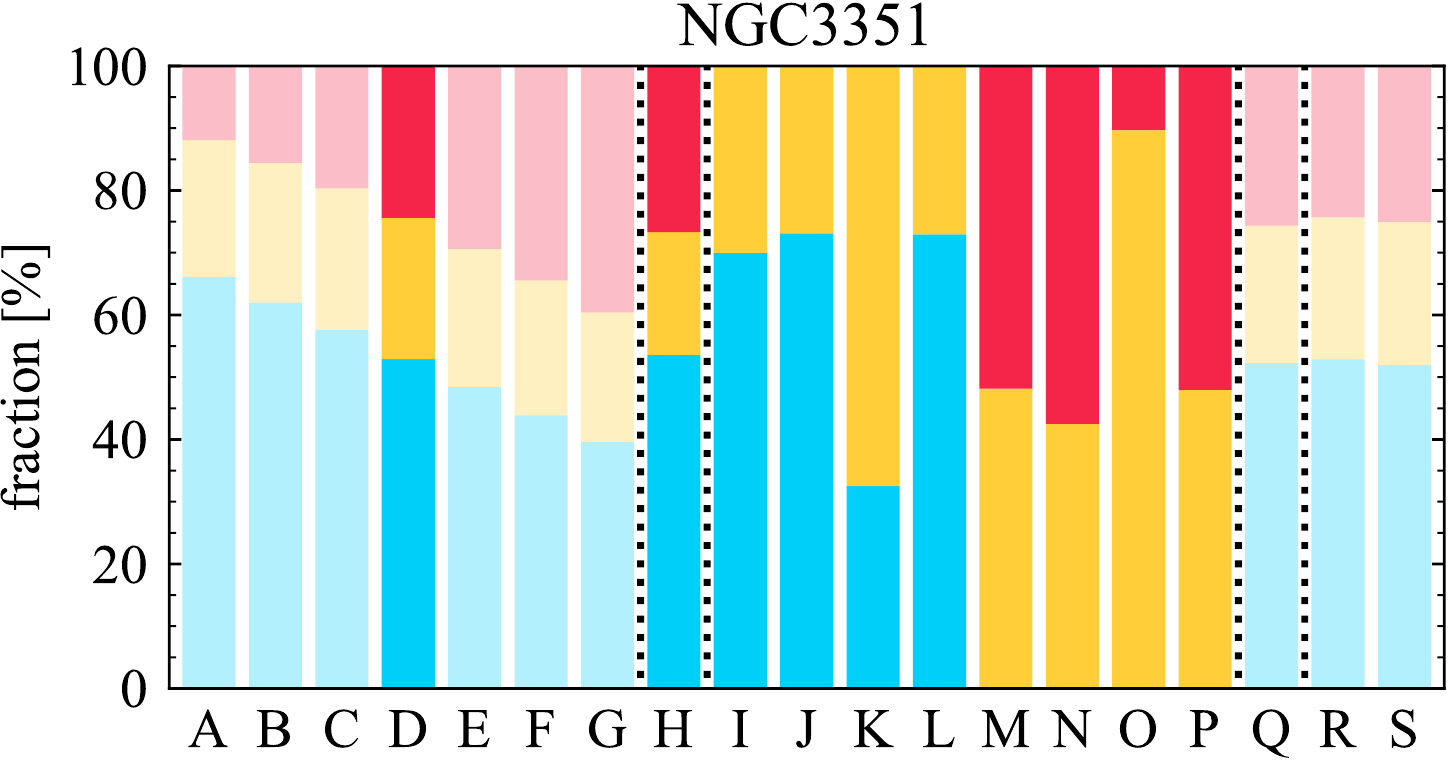}
	\includegraphics[scale=0.52]{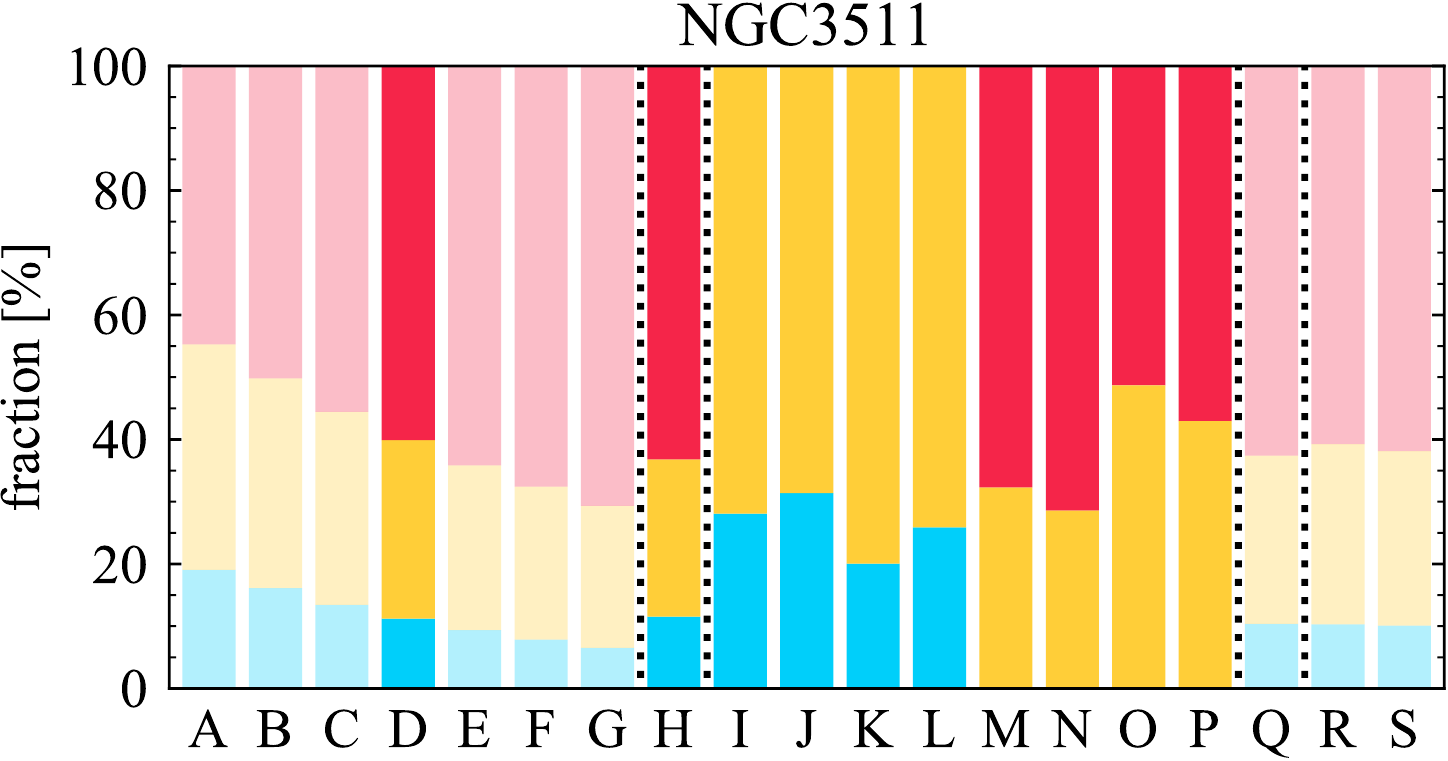}\\
	\vspace{10pt}
	\includegraphics[scale=0.52]{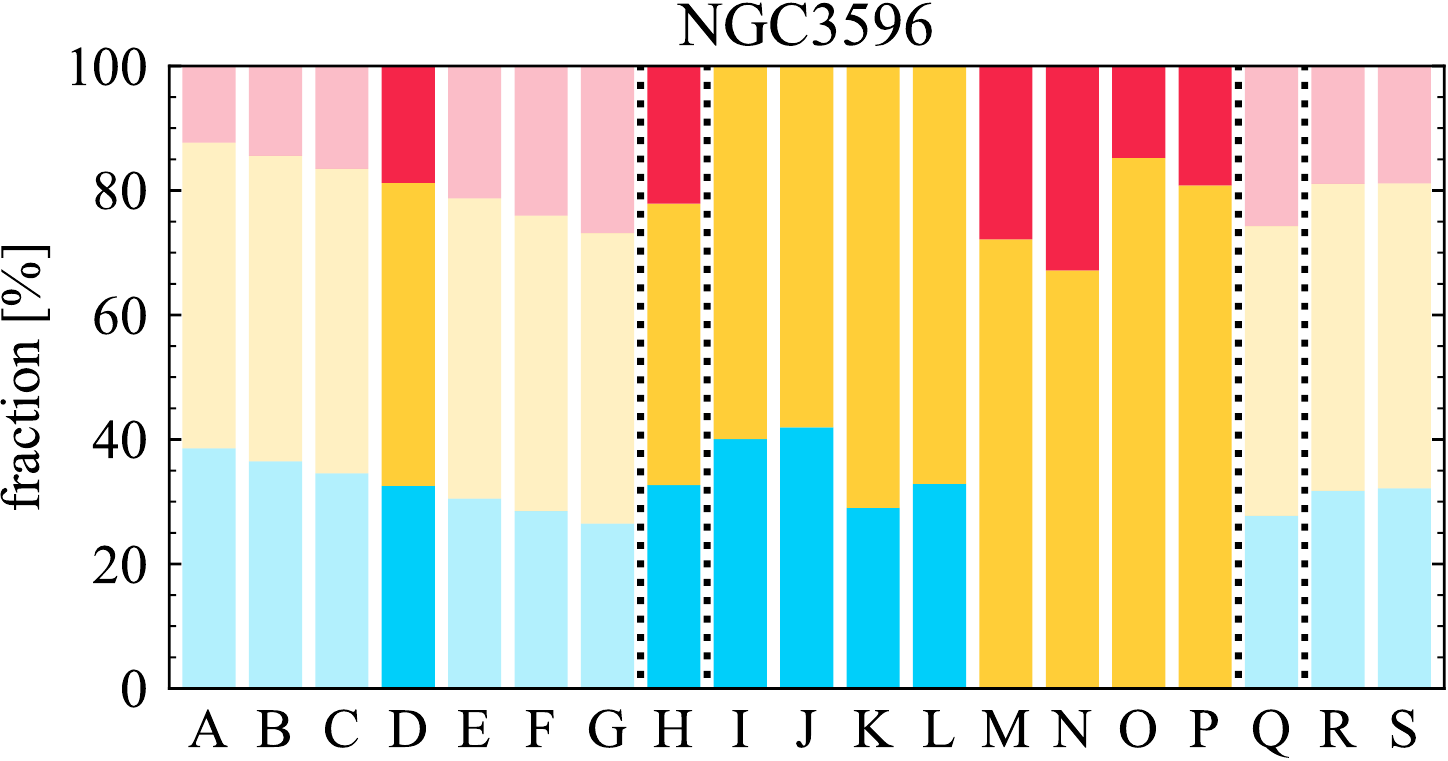}
	\includegraphics[scale=0.52]{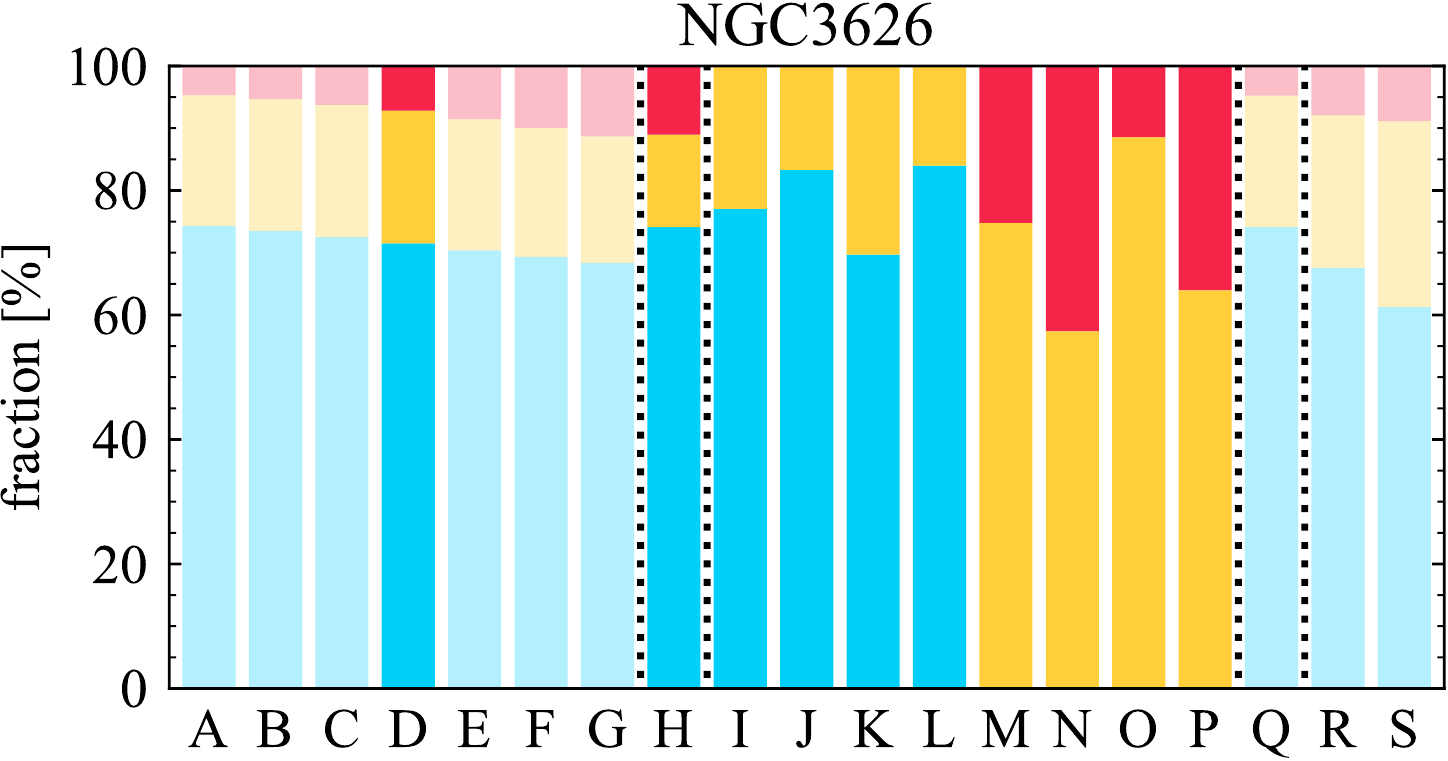}\\
	\vspace{10pt}
	\includegraphics[scale=0.52]{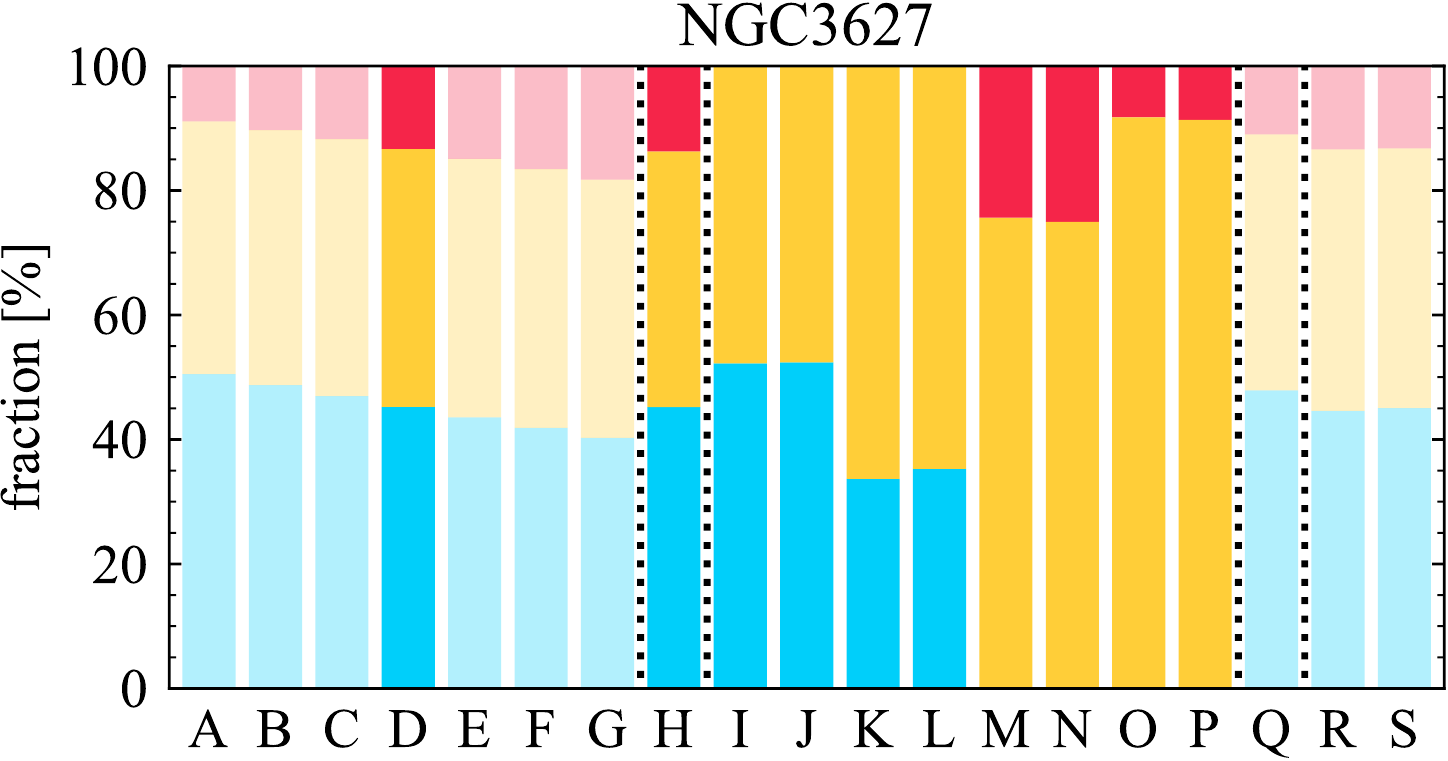}
	\includegraphics[scale=0.52]{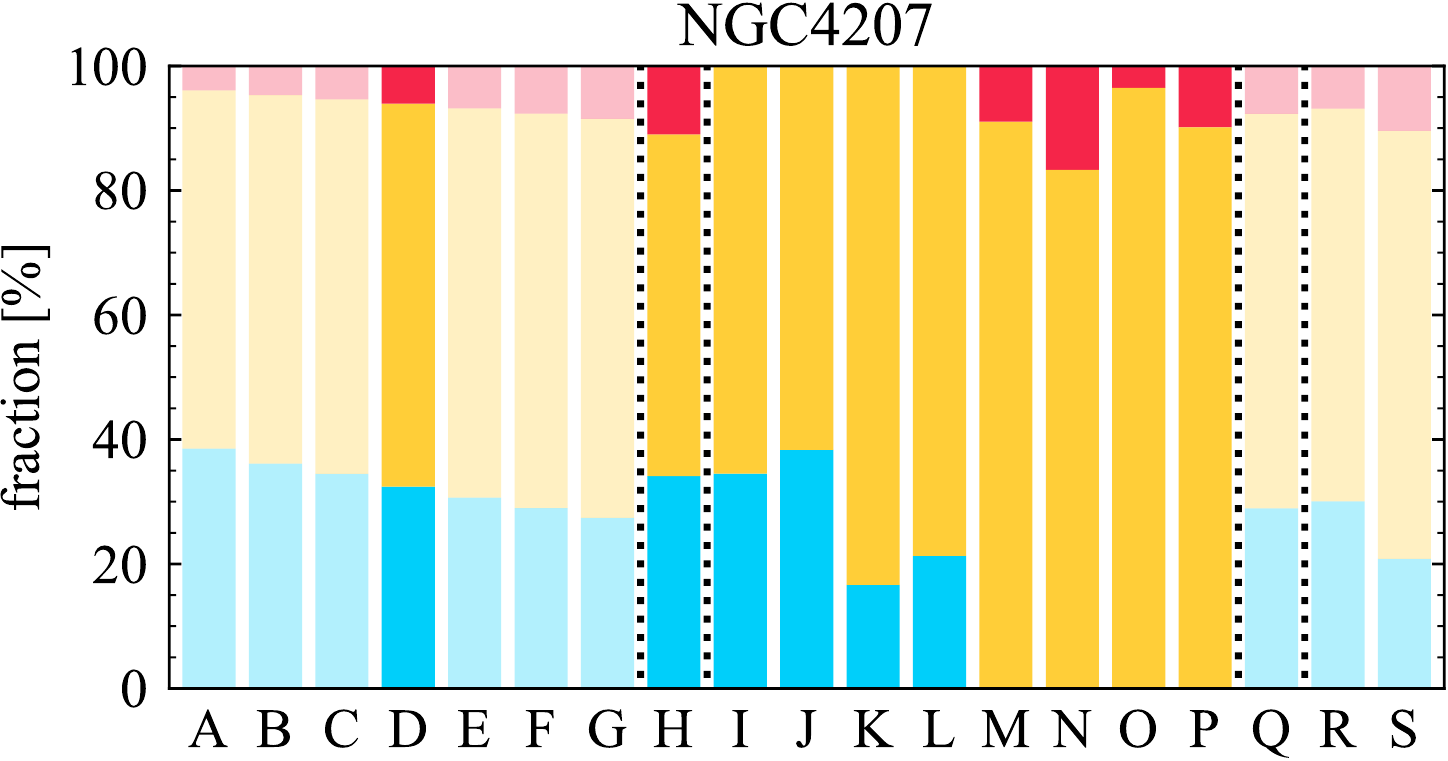}\\
	\vspace{10pt}
	\includegraphics[scale=0.52]{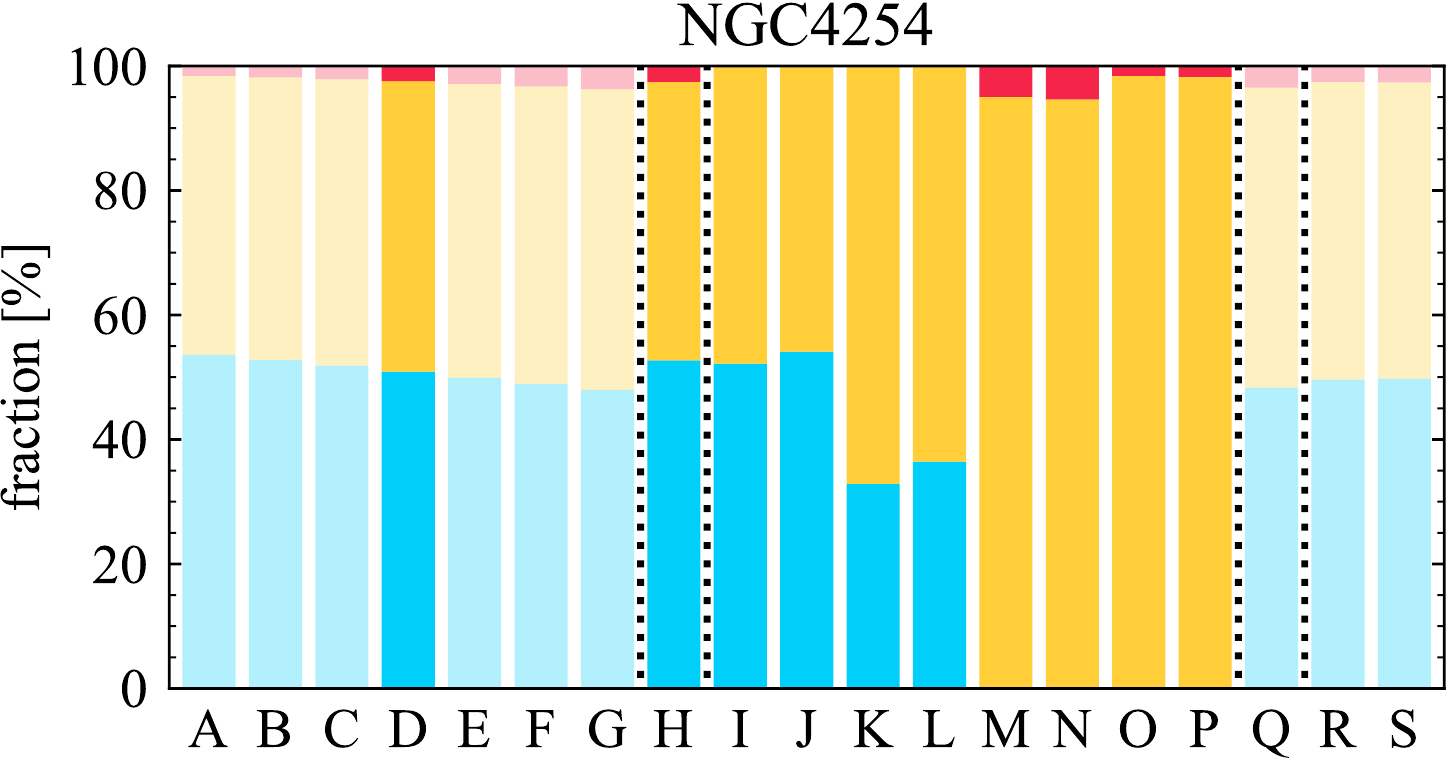}
	\includegraphics[scale=0.52]{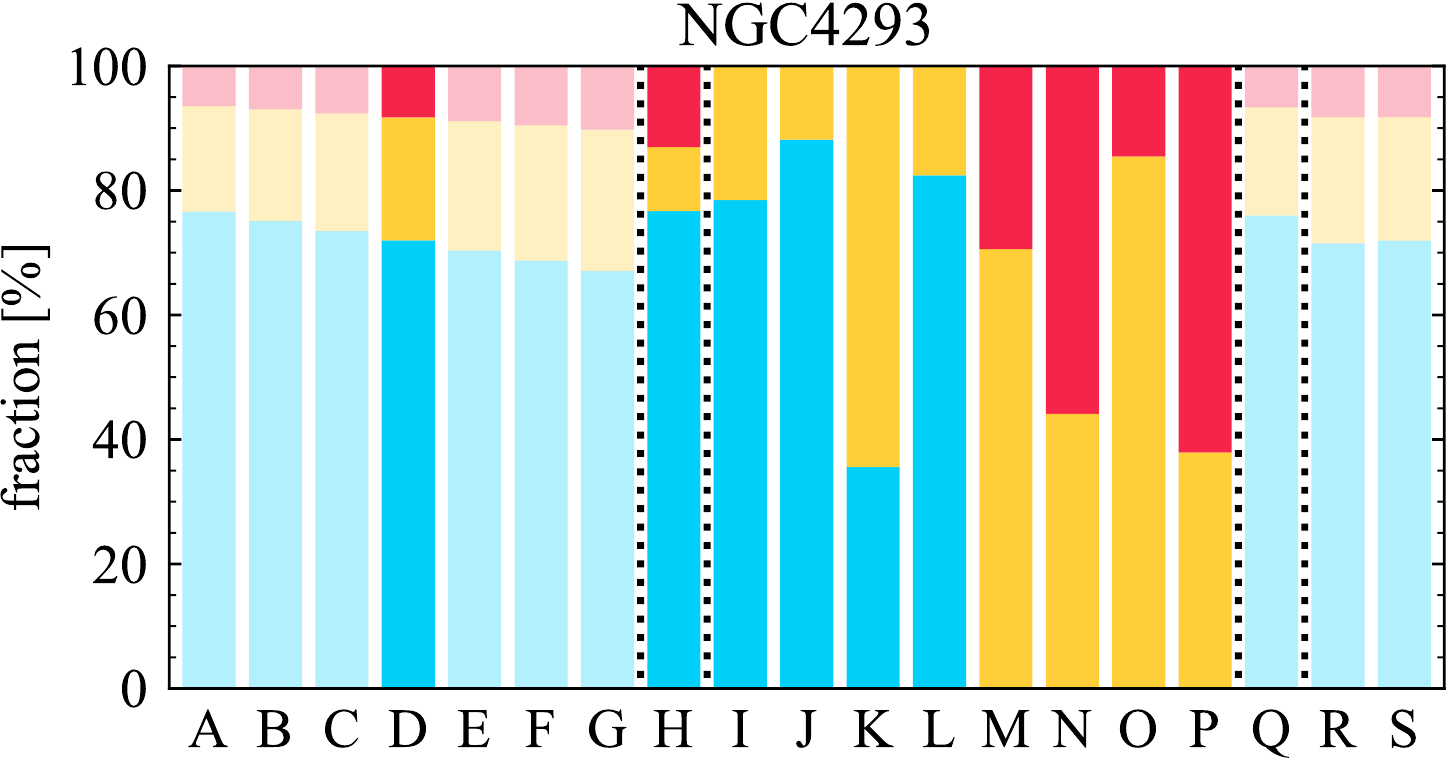}\\
	\vspace{10pt}
	\includegraphics[scale=0.52]{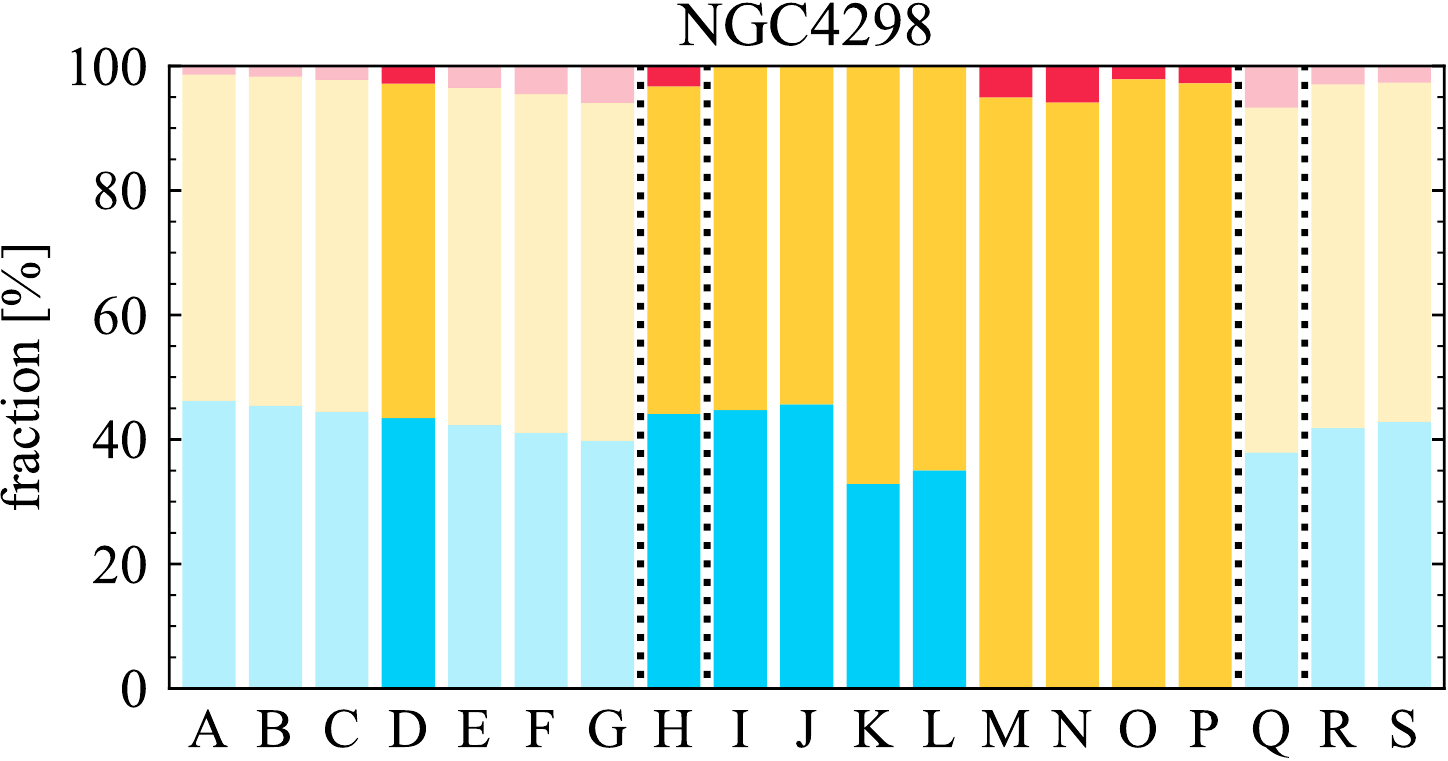}
	\includegraphics[scale=0.52]{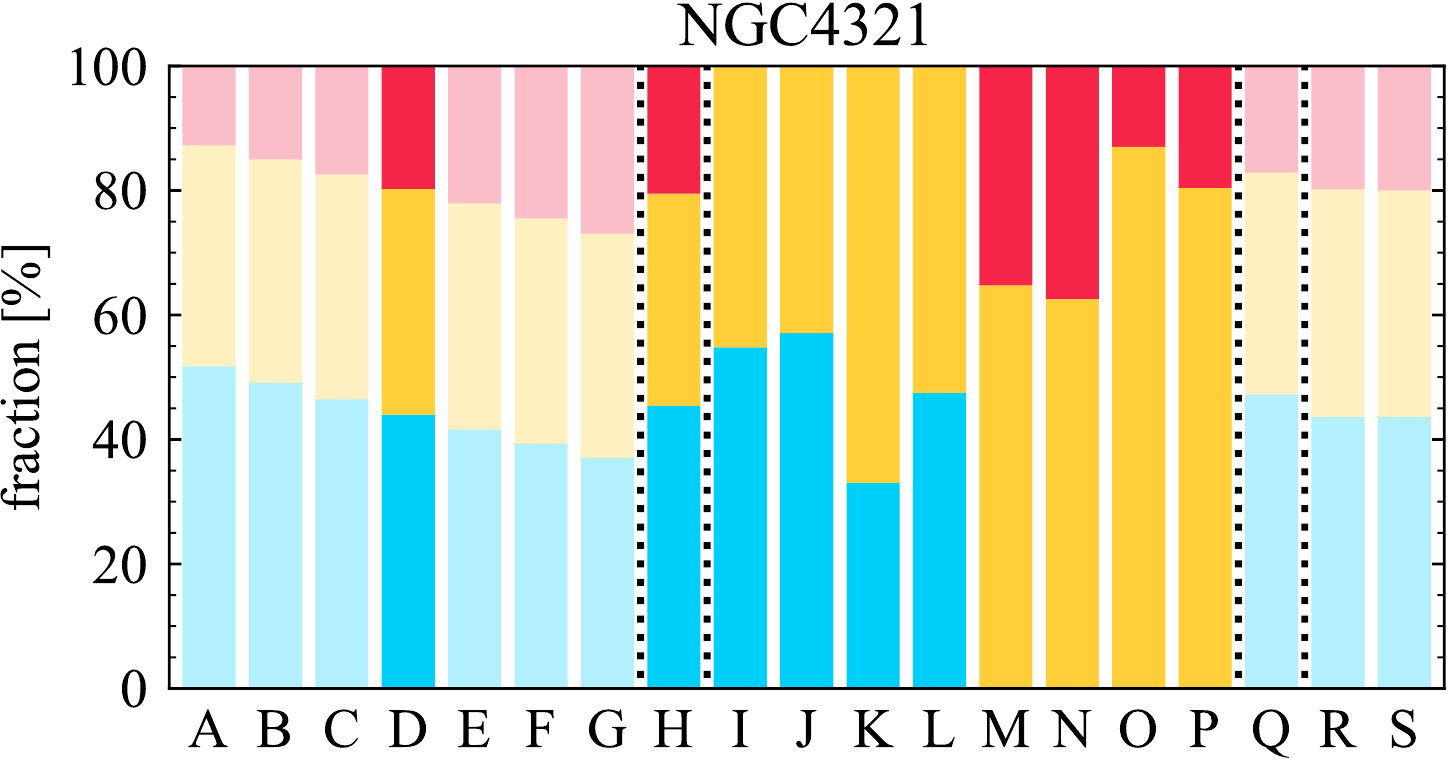}
	\caption{Continued.} 
\end{figure*}

\addtocounter{figure}{-1}
\begin{figure*}
	\centering
	\includegraphics[scale=0.52]{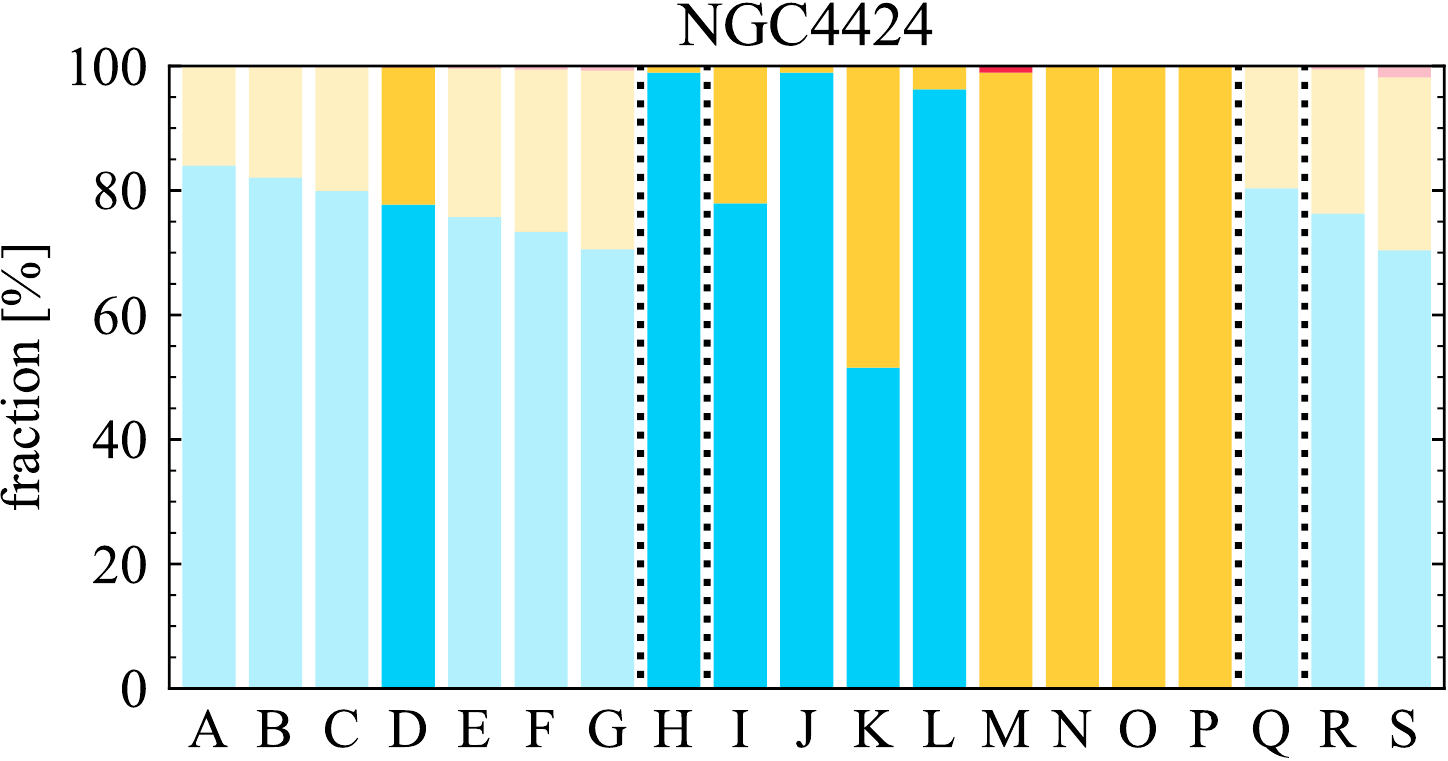}
	\includegraphics[scale=0.52]{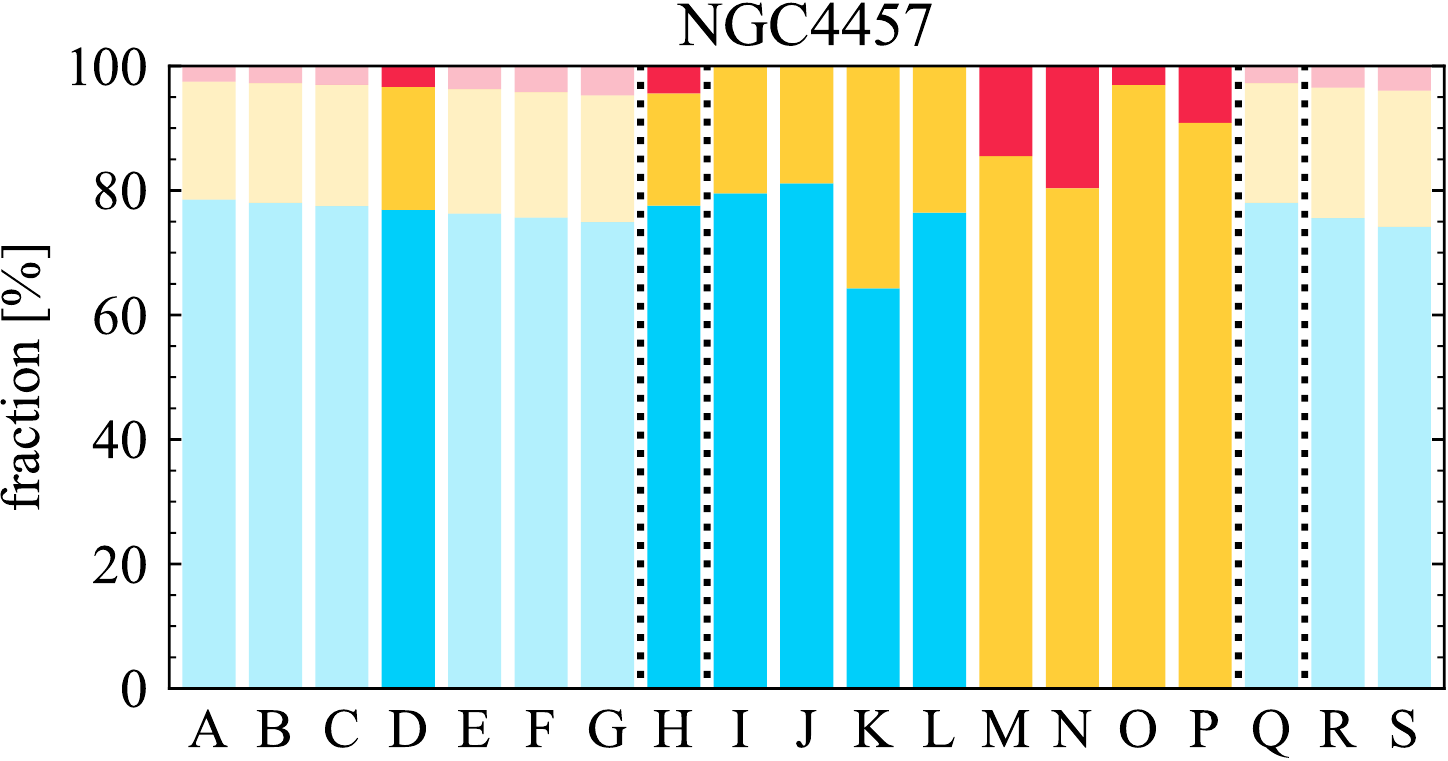}\\
	\vspace{10pt}
	\includegraphics[scale=0.52]{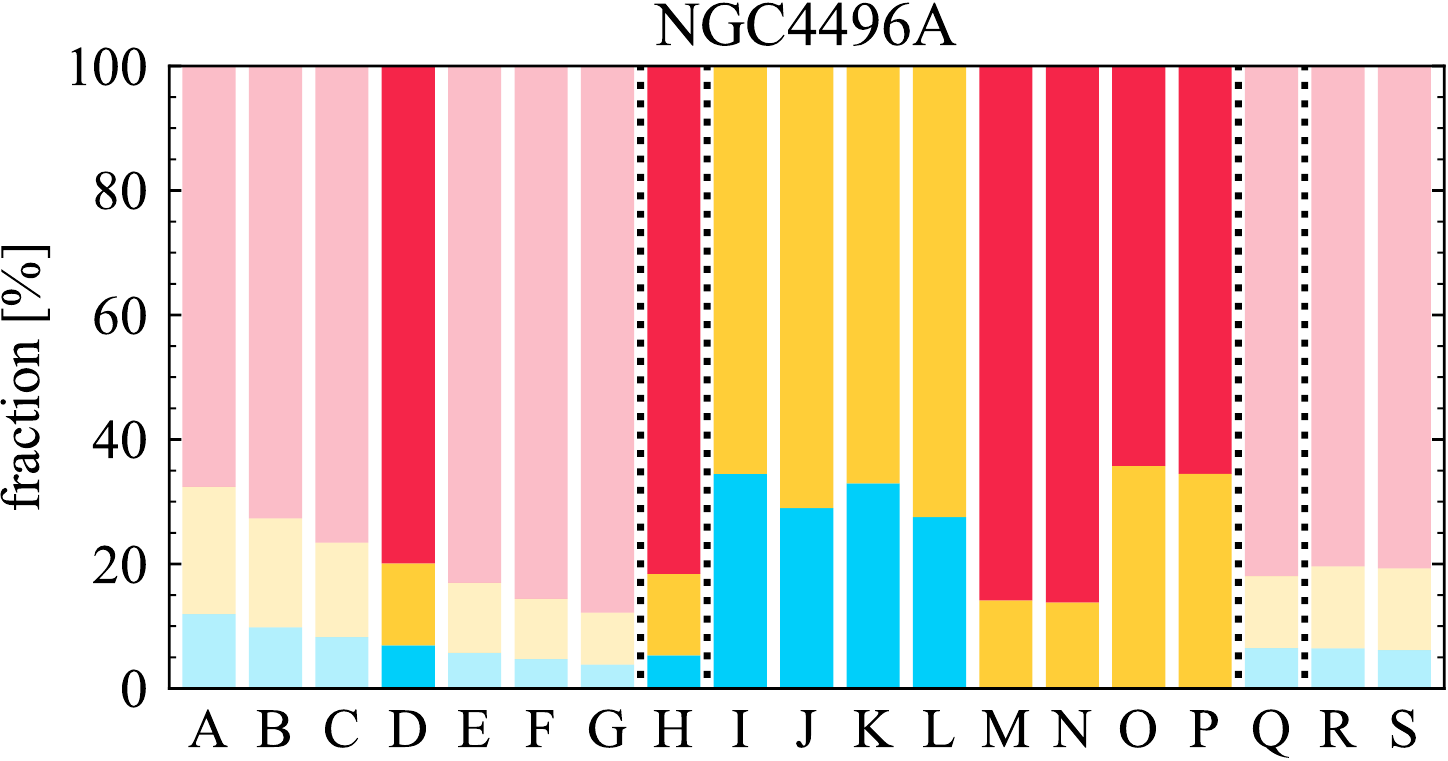}
	\includegraphics[scale=0.52]{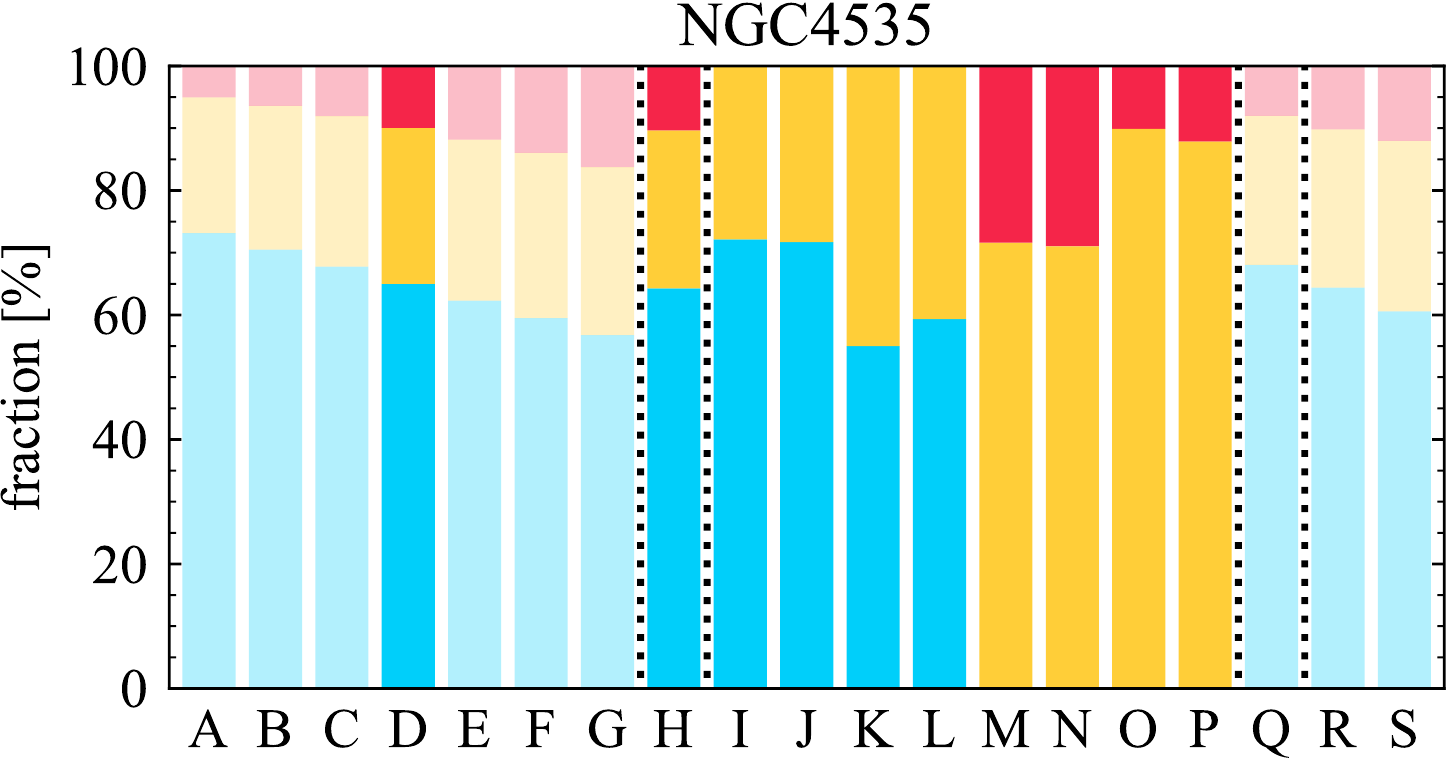}\\
	\vspace{10pt}
	\includegraphics[scale=0.52]{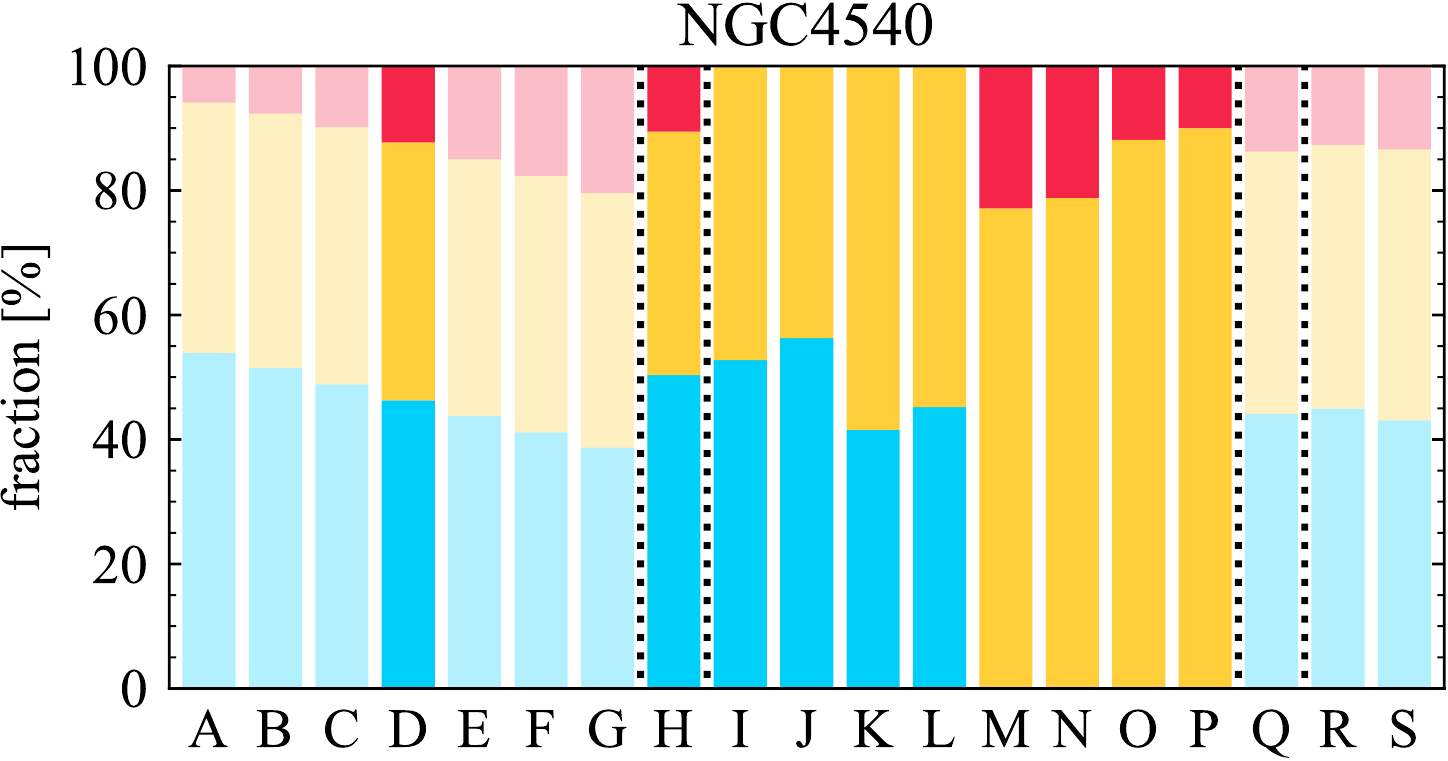}
	\includegraphics[scale=0.52]{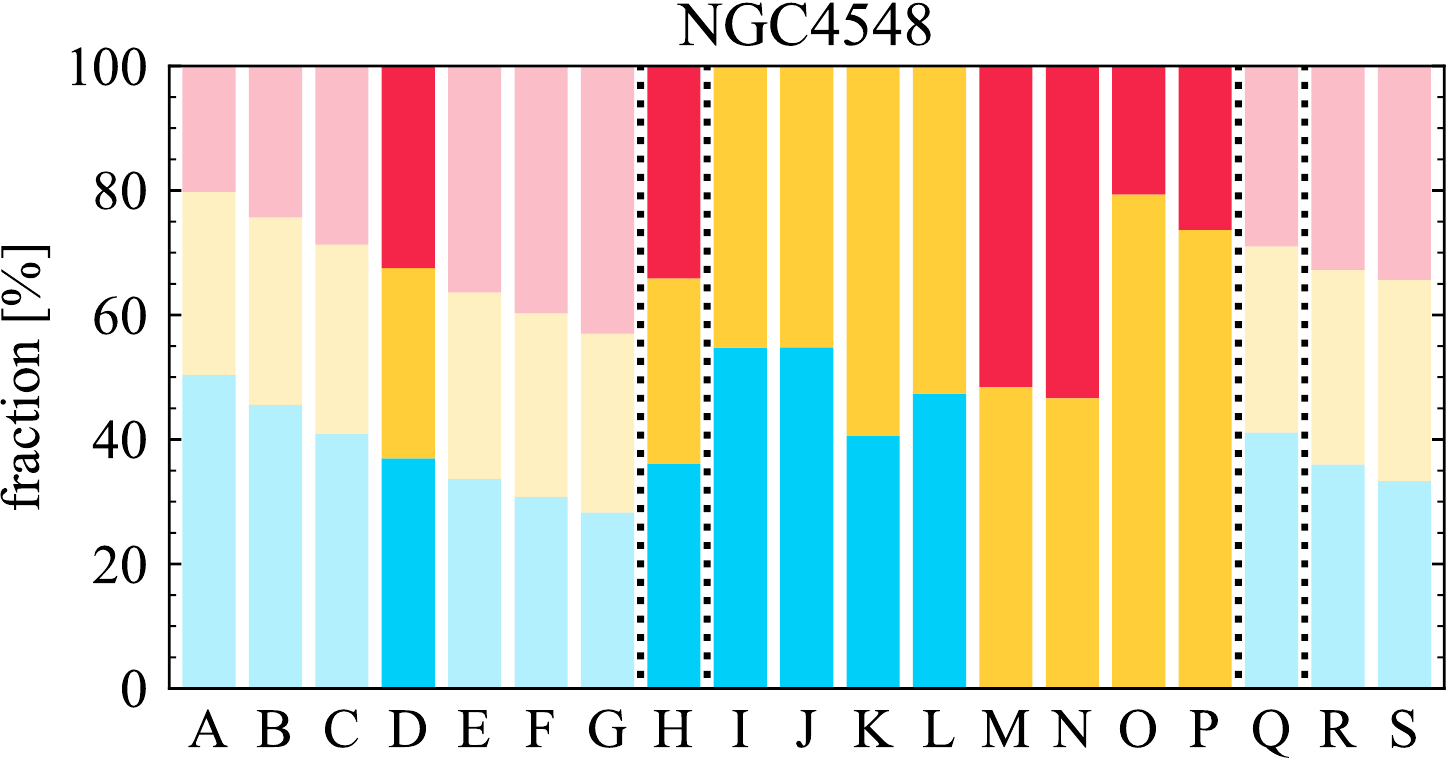}\\
	\vspace{10pt}
	\includegraphics[scale=0.52]{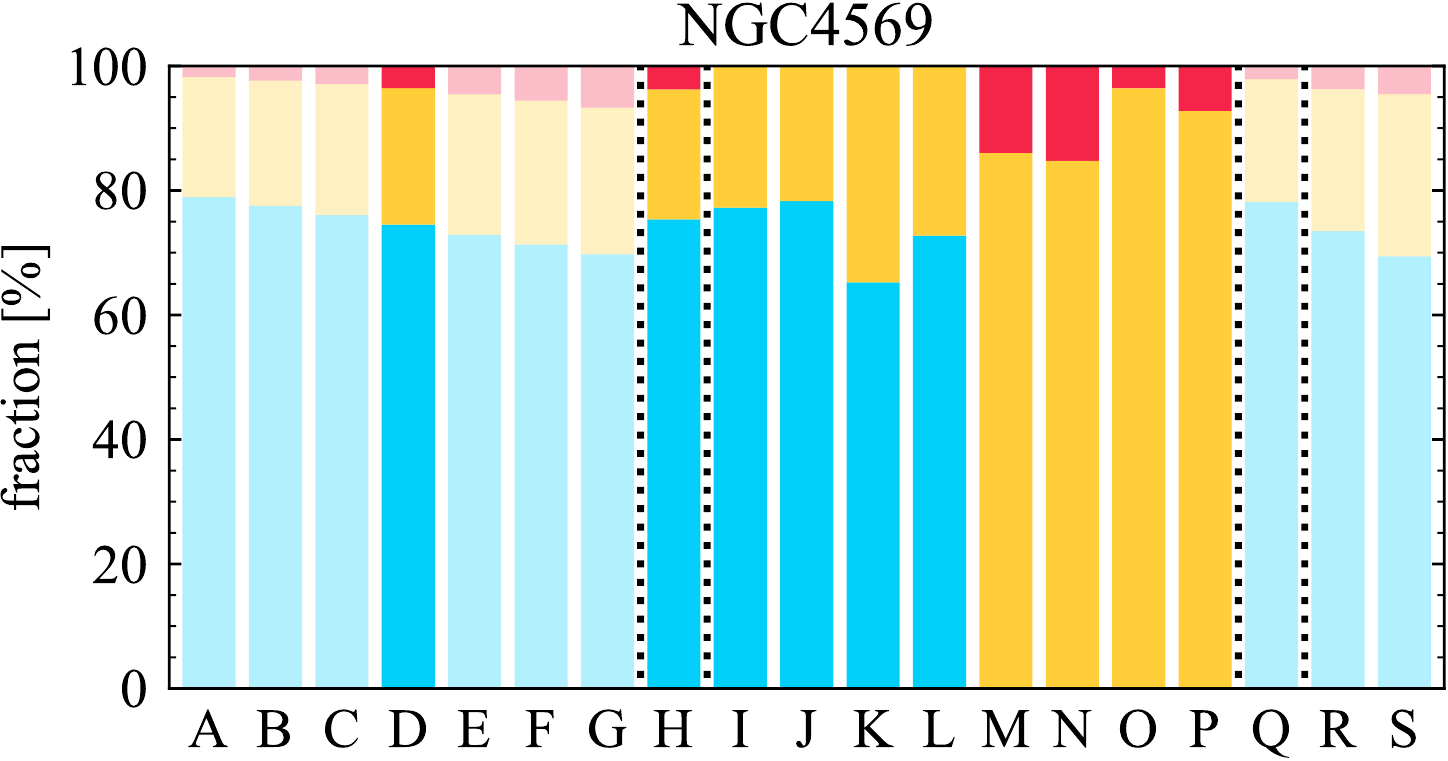}
	\includegraphics[scale=0.52]{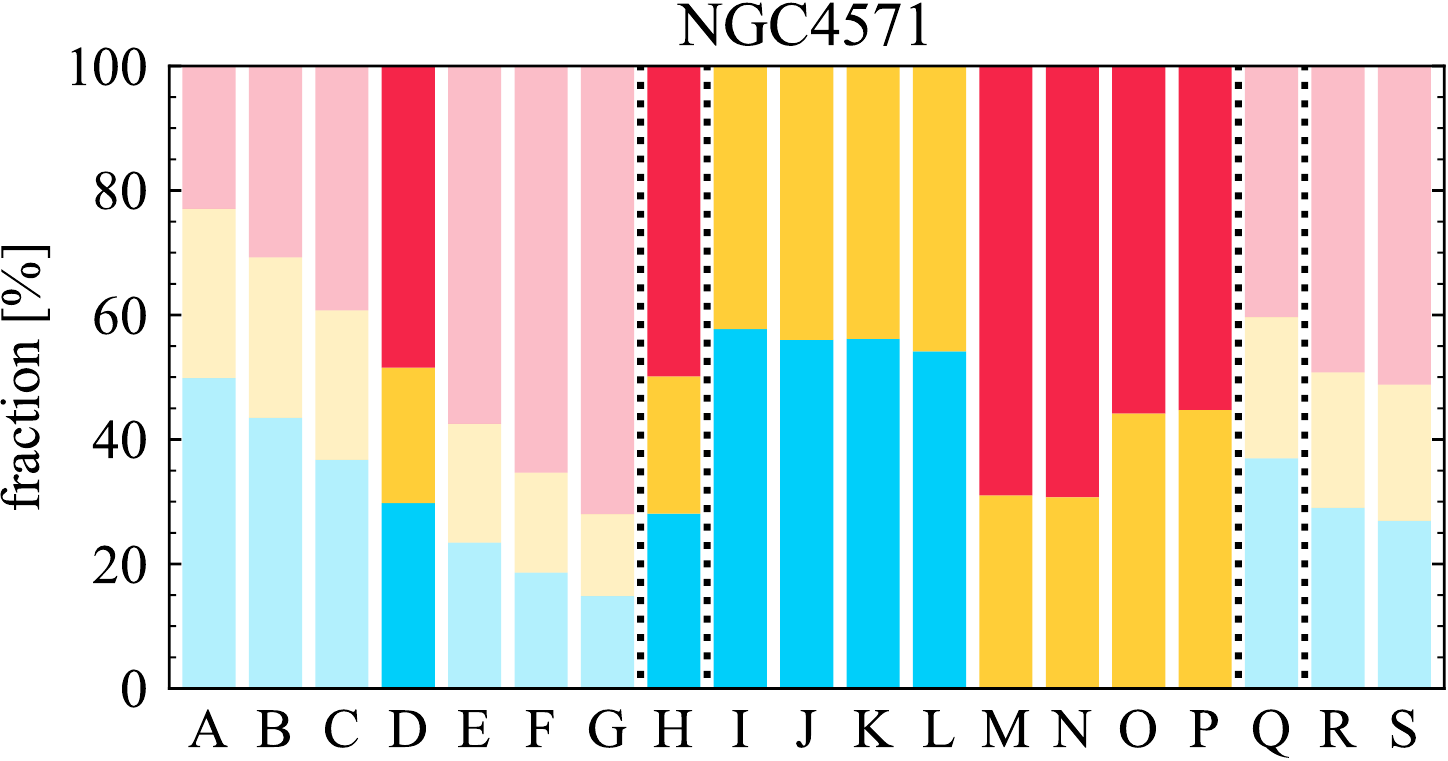}\\
	\vspace{10pt}
	\includegraphics[scale=0.52]{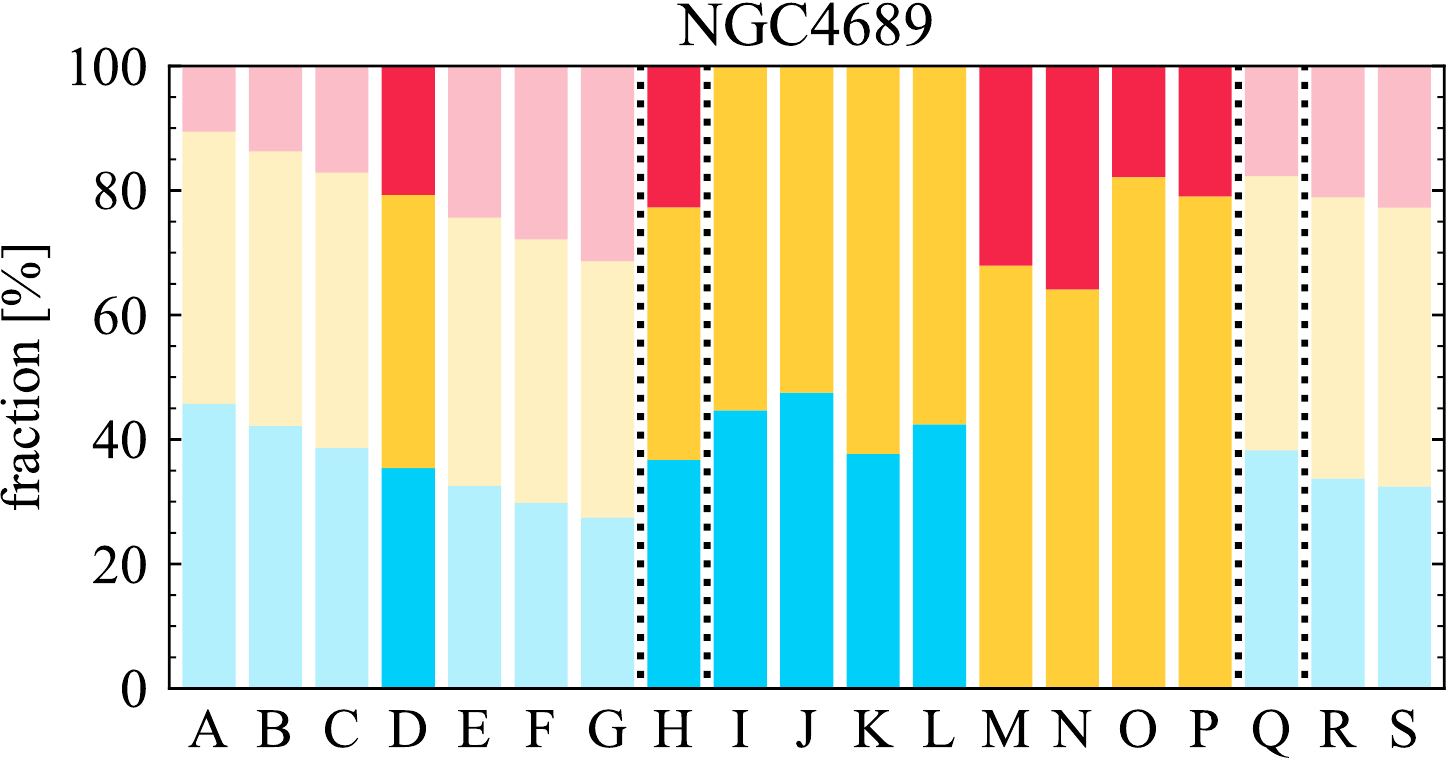}
	\includegraphics[scale=0.52]{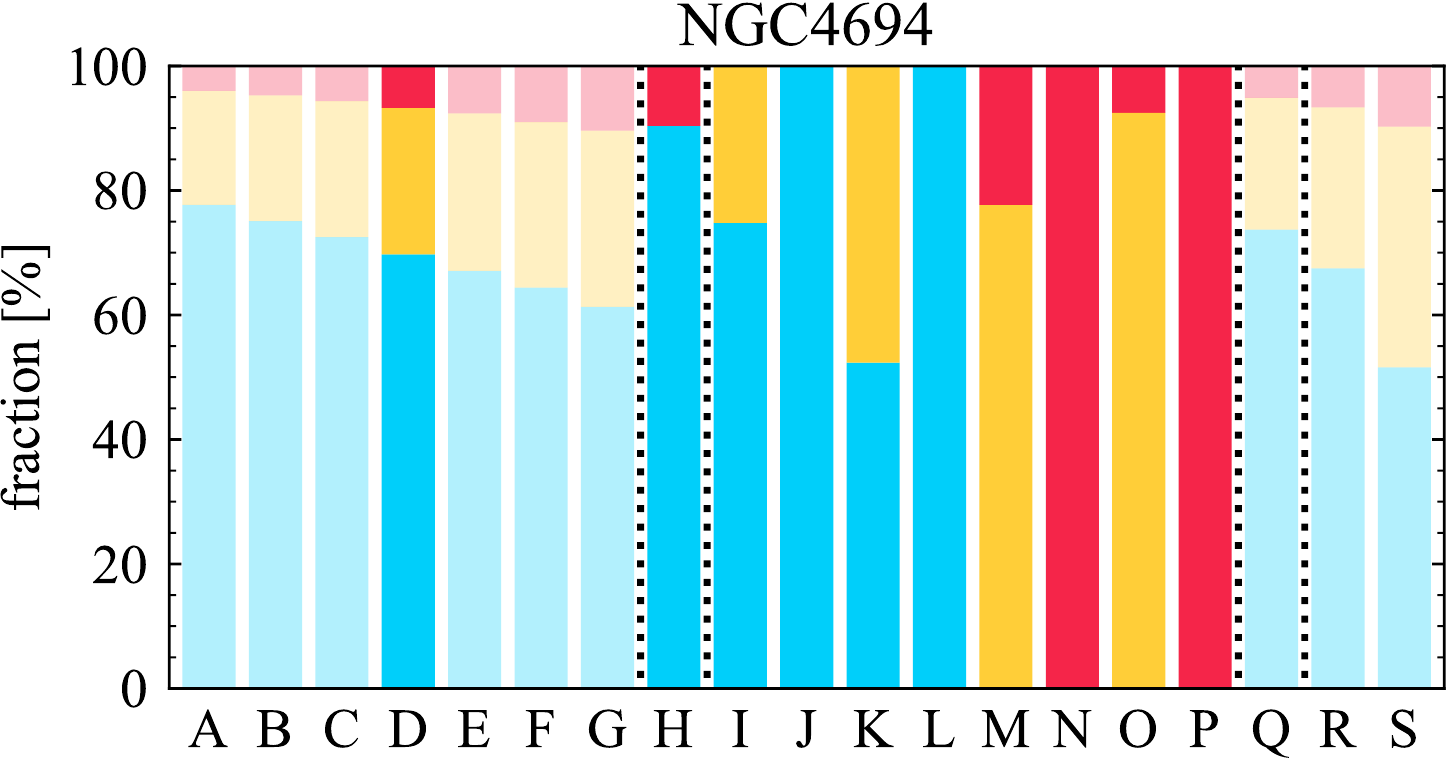}
	\caption{Continued.} 
\end{figure*}

\addtocounter{figure}{-1}
\begin{figure*}
	\centering
	\addtocounter{subfigure}{36}
	\includegraphics[scale=0.52]{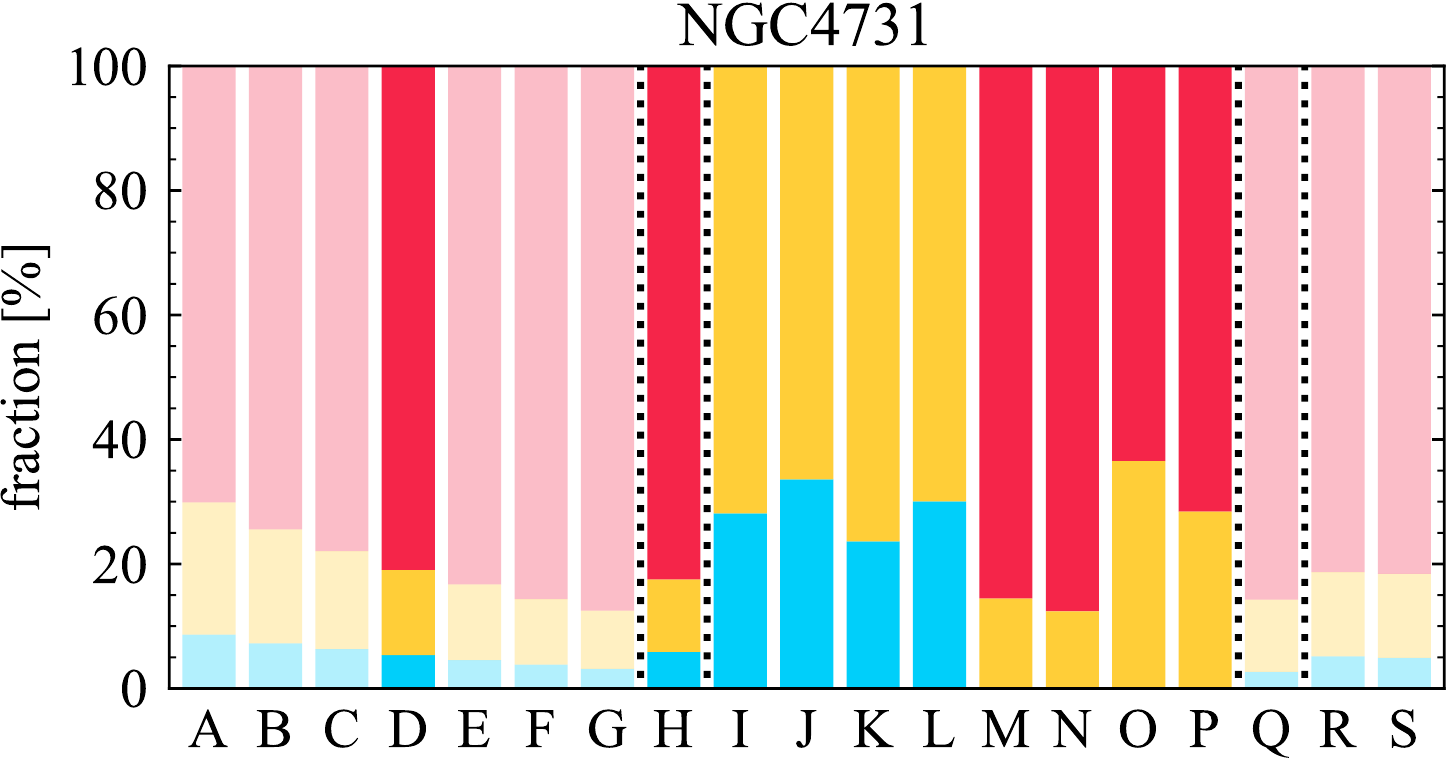}
	\includegraphics[scale=0.52]{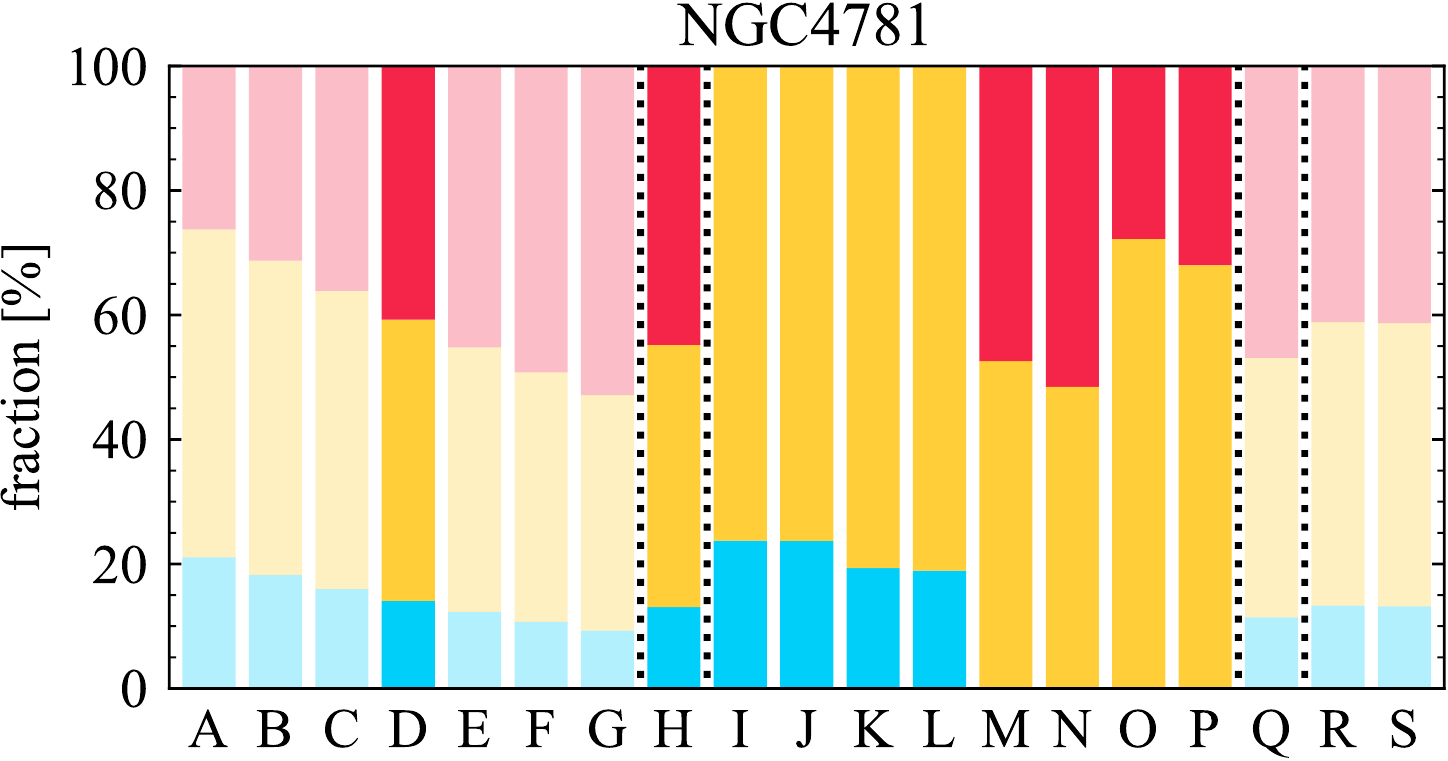}\\
	\vspace{10pt}
	\includegraphics[scale=0.52]{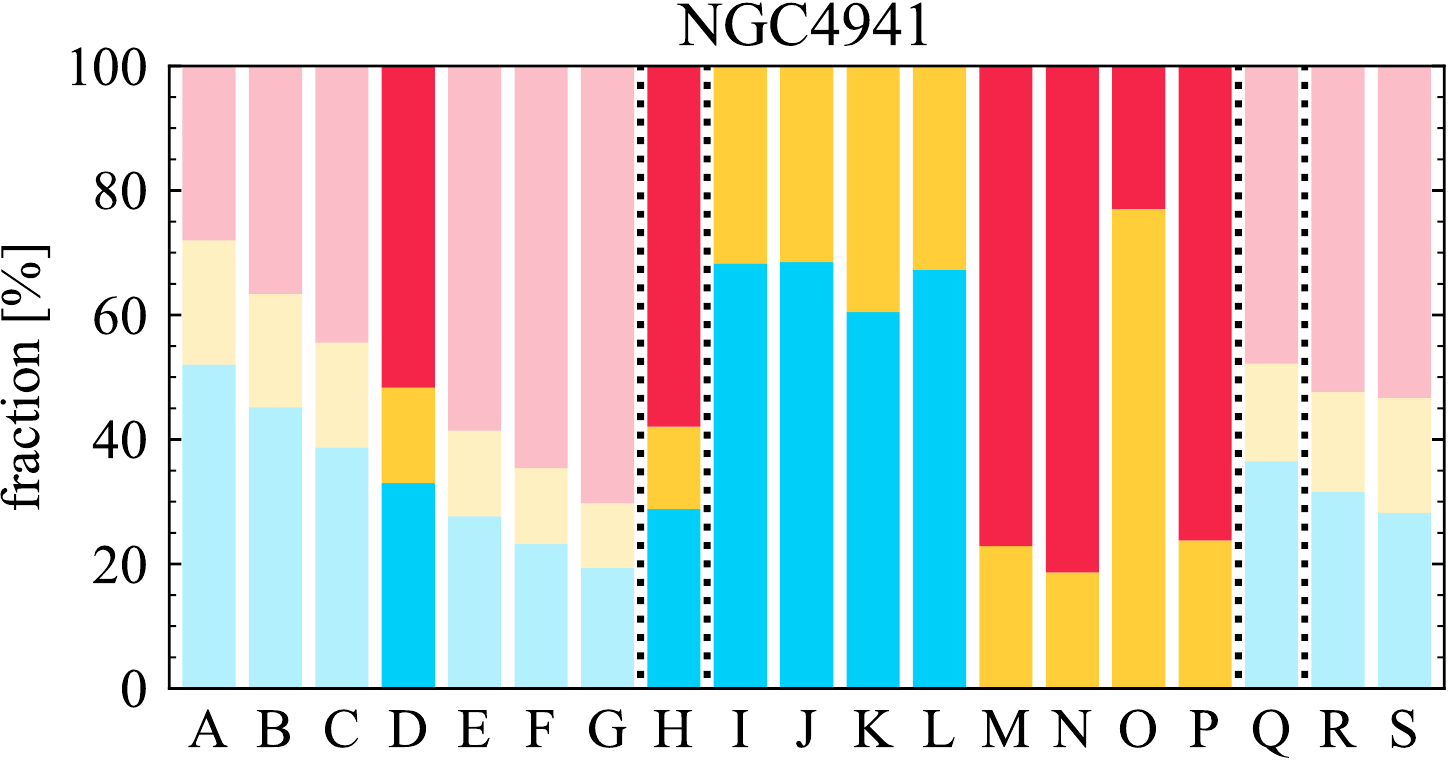}
	\includegraphics[scale=0.52]{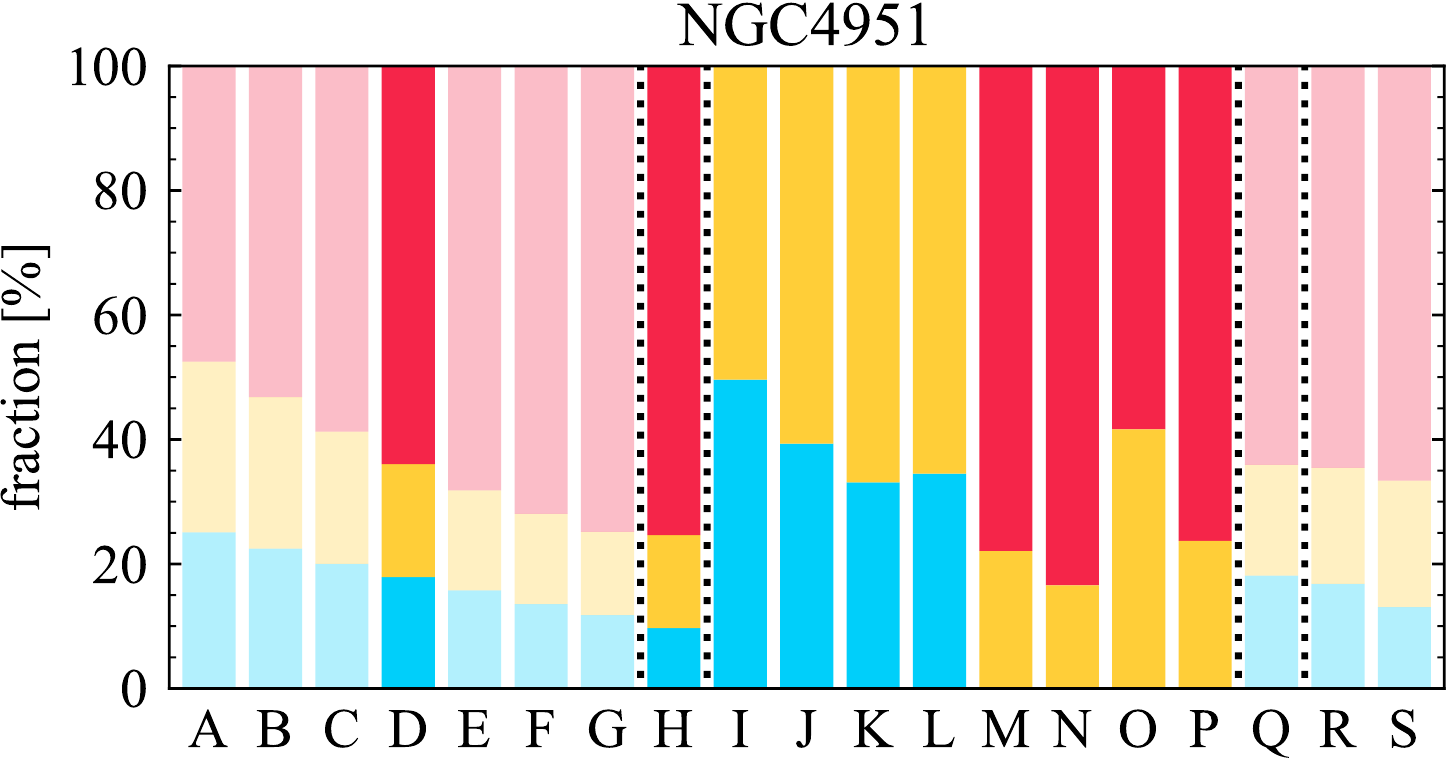}\\
	\vspace{10pt}
	\includegraphics[scale=0.52]{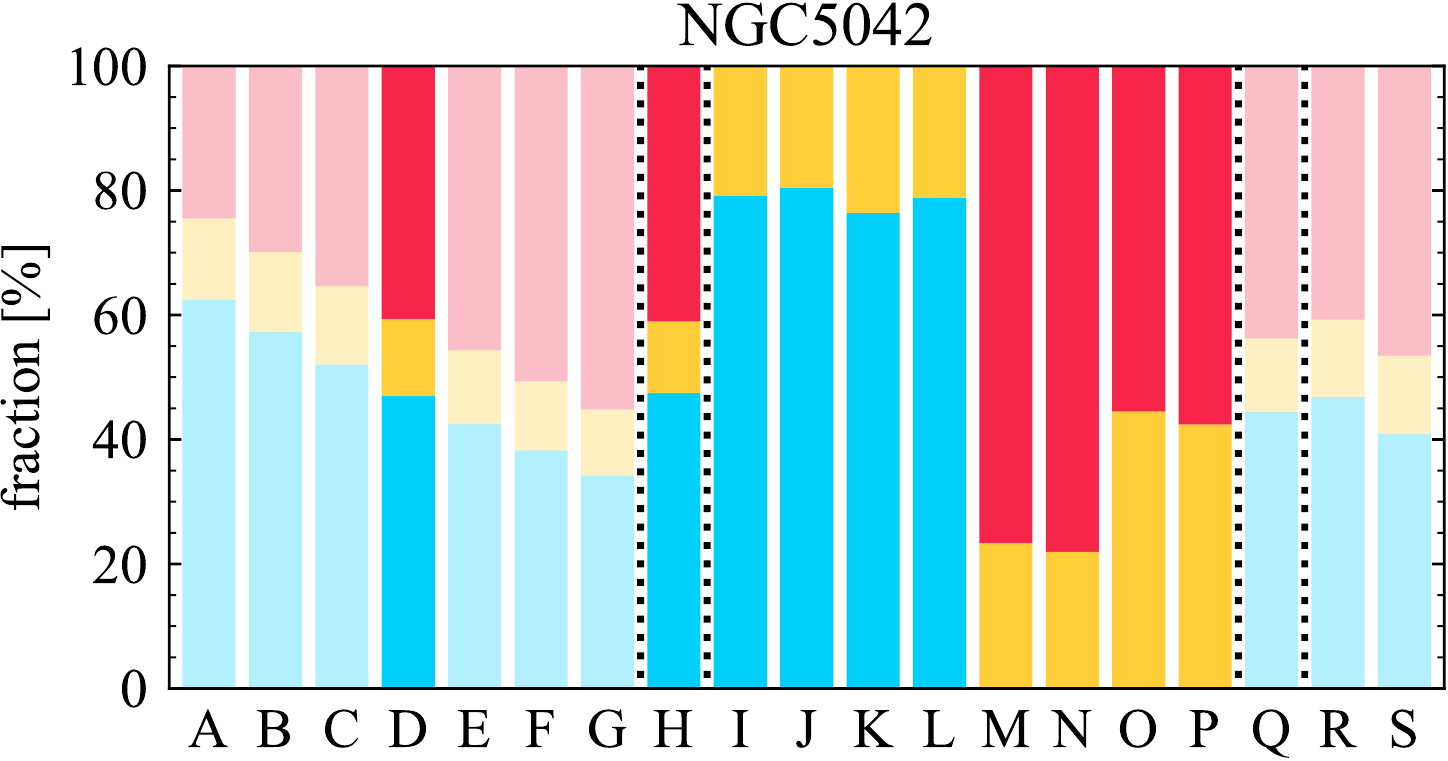}
	\includegraphics[scale=0.52]{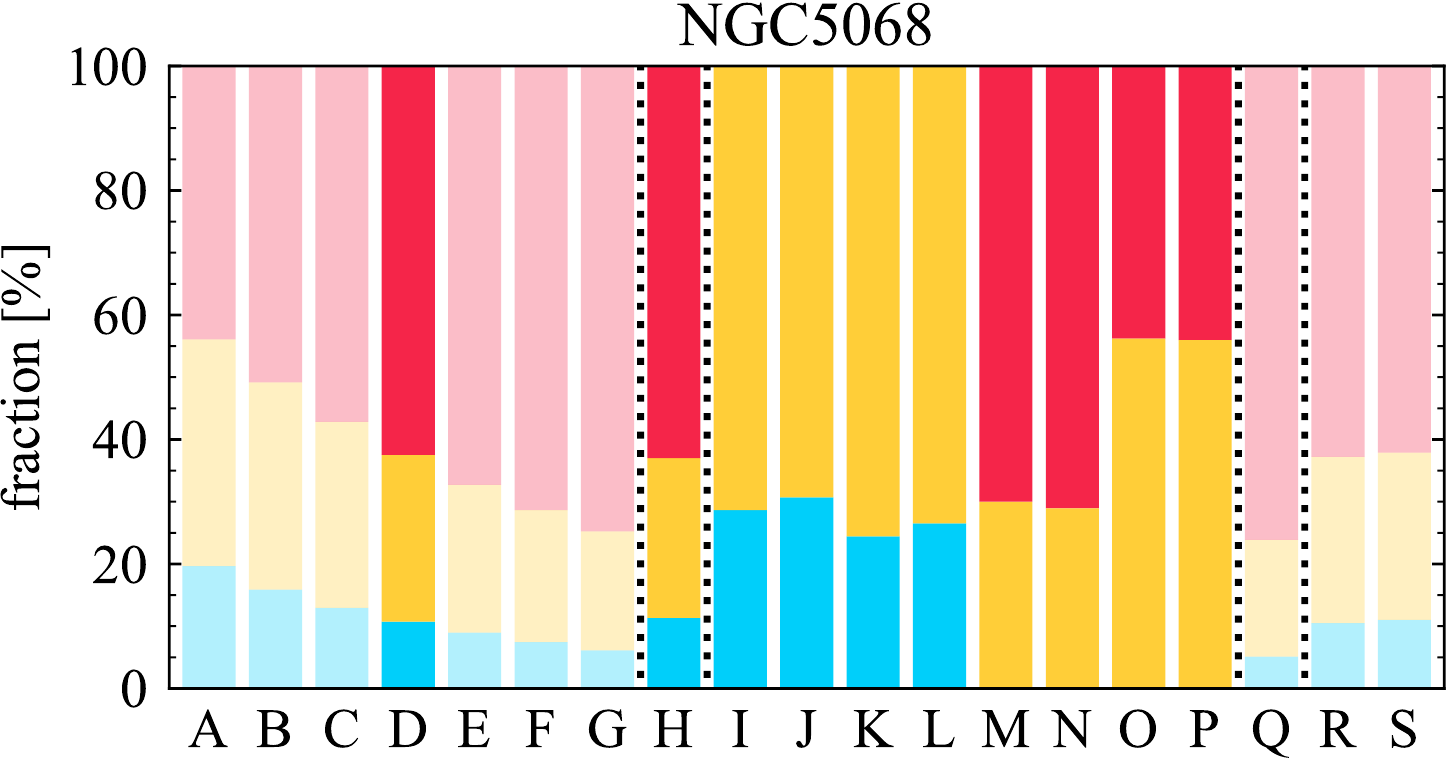}\\
	\vspace{10pt}
	\includegraphics[scale=0.52]{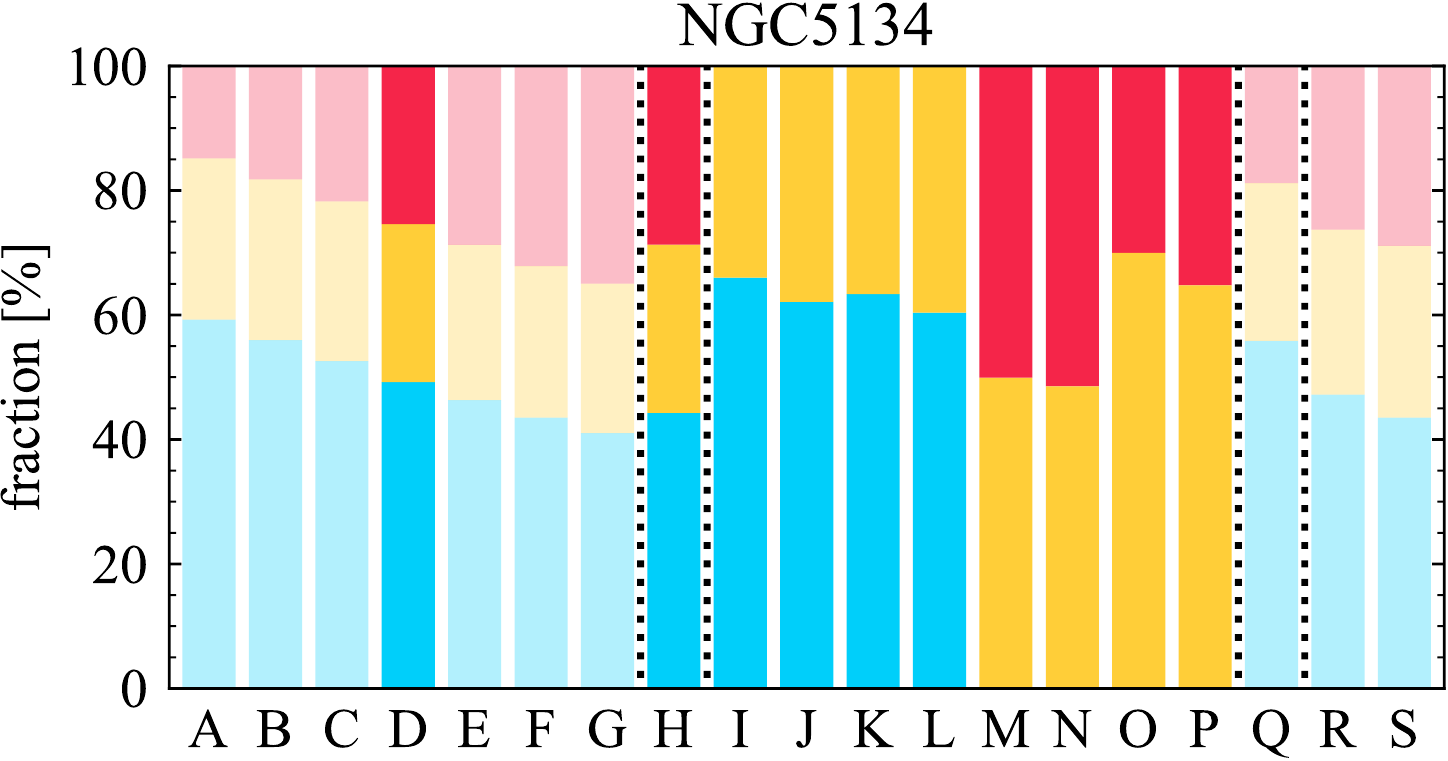}
	\includegraphics[scale=0.52]{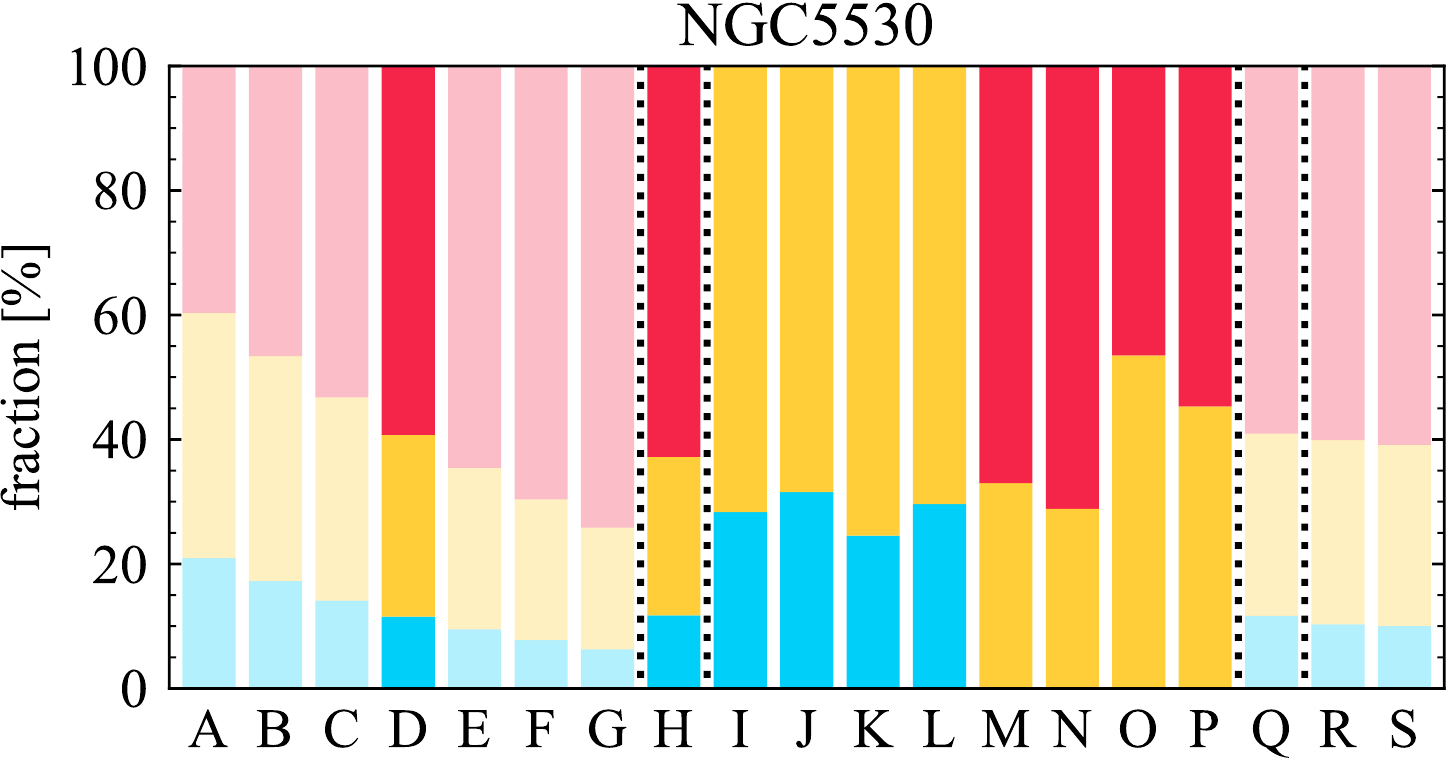}\\
	\vspace{10pt}
	\includegraphics[scale=0.52]{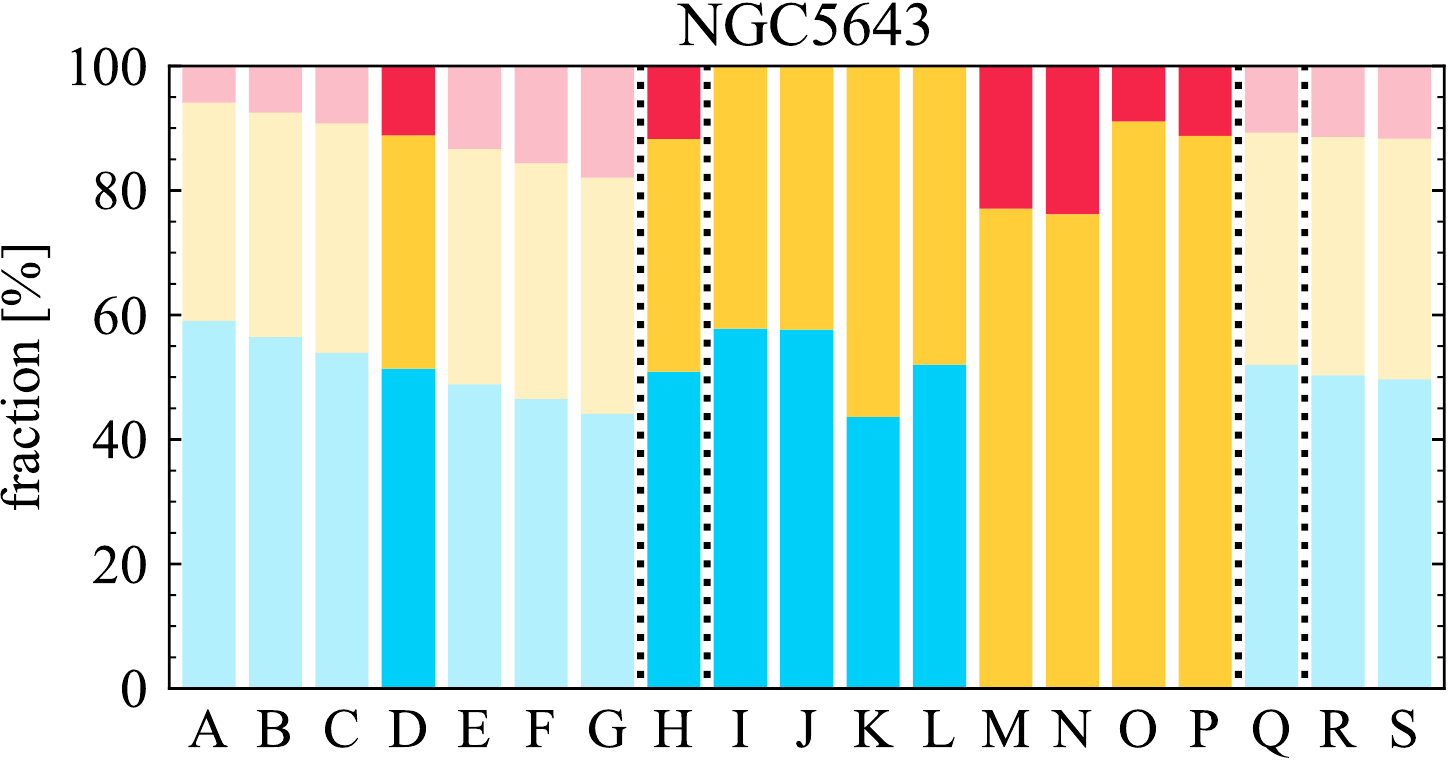}
	\includegraphics[scale=0.52]{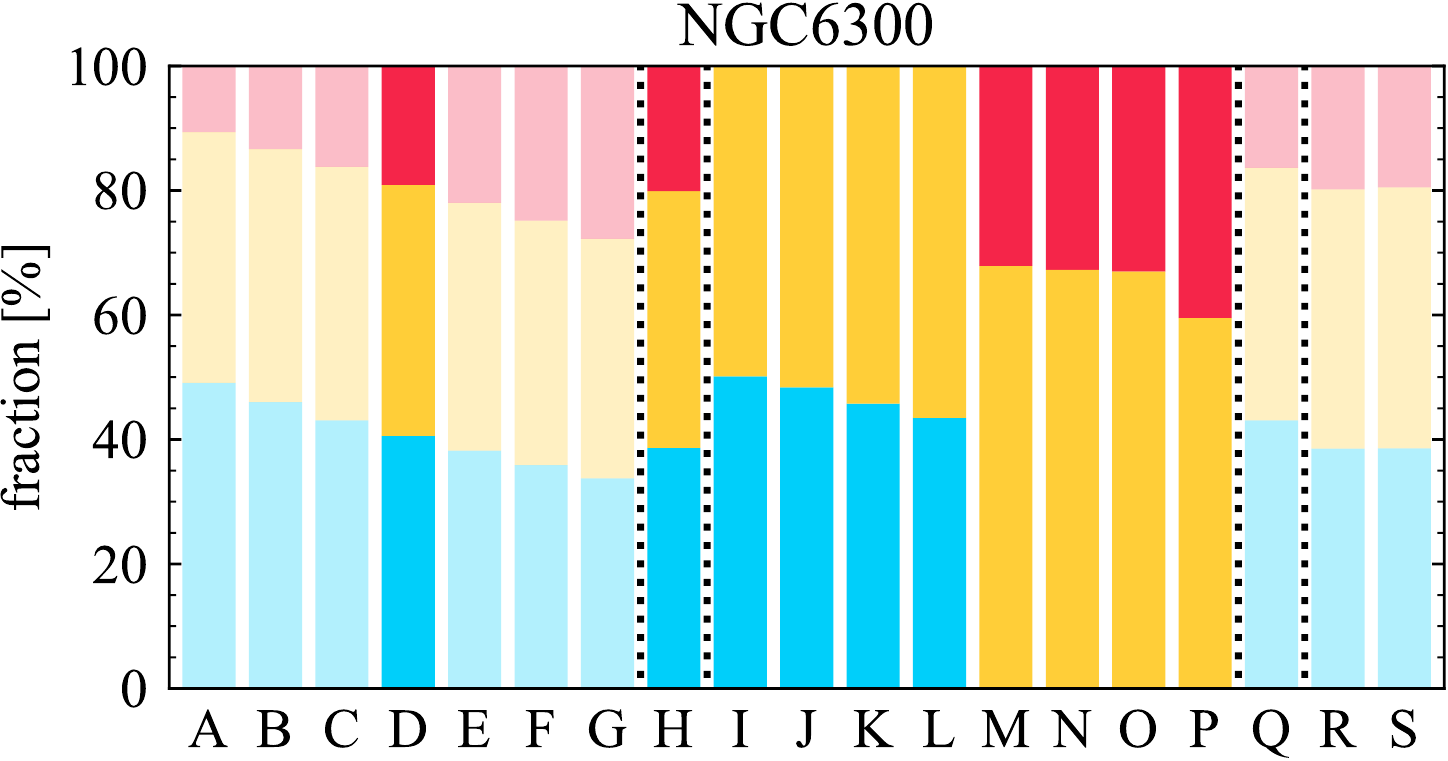}
	\caption{Continued.} 
\end{figure*}

\addtocounter{figure}{-1}
\begin{figure*}
	\centering
	\addtocounter{subfigure}{36}
	\includegraphics[scale=0.52]{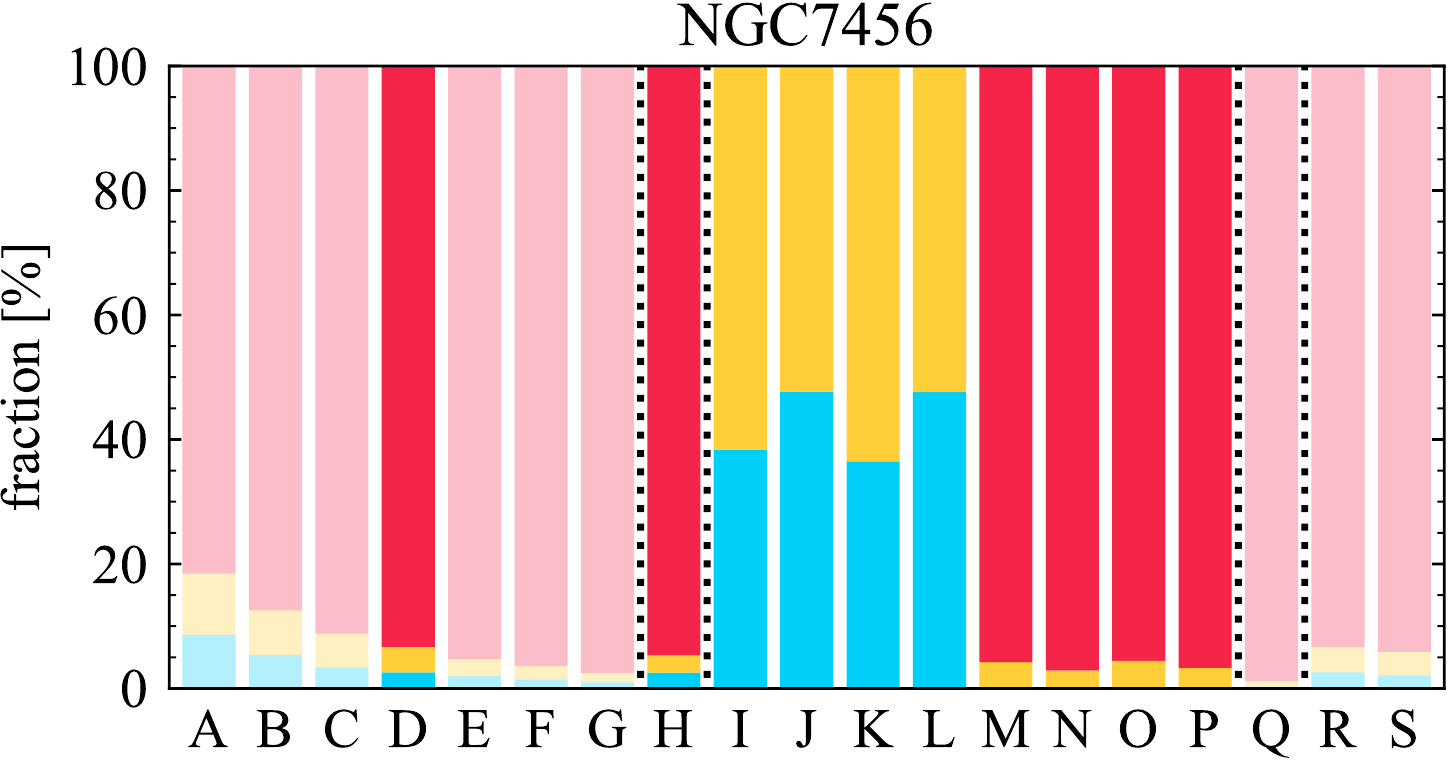}
	\caption{Continued.} 
\end{figure*}

\subsection{H$\alpha$ Threshold}
\label{sec_appendix_ha_sens}
For a point source at the native resolution of our H$\alpha$  data,  the effective sensitivity limits in terms of H$\alpha$ surface brightness threshold applied to the fiducial maps  corresponds to  \textsc{H\,ii} region luminosities ($\log (L_\mathrm{\textsc{H\,ii}\,region}^\mathrm{sensitivity})$) between $36.7$ and $38.4$~erg~s$^{-1}$.
Here we carried out two tests to examine the impact of the varying H$\alpha$ threshold (which originates from the non-uniform H$\alpha$ sensitivity among the sample) on the results.

In the first test, we compare the sight line fractions estimated in the PHANGS-MUSE H$\alpha$ images at 150~pc resolution  to that measured in the default narrowband images at the same resolution. 
The PHANGS-MUSE observations are sensitive enough to probe down to   \textsc{H\,ii} regions with $\log(L)$ $\approx$ 36 erg\,s$^{-1}$, about an order of magnitude deeper than the narrowband images.
We measure the  sight line fractions for the 15 galaxies that have both observations (hereafter overlapping sample). 
 The  MUSE images are  treated by the same method as the narrowband data for the removal of DIG (see Section \ref{sec_ha_filtering}). Namely,  the only difference between the two images is the increased sensitivity of the MUSE data.

\begin{figure*}[h]
	\centering
	\includegraphics[scale=0.67]{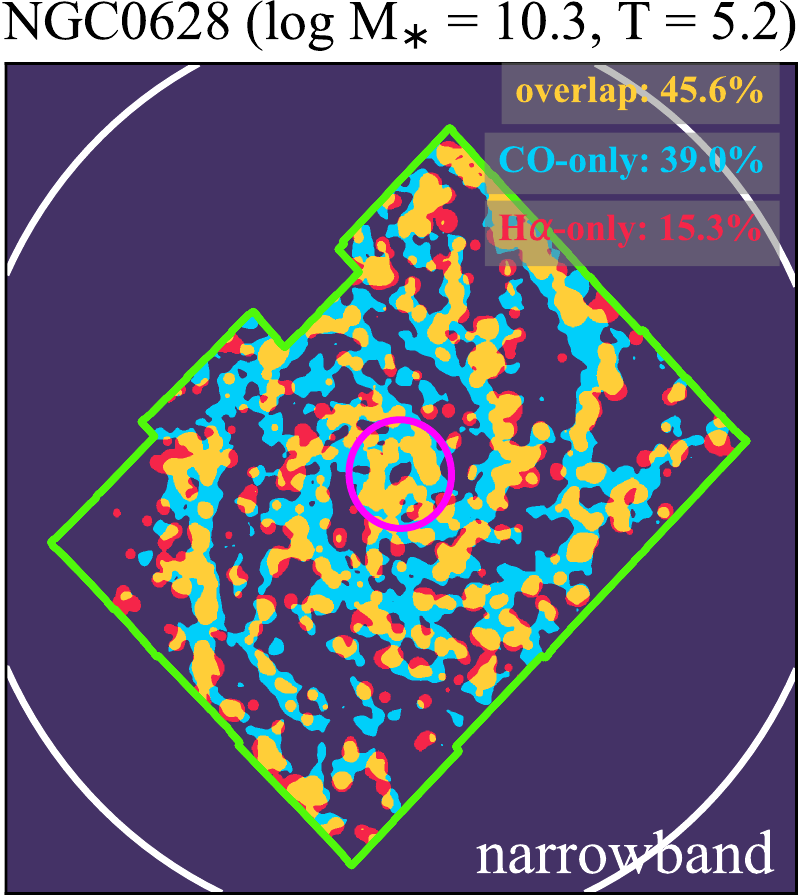}
	\includegraphics[scale=0.67]{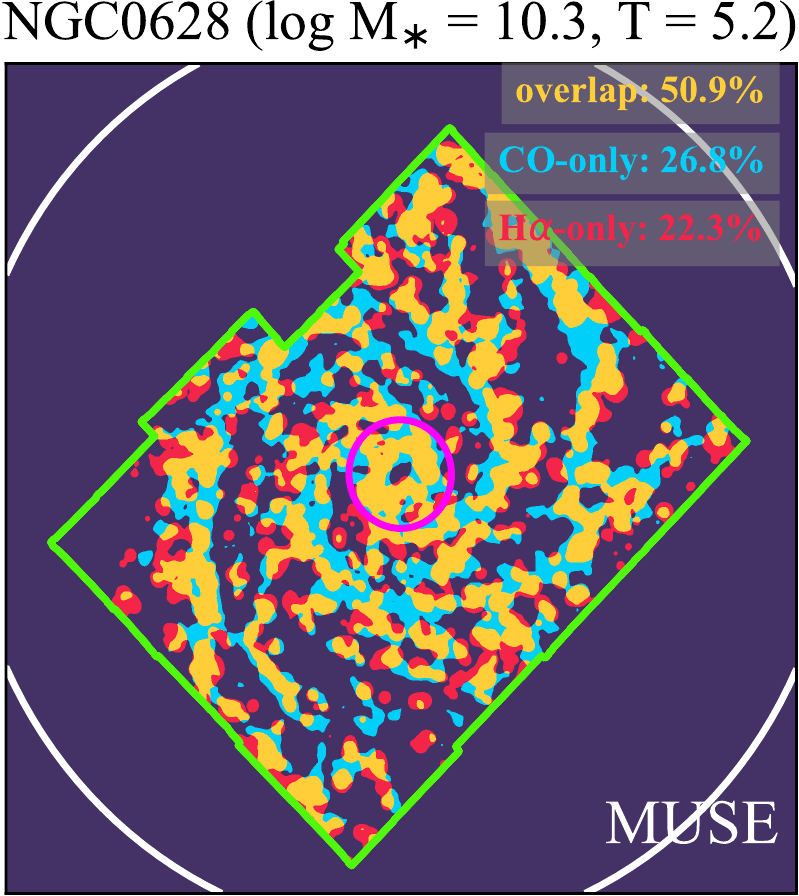}
	\caption{Comparison of the sight line maps of NGC~0628 at 150~pc resolution produced based on  narrowband (left) and MUSE IFS (right) H$\alpha$ images. The symbols are the same as in Figure \ref{fig_pie_example}. The sight line fractions based on both narrowband and MUSE observations are measured using the regions enclosed within the green box (MUSE FoV). Note that the MUSE FoV is smaller than that of  narrowband, so the narrowband sight line fractions measured in Section \ref{sec_appendix_ha_sens} are not necessarily the same as the fractions listed in Table \ref{tab_fractions}.}
	\label{fig_muse_nb_sightlines}
\end{figure*}

\begin{figure*}
	\centering
	\includegraphics[scale=0.52]{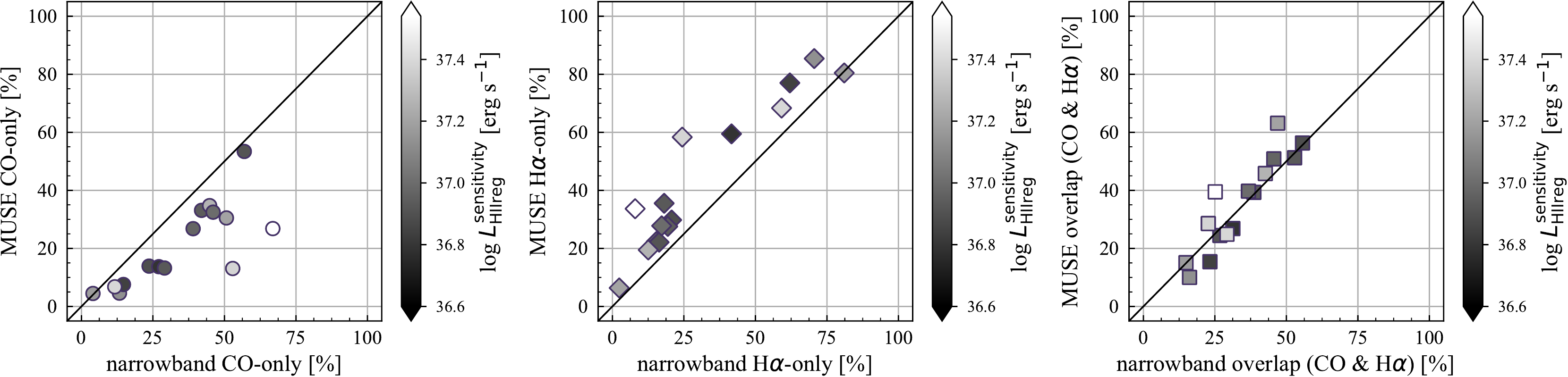}
	\caption{Comparison between the sight line fractions at  150~pc resolution  determined from  the narrowband ($x$-axis) and MUSE ($y$-axis) observations, color-coded by $\log (L_\mathrm{\textsc{H\,ii}\,region}^\mathrm{sensitivity})$. The solid line marks the one-to-one relation. } 
	\label{fig_compare_with_muse}
\end{figure*}

As an example, Figure \ref{fig_muse_nb_sightlines} shows the sight line maps for galaxy NGC~0628 based on the narrowband (left) and MUSE (right) images.
It can be seen directly that the amount of \ha\ sight lines increases in the MUSE map as a result of its higher sensitivity.  
This is accompanied by a decrease in \co\ regions.
A direct comparison is provided in Figure \ref{fig_compare_with_muse}.
While most galaxies show differences within a factor of two, the difference can be up to a factor 2.5 or even more for those galaxies with the highest $\log (L_\mathrm{\textsc{H\,ii}\,region}^\mathrm{sensitivity})$ (i.e., most shallow narrowband observation).
The \overlap\ regions show the least difference between narrowband and MUSE.
This is because the fluxes  of H$\alpha$ emission are higher in \overlap\ regions  compared to \ha\ regions (Section \ref{sec_number_vs_flux}), so the  \overlap\ fraction is less affected by sensitivity.

It is worth noting that the \emph{true} discrepancy between the MUSE and narrowband sight line fractions are likely smaller than what is obtained here. 
The low-H$\alpha$-luminosity regions (on the order of $L_\mathrm{H\alpha}$ $\approx$ 10$^{36 - 37}$~erg~s$^{-1}$) are the main source of the discrepancy between the MUSE and narrowband sight line fractions as they are not present in the  narrowband observations due to the limited sensitivity.	
Previous studies on the nature of H$\alpha$-emitting sources in nearby galaxies show that  regions ionized by non-\textsc{H\,ii} sources (e.g., supernova remnants and planetary nebula) tend to have lower H$\alpha$ luminosity compared to regions powered by \textsc{H\,ii} regions \citep[e.g.,][]{Bel16,Hsi17,Pan18}.
The same characteristic is observed in our MUSE data (e.g., \citealt{San21}; F.~Scheuermann~et~al. submitted),  suggesting that a certain fraction of regions missed by the narrowband observations are not  \textsc{H\,ii}~regions\footnote{About $\sim$ 30\% based on the spectroscopic analysis by \cite{San21}.}. Therefore the discrepancy between the MUSE and narrowband sight line fractions reported here is an upper limit\footnote{To be in line with the analysis of narrowband data, for this comparison we did not apply a spectroscopic classification \citep[e.g., like a Baldwin-Phillips-Telervich (BPT) diagram,][]{Bal81} to the regions identified in the MUSE H$\alpha$ map, so the non-\textsc{H\,ii}~regions remain in our DIG-removed MUSE maps. Such a spectroscopic classification is not possible for our  narrowband data.}.

Further we find no correlation between the slight line fractions and global galaxy properties (\Mstar\ and Hubble type) no matter which H$\alpha$ image is used, presumably due to the low number of galaxies available in the overlapping sample.
Therefore, we carried out a second test to verify whether the trend between the sight line fractions and global galaxy properties (Section \ref{sec_global_galprops}) still holds when narrowband $\log (L_\mathrm{\textsc{H\,ii}\,region}^\mathrm{sensitivity})$ is taken into account.

\begin{figure*}
	\centering
	\includegraphics[scale=0.65]{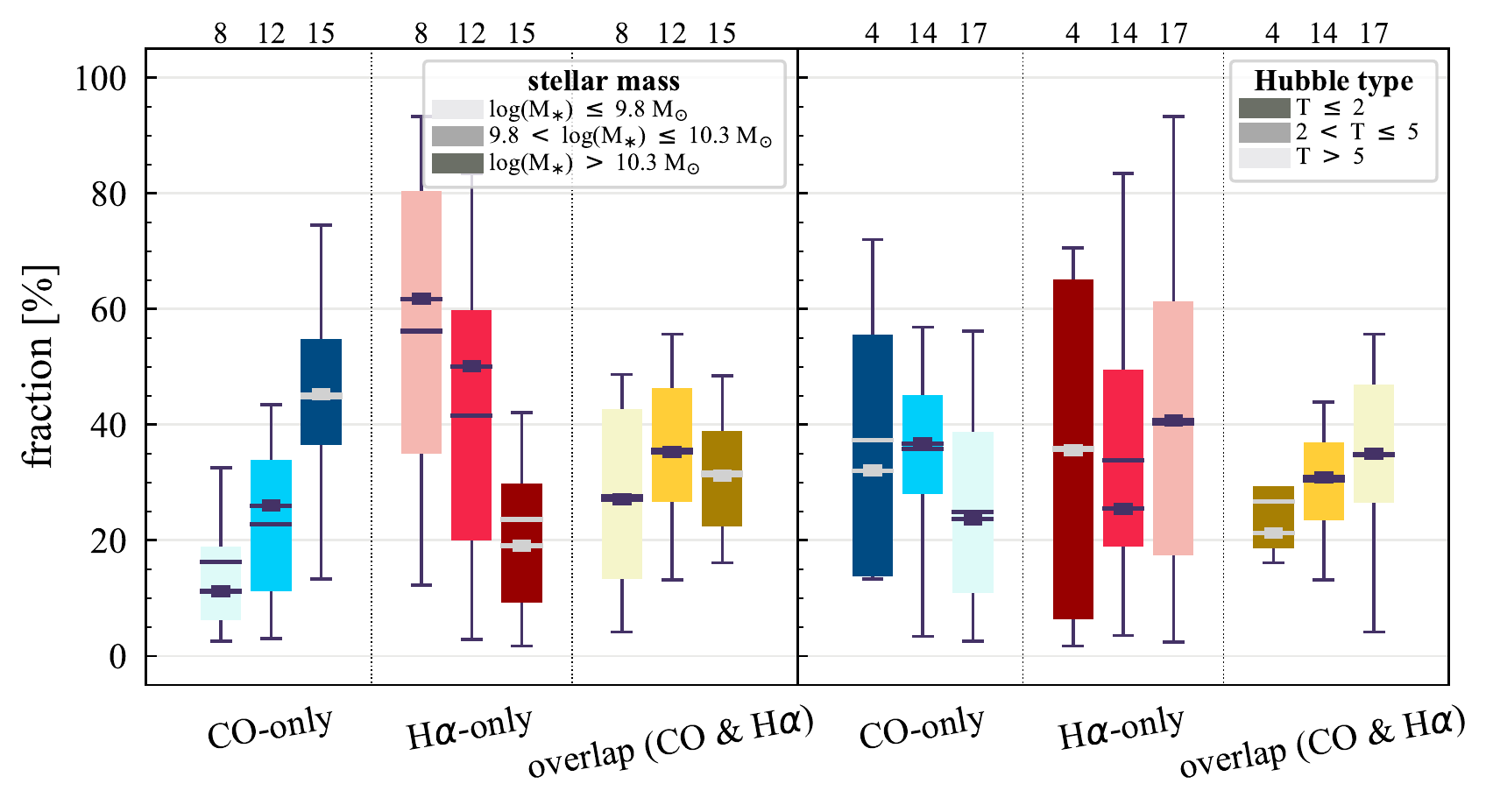}
	\caption{Variations of the global sight line fractions at  150~pc resolution   as a function of \Mstar\ (left) and Hubble type (right). The figure is analogous to Figure \ref{fig_global_boxes}, but only the 35 galaxies with $\log (L_\mathrm{\textsc{H\,ii}\,region}^\mathrm{sensitivity})$~$<$~37.5 erg~s$^{-1}$ are used.}
	\label{fig_global_props_subsample}
\end{figure*}

In the second test, we examine the relation between  the  sight line fractions and global galaxy properties at 150~pc resolution using only galaxies with relatively high narrowband sensitivity (i.e., low $\log (L_\mathrm{\textsc{H\,ii}\,region}^\mathrm{sensitivity})$).
Specifically, we re-plot Figure \ref{fig_global_boxes} using only galaxies with $\log (L_\mathrm{\textsc{H\,ii}\,region}^\mathrm{sensitivity})$~$<$~37.5~erg~s$^{-1}$.
Thirty five  galaxies satisfy this criterion, the result is shown in Figure \ref{fig_global_props_subsample}.
We find good agreement between Figure \ref{fig_global_boxes} (full sample) and Figure \ref{fig_global_props_subsample} for \Mstar\ (left panel), although the median fractions can differ by a factor of a few between two samples.
For Hubble type, the trends with  \co\ and \ha\ fractions is no longer obvious when only using galaxies with relatively high narrowband sensitivity. 
This is partially (or perhaps even completely) due to the low-number statistics in the lowest-$T$ bin as earlier type galaxies in our sample tend to have less sensitive H$\alpha$ images.
In spite of that, the median \co\ and \ha\ fractions of the other two $T$-type bins and the \overlap\ fractions agree well with those derived using the full sample.

In summary, these two tests demonstrate that the H$\alpha$ threshold is  an important factor in the sight line analysis. 
Our results remain qualitatively robust once accounting for H$\alpha$ threshold, but further confirmation with a high sensitivity and large sample is required in the future.

\subsection{CO ($\Sigma_\mathrm{H_{2}}$) Threshold}
\label{sec_appendix_co_threshold}
We clip the CO images at our best-matching resolution of 150~pc using a surface density threshold of  10 M$_{\sun}$~pc$^{-2}$.
We test the effect of this threshold by varying the threshold from 7 to 13 M$_{\sun}$~pc$^{-2}$, $\sim -30$\% to $+30$\% with respect to the fiducial threshold.
The sight line fractions for a threshold of 7--13 M$_{\sun}$~pc$^{-2}$ with an interval of 1~M$_{\sun}$~pc$^{-2}$   are presented in Figure~\ref{fig_test_assumptions} columns A--G, respectively (column D is the default result of this work).

As somewhat expected, the fractions of \co\ sight lines gradually decrease with increasing threshold in all galaxies, while \ha\ sight lines gradually increase.
The fractions of \co\ sight lines show a decrease within a factor $\sim$ 3 (with a few exceptions) when the threshold varies from $-30$\% to $+30$\% with respect to the fiducial threshold.
An opposite trend is observed for \ha\ sight lines, but the magnitude of the change is comparable to that of \co\ sight line.
There is no uniform trend between CO threshold and the  fraction of \overlap\ sight lines, both increasing and decreasing trends are seen in our galaxies.
Nonetheless, the typical magnitude of the change is smaller than for \co\ sight lines, not larger than a factor of~2. 
This is because the flux of CO in the \overlap\ regions tend to be higher than that in the \co\ regions  (and presumably much higher than the applied threshold); therefore they are less sensitive to the choice of the surface density threshold.
Overall, we find that the dependence of sight line fractions on the applied threshold is rather uniform across the sample. For this reason, our results remain qualitatively the same when the CO threshold varies by $\pm$ 30\%.

\section{Sight Line Maps and Fractions for Individual Galaxies at Different Spatial Scales}
\label{sec_appendix_individual_galaxy}
This appendix presents an atlas of the sight line distributions in our 49 galaxies.
Figure~\ref{fig_appendix_spatial_scale} presents galaxy   maps showing  \co\ (blue), \ha\ (red), and \textit{overlapping} CO and H$\alpha$ emission (yellow) at spatial resolutions of 150, 300, 500, 1000, and 1500~pc. 
The  sight line fractions measured within $R$ $<$ 0.6$R_{25}$ are listed in Table~\ref{tab_fractions}.

\begin{figure*}
	\centering
	\includegraphics[scale=0.7]{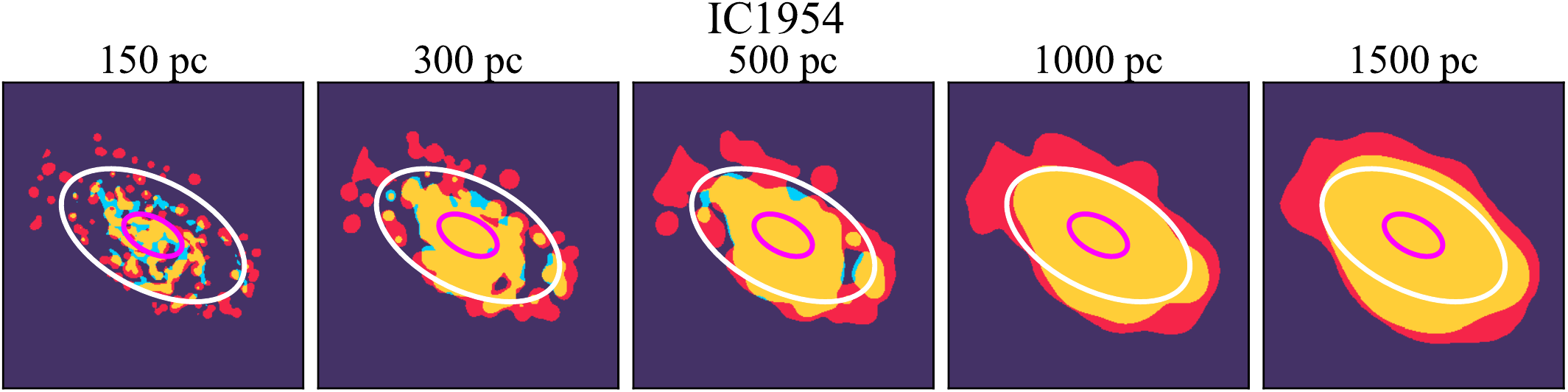}\\
	\vspace{10pt}
	\includegraphics[scale=0.7]{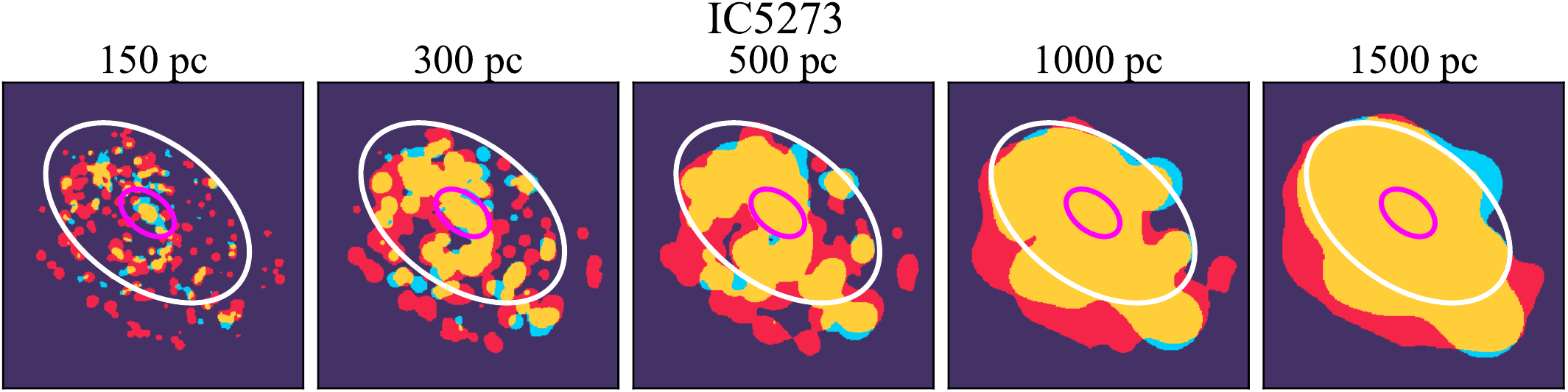}\\
	\vspace{10pt}
	\includegraphics[scale=0.7]{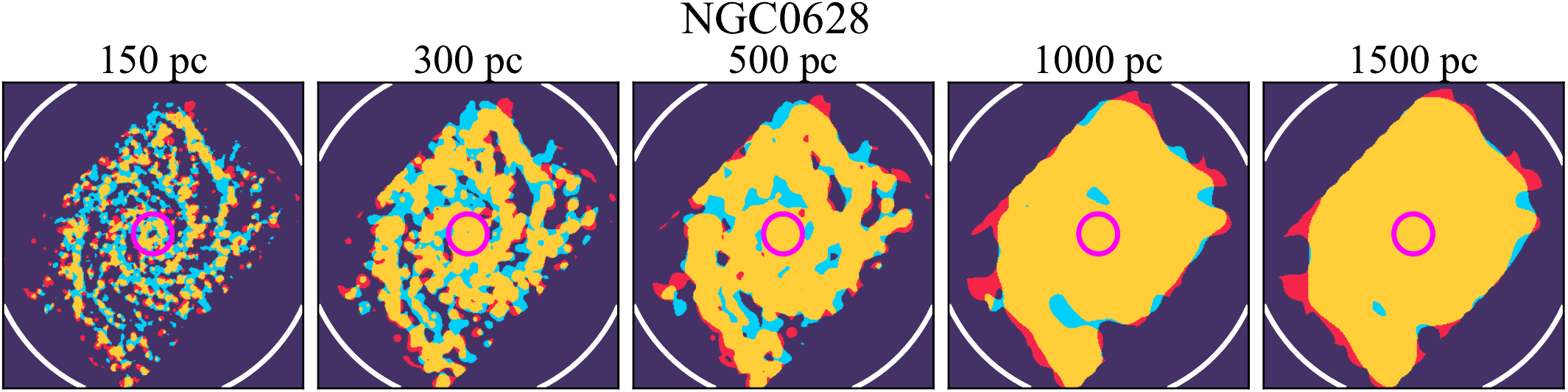}\\
	\vspace{10pt}
	\includegraphics[scale=0.7]{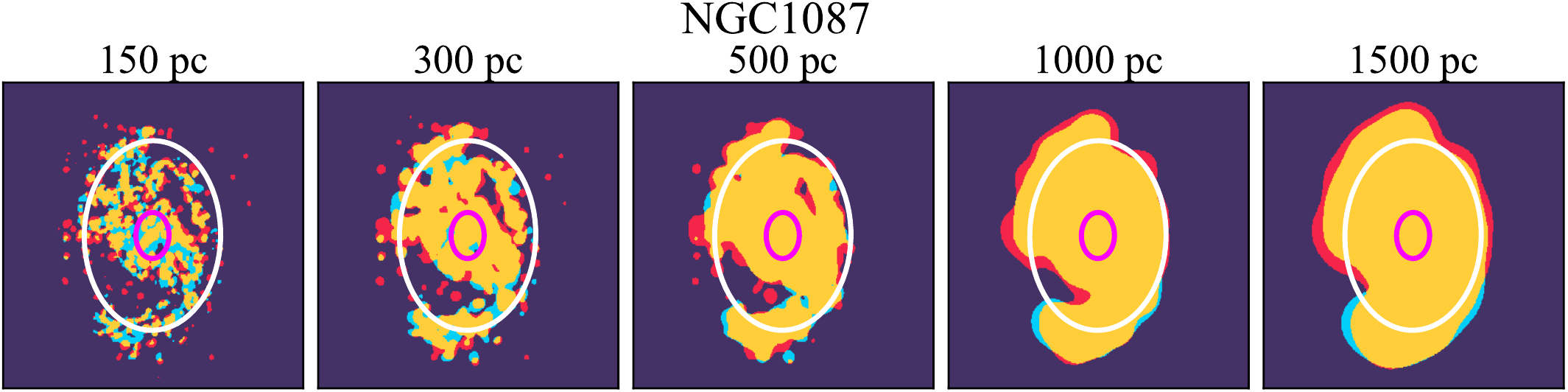}\\
	\vspace{10pt}
	\includegraphics[scale=0.7]{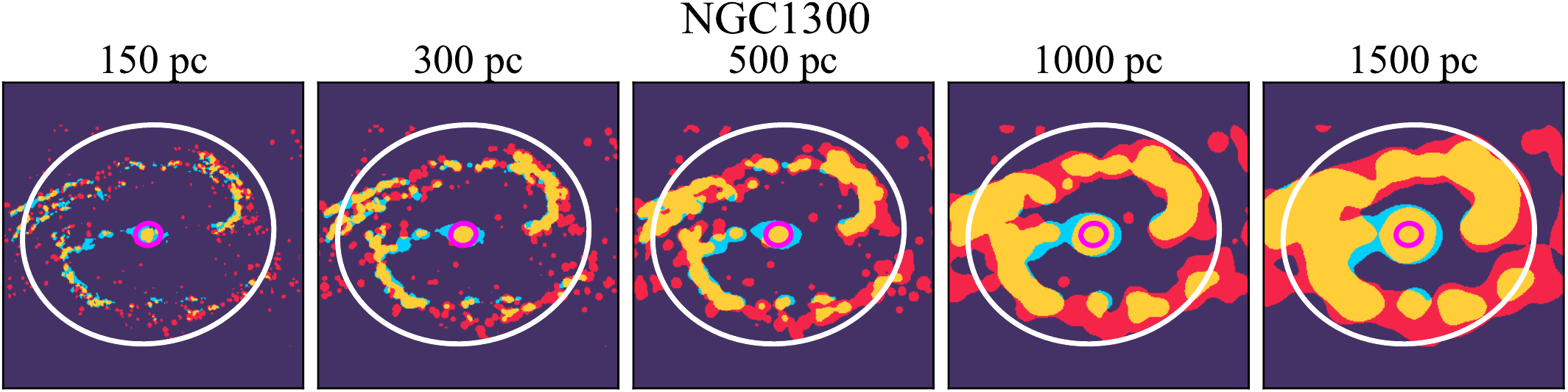}
	
	\caption{Galaxy maps showing regions with \co\ (blue), \ha\ (red), and \textit{overlapping} CO and H$\alpha$ emission (yellow) at  150, 300, 500, 1000, and 1500~pc resolutions.  The inner ellipses (magenta) mark the central region, defined as the central 2~kpc in  diameter. The outer ellipses (white) indicate the 0.6~$R_{25}$ regions where we measure the global sight line fractions.} 
	\label{fig_appendix_spatial_scale}
\end{figure*}

\addtocounter{figure}{-1}
\begin{figure*}
	\centering
	\includegraphics[scale=0.7]{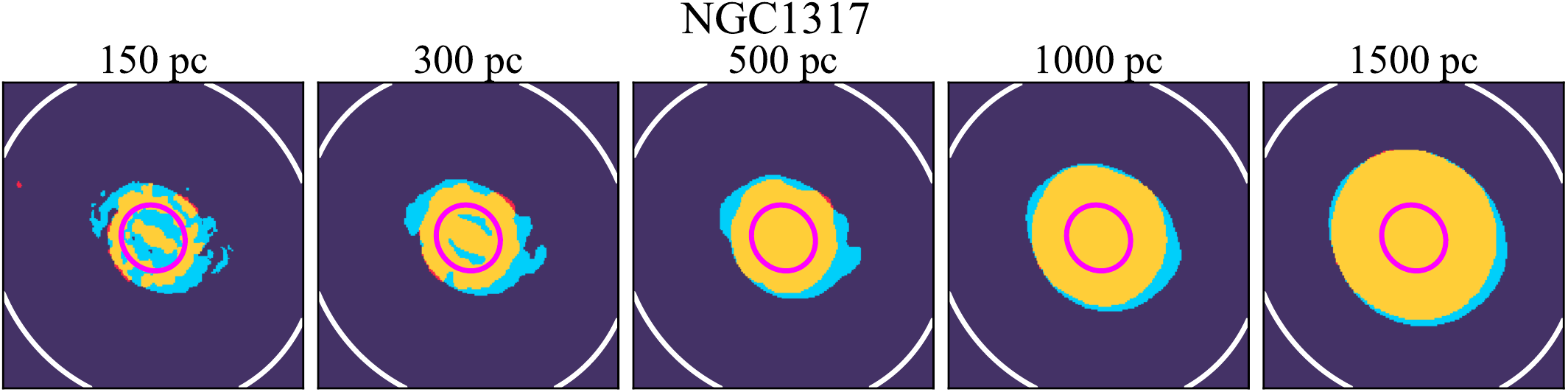}\\
	\vspace{10pt}
	\includegraphics[scale=0.7]{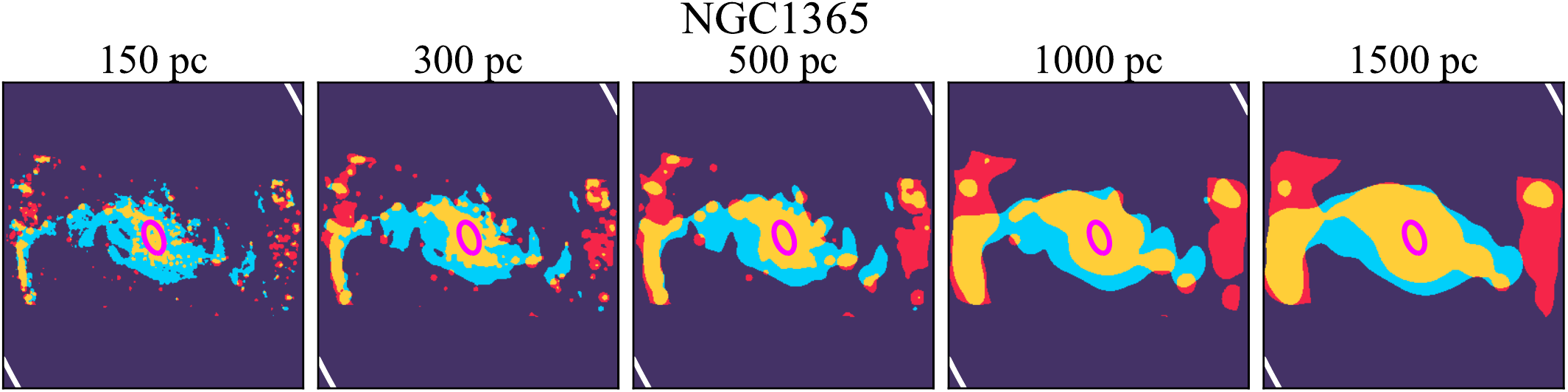}\\
	\vspace{10pt}
	\includegraphics[scale=0.7]{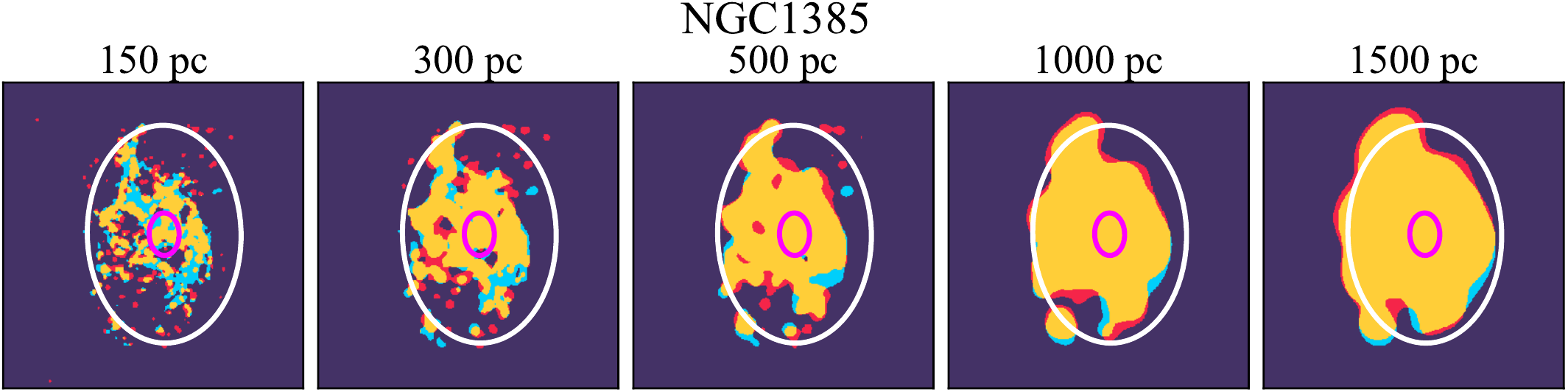}\\
	\vspace{10pt}
	\includegraphics[scale=0.7]{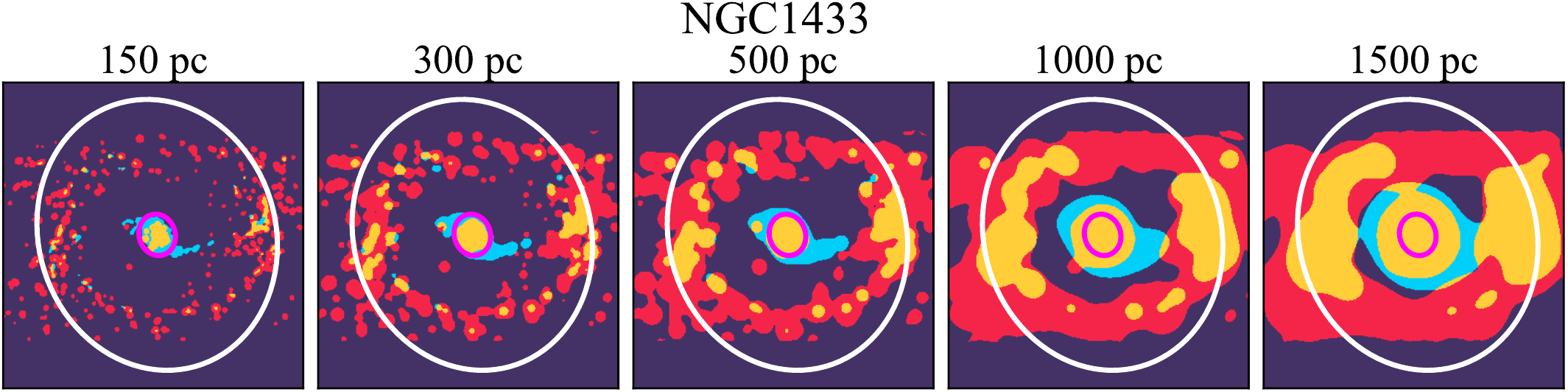}\\
	\vspace{10pt}
	\includegraphics[scale=0.7]{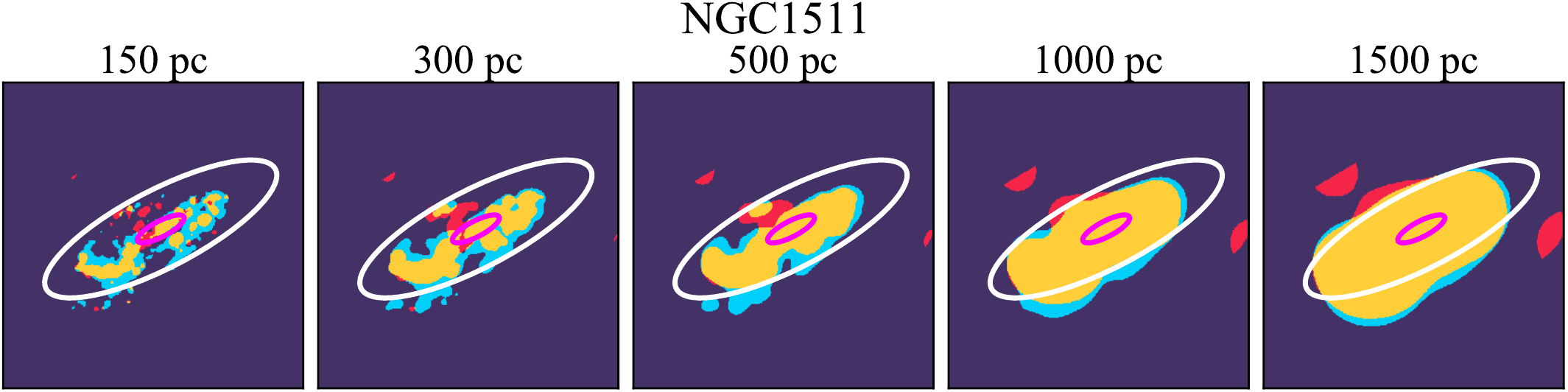}\\
	\caption{Continued.} 
\end{figure*}

\addtocounter{figure}{-1}
\begin{figure*}
	\centering
	\includegraphics[scale=0.7]{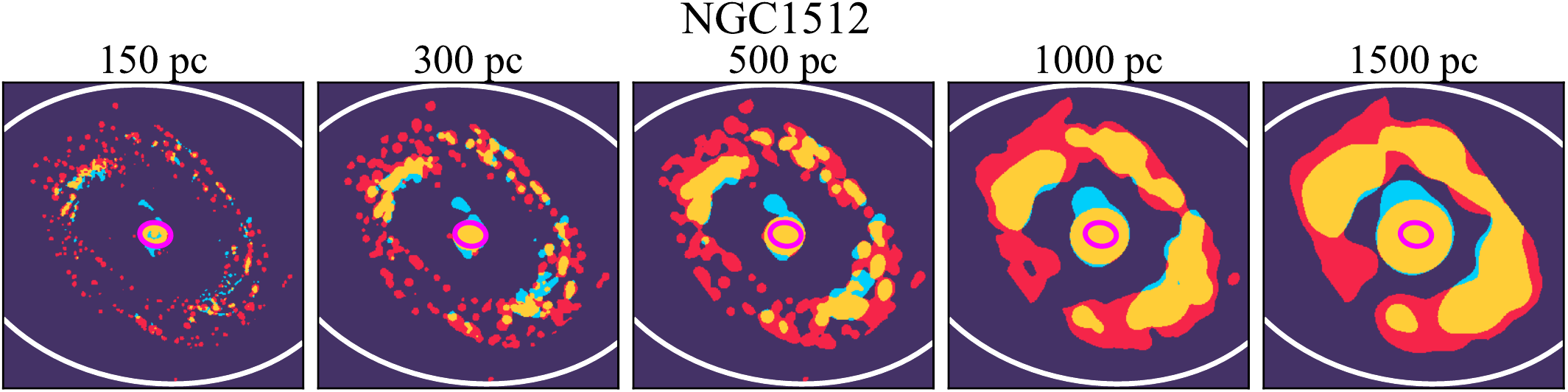}\\
	\vspace{10pt}
	\includegraphics[scale=0.7]{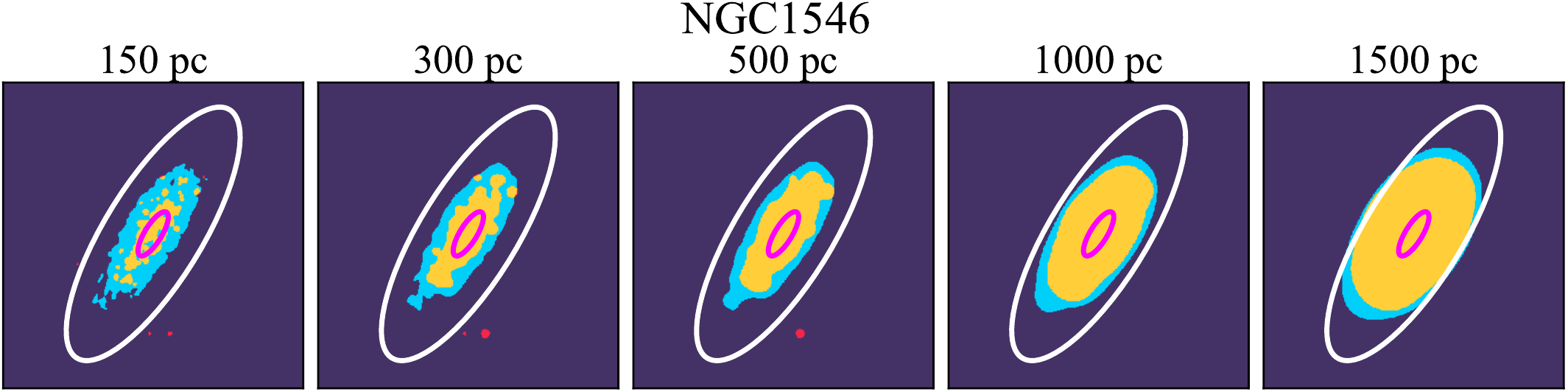}\\
	\vspace{10pt}
	\includegraphics[scale=0.7]{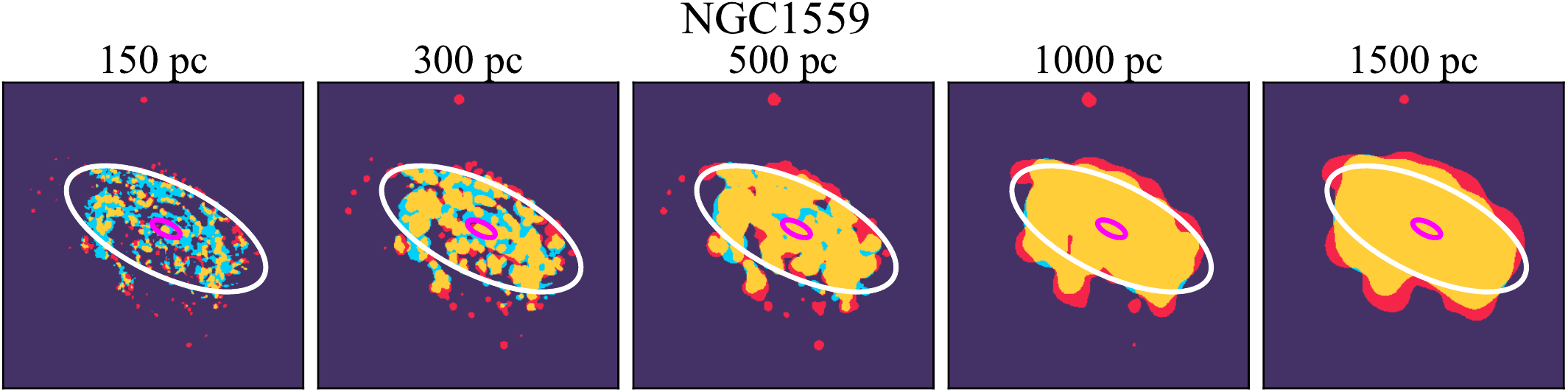}\\
	\vspace{10pt}
	\includegraphics[scale=0.7]{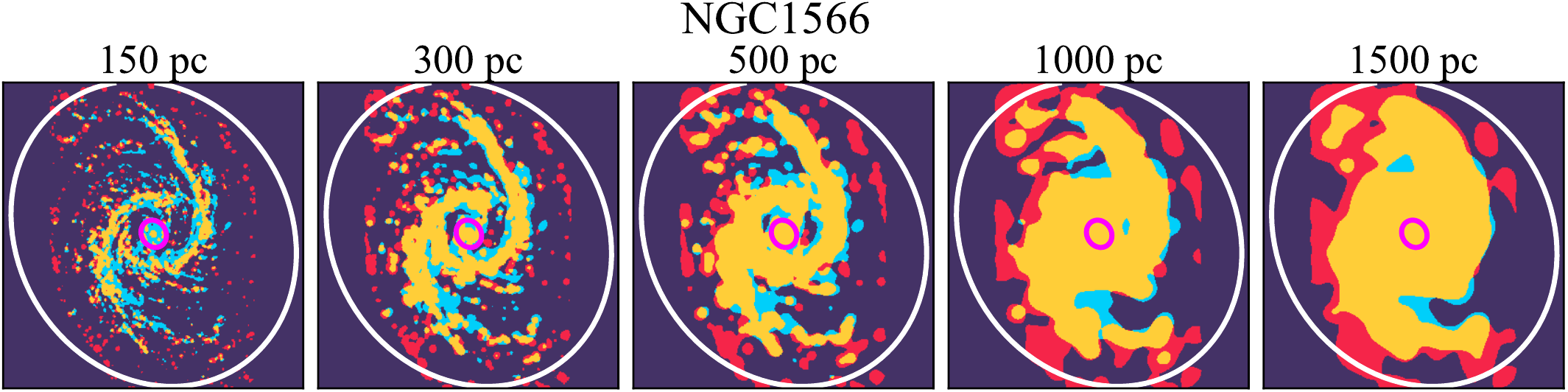}\\
	\vspace{10pt}
	\includegraphics[scale=0.7]{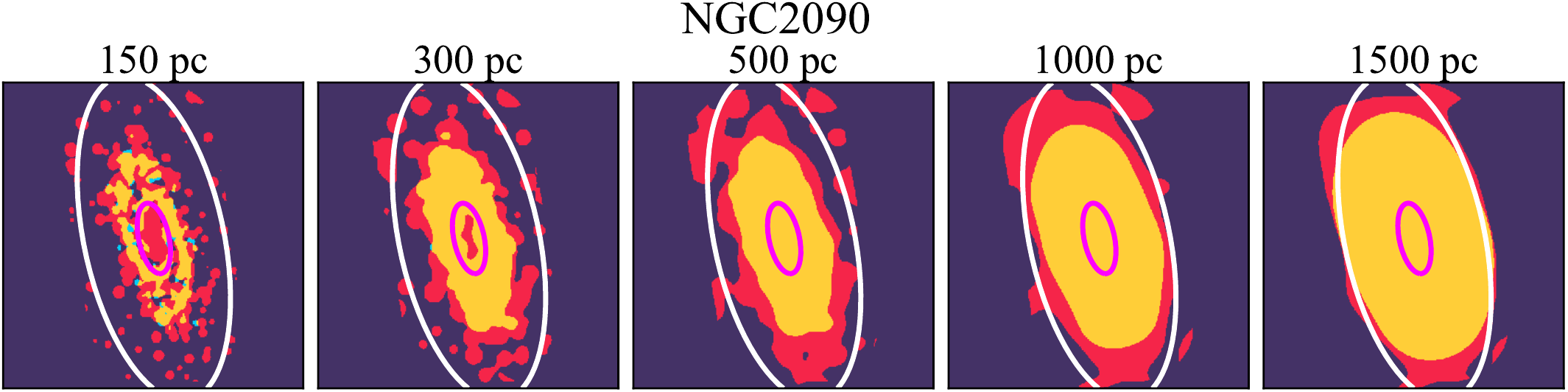}\\
	\caption{Continued.} 
\end{figure*}

\addtocounter{figure}{-1}
\begin{figure*}
	\centering
	\includegraphics[scale=0.7]{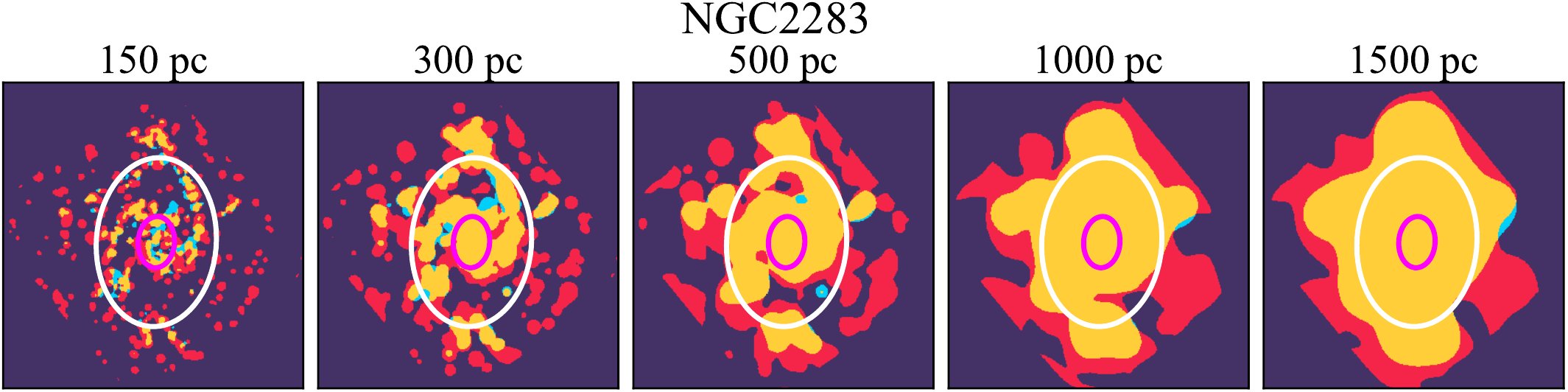}\\
	\vspace{10pt}
	\includegraphics[scale=0.7]{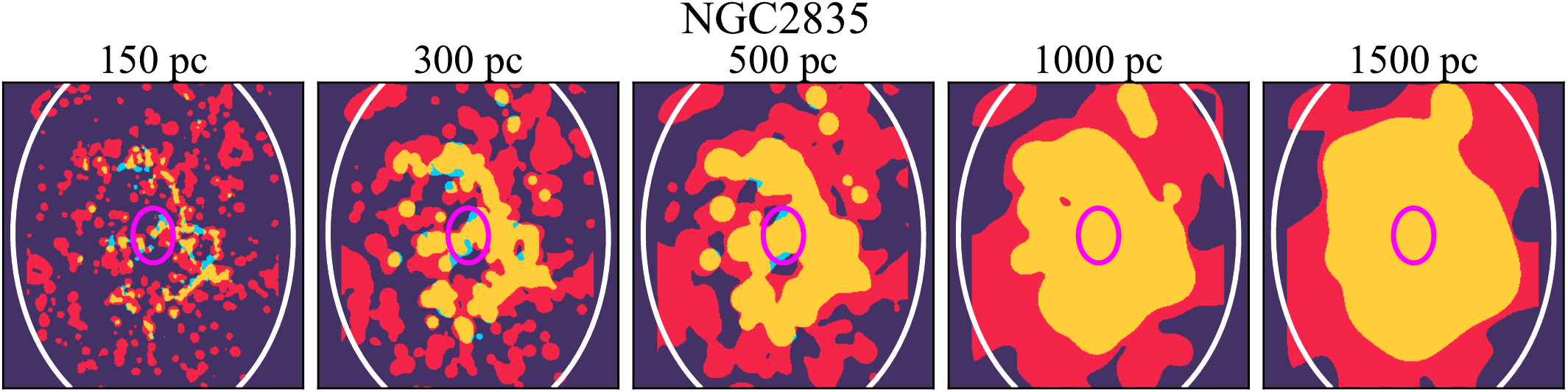}\\
	\vspace{10pt}
	\includegraphics[scale=0.7]{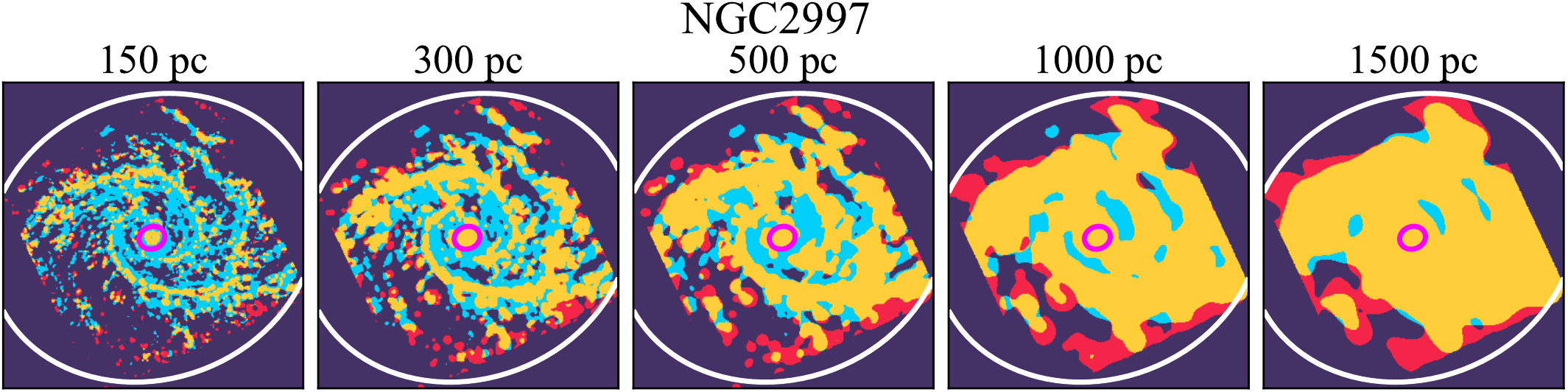}\\
	\vspace{10pt}
	\includegraphics[scale=0.7]{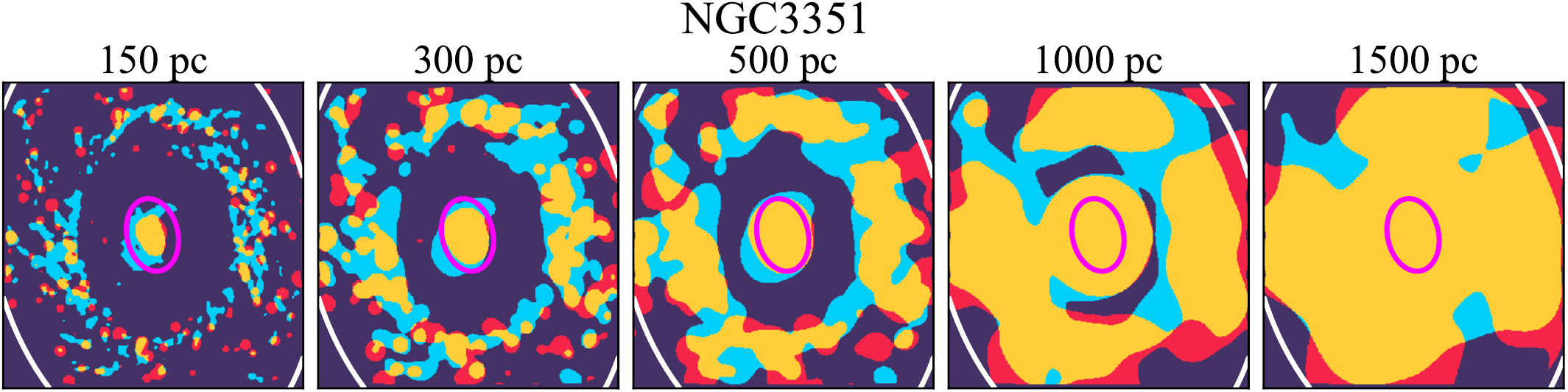}\\
	\vspace{10pt}
	\includegraphics[scale=0.7]{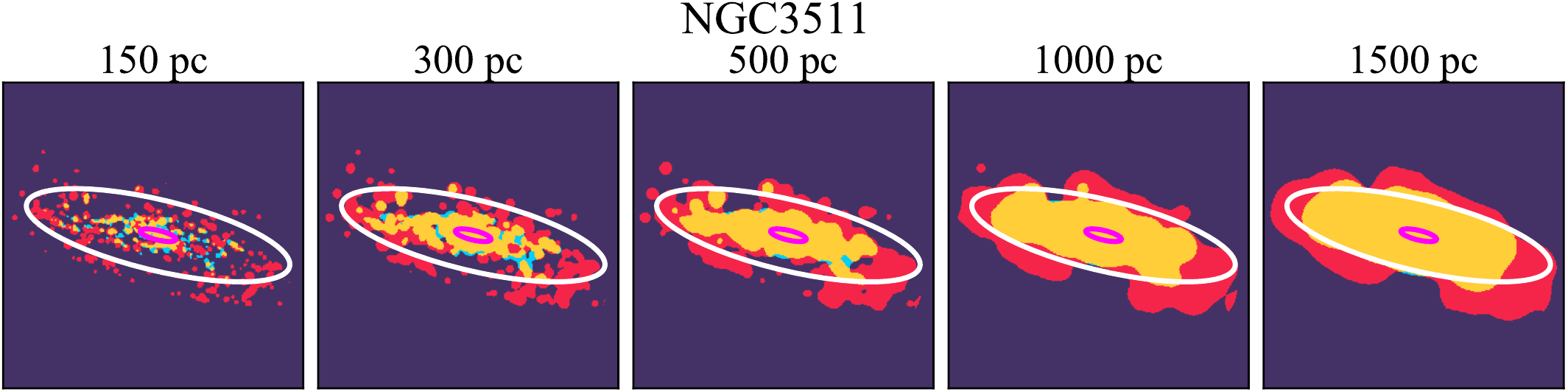}\\
	\caption{Continued.} 
\end{figure*}

\addtocounter{figure}{-1}
\begin{figure*}
	\centering
	\includegraphics[scale=0.7]{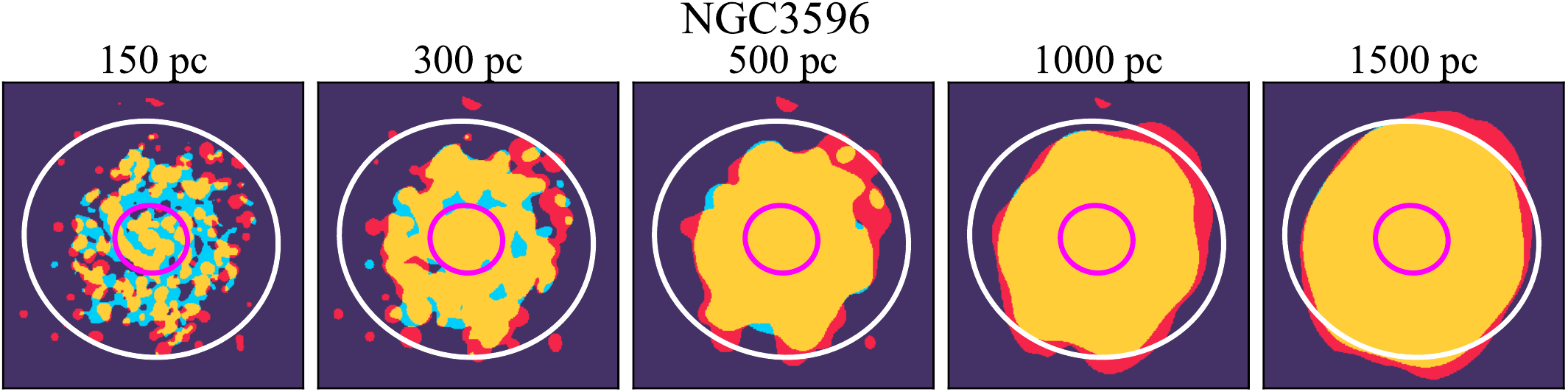}\\
	\vspace{10pt}
	\includegraphics[scale=0.7]{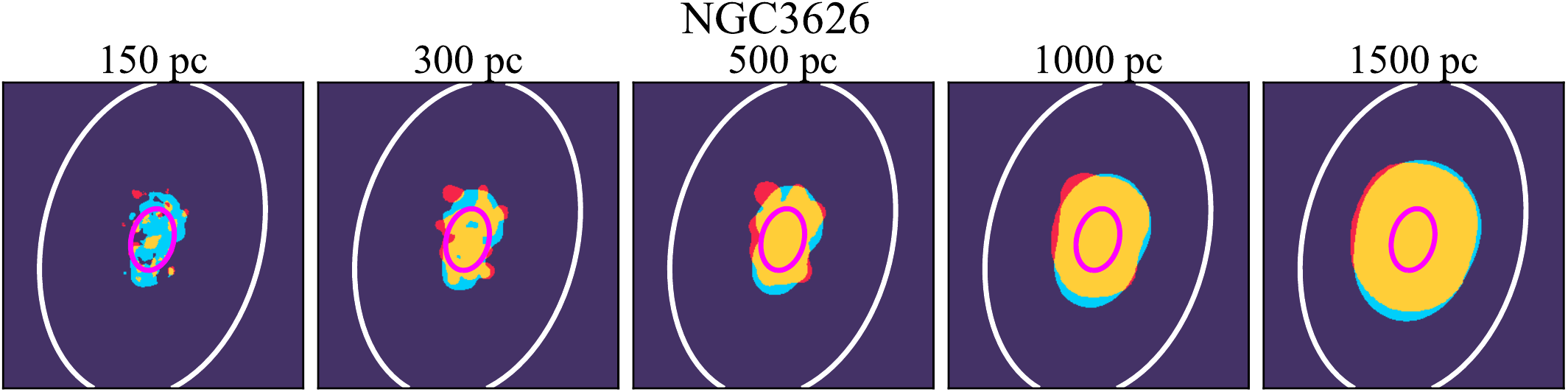}\\
	\vspace{10pt}
	\includegraphics[scale=0.7]{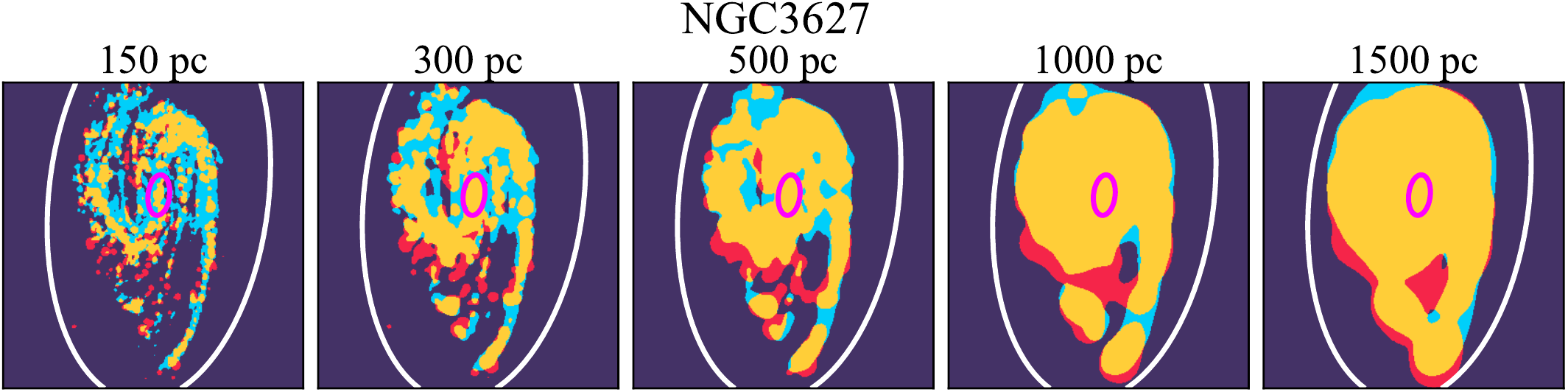}\\
	\vspace{10pt}
	\includegraphics[scale=0.7]{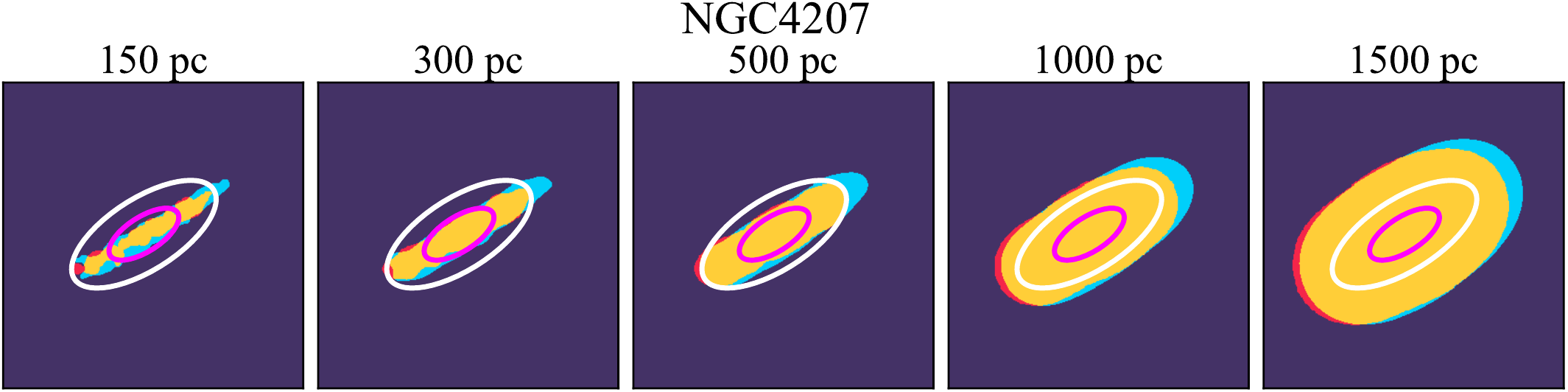}\\
	\vspace{10pt}
	\includegraphics[scale=0.7]{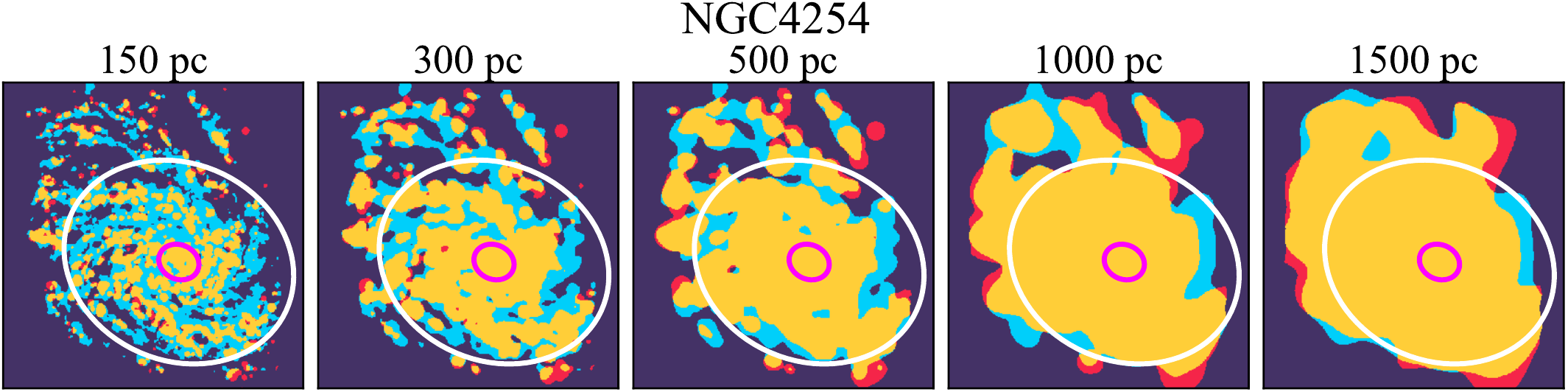}\\
	\caption{Continued.} 
\end{figure*}

\addtocounter{figure}{-1}
\begin{figure*}
	\centering
	\includegraphics[scale=0.7]{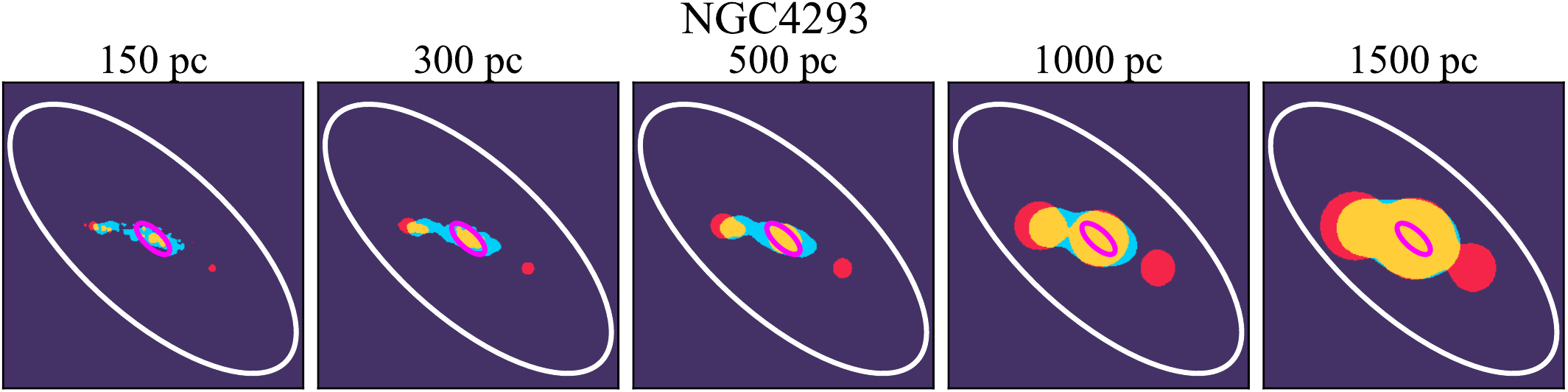}\\
	\vspace{10pt}
	\includegraphics[scale=0.7]{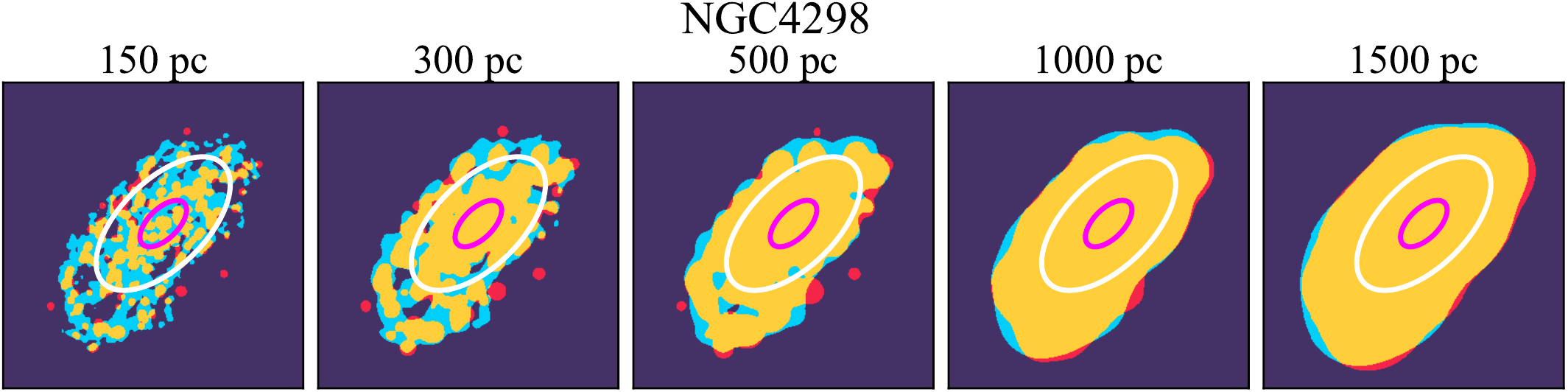}\\
	\vspace{10pt}
	\includegraphics[scale=0.7]{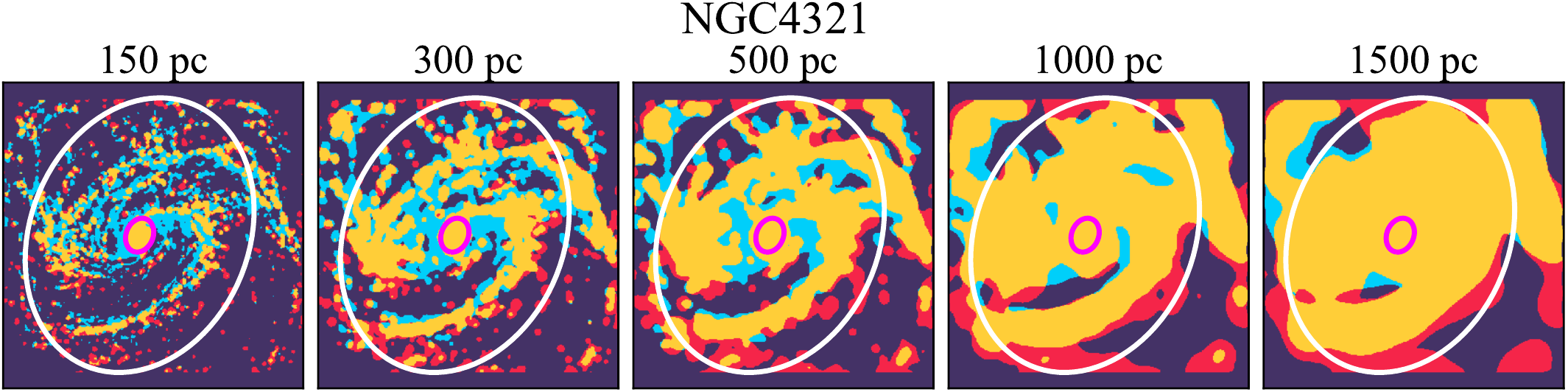}\\
	\vspace{10pt}
	\includegraphics[scale=0.7]{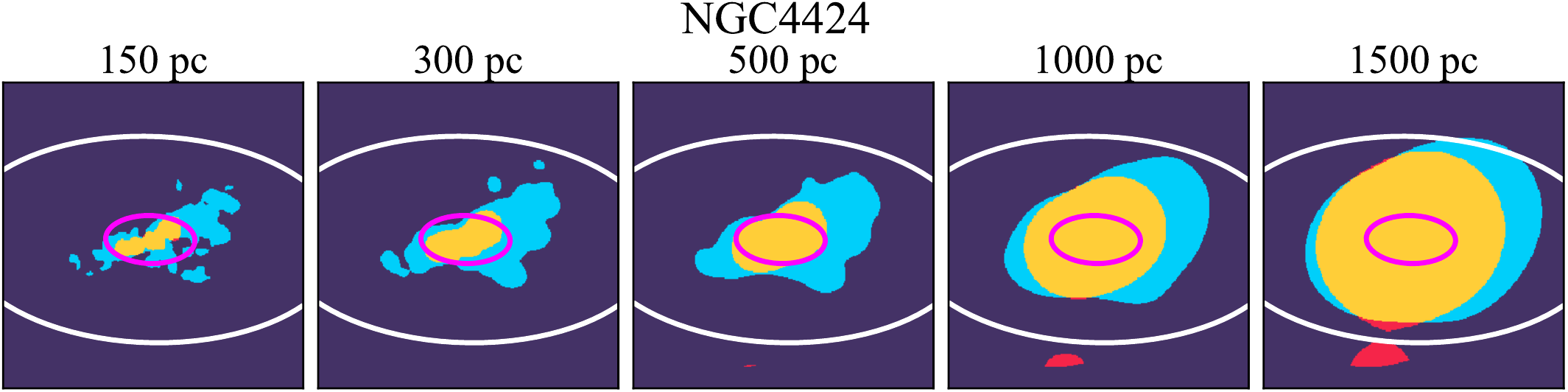}\\
	\vspace{10pt}
	\includegraphics[scale=0.7]{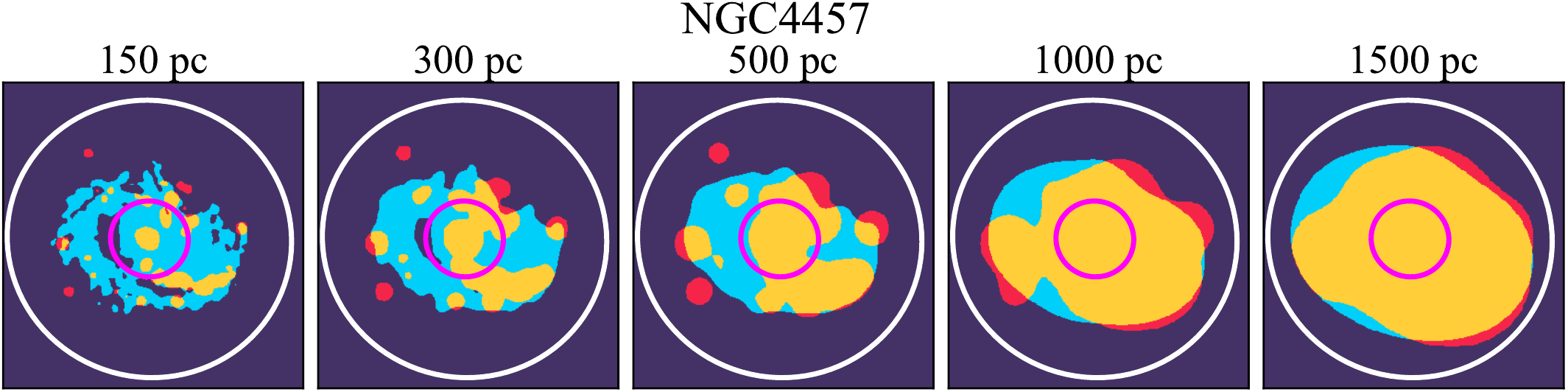}\\
	\caption{Continued.} 
\end{figure*}

\addtocounter{figure}{-1}
\begin{figure*}
	\centering
	\includegraphics[scale=0.7]{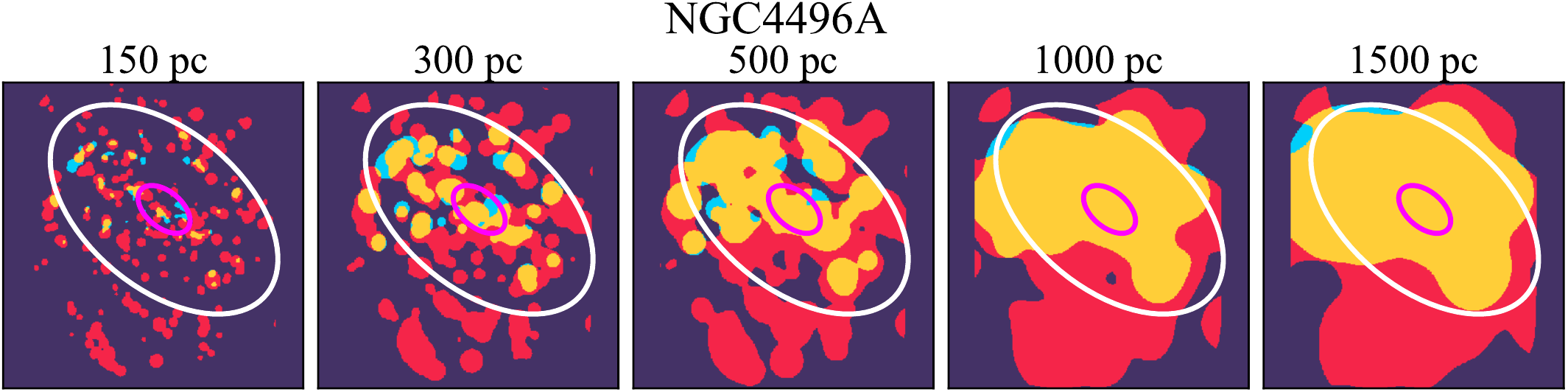}\\
	\vspace{10pt}
	\includegraphics[scale=0.7]{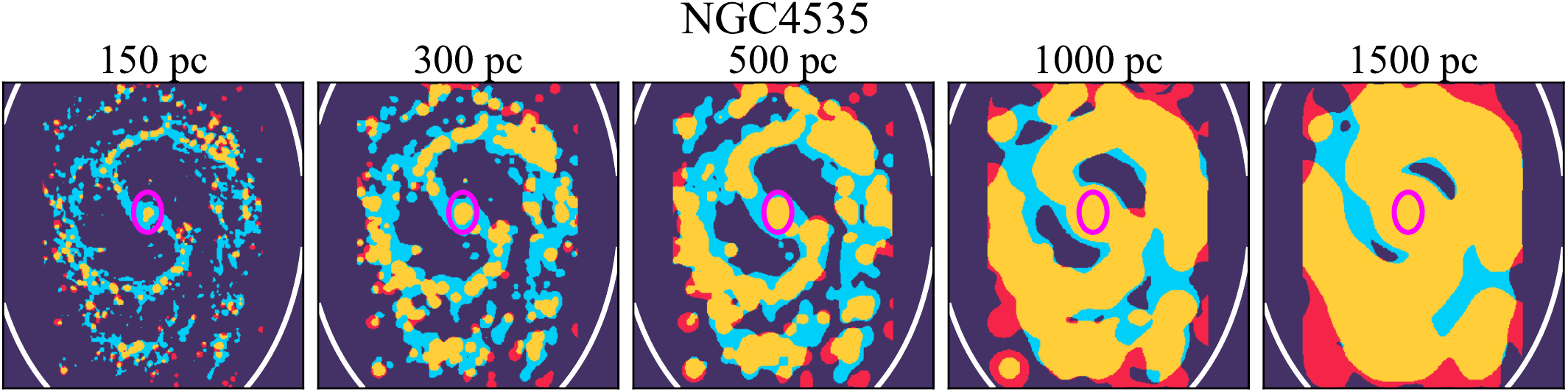}\\
	\vspace{10pt}
	\includegraphics[scale=0.7]{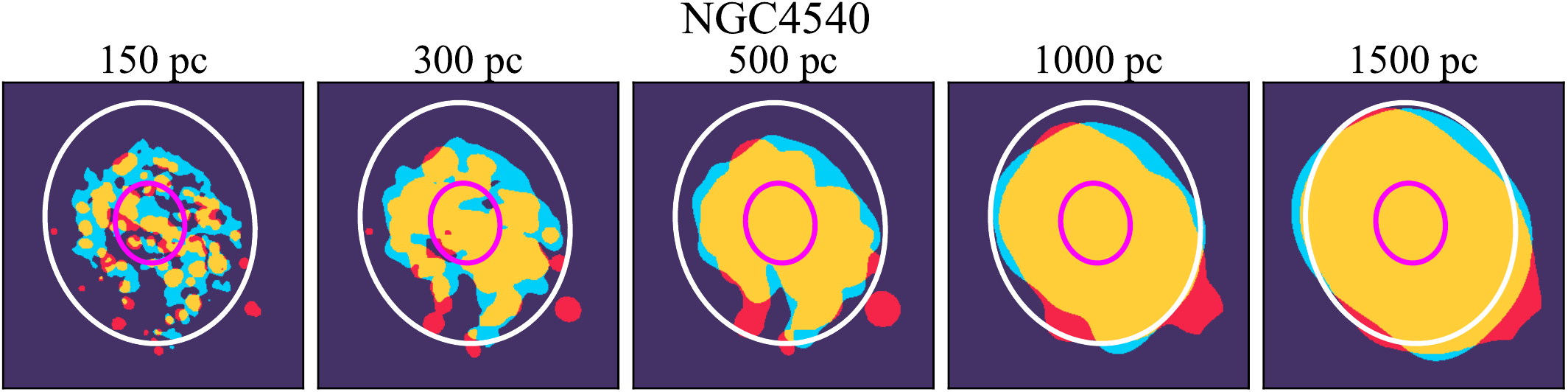}\\
	\vspace{10pt}
	\includegraphics[scale=0.7]{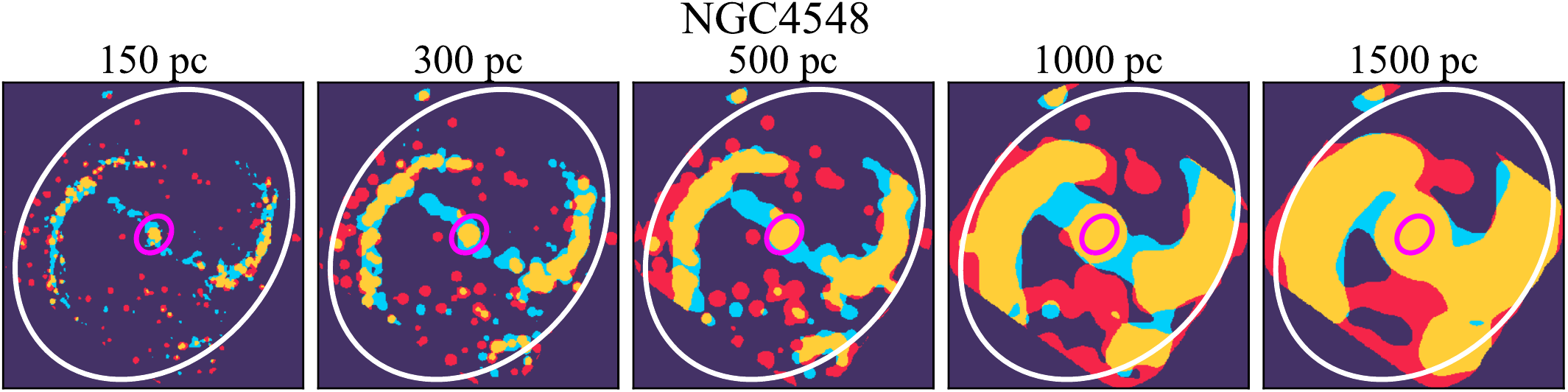}\\
	\vspace{10pt}
	\includegraphics[scale=0.7]{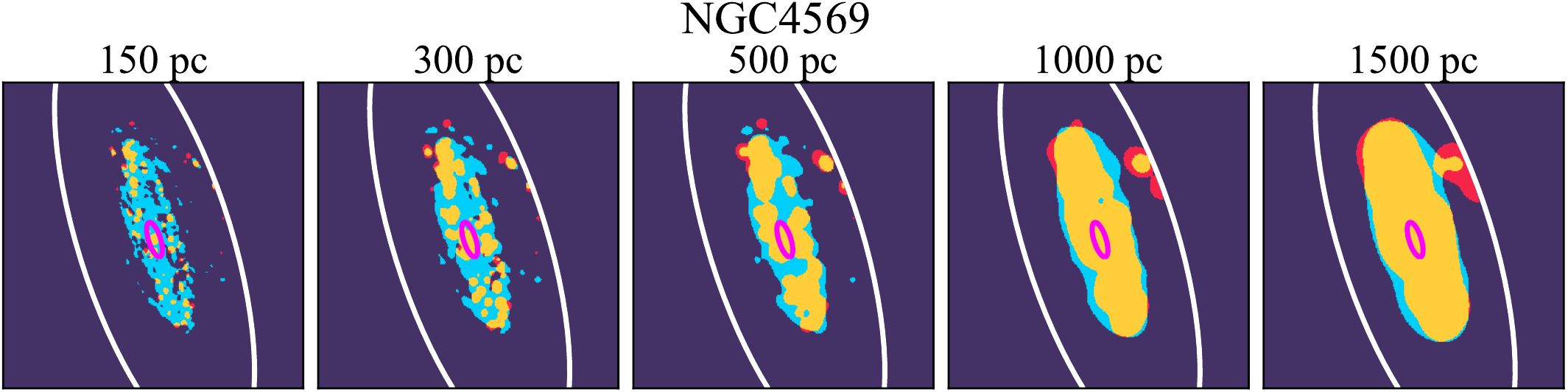}\\
	\caption{Continued.} 
\end{figure*}

\addtocounter{figure}{-1}
\begin{figure*}
	\centering
	\includegraphics[scale=0.7]{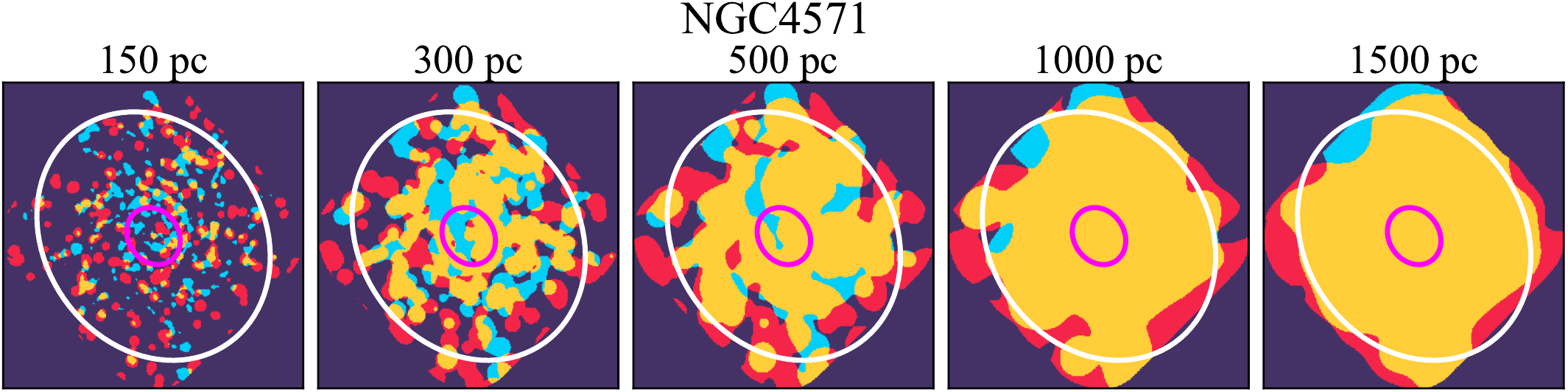}\\
	\vspace{10pt}
	\includegraphics[scale=0.7]{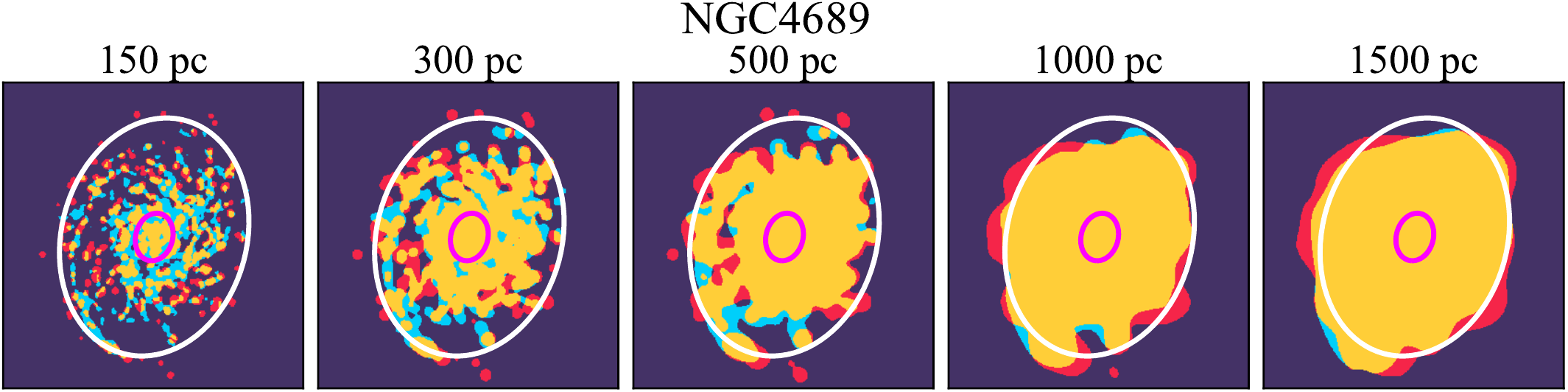}\\
	\vspace{10pt}
	\includegraphics[scale=0.7]{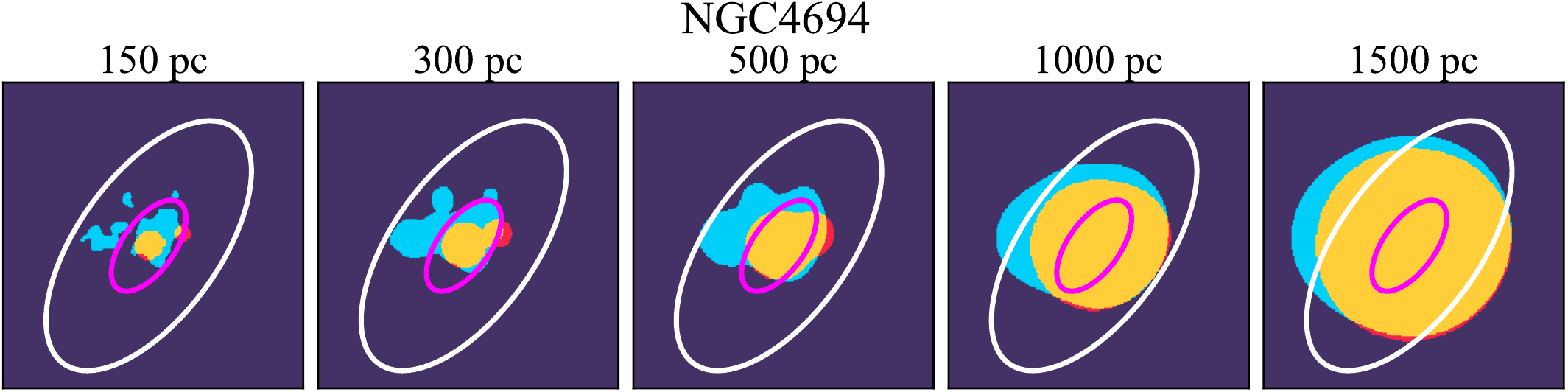}\\
	\vspace{10pt}
	\includegraphics[scale=0.7]{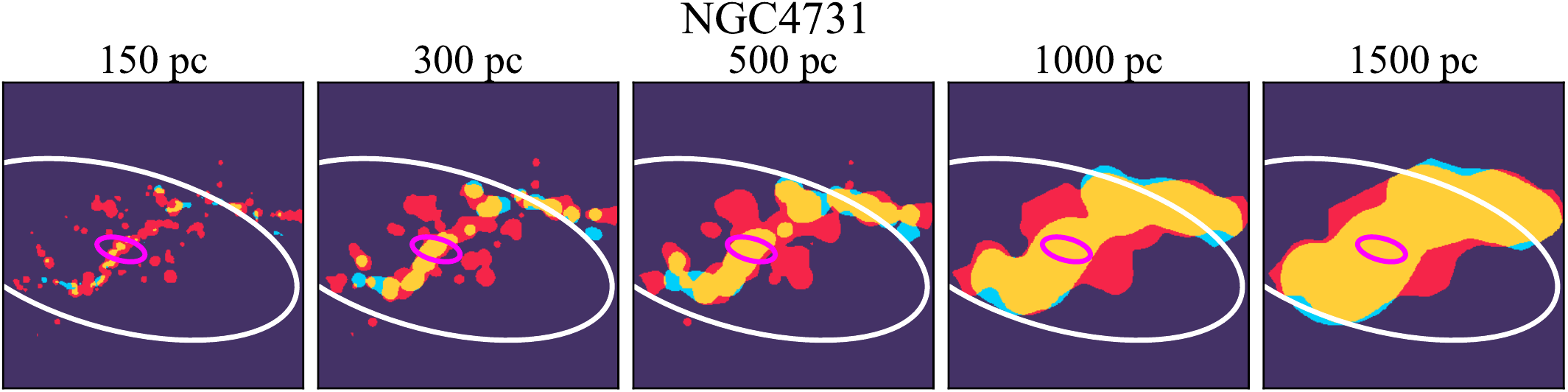}\\
	\vspace{10pt}
	\includegraphics[scale=0.7]{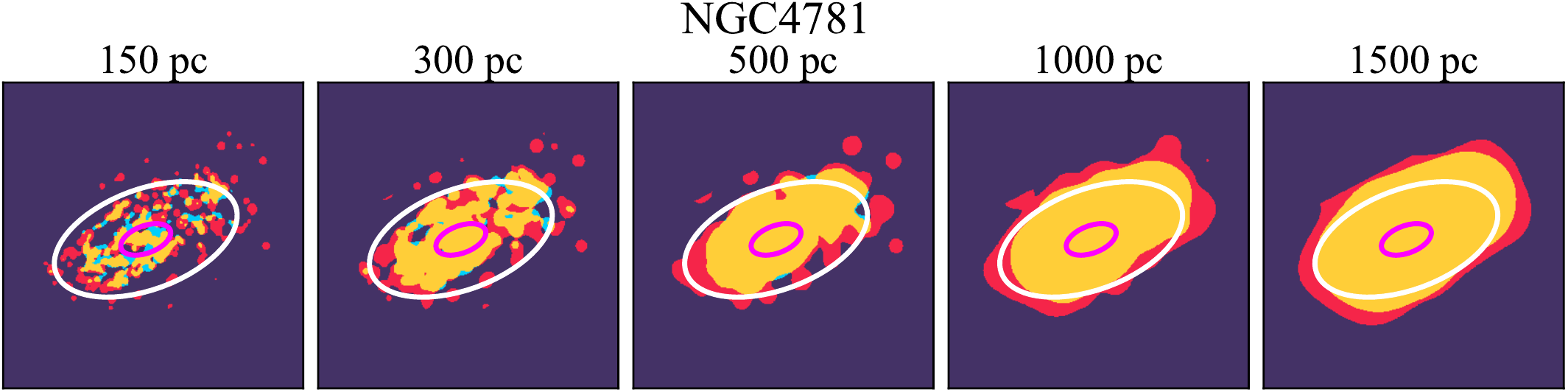}\\
	
	\caption{Continued.} 
\end{figure*}

\addtocounter{figure}{-1}
\begin{figure*}
	\centering
	\includegraphics[scale=0.7]{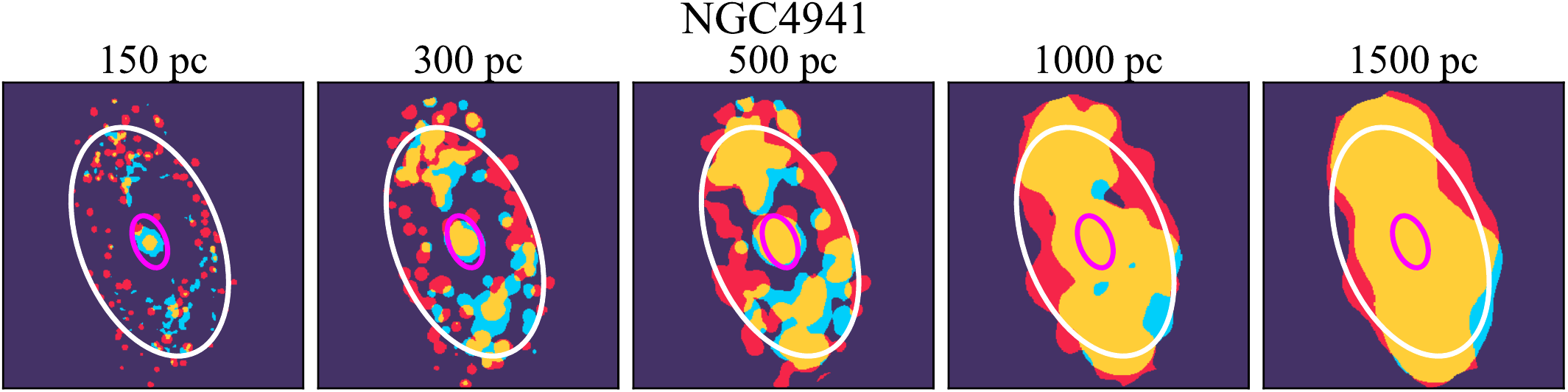}\\
	\vspace{10pt}
	\includegraphics[scale=0.7]{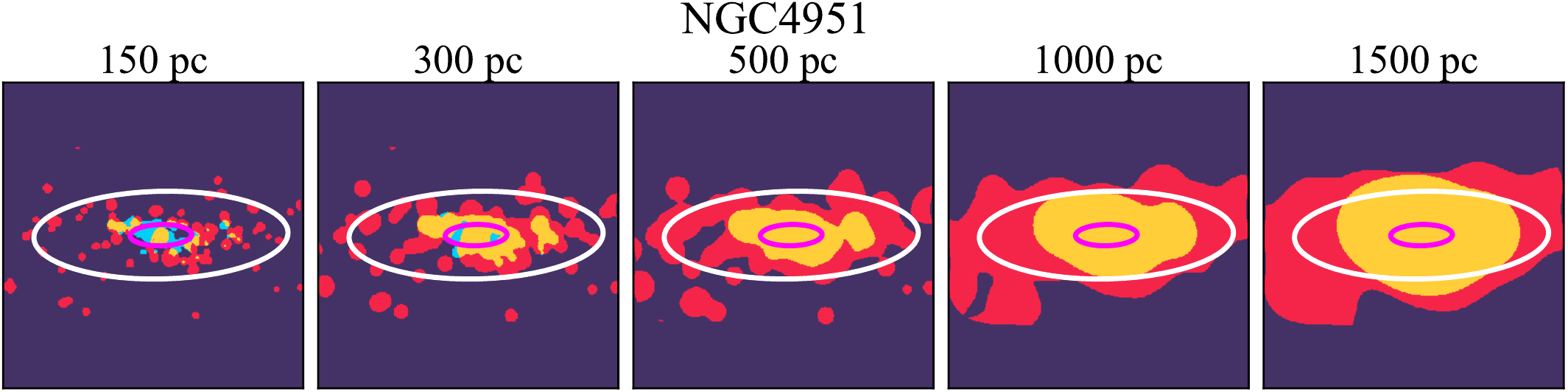}\\
	\vspace{10pt}
	\includegraphics[scale=0.7]{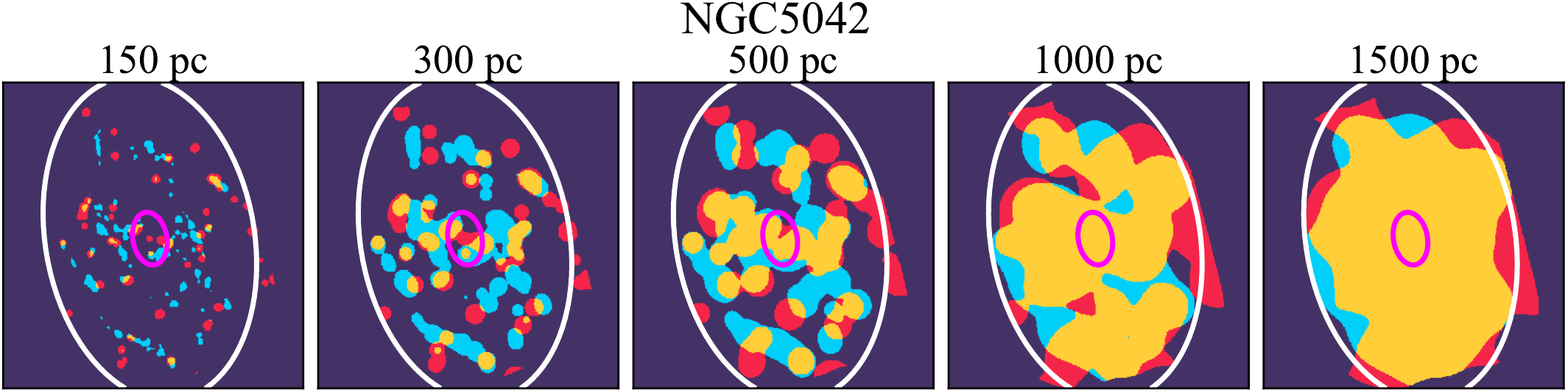}\\
	\vspace{10pt}
	\includegraphics[scale=0.7]{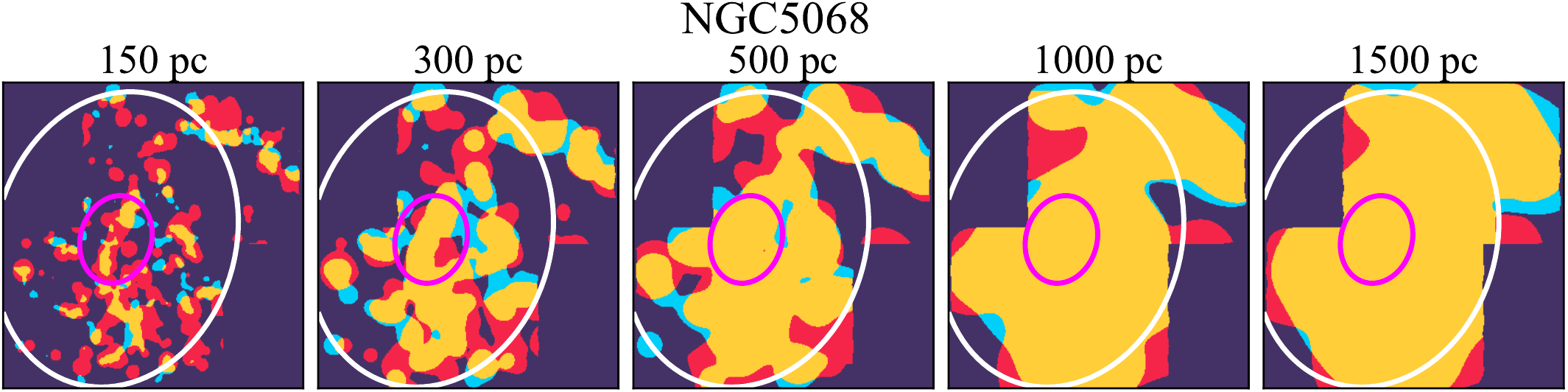}\\
	\vspace{10pt}
	\includegraphics[scale=0.7]{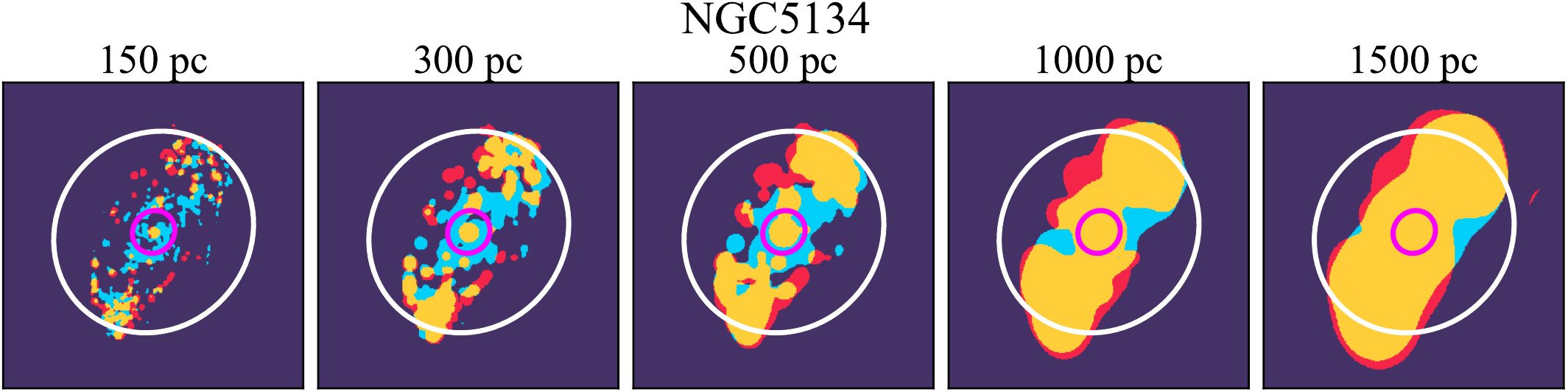}\\
	\caption{Continued.} 
\end{figure*}

\addtocounter{figure}{-1}
\begin{figure*}
	\centering
	\includegraphics[scale=0.7]{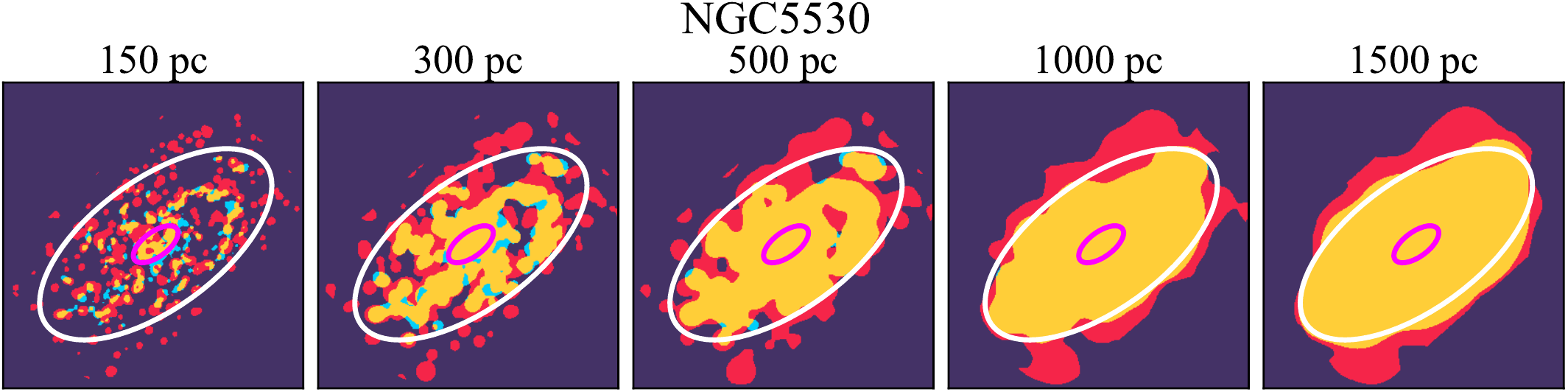}\\
	\vspace{10pt}
	\includegraphics[scale=0.7]{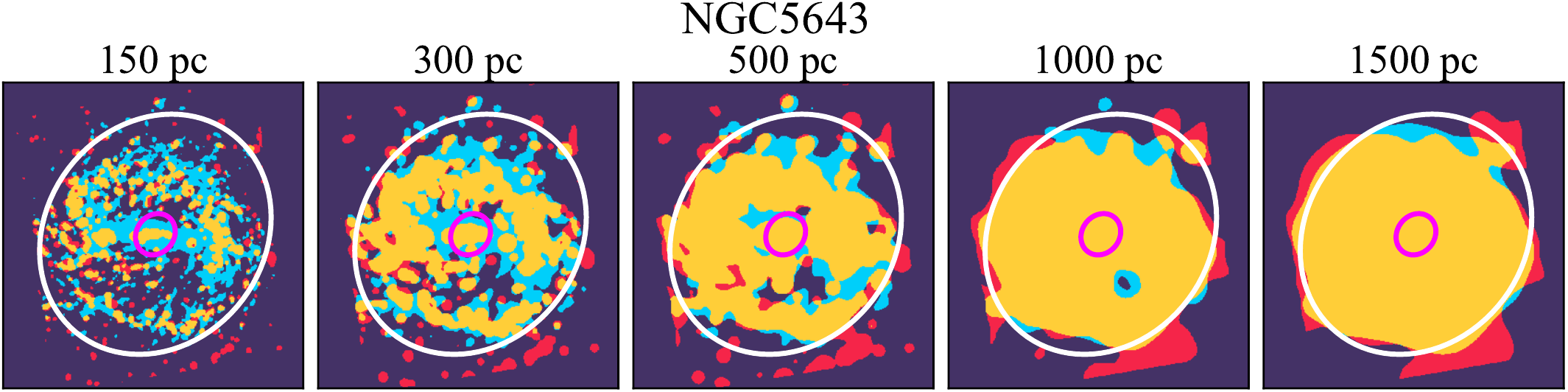}\\
	\vspace{10pt}
	\includegraphics[scale=0.7]{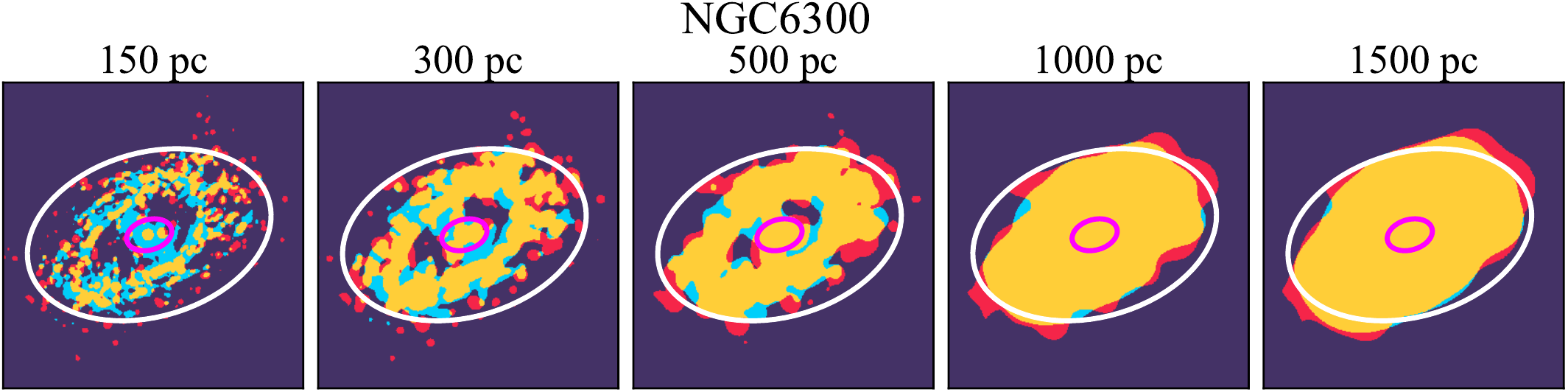}\\
	\vspace{10pt}
	\includegraphics[scale=0.7]{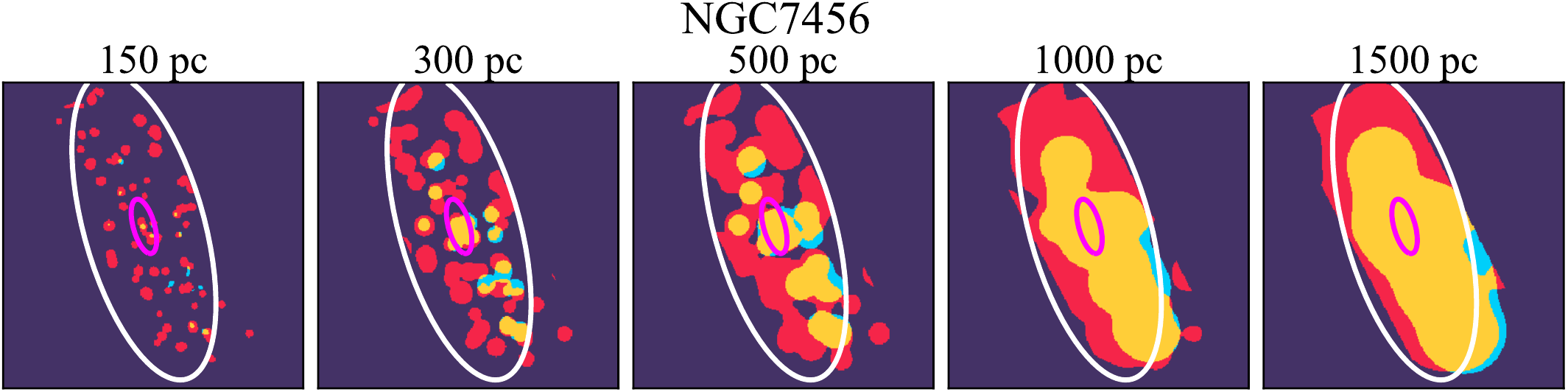}\\
	\caption{Continued.} 
\end{figure*}

\newpage
\begin{longtable*}{llllll}
	\caption{Fractions (\%) of sight lines (\co, \ha, \overlap) at 150, 300, 500, 1000, and 1500~pc  resolutions.}\\
	\hline
	Galaxy	& 150~pc  & 300~pc & 500~pc & 1000~pc & 1500~pc \\
	\hline
	IC1954&(19.3, 34.2, 46.5)&(6.1, 19.2, 74.7)&(3.6, 17.0, 79.3)&(0.0, 4.7, 95.3)&(0.0, 0.0, 100.0)\\
	IC5273&(11.6, 61.0, 27.4)&(8.3, 35.7, 56.1)&(3.9, 23.8, 72.3)&(2.5, 4.9, 92.6)&(1.6, 0.4, 98.1)\\
	NGC0628&(38.5, 17.7, 43.7)&(24.6, 9.6, 65.8)&(13.7, 7.3, 79.0)&(4.8, 7.0, 88.2)&(1.8, 6.6, 91.6)\\
	NGC1087&(23.6, 20.7, 55.6)&(8.1, 12.7, 79.2)&(2.0, 9.3, 88.7)&(0.2, 4.2, 95.6)&(0.1, 1.1, 98.8)\\
	NGC1300&(26.7, 42.1, 31.3)&(14.1, 42.3, 43.6)&(7.8, 41.7, 50.4)&(5.0, 35.8, 59.2)&(4.5, 27.5, 68.0)\\
	NGC1317&(49.8, 1.7, 48.4)&(30.8, 1.2, 68.0)&(23.8, 0.4, 75.8)&(14.4, 0.0, 85.6)&(9.0, 0.1, 90.9)\\
	NGC1365&(56.9, 16.4, 26.7)&(40.7, 23.6, 35.6)&(32.4, 25.9, 41.7)&(24.1, 26.9, 49.0)&(19.8, 26.8, 53.4)\\
	NGC1385&(28.4, 18.7, 52.9)&(12.1, 16.1, 71.8)&(5.7, 12.0, 82.4)&(2.6, 7.8, 89.6)&(2.3, 4.5, 93.3)\\
	NGC1433&(13.3, 70.6, 16.1)&(9.2, 68.1, 22.7)&(8.0, 63.6, 28.4)&(8.4, 51.3, 40.3)&(8.6, 43.0, 48.4)\\
	NGC1511&(48.1, 11.7, 40.1)&(29.8, 10.7, 59.4)&(19.5, 7.6, 73.0)&(6.0, 2.5, 91.6)&(2.4, 0.2, 97.5)\\
	NGC1512&(14.3, 63.0, 22.7)&(11.0, 56.8, 32.2)&(7.2, 54.8, 38.0)&(4.8, 43.6, 51.6)&(4.9, 34.0, 61.1)\\
	NGC1546&(69.7, 0.2, 30.2)&(48.2, 0.2, 51.7)&(36.9, 0.0, 63.0)&(25.1, 0.0, 74.9)&(16.2, 0.0, 83.8)\\
	NGC1559&(50.4, 11.6, 38.0)&(22.9, 10.7, 66.4)&(7.5, 8.8, 83.7)&(0.8, 4.7, 94.4)&(0.2, 2.4, 97.5)\\
	NGC1566&(36.6, 26.5, 36.9)&(22.3, 26.9, 50.8)&(13.7, 28.6, 57.6)&(6.2, 28.9, 64.9)&(3.6, 28.3, 68.1)\\
	NGC2090&(3.0, 61.4, 35.5)&(0.0, 45.8, 54.1)&(0.0, 45.9, 54.1)&(0.0, 33.0, 67.0)&(0.0, 14.7, 85.3)\\
	NGC2283&(11.9, 53.2, 34.9)&(5.2, 29.9, 64.9)&(1.5, 20.1, 78.4)&(0.0, 3.6, 96.4)&(0.0, 0.2, 99.8)\\
	NGC2835&(3.4, 83.4, 13.2)&(2.0, 67.7, 30.3)&(0.7, 61.9, 37.4)&(0.0, 49.9, 50.1)&(0.0, 41.1, 58.9)\\
	NGC2997&(56.2, 10.8, 33.0)&(35.7, 13.2, 51.1)&(22.3, 15.0, 62.7)&(9.5, 16.0, 74.5)&(3.7, 14.6, 81.8)\\
	NGC3351&(52.9, 24.4, 22.7)&(42.3, 17.9, 39.8)&(28.6, 15.7, 55.7)&(16.5, 12.4, 71.1)&(8.2, 8.9, 82.9)\\
	NGC3511&(11.2, 60.1, 28.7)&(5.3, 47.2, 47.4)&(1.5, 41.9, 56.6)&(0.1, 25.1, 74.8)&(0.0, 12.9, 87.1)\\
	NGC3596&(32.5, 18.8, 48.7)&(9.7, 16.9, 73.5)&(1.6, 15.4, 83.0)&(0.1, 11.6, 88.3)&(0.1, 4.1, 95.8)\\
	NGC3626&(71.5, 7.2, 21.3)&(22.0, 14.8, 63.2)&(11.4, 12.3, 76.3)&(8.8, 6.3, 85.0)&(8.2, 3.1, 88.7)\\
	NGC3627&(45.2, 13.3, 41.4)&(27.8, 13.3, 58.9)&(16.1, 12.5, 71.4)&(6.8, 11.2, 82.0)&(4.8, 9.9, 85.2)\\
	NGC4207&(32.4, 6.0, 61.5)&(16.1, 4.0, 79.9)&(7.4, 1.9, 90.7)&(0.0, 0.0, 100.0)&(0.0, 0.0, 100.0)\\
	NGC4254&(50.9, 2.4, 46.7)&(29.5, 1.7, 68.8)&(18.0, 1.1, 80.8)&(8.0, 0.9, 91.2)&(4.8, 0.4, 94.9)\\
	NGC4293&(72.0, 8.2, 19.8)&(52.5, 14.1, 33.4)&(36.5, 20.3, 43.1)&(11.2, 29.0, 59.8)&(1.7, 29.3, 69.1)\\
	NGC4298&(43.4, 2.9, 53.7)&(14.0, 0.6, 85.4)&(2.0, 0.0, 98.0)&(0.0, 0.0, 100.0)&(0.0, 0.0, 100.0)\\
	NGC4321&(44.0, 19.7, 36.3)&(29.1, 17.1, 53.8)&(17.2, 16.7, 66.1)&(5.7, 14.2, 80.1)&(1.8, 11.7, 86.6)\\
	NGC4424&(77.7, 0.2, 22.0)&(74.1, 0.0, 25.9)&(63.6, 0.0, 36.4)&(40.0, 0.1, 59.9)&(24.5, 1.2, 74.3)\\
	NGC4457&(76.9, 3.3, 19.8)&(55.1, 6.4, 38.6)&(35.7, 10.7, 53.5)&(12.7, 10.2, 77.2)&(8.8, 5.8, 85.3)\\
	NGC4496A&(6.9, 79.9, 13.2)&(7.9, 54.4, 37.7)&(4.3, 37.5, 58.2)&(1.1, 21.1, 77.8)&(1.0, 13.0, 86.0)\\
	NGC4535&(65.0, 9.9, 25.1)&(50.0, 8.0, 42.0)&(34.1, 8.9, 57.0)&(15.5, 11.9, 72.6)&(10.1, 12.8, 77.1)\\
	NGC4540&(46.3, 12.3, 41.5)&(26.0, 6.7, 67.3)&(14.8, 6.2, 79.0)&(8.6, 4.2, 87.2)&(4.1, 1.9, 94.0)\\
	NGC4548&(37.0, 32.5, 30.5)&(29.4, 28.6, 42.1)&(20.4, 30.5, 49.1)&(12.4, 27.0, 60.6)&(5.2, 19.4, 75.4)\\
	NGC4569&(74.5, 3.6, 21.9)&(52.0, 5.4, 42.7)&(34.3, 5.2, 60.5)&(16.6, 6.9, 76.5)&(10.2, 7.9, 81.9)\\
	NGC4571&(29.7, 48.5, 21.8)&(22.7, 22.9, 54.4)&(10.7, 18.6, 70.7)&(3.5, 8.5, 88.0)&(3.2, 3.9, 92.9)\\
	NGC4689&(35.4, 20.7, 43.9)&(16.5, 14.3, 69.2)&(8.1, 10.2, 81.7)&(1.6, 5.5, 92.9)&(0.3, 2.7, 97.0)\\
	NGC4694&(69.7, 6.8, 23.5)&(62.6, 4.1, 33.3)&(48.8, 3.1, 48.2)&(19.4, 1.4, 79.3)&(6.0, 1.0, 93.0)\\
	NGC4731&(5.4, 81.0, 13.7)&(6.8, 58.6, 34.6)&(6.7, 49.4, 43.9)&(4.0, 32.4, 63.7)&(2.4, 19.1, 78.4)\\
	NGC4781&(14.1, 40.8, 45.2)&(3.9, 27.0, 69.1)&(0.7, 21.4, 77.9)&(0.0, 9.4, 90.6)&(0.0, 0.5, 99.5)\\
	NGC4941&(33.0, 51.7, 15.3)&(27.6, 36.5, 35.9)&(18.5, 30.1, 51.4)&(5.2, 18.6, 76.2)&(2.0, 9.5, 88.5)\\
	NGC4951&(17.9, 64.0, 18.2)&(3.3, 60.7, 36.0)&(0.0, 60.8, 39.1)&(0.0, 39.9, 60.1)&(0.0, 21.9, 78.1)\\
	NGC5042&(47.0, 40.6, 12.4)&(47.9, 26.1, 26.1)&(32.6, 24.1, 43.3)&(12.3, 17.8, 69.9)&(6.4, 11.1, 82.6)\\
	NGC5068&(10.7, 62.5, 26.8)&(12.9, 30.3, 56.7)&(6.4, 17.9, 75.6)&(2.4, 8.2, 89.4)&(0.5, 5.0, 94.5)\\
	NGC5134&(49.2, 25.4, 25.4)&(40.1, 17.3, 42.6)&(26.6, 16.8, 56.6)&(9.1, 13.5, 77.4)&(4.1, 10.4, 85.5)\\
	NGC5530&(11.5, 59.3, 29.2)&(6.3, 33.2, 60.5)&(1.2, 22.5, 76.2)&(0.1, 8.3, 91.7)&(0.0, 1.2, 98.8)\\
	NGC5643&(51.4, 11.2, 37.5)&(30.6, 8.5, 60.9)&(15.6, 6.4, 78.0)&(5.9, 3.5, 90.6)&(3.5, 2.3, 94.2)\\
	NGC6300&(40.6, 19.1, 40.4)&(19.8, 16.4, 63.9)&(8.5, 15.4, 76.1)&(1.7, 8.0, 90.3)&(1.0, 3.4, 95.5)\\
	NGC7456&(2.6, 93.3, 4.1)&(5.6, 72.3, 22.1)&(4.4, 63.4, 32.1)&(1.8, 43.4, 54.8)&(0.6, 29.0, 70.4)\\
	\hline
	\label{tab_fractions}
\end{longtable*}

\newpage
\section{Sight line Fractions versus  Properties of Galaxies and Observations}
\label{sec_appendix_obs_impact}
\restartappendixnumbering
Here we present the scatter plots of galaxy and observational properties against the three sight line fractions in Figure~\ref{fig_test_obs_gal_props}.
The properties we explore are (a) stellar mass, (b) Hubble type, (c) galaxy distance, (d)  optical size indicated by $R_{25}$, (e) disk inclination, (f) effective \textsc{H\,ii} region sensitivity (log($L_\mathrm{\textsc{H\,ii}\,region}^\mathrm{sensitivity}$), Section~\ref{sec_ha_filtering}, (g) native resolutions of H$\alpha$ observation, (h)  DIG fraction, (i) effective sensitivity of CO observation (1$\sigma$ sensitivity in $\Sigma_\mathrm{H_{2}}$ at 150~pc resolution), (j) native resolution of CO observation, (k) specific star formation rate sSFR, and (l) offset from the star-forming main sequence $\Delta$MS. 
Discussion can be found in the main text in Section~\ref{sec_global_galprops}.
Correlation coefficients  of each relation shown in  Figure~\ref{fig_test_obs_gal_props} are provided in Table \ref{tab_sight line_cc_galprops}  in the main text.

\begin{figure*}[h]
	\centering
	\includegraphics[scale=0.53]{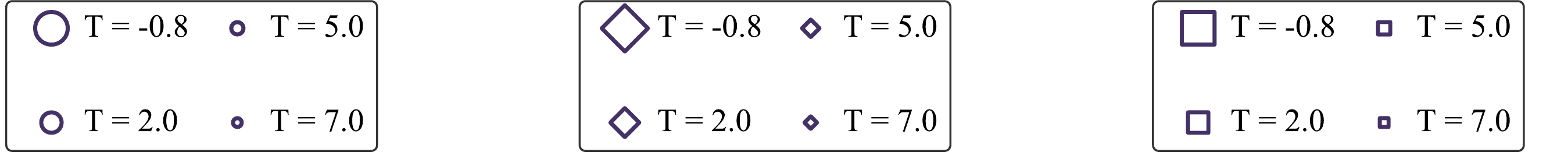}\\
	\vspace{10pt}
	\subfigure[]{\label{fig_frac_mstar}\includegraphics[scale=0.53]{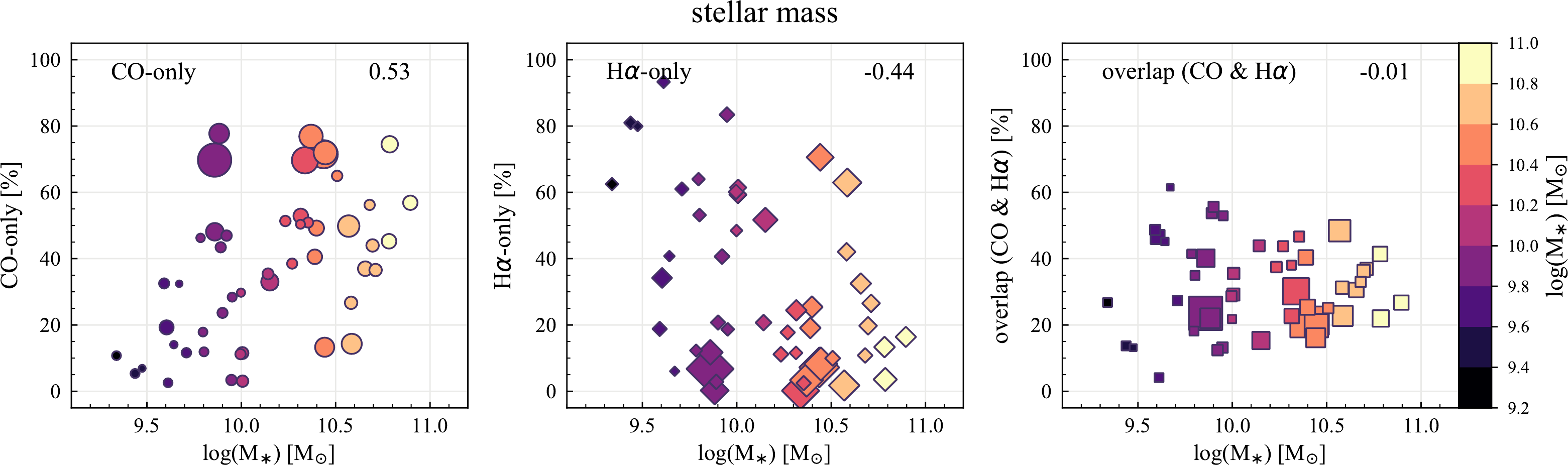}}
	\subfigure[]{\label{fig_frac_t}\includegraphics[scale=0.53]{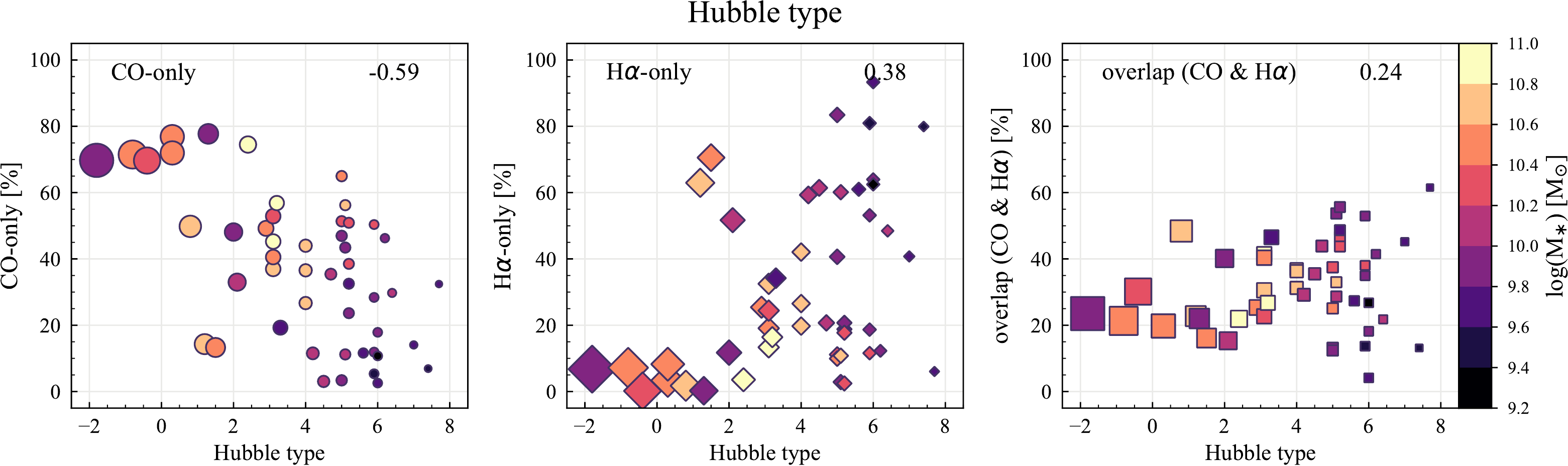}}
	\subfigure[]{\label{fig_frac_dist}\includegraphics[scale=0.53]{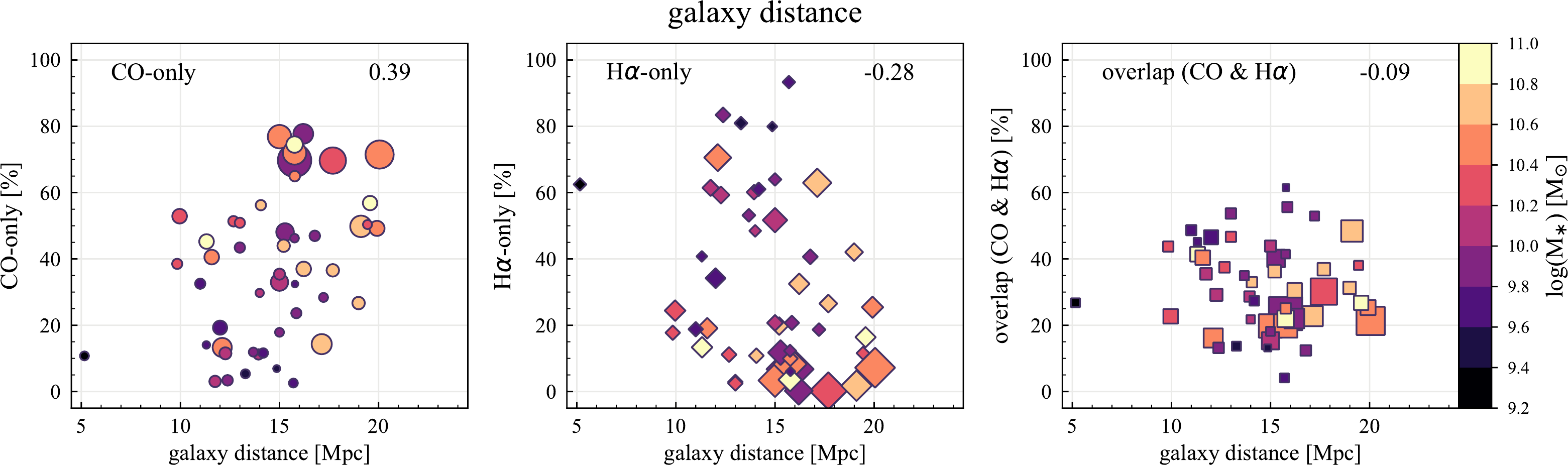}}

	\caption{Sight line fractions as a function of (a) stellar mass, (b) Hubble type,  (c) galaxy distance, (d)  optical size indicated by $R_{25}$, (e) disk inclination, (f) effective \textsc{H\,ii} region sensitivity (log($L_\mathrm{\textsc{H\,ii}\,region}^\mathrm{sensitivity}$); Section~\ref{sec_ha_filtering}, (g)  DIG fraction, (h) native resolutions of the H$\alpha$ observation,  (i) effective sensitivity of the CO observation (1$\sigma$ sensitivity in $\Sigma_\mathrm{H_{2}}$ at 150~pc resolution), (j) native resolution of the CO observation, (k) specific SFR, and (l) offset from the star-forming main sequence $\Delta$MS. Galaxies are color coded by \Mstar. The symbol size indicates Hubble type, with larger symbols for earlier types. Examples of symbol  sizes for different Hubble types are given in the top rows. The correlation coefficient of each pair of properties is given in the upper-right corner of each plot.} 
	\label{fig_test_obs_gal_props}
\end{figure*}

\addtocounter{figure}{-1}
\begin{figure*}
	\addtocounter{subfigure}{3}
	\centering
	\includegraphics[scale=0.53]{global_frac_legends_0150pc.pdf}\\
	\vspace{10pt}
	\subfigure[]{\label{fig_frac_r25}\includegraphics[scale=0.53]{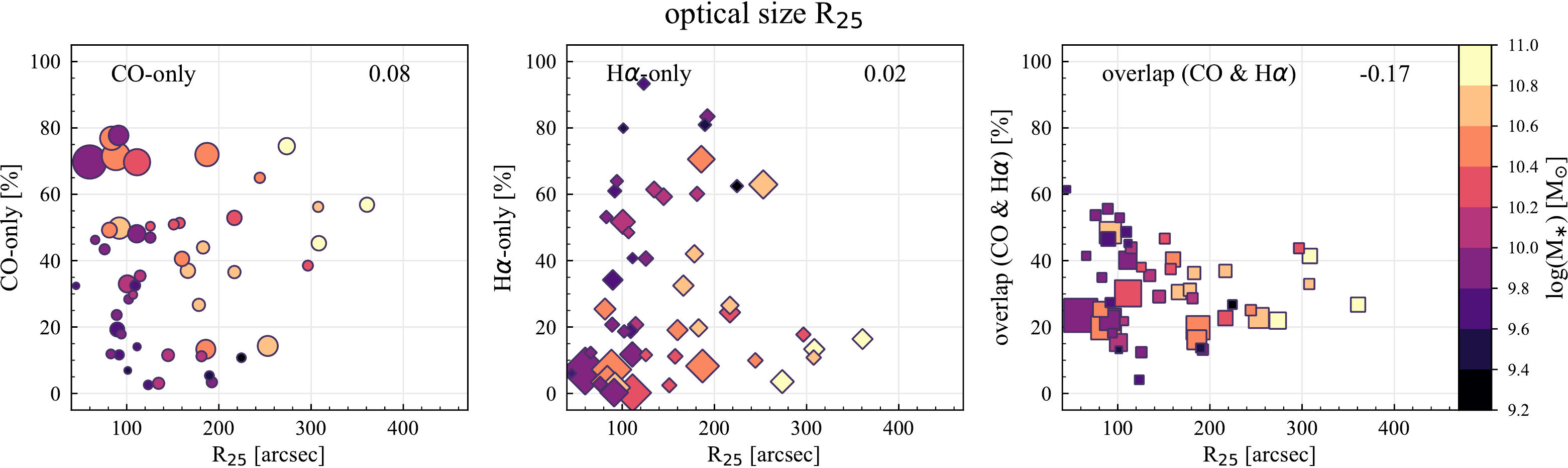}}
	\subfigure[]{\label{fig_frac_incl}\includegraphics[scale=0.53]{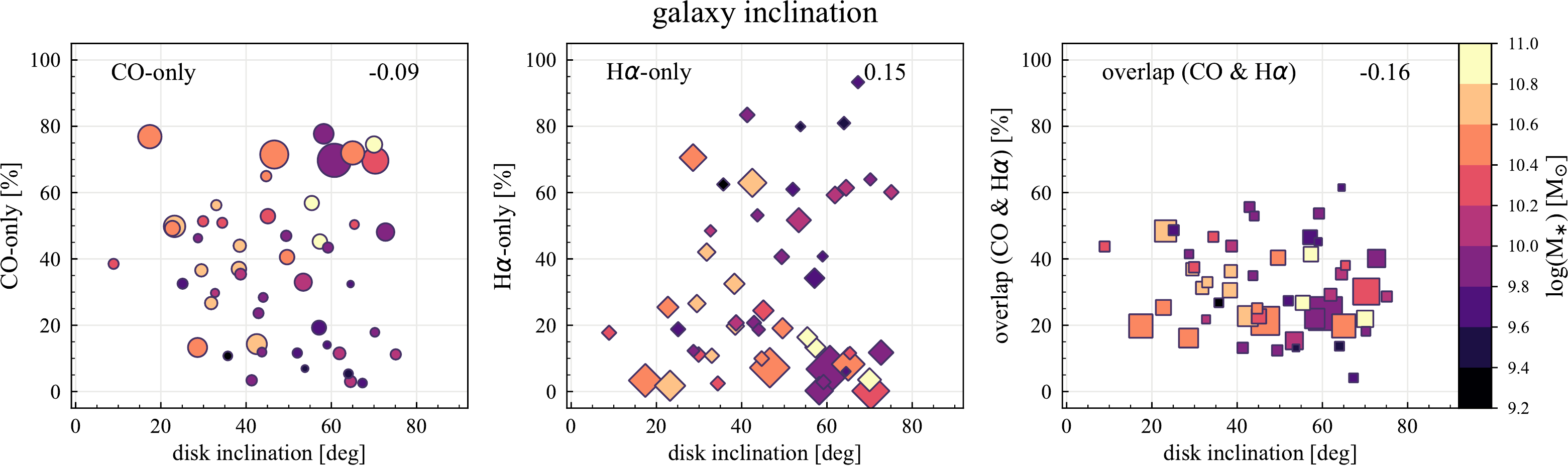}}		
\subfigure[ ]{\label{fig_frac_dig}\includegraphics[scale=0.53]{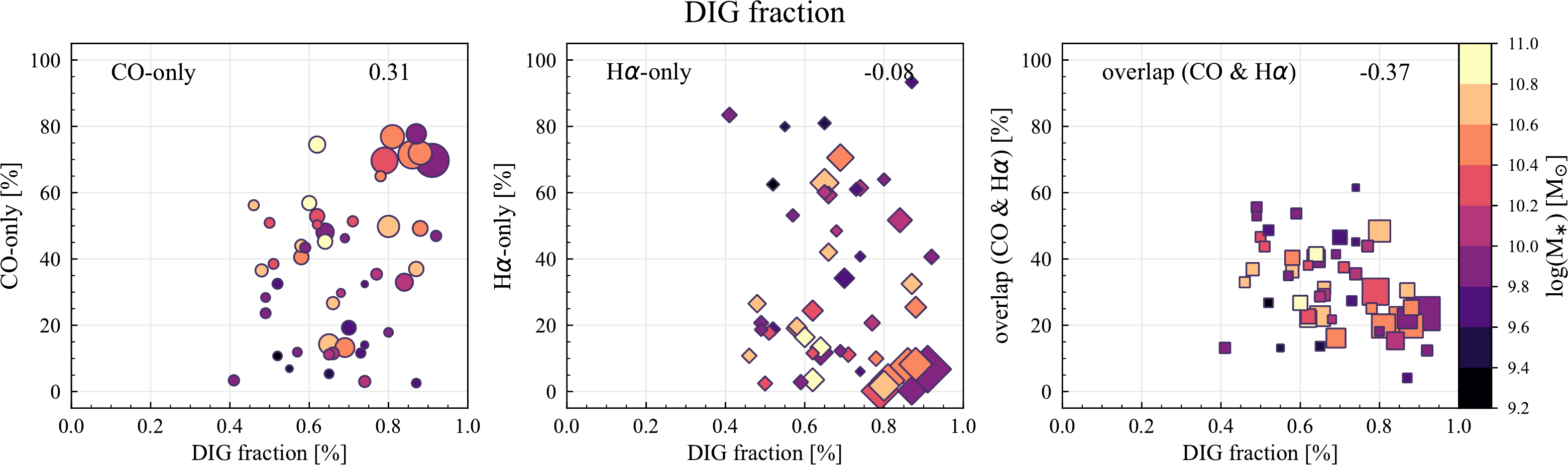}}

	\caption{Continued.} 
\end{figure*}

\addtocounter{figure}{-1}
\begin{figure*}
	\addtocounter{subfigure}{6}
	\centering
	\includegraphics[scale=0.53]{global_frac_legends_0150pc.pdf}\\
	\vspace{10pt}
	\subfigure[]{\label{fig_frac_hasen}\includegraphics[scale=0.53]{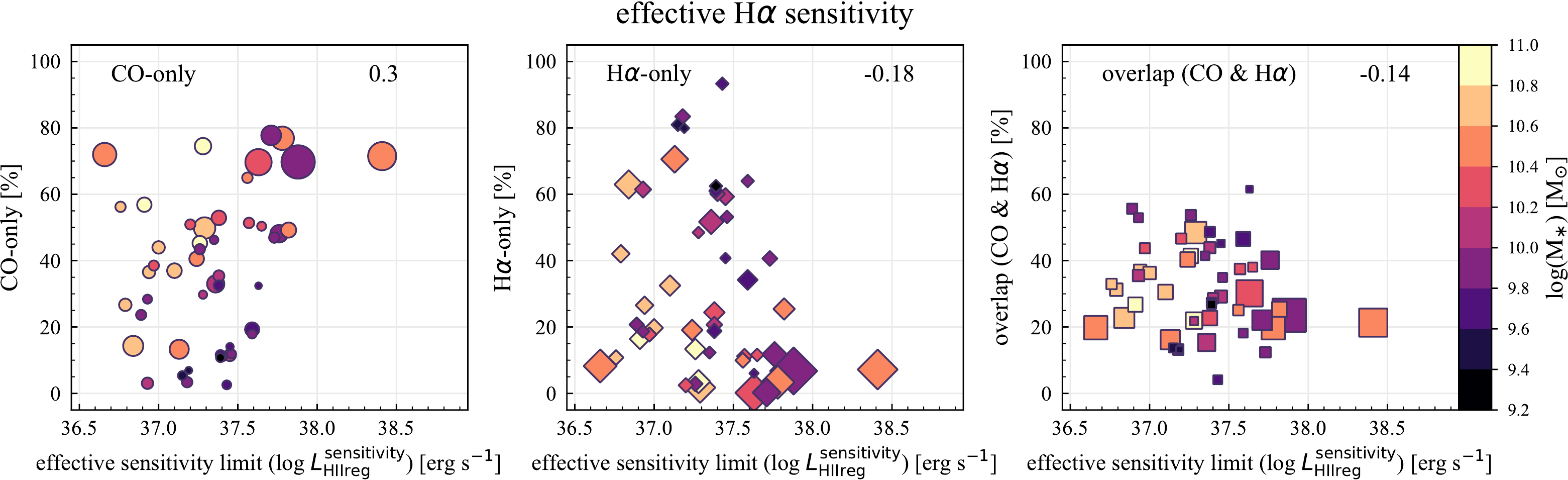}}
\subfigure[]{\label{fig_frac_hares}\includegraphics[scale=0.53]{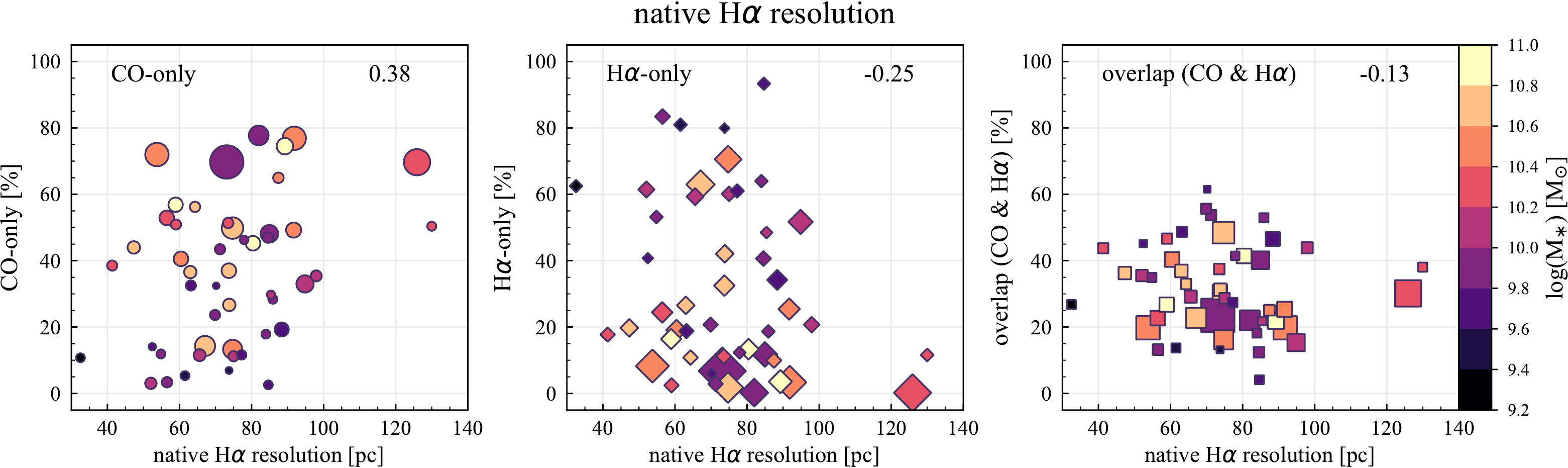}}	
	\subfigure[]{\label{fig_frac_cosen}\includegraphics[scale=0.53]{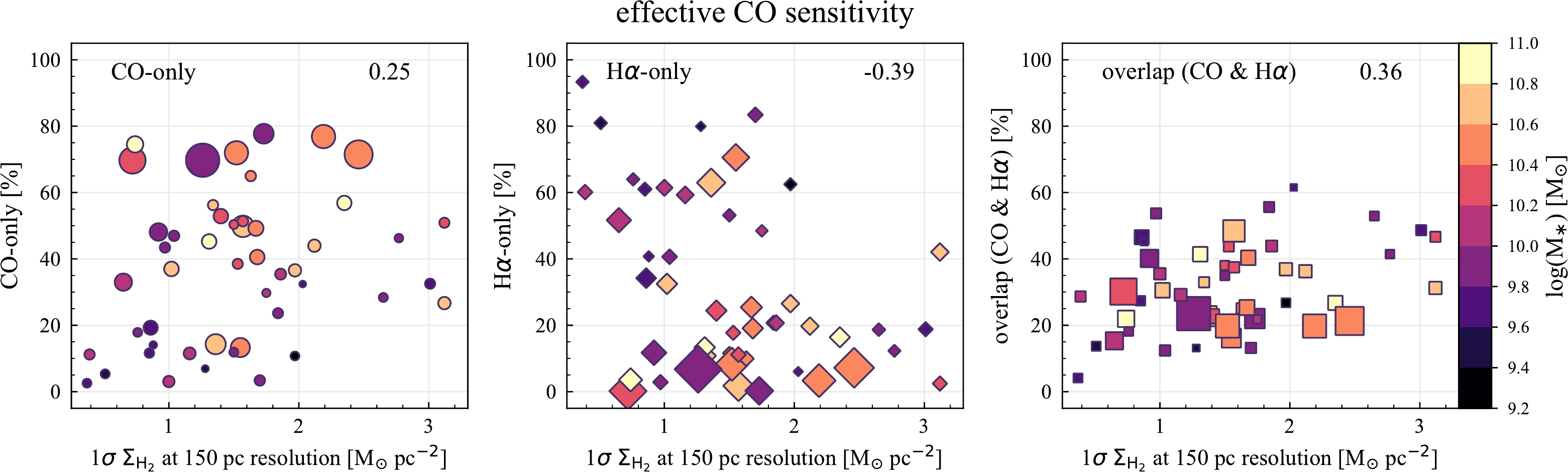}}
	
	\caption{Continued.} 
\end{figure*}

\addtocounter{figure}{-1}
\begin{figure*}
	\addtocounter{subfigure}{9}
	\centering
	\includegraphics[scale=0.53]{global_frac_legends_0150pc.pdf}\\
	\vspace{10pt}
	\subfigure[]{\label{fig_frac_cores}\includegraphics[scale=0.53]{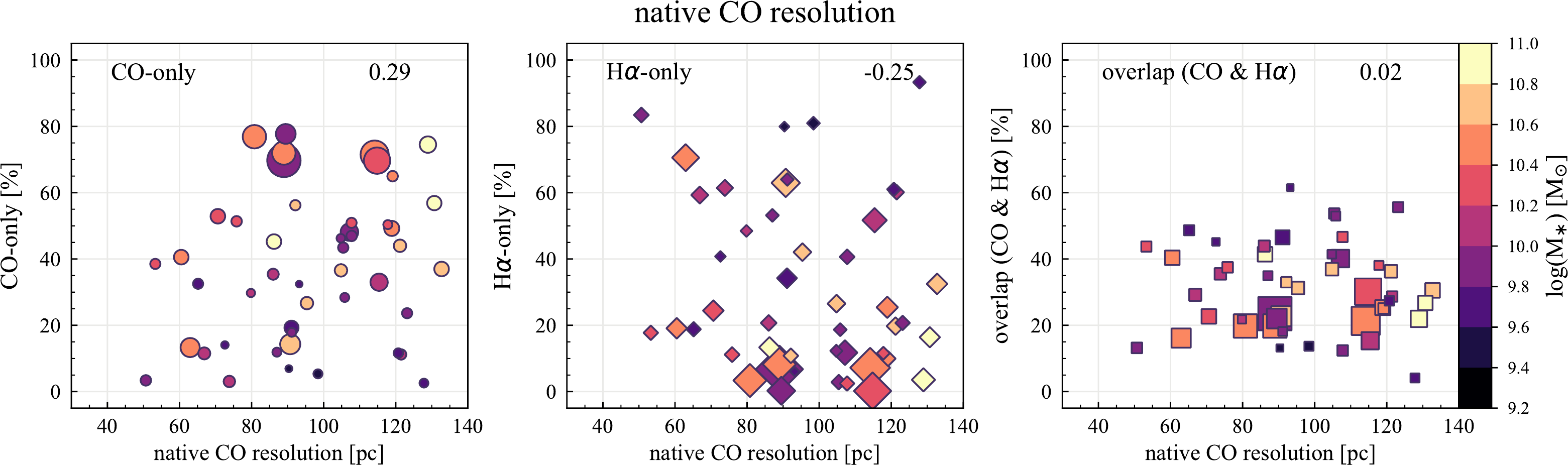}}	
	\subfigure[]{\label{fig_frac_dig}\includegraphics[scale=0.53]{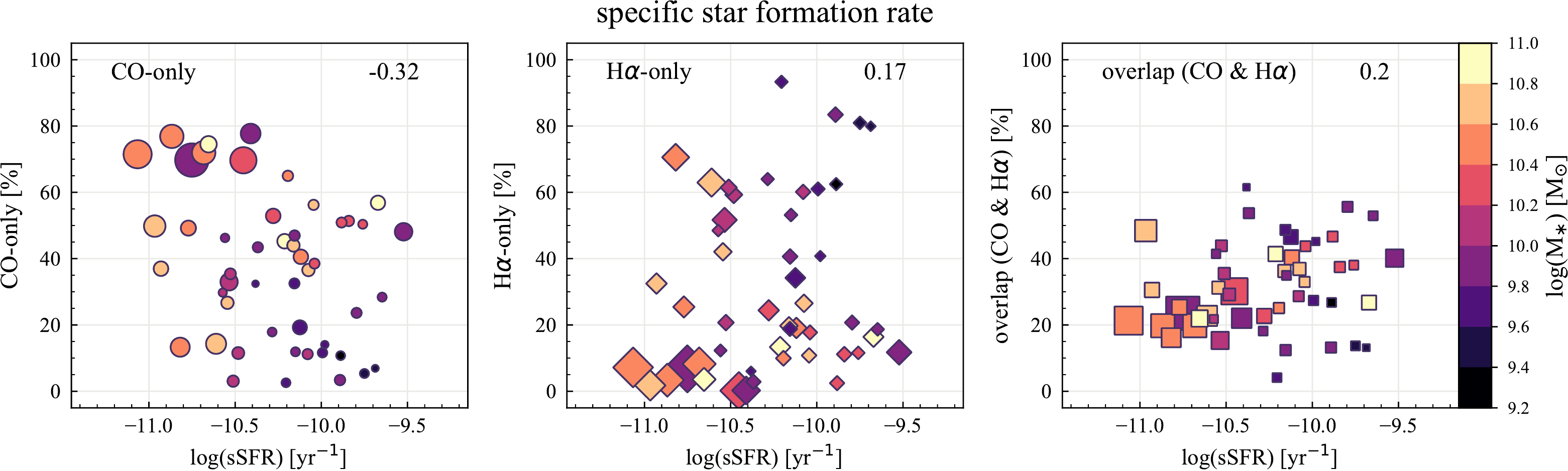}}	
	\subfigure[]{\label{fig_frac_dig}\includegraphics[scale=0.53]{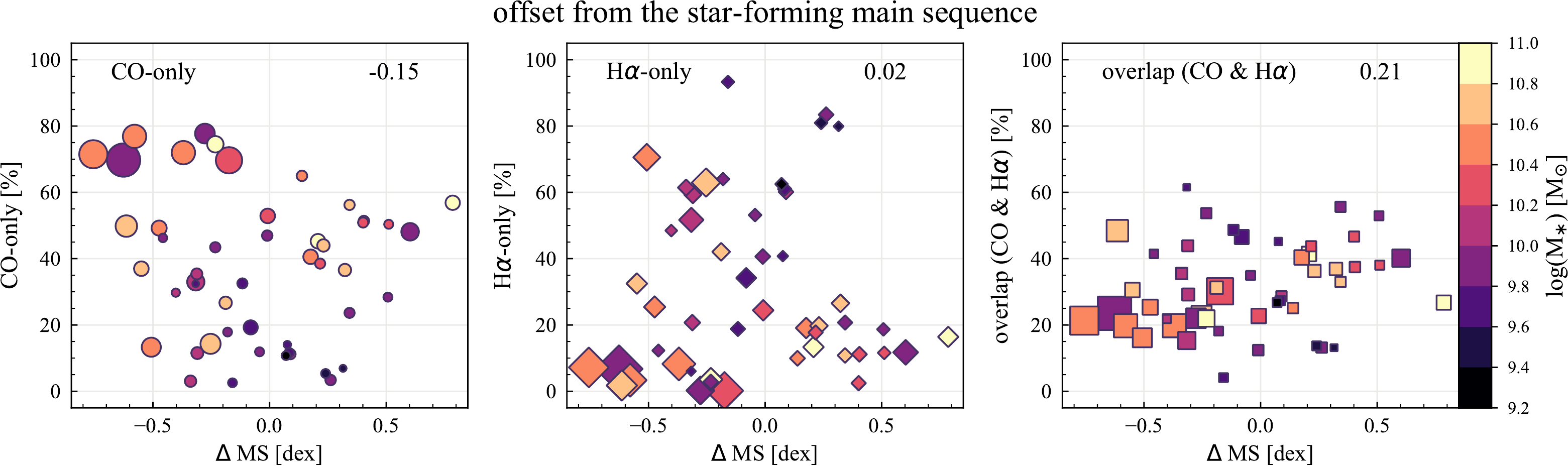}}	
	
	\caption{Continued.} 
\end{figure*}

\section{Comparison  of Cloud Visibility time with Chevance et al.(2020)}
\begin{figure*}
	\centering
	\includegraphics[scale=0.5]{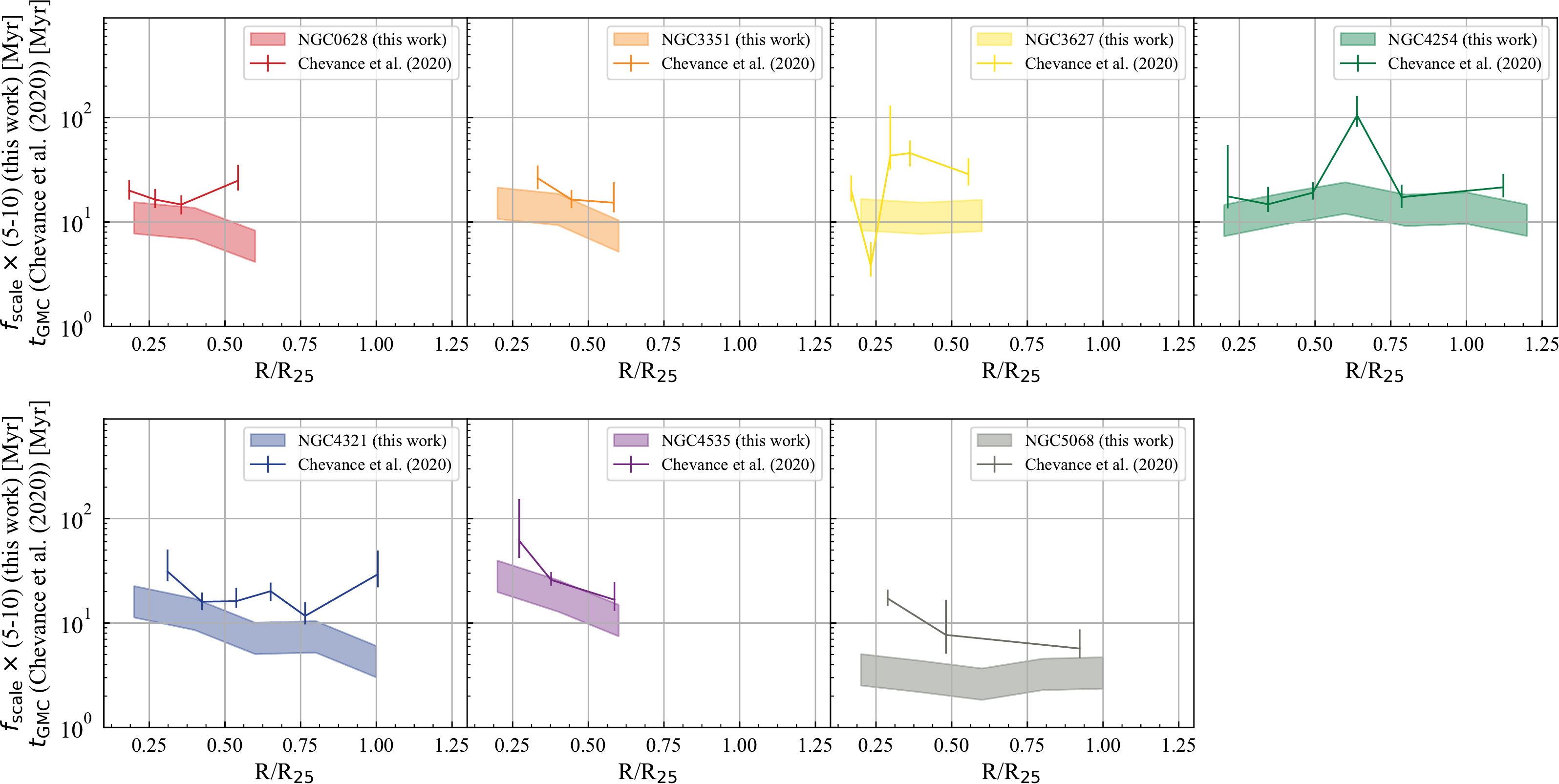}
	\caption{ Comparison of our radial cloud visibility time ($t_\mathrm{gas}$  $=$ $f_\mathrm{scale}$ $\times$ $t_\mathrm{H\alpha}$) (thick shaded curve) with the GMC lifetime during which CO is visible ($t_\mathrm{GMC}$) measured by \cite{Che20} (thin curve with errorbars). In this work, we assume the fiducial timescale $t_\mathrm{H\alpha}$ to be 5 -- 10 Myr.   Note that the radial bins adopted by the two studies are slightly different.  }
	\label{fig_radial_tgmc}
\end{figure*}

\restartappendixnumbering
\label{sec_appendix_comparison_Che20}
Here we directly compare the radial variation in the cloud visibility time derived from our analysis ($t_\mathrm{gas}$) with the GMC lifetime during which CO is visible ($t_\mathrm{GMC}$) measured by \cite{Che20} for the seven galaxies analyzed in both studies.
The results from the two studies are compared  in Figure \ref{fig_radial_tgmc}.
Overall, the two studies show  similar results, particularly in regard to the absolute $t_\mathrm{gas}$ (or $t_\mathrm{GMC}$) for most radial bins. 
The relative $t_\mathrm{gas}$ (or $t_\mathrm{GMC}$) among different galaxies also shows reasonable agreement.
The robustness to the adopted methodology suggests that both results indeed reflect the  visibility time  of molecular cloud traced by CO emission.
Larger discrepancies are seen in regions where the emission (either CO or H$\alpha$) are relatively sparse and/or weak, such as the larger radii and lower \Mstar\ galaxy (NGC~5068), suggesting that the sensitivity and completeness of observation are  important factors in estimating cloud visibility time.

\end{CJK}
\end{document}